\documentclass[showpacs,reprint,groupedaddress,amsmath,amssymb,aps,pra]{revtex4-1}

\usepackage{amsmath,amssymb}
\usepackage{booktabs}
\usepackage[hiresbb,pdftex]{graphicx}		
\usepackage{bm}
\usepackage{braket}
\usepackage{float}
\usepackage{latexsym}
\usepackage{longtable}
\usepackage{multirow}
\usepackage{textcomp}
\usepackage{colortbl}
\usepackage[pdftex,
draft=false,
bookmarks=true,
bookmarksnumbered=false,
bookmarksopen=false,
bookmarksopenlevel=4,
colorlinks=true,
anchorcolor=black,
citecolor=black,
filecolor=black,
linkcolor=black,
linkbordercolor={1 0 0},
urlcolor=black,
pdfborder={0 0 1},
pdftitle={},
pdfauthor={},
pdfsubject={},
pdfkeywords={},
pdfpagemode=UseThumbs,
]{hyperref}

\newcommand{\mr}[2]{#1 \, {\mathrm{ \,  #2 \,}}}
\newcommand{\aizu}[2]{ #1 \, \scalebox{0.8}{\textit{F}} \, #2}

\begin{document}

\title{Group-theoretical classification of multipole order: \\emergent responses and candidate materials}
\author{Hikaru Watanabe}
\email[]{watanabe.hikaru.43n@st.kyoto-u.ac.jp}
\author{Youichi Yanase}
\affiliation{Department of Physics, Graduate School of Science, Kyoto University, Kyoto 606-8502, Japan}
\date{\today}

\begin{abstract}
The multipole moment is an established concept of electrons in solids. Entanglement of spin, orbital, and sublattice degrees of freedom is described by the multipole moment, and spontaneous multipole order is a ubiquitous phenomenon in strongly correlated electron systems. 
In this paper, we present group-theoretical classification theory of multipole order in solids. Intriguing duality between the real space and momentum space properties is revealed for odd-parity multipole order which spontaneously breaks inversion symmetry. Electromagnetic responses in odd-parity multipole states are clarified on the basis of the classification theory. A direct relation between the multipole moment and the magnetoelectric effect, Edelstein effect, magnetopiezoelectric effect, and dichromatic electron transport is demonstrated. More than 110 odd-parity magnetic multipole materials are identified by the group-theoretical analysis. Combining the list of materials with the classification tables of multipole order, we predict emergent responses of the candidate materials.
\end{abstract}
\maketitle
\twocolumngrid

\section{Introduction} \label{Section1}
 Physical phenomena originating from spin-orbit coupling in parity-violated systems are attracting growing interest in condensed matter physics. The topics include topological phases of matter~\cite{Hasan2010,Qi2011,Witczak-Krempa2014}, unconventional superconductivity~\cite{bauer2012non,Sato2017topological,Smidman2016super}, multiferroics~\cite{Katsura2005,Arima2007,Kimura2003,Fiebig2005b,Tokura2014multiferroic,Schmid2008}, and spintronics~\cite{Murakami2003spinhall,Sinova2004,Manchon2008,Manchon2009,Garate2009}. It has been shown that these quantum phases exhibit nontrivial electromagnetic responses characterized by peculiar symmetry. A concept of multipole may interconnect the broad research fields because the electromagnetic charge and current distributions are described by the multipole expansion in a unified way. 

So far the multipole physics has mainly been investigated with focus on the spontaneous multipole ordering of localized $d$ and $f$ electrons~\cite{Kuramoto2009Review,Santini2009}. The atomic multiplet formed by strong Coulomb interaction and spin-orbit coupling is characterized by the multipole moments, and inter-multipole interactions may lead to spontaneous multipole order. Furthermore, the multipole degree of freedom gives rise to intriguing quantum phases; \textit{e.g.} superconductivity mediated by multipole fluctuations~\cite{Koga2006,Matsubayachi2012MultipolarSC}, multipole Kondo effects~\cite{cox1998exotic,Onimaru2016QuadKondo,Yamane2018QuadKondo}, and so on. 

Recent studies shed light on odd-parity multipole order, where the inversion symmetry is spontaneously broken, although the studies of atomic multipole physics were restricted to even-parity order. In addition to the electric dipole order, which has been well-known as ferroelectric order, electric octupole~\cite{Hitomi2014,Hitomi2016},  magnetic monopole~\cite{Sumita2016,Spaldin2013,Th2016}, magnetic quadrupole~\cite{Yanase2014,Sumita2016,Sumita2017,Fechner2016,DiMatteo2017b}, magnetic toroidal dipole~\cite{Clin1991,Popov1998,Arima2005,VanAken2007observation,Zimmermann2014a,Hayami2014b,Ederer2007,Spaldin2008,Saito2018CurrentInduced}, and magnetic hexadecapole order~\cite{hikaruwatanabe2017} have been investigated. 

The key to the odd-parity multipole order is \textit{a locally-noncentrosymmetric crystal structure}, where an inversion center cannot be taken at atomic sites although the inversion symmetry is globally preserved. Then, there are at least two sublattices, and the parity operation is accompanied by site permutation. For instance, a zigzag chain contains two nonequivalent sites which are interchanged by the parity operation~\cite{Yanase2014,Sumita2016}. When an atomic even-parity multipole, such as spin, is antiferroically-ordered between the two sublattices, the system breaks the inversion symmetry instead of the translational symmetry. Then, the seemingly antiferroic states, indeed, have zero N\'eel vector $\bm{Q}=\bm{0}$, and they are characterized by ferroic odd-parity multipole moments. In fact, all the candidates for odd-parity magnetic multipole ordered materials in Sec.~\ref{Section4_candidates} show the locally-noncentrosymmetric property in the magnetic sites.

Furthermore, the sublattice degrees of freedom can be replaced with subsectors such as layers~\cite{Hitomi2014,Hitomi2016} or clusters consisting of several atoms~\cite{DiMatteo2017b}. The above idea is applicable when each subsector has no local inversion symmetry as the tetrahedral network of Cd atoms does in Cd$_2$Re$_2$O$_7$. The multipole degrees of freedom formed by several atoms, called as {\it augmented} multipole, ubiquitously exist in solids, and candidates for the odd-parity multipole ordered systems may be found widely in crystalline materials with broken local inversion symmetry. This is in sharp contrast to the conventional multipole order of atomic scale whose candidates are limited to highly symmetric crystals such as cubic systems~\cite{Kuramoto2009Review,Santini2009}. Thus, owing to the unavoidable reason, candidate materials for odd-parity multipole order may have complex crystal structures. Therefore, it is desirable to develop a theoretical tool to identify multipole order parameter in a systematic way.

Compounds with ferroic odd-parity magnetic multipole order are also called magnetoelectric materials, since the electromagnetic crossed correlation appears. 
The magnetoelectricity has been extensively studied in the research field of multiferroic materials, and then the candidates should be insulating to hold well-defined electric polarization reversible by external electric fields~\cite{Kimura2003,Fiebig2005b,Tokura2014multiferroic}. On the other hand, the odd-parity multipole order also leads to nontrivial electronic structures such as spontaneous emergence of spin-momentum locking~\cite{Hitomi2014,Hitomi2016,Fu2015} and asymmetric band distortions in itinerant systems~\cite{Yanase2014,Hayami2014b,Sumita2016,Sumita2017}. Accordingly, characteristic electromagnetic responses may occur in metallic states~\cite{Yanase2014}. Intriguing phenomena in odd-parity multipole states, however, have not been fully explored. Thus, to investigate emergent property, systematic studies are required.

In this paper, we report group-theoretical classification of multipole order, by which the relevant multipole moment in real materials is identified and the connection to the emergent electromagnetic responses is clarified. In Sec.~\ref{Section2}, we classify electric and magnetic multipole moments by high-symmetry crystal point groups. The tables in Sec.~\ref{Section2_multipole} list multipole moments up to rank-$4$. The results of classification are summarized in Sec.~\ref{Section2_basisfunctions}. We can identify the nontrivial properties of itinerant odd-parity multipole ordered states by the basis in the momentum space. In Sec.~\ref{Section3}, we discuss emergent responses arising from odd-parity multipole order such as magnetoelectric effect, Edelstein effect, magnetopiezoelectric effect, and dichromatic electron transport. We demonstrate that our representation analysis predicts these electromagnetic responses in an intuitive way. Application to odd-parity multipole materials is discussed by taking Cd$_2$Re$_2$O$_7$, Ba$_{1-x}$K$_x$Mn$_2$As$_2$, and GdB$_4$ as examples. We also show more than 110 candidate materials in Sec.~\ref{Section4_candidates} with use of the group-theoretical analysis. A brief summary is given in Sec.~V.

\section{Classification theory of multipole order} \label{Section2}

In the classical electromagnetism, static electromagnetic fields are characterized by electric and magnetic multipole moments. The multipole expansion of scalar and vector gauge potentials is given by
		\begin{align}
		\phi (\bm{r})= \sum_{l,m} a_l Q_{lm} \frac{Y_{lm}\left(\hat{r} \right)}{r^{l+1}},\label{electric_multipole}\\
		\bm{A} (\bm{r})= \sum_{l,m}b_l M_{lm} \frac{\bm{Y}_{lm}^l\left(\hat{r} \right)}{r^{l+1}},\label{magnetic_multipole}
		\end{align} 
where the electric multipole moment $Q_{lm}$ and magnetic multipole moment $M_{lm}$ are introduced, and $Y_{lm}$ and $\bm{Y}_{lm}^{l}$ are the spherical harmonics and rank-$l$ vector spherical harmonics, respectively. These spherical harmonics are defined with the Condon-Shotley phase, $Y_{lm}^\ast = (-1)^m Y_{l-m}$~\cite{varshalovich1988quantum}. Here, we adopt the Coulomb gauge $\nabla \cdot \bm{A}=0$ and the position $\bm{r}$ is set outside charge-current source.

The electric and magnetic multipole moments are quantum-mechanically written as
	\begin{align}
	&\hat{Q}_{lm} =-e \sum_{i}  \hat{r}_i^l  \sqrt{\frac{4\pi}{2l+1}  } Y_{lm}^{\ast}  (\hat{\theta}_i ,\hat{\phi}_i ), \label{elemulti}\\
	&\hat{M}_{lm} =\sum_{i} \bm{M}\cdot \nabla_i \left( 	\hat{r}_i^l  \sqrt{\frac{4\pi}{2l+1}  } Y_{lm}^{\ast} (\hat{\theta}_i ,\hat{\phi}_i ) 	\right), \label{magmulti}
	\end{align}
where the summation $i$ labels electrons and we define $\bm{M} = \left(  \hat{\bm{x}},\hat{\bm{y}},\hat{\bm{z}}\right) \equiv \mu_\mathrm{B} \left( 2\bm{l}/(l+1) +2\bm{s} \right)$ with $\bm{l}$ and $\bm{s}$ being orbital and spin angular momentum. Following the definitions in Eqs.~\eqref{elemulti} and \eqref{magmulti}, we have coefficients $a_l=\epsilon_0^{-1} \left[ 4 \pi \left(2 l+1 \right)  \right]^{-1/2}$ and $b_l= i \mu_0 \left(l+1 \right) \left[ 4 \pi \left(2 l+1 \right)  \right]^{-1/2}$ by using the vacuum permittivity $\epsilon_0$ and permeability $\mu_0$. $(r_i,\theta_i,\phi_i)$ denotes spherical coordinates of the $i$-th electron. Thermodynamical and quantum-mechanical expectation values of these quantum operators, that are classical variables, are treated  in our classification theory of multipole ordered phases. Then, $\Braket{\bm{M}}$ is a classical axial vector. We take the unit $e=1$. 

The space inversion parity is odd for odd-$l$ (even-$l$) electronic (magnetic) multipole moment, while it is even for even-$l$ (odd-$l$) electronic (magnetic) multipole moment. 
Later we show that the odd-parity multipole states exhibit intriguing properties distinct from even-parity states. Polar and time-reversal odd multipole moments, magnetic toroidal multipole moments, are not introduced in Eq.~\eqref{magnetic_multipole} in contrast to the previous studies~\cite{Spaldin2008,Dubovik1990a,Hayami2017MultipoleExpansion}. In the electrostatic case, however, we can perform the gauge transformation to erase toroidal multipole moments in Eqs.~\eqref{electric_multipole} and \eqref{magnetic_multipole} with the Coulomb gauge condition kept. For this reason, we focus on the electric and magnetic multipoles. On the other hand, our classification includes the magnetic toroidal dipole moment and the magnetic monopole moment because recent studies clarified a quantitative relation between these multipole moments and the magnetoelectric effect~\cite{Gao2018spin,Shitade2018theory,gao2018theory}.

\subsection{Classification of electric and magnetic multipole moment}\label{Section2_multipole}

The multipole moments $\left(Q_{lm},M_{lm}\right)$ are basis in the spherical space and characterized by angular momentum quantum numbers $l$ and $m$. On the other hand, the crystal symmetry restricts allowed symmetry operations, and therefore, irreducible representations in the rotation group become reducible. Thus, the multipole moment in crystals is classified by the compatibility relation with irreducible representations in a given point group. First, we show the classification of multipole moments up to rank $l= 4$.  For this purpose, we should adopt the solid representation~\cite{Kusunose2008}, 
		\begin{equation}
		\begin{split}
			Q_{lm}^+ = \frac{(-1)^m}{\sqrt{2}} \left( Q_{lm} + Q_{lm}^\ast \right), \\
			Q_{lm}^{-} = \frac{(-1)^m}{i\sqrt{2}} \left( Q_{lm} - Q_{lm}^\ast \right),
		\end{split}\label{elecubic}
		\end{equation}
and
		\begin{equation}
		\begin{split}
			M_{lm}^+ = \frac{(-1)^m}{\sqrt{2}} \left( M_{lm} + M_{lm}^\ast \right), \\
			M_{lm}^{-} = \frac{(-1)^m}{i\sqrt{2}} \left( M_{lm} - M_{lm}^\ast \right),
		\end{split}\label{magcubic}
		\end{equation}
for $0< l$ and $0 < m \leq l$. We deal with $Q_{l0}$ and $M_{l0}$ for $m=0$. The derivation of Eqs.~\eqref{electric_multipole}-\eqref{magcubic} is described in Refs.~\cite{Schwartz1957,Kusunose2008}.

In the classification Tables~\ref{electric_multipole_cubic}-\ref{magnetic_multipole_tetragonal_hexagonal}, we take three high-symmetry point groups, $O_h$, $D_{4h}$, and $D_{6h}$, following the classification theory of unconventional superconductivity by Sigrist and Ueda~\cite{Sigrist1991}. All the other crystal point groups are descended from these point groups, and the classification in the low-symmetry point groups can be obtained from Tables~\ref{electric_multipole_cubic}-\ref{magnetic_multipole_tetragonal_hexagonal} by the compatibility relation~\cite{inui1990group}. Thus, the classification of multipole moment in crystals has been completed, although previous results were limited to even-parity multipoles~\cite{Shiina1997b} or tetragonal systems~\cite{hikaruwatanabe2017}. Similar classification was performed for orientational ordering tensors by a gauge-theoretical approach~\cite{Liu2016GaugeTheoreticClassification}.

%
%
%
%
%
%
\onecolumngrid

	\begin{table}[H]
\caption{Classification of electric multipole moment up to rank $l=4$ in the cubic point group $O_h$. IR is an abbreviation of ``irreducible representation''. Since some electric/magnetic multipole moments, such as $Q_{31}^{+}$, are not basis of an irreducible representation in $O_h$, we divide such multipole moments to sum of basis of two irreducible representations (See also Table~\ref{magnetic_multipole_cubic}). The basis for multi-dimensional irreducible representations, $E_{g(u)}$, $T_{1g(1u)}$, and $T_{2g(2u)}$, are labeled by the convention listed in Table~\ref{basis_conventions}.}
 		 \label{electric_multipole_cubic}
\centering
{\renewcommand\arraystretch{1.4}
           \begin{tabular}{llll}
 		 \hline \hline
		 $l$&$Q_{lm}$&IR& basis function  \\ \hline 
		$l=0$&$Q_{00}	$&$ A_{1g}^+$&$1\left( =e\right)$\\ \hline
		$l=1$&$Q_{10}	$&$ T_{1u}^+\left(z \right)$&$z		$\\
		&$Q_{11}^{+}	$&$ T_{1u}^+\left(x \right)$&$x 		$\\
		&$Q_{11}^{-}	$&$ T_{1u}^+\left(y \right)$&$y		$\\ \hline
		$l=2$&$Q_{20}		$&$ E_{g}^+\left(u \right)$&$   \frac{1}{2} \left( 3z^2-r^2\right)$\\
		&$Q_{21}^{+}		$&$ T_{2g}^+\left(\eta \right)$&$\sqrt{3} zx		$\\
		&$Q_{21}^{-}		$&$ T_{2g}^+\left(\xi \right)$&$\sqrt{3} yz		$\\
		&$Q_{22}^{+}		$&$ E_{g}^+\left(v \right)$&$\frac{\sqrt{3}}{2}  \left( 	 x^2-   y^2	\right)		$\\
		&$Q_{22}^{-}		$&$ T_{2g}^+\left(\zeta \right)$&$\sqrt{3}xy $\\ \hline
		$l=3$&$Q_{30}		$&$ T_{1u}^+\left(	z \right)$&$\frac{1}{2}    \left( 5z^2 -3r^2 \right)z $\\
		&$Q_{31}^{+}		$&$ T_{1u}^+\left(x \right)$&$ \frac{\sqrt{6}}{8} x(  3r^2 -5x^2) $\\
		&&$T_{2u}^+\left( \alpha \right)$&\ \ $-\frac{5\sqrt{6}}{8}x ( y^2-z^2 )$\\
		&$Q_{31}^{-}		$&$ T_{1u}^+\left(y  \right)$&$\frac{\sqrt{6}}{8} y \left(3 r^2-5 y^2 \right)$\\
		&&$ T_{2g}^+\left(\beta \right)$&\ \ $+\frac{5\sqrt{6}}{8}  y \left(z^2-x^2\right)$\\
		&$Q_{32}^{+}		$&$ T_{2u}^+ \left(\gamma \right)$&$ \frac{\sqrt{15}}{2} z\left(x^2-y^2 \right)	$\\
		&$Q_{32}^{-}		$&$ A_{2u}^+$&$ \sqrt{15} xyz	$\\
		&$Q_{33}^{+}		$&$ T_{1u}^+\left(x \right)$&$\frac{\sqrt{10}}{8}  x \left(5 x^2-3r^2\right)$\\
		&&$ T_{2u}^+\left(\alpha \right)$&\ \ $-\frac{3\sqrt{10}}{8} x \left(y^2-z^2\right)$\\
		&$Q_{33}^{-}		$&$ T_{1u}^+\left(y  \right)$&$\frac{\sqrt{10}}{8}  y \left(3 r^2-5 y^2\right)$\\ 
		&&$T_{2u}^+\left(\beta \right)$&\ \ $-\frac{3\sqrt{10}}{8}  y \left(z^2-x^2\right)$\\ \hline
		$l=4$&$Q_{40}		$&$ A_{1g}^+$&$\frac{7}{12} \left(x^4+y^4+z^4\right)-\frac{7}{4} \left(  x^2 y^2+y^2 z^2+ z^2x^2\right)$\\
		&&$ E_g^+\left(u \right)$&\ \ $+\frac{5}{24} \left( 2 z^4-x^4 -y^4\right)+\frac{5}{4} \left(2x^2 y^2 - y^2 z^2- z^2 x^2\right)$\\
		&$Q_{41}^{+}		$&$ T_{1g}^+\left(\hat{\bm{y}} \right)$&$\frac{7\sqrt{10}}{8} zx \left(z^2-x^2\right)$\\
			&&$ T_{2g}^+\left( \eta \right)$&\ \ $+\frac{\sqrt{10}}{8}  zx \left(r^2-7 y^2\right)$\\
		&$Q_{41}^{-}		$&$ T_{1g}^+\left( \hat{\bm{x}} \right)$&$-\frac{7\sqrt{10}}{8}  y z \left(y^2-z^2\right)$\\
			&&$ T_{2g}^+\left(\xi \right)$&\ \ $+\frac{\sqrt{10}}{8} yz \left(r^2-7 x^2\right)$\\
		&$Q_{42}^{+}		$&$E_{g}^+\left(v \right)$&$\frac{\sqrt{5}}{4}  \left(x^2-y^2\right) \left(7 z^2-r^2\right)$\\
		&$Q_{42}^{-}		$&$ T_{2g}^+\left(\zeta \right)$&$ \frac{\sqrt{5}}{2}  x y \left(7 z^2-r^2\right)$\\
		&$Q_{43}^{+}		$&$ T_{1g}^+\left( \hat{\bm{y}} \right)$&$-\frac{\sqrt{70}}{8}  zx \left(z^2-x^2\right)$\\
		&&$ T_{2g}^+\left(\eta \right)$&\ \ $+\frac{\sqrt{70}}{8}zx \left(r^2-7 y^2\right)$\\
		&$Q_{43}^{-}		$&$ T_{1g}^+\left(\hat{\bm{x}} \right)$&$-\frac{\sqrt{70}}{8}  y z \left(y^2-z^2\right)$\\
		&&$ T_{2g}^+\left(\xi \right)$&\ \ $-\frac{\sqrt{70}}{8} y z \left(r^2 -7 x^2\right)$\\
		&$Q_{44}^{+}		$
			&$ A_{1g}^+$&$\frac{\sqrt{35}}{12} \left(x^4+y^4+z^4\right)-\frac{\sqrt{35}}{4} \left( x^2  y^2+y^2  z^2+z^2  x^2\right)$\\
			&&$ E_g^+\left(u \right) $&\ \ $-\frac{\sqrt{35}}{24}  \left(2 z^4-x^4-y^4\right)-\frac{\sqrt{35}}{4}  \left(2 x^2y^2 -y^2z^2-z^2x^2 \right)$\\
		&$Q_{44}^{-}		$&$ T_{1g}^+\left( \hat{\bm{z}} \right) $&$\frac{\sqrt{35}}{2}  x y \left(x^2-y^2\right)$\\ \hline \hline
           \end{tabular}
}
		\end{table}
	\begin{table}[H]
	\caption{Classification of magnetic multipole moment up to rank $l=4$ in the cubic point group $O_h$. We also show toroidal dipole moment $T_\mu$ and magnetic monopole moment $\bm{r}\cdot \bm{s}$ which are not represented by any linear combination of magnetic multipole moment.}
 		 \label{magnetic_multipole_cubic}
\centering
{\renewcommand\arraystretch{1.4}
           \begin{tabular}{llll}
           \hline \hline
		 $l$&$M_{lm}$&IR& basis function  \\ \hline 
		$l=0$&$\bm{r}\cdot \bm{s}$&$  A_{1u}^-	$&$	 x\hat{\bm{x}}+ y\hat{\bm{y}} + z\hat{\bm{z}}						$\\ \hline
		$l=1$&$M_{10}	$&$ T_{1g}^-\left(\hat{\bm{z}} \right)$&$\hat{\bm{z}}		$\\
		&$M_{11}^{+}	$&$ T_{1g}^-\left(\hat{\bm{x}} \right)$&$ \hat{\bm{x}} 		$\\
		&$M_{11}^{-}	$&$ T_{1g}^-\left(\hat{\bm{y}} \right)$&$ \hat{\bm{y}}		$\\ \cline{2-4}
		&$T_x$&$  T_{1u}^-\left(x \right)	$&$	 y\hat{\bm{z}}-z\hat{\bm{y}}						$\\
		&$T_y$&$  T_{1u}^-\left(y \right)	$&$	 z\hat{\bm{x}}-x\hat{\bm{z}}						$\\
		&$T_z$&$  T_{1u}^-\left(z \right)	$&$	 x\hat{\bm{y}}-y\hat{\bm{x}}						$\\ \hline
		$l=2$&$M_{20}		$&$ E_{u}^-\left(q \right)$&$   2z\hat{\bm{z}} - x\hat{\bm{x}} -  y\hat{\bm{y}}$\\
		&$M_{21}^{+}		$&$ T_{2u}^-\left(\beta \right)$&$\sqrt{3}  \left( 	 x\hat{\bm{z}}+  z\hat{\bm{x}} 	\right)		$\\
		&$M_{21}^{-}		$&$ T_{2u}^-\left(\alpha \right)$&$\sqrt{3}  \left( 	 y\hat{\bm{z}}+  z\hat{\bm{y}} 	\right)		$\\
		&$M_{22}^{+}		$&$ E_{u}^-\left(p \right)$&$\sqrt{3}  \left( 	 x\hat{\bm{x}}-   y\hat{\bm{y}} 	\right)		$\\
		&$M_{22}^{-}		$&$ T_{2u}^-\left(\gamma \right)$&$\sqrt{3}  \left( 	  y\hat{\bm{x}}+  x\hat{\bm{y}} 	\right)		$\\ \hline
		$l=3$&$M_{30}		$&$ T_{1g}^-\left(\hat{\bm{z}}  \right)$&$\frac{3}{2}    \left( 3z^2 -r^2 \right) \hat{\bm{z}} -  3z  \left( 	 x \hat{\bm{x}}+  y\hat{\bm{y}} 	\right)		$\\
		&$M_{31}^{+}		$&$ T_{1g}^-\left(\hat{\bm{x}} \right)$&$\frac{3\sqrt{6}}{8} \left[  (r^2-3x^2) \hat{\bm{x}} +2xy\hat{\bm{y}}+2zx\hat{\bm{z}}  \right] $\\
		&&$T_{2g}^-\left(\xi \right)$&\ \ $-\frac{5\sqrt{6}}{8} \left[ ( y^2-z^2 ) \hat{\bm{x}}  +2 x y  \hat{\bm{y}} -2 zx \hat{\bm{z}} \right]  $\\
		&$M_{31}^{-}		$&$ T_{1g}^-\left(\hat{\bm{y}}  \right)$&$\frac{3\sqrt{6}}{8} \left[ 2 xy\hat{\bm{x}} +(r^2-3y^2) \hat{\bm{y}} +2yz\hat{\bm{z}} \right]$\\
		&&$ T_{2g}^-\left(\eta \right)$&\ \ $-\frac{5\sqrt{6}}{8} \left[ 2 x y  \hat{\bm{x}}+( x^2-z^2 ) \hat{\bm{y}}   -2 yz \hat{\bm{z}} \right]  $\\
		&$M_{32}^{+}		$&$ T_{2g}^-\left(\zeta \right)$&$\frac{\sqrt{15}}{2} \left(	x^2-y^2	\right)\hat{\bm{z}} +\sqrt{15} z \left(	  x \hat{\bm{x}} - y \hat{\bm{y}}  	\right) 	$\\
		&$M_{32}^{-}		$&$ A_{2g}^-$&$ \sqrt{15}  \left(	 y z \hat{\bm{x}}  + z x \hat{\bm{y}}	+ xy \hat{\bm{z}}  \right)  		$\\
		&$M_{33}^{+}		$&$ T_{1g}^-\left(\hat{\bm{x}}  \right)$&$ \frac{3\sqrt{10}}{8} 	 \left[	( 3x^2-r^2) \hat{\bm{x}}-2 xy \hat{\bm{y}}  -2 zx \hat{\bm{z}}\right] $\\
		&&$ T_{2g}^-\left(\xi \right)$&\ \ $  - \frac{3\sqrt{10}}{8} 	 \left[ (y^2-z^2) \hat{\bm{x}}+2  x y\hat{\bm{y}}- 2 zx \hat{\bm{z}} 	\right] $\\
		&$M_{33}^{-}		$&$ T_{1g}^-\left(\hat{\bm{y}}  \right)$&$ \frac{3\sqrt{10}}{8} 	 \left[	2 xy \hat{\bm{x}} +(r^2 -3 y^2) \hat{\bm{y}}+2 yz \hat{\bm{z}} )\right] $\\ 
		&&$T_{2g}^-\left(\eta \right)$&\ \ $ - \frac{3\sqrt{10}}{8} 	 \left[ -2x y\hat{\bm{x}} +(z^2-x^2) \hat{\bm{y}} +2yz \hat{\bm{z}} \right]$\\ \hline
		$l=4$&$M_{40}		$&$ A_{1u}^-$&$\frac{7}{6} \left[ \left(5x^2-3r^2\right)x \hat{\bm{x}}+\left(5y^2-3r^2\right)y \hat{\bm{y}}+\left(5z^2-3r^2\right)z \hat{\bm{z}} \right]$\\
		&&$ E_u^-\left(q \right)$&\ \ $+\frac{5}{6} \left[ - \left(x^2+ 3 z^2-6 y^2\right) x\hat{\bm{x}}-  \left(y^2+ 3 z^2-6 x^2\right)y\hat{\bm{y}} +\left(5 z^2-3 r^2 \right) z\hat{\bm{z}}\right]$\\
		&$M_{41}^{+}		$&$ T_{1u}^-\left(y \right)$&$-\frac{7\sqrt{10}}{8}\left[  \left(3 zx^2 -z^3\right) \hat{\bm{x}}+  \left(x^3-3  z^2x\right)\hat{\bm{z}} \right]$\\
		&&$ T_{2u}^-\left( \beta \right)$&\ \ $+\frac{\sqrt{10}}{8}  \left[  \left(  z^2-6 y^2+3 x^2 \right) z\hat{\bm{x}} -12 x y z \hat{\bm{y}} +\left(x^2-6  y^2+3  z^2\right) x\hat{\bm{z}}\right]$\\
		&$M_{41}^{-}		$&$ T_{1u}^-\left(x \right)$&$-\frac{7\sqrt{10}}{8}  \left[   \left(3 y^2 z-z^3\right) \hat{\bm{y}}+ \left(y^3-3 y z^2\right)\hat{\bm{z}} \right]$\\
		&&$ T_{2u}^-\left(\alpha \right)$&\ \ $+\frac{\sqrt{10}}{8}  \left[ -12 x  y z\hat{\bm{x}} + \left(-6 x^2 +3 y^2 +z^2\right) z\hat{\bm{y}}+ \left(-6 x^2 +y^2+3 z^2\right) y\hat{\bm{z}}\right]$\\
		&$M_{42}^{+}		$&$E_{u}^-\left(p \right)$&$- \sqrt{5} \left[    \left(x^3-3 x z^2\right) \hat{\bm{x}}+   \left(3 y z^2-y^3\right) \hat{\bm{y}}+\left(3 y^2 z-3 x^2 z\right) \hat{\bm{z}} \right]$\\
		&$M_{42}^{-}		$&$ T_{2u}^-\left(\gamma \right)$&$-\frac{\sqrt{5}}{6} \left[     \left(3 x^2 + y^2-6 z^2\right)y\hat{\bm{x}} +\left(x^2+3 y^2-6 z^2\right) x\hat{\bm{y}}-12 x y z \hat{\bm{z}} \right]$\\
		&$M_{43}^{+}		$&$ T_{1u}^-\left(y \right)$&$\frac{\sqrt{70}}{8} \left[   \left(3 x^2 z-z^3\right) \hat{\bm{x}} + \left(x^3-3 x z^2\right) \hat{\bm{z}}\right]$\\
		&&$ T_{2u}^-\left(\beta \right)$&\ \ $+\frac{\sqrt{70}}{8} \left[  \left(z^2-6 y^2 +3 x^2   \right) z\hat{\bm{x}}-12 x y  z \hat{\bm{y}} +\left(x^2-6  y^2+3  z^2\right) x\hat{\bm{z}}\right]$\\
		&$M_{43}^{-}		$&$ T_{1u}^-\left(x \right)$&$\frac{\sqrt{70}}{8} \left[   \left(z^3-3 y^2 z\right)\hat{\bm{y}} + \left(3 y z^2-y^3\right) \hat{\bm{z}}\right]$\\
		&&$ T_{2u}^-\left(\alpha \right)$&\ \ $	+\frac{\sqrt{70}}{8}  \left[ 12 x y z\hat{\bm{x}}  + \left(6 x^2 -3 y^2 -z^2\right) z\hat{\bm{y}} +\left(6 x^2 -y^2-3  z^2\right)y \hat{\bm{z}}\right]$\\
		&$M_{44}^{+}		$&$ A_{1u}^-$&$\frac{\sqrt{35}}{6}  \left[  \left(5 x^2- 3 r^2 \right) x \hat{\bm{x}} +\left(5 y^2- 3 r^2 \right) y \hat{\bm{y}}+\left(5 z^2- 3 r^2 \right) z \hat{\bm{z}} \right]$\\
							&&$ E_u^-\left(q \right) $&\ \ $+\frac{\sqrt{35}}{6}  \left[  \left(x^2+3 z^2-6 y^2\right) x\hat{\bm{x}} + \left( y^2+3  z^2-6 x^2\right)y\hat{\bm{y}} +\left(3 r^2-5 z^2\right) z \hat{\bm{z}} \right]$\\
		&$M_{44}^{-}		$&$ T_{1u}^-\left(z \right) $&$\frac{\sqrt{35}}{2} \left[ \left(3 x^2 y-y^3\right) \hat{\bm{x}}+  \left(x^3-3 x y^2\right)\hat{\bm{y}}   \right]$\\ \hline \hline
           \end{tabular}
}
		\end{table}
\twocolumngrid
%
%
%
%
%
%
\onecolumngrid 

	\begin{table}[H]
		\caption{Classification of electric multipole moment up to rank $l=4$ in the tetragonal point group $D_{4h}$ and in the hexagonal point group $D_{6h}$. 
 Irreducible representations in $D_{4h}$ and $D_{6h}$ are illustrated in the third and fourth columns by IR(T) and IR(H), respectively. 
}
 		 \label{electric_multipole_tetragonal_hexagonal}
\centering
{\renewcommand\arraystretch{1.4}
           \begin{tabular}{llllllllll}
		 \hline\hline
		 $l$&$Q_{lm}$&IR$\left(  \mathrm{T}\right)$&IR$ \left( \mathrm{H}\right)$& basis function \\ \hline
		$l=0$&$Q_{00}	$&$ A_{1g}^+$&$ A_{1g}^+$&$1\left( =e\right)$&
		$l=4$&$Q_{40}$
			&$ A_{1g}^+$&$ A_{1g}^+$&$\frac{1}{8} \left( 35z^4 -30z^2r^2 +3r^4 \right)$\\ \cline{1-5}
		$l=1$&$Q_{10}	$&$ A_{2u}^+$&$ A_{2u}^+$&$z		$&
		&$Q_{41}^{+}$&$ E_{g}^+\left(\hat{\bm{y}} \right)$&$ E_{1g}^+\left(\hat{\bm{y}} \right)$&$\frac{\sqrt{10}}{4} \left(7z^2-3r^2\right)zx$\\
		&$Q_{11}^{+}	$&$ E_{u}^+\left(x \right)$&$ E_{1u}^+\left(x \right)$&$x 		$&
		&$Q_{41}^{-}		$&$ E_{g}^+\left( \hat{\bm{x}} \right)$&$ E_{1g}^+\left( \hat{\bm{x}} \right)$&$\frac{\sqrt{10}}{4} \left(7z^2-3r^2\right) yz$\\
		&$Q_{11}^{-}	$&$ E_{u}^+\left(y \right)$&$ E_{1u}^+\left(y \right)$&$y		$&
		&$Q_{42}^{+}		$&$B_{1g}^+$&$E_{2g}^+ \left(v \right)$&$\frac{\sqrt{5}}{4}  \left(x^2-y^2\right) \left(7 z^2-r^2\right)$\\\cline{1-5}
		$l=2$&$Q_{20}		$&$ A_{1g}^+$&$ A_{1g}^+$&$   \frac{1}{2} \left( 3z^2-r^2\right)$&
		&$Q_{42}^{-}		$&$ B_{2g}^+$&$ E_{2g}^+ \left(u \right)$&$ \frac{\sqrt{5}}{2}  \left(7 z^2-r^2\right) xy$\\
		&$Q_{21}^{+}		$&$ E_{g}^+\left( \hat{\bm{y}} \right)$&$ E_{1g}^+\left(\hat{\bm{y}} \right)$&$\sqrt{3} zx		$&
		&$Q_{43}^{+}		$&$ E_{g}^+\left( \hat{\bm{y}} \right)$&$ B_{2g}^+$&$\frac{\sqrt{70}}{4} \left(x^2-3y^2\right)zx$\\
		&$Q_{21}^{-}		$&$ E_{g}^+\left(\hat{\bm{x}} \right)$&$ E_{1g}^+\left(\hat{\bm{x}} \right)$&$\sqrt{3} yz		$&
		&$Q_{43}^{-}		$&$ E_{g}^+\left(\hat{\bm{x}} \right)$&$ B_{1g}^+$&$\frac{\sqrt{70}}{8}  \left(3x^2-y^2\right) yz$\\
		&$Q_{22}^{+}		$&$ B_{1g}^+$&$ E_{2g}^+ \left(v \right)$&$\frac{\sqrt{3}}{2}  \left( 	 x^2-   y^2	\right)		$&
		&$Q_{44}^{+}		$&$ A_{1g}^+$&$ E_{2g}^+ \left(v \right)$&$\frac{\sqrt{35}}{8} \left(x^4 -6x^2y^2+z^4\right)$\\ 
		&$Q_{22}^{-}		$&$ B_{2g}^+$&$ E_{2g}^+ \left(u \right)$&$\sqrt{3}xy $& 
		&$Q_{44}^{-}		$&$ A_{2g}^+$&$ E_{2g}^+ \left(u \right)$&$\frac{\sqrt{35}}{2}  \left(x^2-y^2\right)x y $\\ \cline{1-5} 
		$l=3$&$Q_{30}		$&$ A_{2u}^+$&$ A_{2u}^+$&$\frac{1}{2}    \left( 5z^2 -3r^2 \right)z $&\\
		&$Q_{31}^{+}		$&$ E_{u}^+\left(x \right)$&$ E_{1u}^+\left(x \right)$&$ \frac{\sqrt{6}}{4} (  5z^2 -r^2)x $\\
		&$Q_{31}^{-}		$&$ E_{u}^+\left(y  \right)$&$ E_{1u}^+\left(y  \right)$&$\frac{\sqrt{6}}{4}  \left(5 z^2-r^2 \right) y $\\
		&$Q_{32}^{+}		$&$ B_{2u}^+$&$ E_{2u}^+ \left(p \right)$&$ \frac{\sqrt{15}}{2} \left(x^2-y^2 \right)z	$\\
		&$Q_{32}^{-}		$&$ B_{1u}^+$&$ E_{2u}^+ \left(q \right)$&$ \sqrt{15} xyz	$\\
		&$Q_{33}^{+}		$&$ E_{u}^+\left(x \right)$&$B_{1u}^+$&$\frac{\sqrt{10}}{4}  \left(x^2-3y^2\right)x$\\
		&$Q_{33}^{-}		$&$ E_{u}^+\left(y  \right)$&$B_{2u}^+$&$\frac{\sqrt{10}}{4}   \left(3 x^2-y^2\right)y$\\ \hline\hline
           \end{tabular}
}
		\end{table}
\twocolumngrid
\onecolumngrid

	\begin{table}[H]
	\caption{Classification of magnetic multipole moment up to rank $l=4$  in the tetragonal point group $D_{4h}$ and in the hexagonal point group $D_{6h}$. Irreducible representations in $D_{4h}$ and $D_{6h}$ are represented by IR(T) and IR(H), respectively. We also show toroidal dipole moment $T_\mu$ and magnetic monopole moment $\bm{r}\cdot \bm{s}$.}
 		 \label{magnetic_multipole_tetragonal_hexagonal}
\centering
{\renewcommand\arraystretch{1.4}
           \begin{tabular}{lllll}
		 \hline\hline
		 $l$&$M_{lm}$&IR$\left(  \mathrm{T}\right)$&IR$ \left( \mathrm{H}\right)$& basis function  \\ \hline 
		$l=0$&$\bm{r}\cdot \bm{s}$&$  A_{1u}^-	$&$  A_{1u}^-	$&$	 x\hat{\bm{x}}+ y\hat{\bm{y}} + z\hat{\bm{z}}$\\ \hline
		$l=1$&$M_{10}	$&$ A_{2g}^-$&$ A_{2g}^-$&$\hat{\bm{z}}		$\\
		&$M_{11}^{+}	$&$ E_{g}^- \left( \hat{\bm{x}}\right)$&$ E_{1g}^- \left( \hat{\bm{x}}\right)$&$ \hat{\bm{x}} 		$\\
		&$M_{11}^{-}	$&$ E_{g}^-\left( \hat{\bm{y}}\right)$&$ E_{1g}^-\left( \hat{\bm{y}}\right)$&$ \hat{\bm{y}}		$\\ \cline{2-5} 
		&$T_x$&$  E_{u}^-	\left( x\right)$&$ E_{1u}^-\left( x\right)$&$	 y\hat{\bm{z}}-z\hat{\bm{y}}						$\\
		&$T_y$&$  E_{u}^-	\left( y\right)$&$ E_{1u}^-\left( y\right)$&$	 z\hat{\bm{x}}-x\hat{\bm{z}}						$\\
		&$T_z$&$  A_{2u}^-	$&$  A_{2u}^-	$&$	 x\hat{\bm{y}}-y\hat{\bm{x}}						$\\ \hline
		$l=2$&$M_{20}		$&$ A_{1u}^-$&$ A_{1u}^-$&$     2z\hat{\bm{z}} - x\hat{\bm{x}} 	 -  y\hat{\bm{y}} 	$\\
		&$M_{21}^{+}		$&$ E_{u}^-\left( y\right)$&$ E_{1u}^-\left( y\right)$&$\sqrt{3}  \left( 	 x\hat{\bm{z}}+  z\hat{\bm{x}} 	\right)		$\\
		&$M_{21}^{-}		$&$ E_{u}^-\left( x\right)$&$ E_{1u}^-\left( x\right)$&$\sqrt{3}  \left( 	 y\hat{\bm{z}}+  z\hat{\bm{y}} 	\right)		$\\
		&$M_{22}^{+}		$&$ B_{1u}^-$&$ E_{2u}^-\left( q\right)$&$\sqrt{3}  \left( 	 x\hat{\bm{x}}-   y\hat{\bm{y}} 	\right)		$\\
		&$M_{22}^{-}		$&$ B_{2u}^-$&$ E_{2u}^-\left( p\right)$&$\sqrt{3}  \left( 	  y\hat{\bm{x}}+  x\hat{\bm{y}} 	\right)		$\\ \hline

		$l=3$&$M_{30}		$&$ A_{2g}^-$&$ A_{2g}^-$&$-3 x z \hat{\bm{x}}-3 y z\hat{\bm{y}}-\frac{3}{2}  \left(x^2+y^2-2 z^2\right)\hat{\bm{z}}$\\
		&$M_{31}^{+}		$&$ E_{g}^-\left( \hat{\bm{x}}\right)$&$ E_{1g}^-\left( \hat{\bm{x}}\right)$&$-\frac{\sqrt{6}}{4} \left[ \left(3 x^2+y^2-4z^2 \right) \hat{\bm{x}}+2 x y \hat{\bm{y}}-8 x z \hat{\bm{z}}\right]$\\
		&$M_{31}^{-}		$&$ E_{g}^-\left( \hat{\bm{y}}\right)$&$ E_{1g}^-\left( \hat{\bm{y}}\right)$&$-\frac{\sqrt{6}}{4}  \left[ 2 x  y\hat{\bm{x}} + \left(x^2+3y^2 -4z^2 \right) \hat{\bm{y}}-8 y z \hat{\bm{z}}\right]$\\
		&$M_{32}^{+}		$&$ B_{2g}^-	$&$ E_{2g}^- \left( u \right)	$&$\frac{\sqrt{15}}{2}  \left[ 2 x  z \hat{\bm{x}}-2 y  z \hat{\bm{y}} +\left(x^2-y^2\right) \hat{\bm{z}}\right]$\\
		&$M_{32}^{-}		$&$ B_{1g}^-	$&$ E_{2g}^- \left( v \right)	$&$\sqrt{15} (y z \hat{\bm{x}} + x z \hat{\bm{y}} + x y \hat{\bm{z}})$\\
		&$M_{33}^{+}		$&$ E_{g}^-\left( \hat{\bm{x}}\right)$&$B_{1g}^-$&$\frac{3\sqrt{10}}{4}  \left[ (x^2 - y^2)\hat{\bm{x}} - 2 x y \hat{\bm{y}} \right]$\\
		&$M_{33}^{-}		$&$ E_{g}^- \left( \hat{\bm{y}}\right)$&$B_{2g}^-$&$\frac{3\sqrt{10}}{4} \left[ 2 x  y  \hat{\bm{x}}+ \left(x^2-y^2\right) \hat{\bm{y}}\right]$\\ \hline
		$l=4$&$M_{40}		$&$ A_{1u}^-$&$ A_{1u}^-$&$ \frac{1}{2} \left[ 3\left(  x^2+  y^2-4  z^2\right)x \hat{\bm{x}}+ 3 \left( x^2 + y^2-4 z^2\right) y \hat{\bm{y}}+ 4 \left(-3 x^2 -3 y^2 +2 z^2\right) z \hat{\bm{z}} \right]$\\
		&$M_{41}^{+}		$&$ E_{u}^-\left( y\right)$&$ E_{1u}^-\left( y\right)$&$\frac{\sqrt{10}}{4} 	\left[ \left(-9 x^2 - 3 y^2+4 z^2 \right) z \hat{\bm{x}} -6 x y  z \hat{\bm{y}} - 3 \left(  x^2+  y^2- 4 z^2\right) x\hat{\bm{z}} \right] $\\
		&$M_{41}^{-}		$&$ E_{u}^-\left( x\right)$&$ E_{1u}^-\left( x\right)$&$\frac{\sqrt{10}}{4} 	\left[-6 x  y z\hat{\bm{x}}  + \left(-3 x^2 -9 y^2 +4 z^2\right) z\hat{\bm{y}} - 3\left( x^2 + y^2-4 z^2\right) y\hat{\bm{z}} \right]$\\
		&$M_{42}^{+}		$&$B_{1u}^-	$&$E_{2u}^- \left(q \right)	$&$ \sqrt{5} \left[ \left(3 z^2-x^2 \right)x  \hat{\bm{x}}+\left(y^2-3  z^2\right) y \hat{\bm{y}}+ 3\left( x^2 - y^2 \right) z\hat{\bm{z}} \right]$\\
		&$M_{42}^{-}		$&$ B_{2u}^-	$&$E_{2u}^- \left(p \right)	$&$\frac{\sqrt{5}}{2}\left[  \left(-3 x^2 -  y^2+6 z^2 \right) y\hat{\bm{x}}+  \left(-x^2-3  y^2+6  z^2\right)x \hat{\bm{y}}+12 x y z \hat{\bm{z}} \right]$\\
		&$M_{43}^{+}		$&$ E_{u}^- \left( y\right)$&$B_{2u}^-$&$ \frac{\sqrt{70}}{4}  \left[ 3 ( x^2  - y^2)z   \hat{\bm{x}}-6 x y z \hat{\bm{y}} + (x^3 - 3 xy^2) \hat{\bm{z}} \right] $\\
		&$M_{43}^{-}		$&$ E_{u}^- \left( x\right) $&$B_{1u}^-$&$ \frac{\sqrt{70}}{2}  \left[6 x y z \hat{\bm{x}} +  3( x^2 -  y^2 )z \hat{\bm{y}}+ (3 x^2y  - y^3)  \hat{\bm{z}} \right] $\\
		&$M_{44}^{+}		$&$ A_{1u}^-$&$E_{2u}^- \left( q \right)$&$  \frac{\sqrt{35}}{2} \left[  (x^3 - 3 xy^2)   \hat{\bm{x}} + (-3 x^2y  + y^3)   \hat{\bm{y}} \right] $\\
		&$M_{44}^{-}		$&$ A_{2u}^- $&$E_{2u}^- \left( p \right)$&$  \frac{\sqrt{35}}{2} \left[ (3 x^2y  - y^3)  \hat{\bm{x}} + (x^3 - 3 xy^2) \hat{\bm{y}} \right]$\\ \hline \hline
           \end{tabular}
}
		\end{table}

\twocolumngrid

Here, we give some comments on the notation in Tables~\ref{electric_multipole_cubic}-\ref{magnetic_multipole_tetragonal_hexagonal}. 
Needless to say, the parity under the time reversal operation is an essential property of physical quantities and responses. We denote irreducible representations with even time-reversal parity by $\Gamma^+$, and those with odd time-reversal parity by $\Gamma^-$. Thus, electric multipole moments listed in Tables~\ref{electric_multipole_cubic} and \ref{electric_multipole_tetragonal_hexagonal} are represented by  $\Gamma^+$, while magnetic multipole moments in Tables~\ref{magnetic_multipole_cubic} and \ref{magnetic_multipole_tetragonal_hexagonal} are represented by $\Gamma^-$. 
Tables~\ref{magnetic_multipole_cubic} and \ref{magnetic_multipole_tetragonal_hexagonal} include the magnetic monopole moment $\bm{r}\cdot \bm{s}$ and the magnetic toroidal moment $\bm{T} \equiv (T_x, T_y, T_z)$ in addition to the magnetic multipole moment $M_{lm}$.

For multi-dimensional irreducible representations, there is an ambiguity in the choice of basis functions; we can choose arbitrary linear combinations of multi-component basis functions. It is useful to specify the basis, since the multi-dimensional basis functions may be separated into lower-dimensional irreducible representations in subgroups. For example, the two-dimensional $E_u$ irreducible representation of $D_{4h}$ point group has basis functions $\{x,y\}$. In $D_{2h}$, a subgroup of $D_{4h}$, the $E_u$ representation is reduced to $B_{2u}$ and $B_{3u}$ representations. Accordingly, $x$ and $y$ are basis functions for $B_{3u}$ and $B_{2u}$, respectively. For the purpose to simplify such compatibility relations, we specify the basis of multi-dimensional irreducible representations, as listed in Table~\ref{basis_conventions}.  
Convention such as $E_u\left( x \right)$ is used in the classification tables~\ref{electric_multipole_cubic}-\ref{magnetic_multipole_tetragonal_hexagonal}. 
For example, a magnetic quadrupole moment $y \hat{\bm{z}} + z \hat{\bm{y}}$ in $D_{4h}$ is classified into $E_u\left(x \right)$ since it is symmetry-adapted to $x$.

		\begin{table}[htbp] 
		\centering
{\renewcommand\arraystretch{1.3}
		\caption{Conventions for the basis of multi-dimensional irreducible representations in the point group $O_h$, $D_{4h}$, and $D_{6h}$.}
		\label{basis_conventions}
		\begin{tabular}{lll} \hline \hline
		$\bm{G}$&IR&Basis\\ \hline
		$O_h$
		&$E_{g}$&$\{u, v\} \equiv \{ 2z^2-x^2-y^2, x^2-y^2 \}$\\ 
		&$T_{1g}$&$\{ \hat{\bm{x}},\hat{\bm{y}},\hat{\bm{z}} \}$\\
		&$T_{2g}$&$\{ \xi,\eta,\zeta \}\equiv \{ yz,zx,xy \}$\\
		&$E_{u}$&$\{ p,q \} \equiv \{ xyz\left( 2z^2-x^2-y^2\right),xyz\left( x^2-y^2\right) \}$\\
		&$T_{1u}$&$\{ x,y,z \}$\\
		&$T_{2u}$&$\{ \alpha,\beta,\gamma \} \equiv \{ x\left(y^2-z^2 \right),y\left(z^2-x^2 \right),z\left(x^2-y^2 \right) \}$ \vspace{1mm}\\ 
		$D_{4h}$
		&$E_{g}$&$\{\hat{\bm{x}},\hat{\bm{y}} \}$\\
		&$E_{u}$&$\{x,y \}$\vspace{1mm}\\
		$D_{6h}$
		&$E_{1g}$&$\{\hat{\bm{x}},\hat{\bm{y}} \}$\\
		&$E_{2g}$&$\{ u,v \} \equiv \{ xy, x^2-y^2 \}$\\
		&$E_{1u}$&$\{x,y \}$\\
		&$E_{2u}$&$\{p,q \} \equiv  \{ z\left(x^2-y^2 \right), xyz \}$\\
		\hline \hline
		\end{tabular}
}
		\end{table}

\subsection{Basis functions} \label{Section2_basisfunctions}
	
In Sec.~\ref{Section2_multipole}, we have classified the multipole moment in crystal point groups. It is practically more useful to summarize the classification of multipole moments with respect to each irreducible representation of a given point group, because the symmetry analysis of ordered states identifies irreducible representations and the candidate for multipole order parameter is determined by the obtained irreducible representation. Furthermore, we may obtain an intuitive and precise understanding of emergent electromagnetic responses in the multipole states by referring to the list of basis functions in both real space and momentum space (See Sec.~\ref{Section3} for details).

For this purpose we show the list of irreducible representations in point groups and tabulate multipole moments as order parameters characterizing each irreducible representation.
In the classification tables~\ref{electric_basis_cubic}-\ref{magnetic_basis_hexagonal}, the lowest-order basis functions in the real-space coordinates $\bm{r} = (x,y,z)$ and those in the momentum-space coordinates $\bm{k} =(k_x,k_y,k_z)$ are shown. Some of the real-space basis functions do not correspond to the multipole moments, although in most cases the multipole moment appears as the lowest order basis in the real space.

We immediately notice the characteristic correspondence of real space and momentum space representations. For the even-parity electric/magnetic multipole order, we can obtain basis functions in the momentum space from those in the real space by replacing $\bm{r} \rightarrow \bm{k}$. For instance, the real space basis function of $B_{2g}^+$ in $D_{4h}$ is $xy$, and the corresponding basis function in the momentum space is $k_xk_y$. The basis function $xy$ represents an electric quadrupole moment in the real space, which can be induced by ferroic orbital order. The momentum space basis $k_xk_y$ may correspond to the $d$-wave Pomeranchuk instability. Indeed, nematic order in 122-type iron-based superconductors belongs to the $B_{2g}^+$ irreducible representation, although the origin of the nematic order is still under debate in spite of intensive studies~\cite{Chu2012,Fernandes2014}.

In contrast to the even-parity multipole order, the real space and momentum space representations of odd-parity multipole order are quite different~\cite{hikaruwatanabe2017}. In the odd-parity electric multipole states, the basis in the momentum space is spin-dependent, indicating a spin-momentum coupling~\cite{Hitomi2014,Hitomi2016,Fu2015}. For instance, an electric octupole order with $Q_{32}^{-} \propto xyz$ in a cubic system is classified into the $A_{2u}^+$ representation. Then, the momentum-space representation $k_x (k_y^2-k_z^2)\hat{\bm{x}}+k_y (k_z^2-k_x^2)\hat{\bm{y}}+k_z (k_x^2-k_y^2)\hat{\bm{z}}  $ indicates the Dresselhaus-type spin-orbit coupling. 

On the other hand, the basis for odd-parity magnetic multipole order in the momentum space is spin-independent, which gives rise to anti-symmetric dispersion of electrons and magnons~\cite{hikaruwatanabe2017,Yanase2014,Hayami2014b,Hayami2016,Sumita2016,Sumita2017}.  For example, the magnetic quadrupole order with $M_{22}^{+} \propto x\hat{\bm{x}} - y\hat{\bm{y}}$ belongs to the $B_{1u}^-$ representation in the tetragonal $D_{4h}$ point group. The corresponding momentum-space representation is $k_x k_y k_z$ which indicates the tetrahedral distortion in the band structure~\cite{hikaruwatanabe2017}. Later we show many candidate materials of the $B_{1u}^-$ multipole order (Sec.~\ref{Section4_candidates}). 
From the classification tables~\ref{electric_basis_cubic}-\ref{magnetic_basis_hexagonal}, we may understand the one-to-one correspondence between the symmetry of multipole order and the unusual electronic structure. The nontrivial duality between the real space and the momentum space is a characteristic and intriguing property of parity-violating order in crystalline systems, while the $l$-rank multipole order in the real-space corresponds to the $l$-rank basis in the momentum space in an isotropic system.

\onecolumngrid

%
%
%
%
%
	\begin{table}[H]
		\caption{Time-reversal even basis functions for irreducible representations in the $O_h$ point group. Electric multipole moments $Q_{lm}$ are shown as a candidate for an order parameter. Basis functions in the real and momentum space are also listed. Some irreducible representations do not have a basis function within the rank $l\leq 4$ electric multipole moment, and then, we show higher-rank electric multipole moment. Note that some multipole moments are reducible in $O_h$; \textit{e.g.} $Q_{62}^+$ is divided into the basis of $A_{2g}^+$ and $E_g^+ \left( v \right)$.}
		\label{electric_basis_cubic}
		\centering
{\renewcommand\arraystretch{1.3}
		\begin{tabular}{llll}\hline\hline
		IR &$Q_{lm}$&Basis in real space&Basis in momentum space	 \\  \hline
		$A_{1g}^+$&\multirow{1}{*}{$Q_{00},Q_{40},Q_{44}^+$}&$x^2+y^2+z^2 $& \\ 
		$A_{2g}^+$&$Q_{62}^+,Q_{66}^+$&$ ( x^2-y^2 )( y^2-z^2 ) ( z^2-x^2 )	$& \\ 
		$E_{g}^+$&\{$Q_{20},Q_{22}^+$\},&\{$ 2z^2-x^2-y^2, x^2-y^2$\}& \\ 
		&$Q_{40},Q_{42}^+,Q_{44}^+$&&\multicolumn{1}{c}{($\bm{r} \rightarrow \bm{k}$)} \\ 
		$T_{1g}^+$&$Q_{41}^\pm, Q_{43}^\pm, Q_{44}^-$&\{$ y z (y^2-z^2), z x (z^2-x^2),x y (x^2-y^2)$\}& \\ 
		$T_{2g}^+$ &\{$Q_{21}^-,Q_{21}^+, Q_{22}^-$\},&\{$ y z , z x, x y $\}& \\ 
		&$Q_{41}^\pm, Q_{42}^-, Q_{43}^\pm$&&  \\ \hline
		$A_{1u}^+$&$Q_{94}^-,Q_{98}^-$&$ x y z (x^2-y^2)(y^2-z^2)(z^2-x^2)$&$k_x\hat{\bm{x}}+k_y\hat{\bm{y}}+k_z\hat{\bm{z}}$ \\ 
		$A_{2u}^+$	&$Q_{32}^-$&$ x y z 	$&$k_x (k_y^2-k_z^2)\hat{\bm{x}}+k_y (k_z^2-k_x^2)\hat{\bm{y}}+k_z (k_x^2-k_y^2)\hat{\bm{z}}  $ \\ 
		$E_u^+$&$\{ Q_{52}^-, Q_{54}^-\}$&\{$x y z (2 z^2 -x^2-y^2 ),x y z ( x^2-y^2 )$\}&\{$k_x \hat{\bm{x}}- k_y \hat{\bm{y} },2 k_z \hat{\bm{z}} - k_x \hat{\bm{x}}- k_y \hat{\bm{y}}$\}  \\ 
		$T_{1u}^+$&\{$Q_{11}^+,Q_{11}^-, Q_{10}$\},&\{$x, y, z$\}&\multirow{1}{*}{\{$k_y \hat{\bm{z} }-k_z\hat{\bm{y} }, k_z \hat{\bm{x} }-k_x\hat{\bm{z} },k_x \hat{\bm{y} }-k_y\hat{\bm{x} }$\}  }\\ 
			&$Q_{30},Q_{31}^\pm,Q_{33}^\pm$&& \\ 
		$T_{2u}^+$&$Q_{31}^\pm,Q_{32}^+, Q_{33}^\pm$&\{$x (y^2-z^2), y (z^2-x^2), z (x^2-y^2)$\} & \{$k_y \hat{\bm{z} }+k_z\hat{\bm{y} }, k_z \hat{\bm{x} }+k_x\hat{\bm{z} },k_x \hat{\bm{y} }+k_y\hat{\bm{x} }$\} \\ \hline \hline
		\end{tabular}
}
		\end{table}
	\begin{table}[H]
		\caption{Time-reversal odd basis functions for irreducible representations in the $O_h$ point group. Magnetic multipole moment $M_{lm}$, toroidal moment $T_\mu$, or monopole moment $\bm{r}\cdot \bm{s}$ are shown as a candidate for an order parameter. Basis functions in the real and momentum space are also listed. When order parameters are not described by the magnetic multipole moment up to rank $l\leq 4$, we show higher-rank magnetic multipole moment. Note that some multipole moments are reducible in $O_h$; \textit{e.g.} $M_{94}^-$ is divided into the basis of $A_{1g}^-$ and $E_g^- \left( u \right)$.}
		\label{magnetic_basis_cubic}
		\centering
		{\renewcommand \arraystretch{1.3}
			\begin{tabular}{llll}\hline\hline
			IR &$M_{lm}$&Basis in real space&Basis in momentum space	 \\ \hline 		
			$A_{1g}^-$ &$M_{94}^-,M_{98}^-$&$y z(y^2-z^2)\hat{\bm{x}}+z x(z^2-x^2)\hat{\bm{y}}+xy(x^2-y^2)\hat{\bm{z}}  $&\\ 
			$A_{2g}^-$ &$M_{32}^-$&$ yz\hat{\bm{x}}+zx\hat{\bm{y}}+xy\hat{\bm{z}}$& \\ 
			$E_{g}^-$ &$M_{52}^-,M_{54}^- $&\{$yz \hat{\bm{x}}- zx \hat{\bm{y} }, 2 xy \hat{\bm{z}} - yz \hat{\bm{x}}- zx \hat{\bm{y}} $\} & \\ 
			$T_{1g}^-$ &\{$M_{11}^+,M_{11}^-,M_{10}$\}&\{$\hat{\bm{x} }, \hat{\bm{y} }, \hat{\bm{z} }$\} &\multicolumn{1}{c}{($\bm{r} \rightarrow \bm{k}$)}	\\ 
			  &$M_{30},M_{31}^\pm,M_{33}^\pm$&&\\  
			$T_{2g}^-$&$ M_{31}^\pm ,M_{33}^\pm$&\{$xy  \hat{\bm{y}}-zx  \hat{\bm{z}}, yz \hat{\bm{z}}-xy \hat{\bm{x}}, zx \hat{\bm{x}}-yz \hat{\bm{y}}$\} &\\
			 &&\{$ (y^2-z^2) \hat{\bm{x}},(z^2-x^2) \hat{\bm{y}}, (x^2-y^2) \hat{\bm{z}}  $\} &\\  \hline
			$A_{1u}^-$&$\bm{r}\cdot \bm{s},$&$x\hat{\bm{x}}+y\hat{\bm{y}}+z\hat{\bm{z}}  $&$ k_xk_y k_z (k_x^2-k_y^2)(k_y^2-k_z^2)(k_z^2-k_x^2)$ \\ 
			&$M_{40},M_{44}^+$&& \\  
			$A_{2u}^-$&$M_{62}^+,M_{66}^+ $&$x (y^2-z^2)\hat{\bm{x}}+y (z^2-x^2)\hat{\bm{y}}+z (x^2-y^2)\hat{\bm{z}}  $&$k_x k_y k_z $ \\ 
			$E_u^-$&\{$ M_{22}^+,M_{20}  $\},&\{$x \hat{\bm{x}}- y \hat{\bm{y} },2 z \hat{\bm{z}} - x \hat{\bm{x}}- y \hat{\bm{y}}$\} &\{$k_x k_y k_z (2 k_z^2 -k_x^2-k_y^2 ),k_x k_y k_z ( k_x^2-k_y^2 )   $\}\\
			&$ M_{40},M_{42}^+,M_{44}^+  $&& \\ 
			$T_{1u}^-$ &\{$ T_x,T_y,T_z $\},&\{$ y \hat{\bm{z} }-z\hat{\bm{y} }, z\hat{\bm{x} }-x \hat{\bm{z} },x\hat{\bm{y} }-y \hat{\bm{x} }$\} & \{$k_x, k_y, k_z$\}\\ 
			&$ M_{41}^\pm,M_{43}^\pm, M_{44}^-  $&&\\ 
			$T_{2u}^-$&\{$ M_{21}^- ,M_{21}^+ ,M_{22}^-  $\},&\{$ y \hat{\bm{z} }+z\hat{\bm{y} }, z\hat{\bm{x} }+x \hat{\bm{z} },x\hat{\bm{y} }+y \hat{\bm{x} }$\} &\{$k_x (k_y^2-k_z^2), k_y (k_z^2-k_x^2), k_z (k_x^2-k_y^2) $\}\\ 
			&$ M_{41}^\pm ,M_{42}^-, M_{43}^\pm  $&&\\ \hline  \hline
			\end{tabular}
}
		\end{table}
%
%
%
%
%
%

		\begin{table}[H] 
			\caption{Time-reversal even basis functions for irreducible representations in the $D_{4h}$ point group. Basis functions in the real space, those in the momentum space, and multipole moments are listed. 
}
			\centering
		{\renewcommand \arraystretch{1.3}
		\begin{tabular}{llll} \hline \hline
		IR &$Q_{lm}$&Basis in real space&Basis in momentum space    \\ \hline
		$A_{1g}^+$
			&$Q_{20},  $&$z^2 $& \\    
			&$Q_{40},Q_{44}^+ $&& \\ 
		$A_{2g}^+$    &$Q_{44}^-  $&$xy (x^2-y^2)$& \\      
		$B_{1g}^+$
			&$Q_{22}^+,  $&$ x^2-y^2$& \\ 
			&$Q_{42}^+  $&&\multicolumn{1}{c}{$ \left( \bm{r}\rightarrow\bm{k} \right) $} \\ 
		$B_{2g}^+$
			&$Q_{22}^-, $&$ xy$& \\      
			&$Q_{42}^- $&& \\       
		$E_{g}^+$&\{$Q_{21}^{-},Q_{21}^{+}$\},
			&\{$ yz, zx $\}& \\ 
			&\{$Q_{41}^{-},Q_{41}^{+}$\}, \{$Q_{43}^{-},Q_{43}^{+}$\}&&\\ \hline
		$A_{1u}^+$
			&$Q_{54}^- $&$xyz(x^2-y^2)$&$k_x \hat{\bm{x}}+ k_y \hat{\bm{y}}+k_z \hat{\bm{z}}$\\ 
			&&&$ 2 k_z \hat{\bm{z}}- k_ x \hat{\bm{x}}- k_y \hat{\bm{y}}$\\    
		$A_{2u}^+$
			&$Q_{10},$&$z,$&$ k_x \hat{\bm{y}}-k_y \hat{\bm{x}}  $ \\ 
			&$Q_{30} $&& \\        
		$B_{1u}^+$
			&$Q_{32}^-$&$xyz$&$   k_ x\hat{\bm{x}} -k_y \hat{\bm{y}}   $\\ 
		$B_{2u}^+$
			&$Q_{32}^+ $&$ z (x^2-y^2 )$&$      k_x \hat{\bm{y}}+ k_y \hat{\bm{x}}$\\ 
		$E_{u}^+$
			&\{$Q_{11}^{+},Q_{11}^{-}$\},&\{$ x,   y  $\}&\{$ k_y \hat{\bm{z}} +k_z \hat{\bm{y}}, k_z \hat{\bm{x}}+k_x \hat{\bm{z}} $\}\\ 
			&\{$Q_{31}^{+},Q_{31}^{-}$\}, \{$Q_{33}^{+},Q_{33}^{-}$\}&&\{$ k_y\hat{\bm{z}}-k_z\hat{\bm{y}},  k_z\hat{\bm{x}}-k_x\hat{\bm{z}} $\} \\  \hline \hline
		\end{tabular}
}
		\end{table} 

	\begin{table}[H]
		\caption{Time-reversal odd basis functions for irreducible representations in the $D_{4h}$ point group. Basis functions in the real space, those in the momentum space, and multipole moments are listed.}
		\centering
		\label{magnetic_basis_tetragonal}
          {\renewcommand\arraystretch{1.3}
		\begin{tabular}{llll}\hline \hline 
		IR &$M_{lm}$&Basis in real space&Basis in momentum space	 \\ \hline	
		$A_{1g}^-$
				&$M_{54}^- $&$z (x\hat{\bm{y}}-y\hat{\bm{x}}) $&\\	
		$A_{2g}^-$
				&$M_{10},  $&$\hat{\bm{z}} $& \\  
				&$M_{30}  $&& \\ 
		$ B_{1g}^-$
				&$M_{32}^-  $&$xy \hat{\bm{z}} ,z  \left(	 y \hat{\bm{x}}+x \hat{\bm{y}} 	\right) 	$&\multicolumn{1}{c}{$\left(\bm{r}\rightarrow \bm{k}  \right)$} \\ 
		$ B_{2g}^-$
				&$M_{32}^+ $&$ (	x^2-y^2) \hat{\bm{z}} , z \left(	 x\hat{\bm{x}}- y\hat{\bm{y}}	\right) 	$& \\ 
		$ E_{g}^-$
				&\{$M_{11}^{+},M_{11}^{-}$\},&\{$ \hat{\bm{x}}, \hat{\bm{y}} $\}& \\ 
				&\{$M_{31}^{+},M_{31}^{-}$\}, \{$M_{33}^{+},M_{33}^{-}$\}&&  \\  \hline
		$ A_{1u}^-$
				&$M_{20},\bm{r}\cdot \bm{s}, $&$2 z \hat{\bm{z}}- x \hat{\bm{x}} - y \hat{\bm{y}},x\hat{\bm{x}}+y\hat{\bm{y}}+z\hat{\bm{z}}  	$&$k_xk_yk_z (	k_x^2-k_y^2) $	 \\ 
				&$M_{40}, M_{44}^+$&&\\ 
		$ A_{2u}^-$
				&$T_z, M_{44}^-$&$	 x\hat{\bm{y}}-y\hat{\bm{x}}	$&$k_z $\\ 
		$ B_{1u}^-$
				&$M_{22}^+,M_{42}^+ $&$    x\hat{\bm{x}} -y \hat{\bm{y}} 	$&$k_xk_yk_z    $ \\ 
		$ B_{2u}^-$
				&$M_{22}^-,M_{42}^- $&$   	 y \hat{\bm{x}}+ x \hat{\bm{y}}	   	$&$ k_z (k_x^2-k_y^2)$\\ 
		$ E_{u}^-$
				&\{$M_{21}^{-},M_{21}^{+}$\}, \{$T_x,T_y$\},&\{$ y \hat{\bm{z}} +z \hat{\bm{y}},z \hat{\bm{x}}+x \hat{\bm{z}} $\}, \{$ y\hat{\bm{z}}-z\hat{\bm{y}},  z\hat{\bm{x}}-x\hat{\bm{z}} $\}&\multirow{1}{*}{\{$  k_x,   k_y 		$\}}\\ 
				&\{$M_{41}^{-},M_{41}^{+}$\}, \{$M_{43}^{-},M_{43}^{+}$\}&&\\ \hline \hline
			\end{tabular}
}
			\end{table}

%
%
%
%
%
%

	\begin{table}[H]
		\caption{Time-reversal even basis functions for irreducible representations in the $D_{6h}$ point group. Basis functions in the real space, those in the momentum space, and multipole moments are listed.}
		\centering
		\label{electric_basis_hexagonal}
	 {\renewcommand \arraystretch{1.3}
		\begin{tabular}{llll} \hline \hline
		IR &$Q_{lm}$&Basis in real space&Basis in momentum space	 \\ \hline
		$A_{1g}^+$
				&$Q_{20},$&$ z^2$& \\  
				&$Q_{40}$&& \\ 
		$A_{2g}^+$
				&$ Q_{66}^-$&$ 3x^5 y-10x^3 y^3+3x y^5	$&\\ 
		$B_{1g}^+$
				&$Q_{43}^{-}$&$ ( 3x^2 -y^2 ) y z	$&\\ 
		$B_{2g}^+$
				&$Q_{43}^{+}$&$ ( x^2 -3 y^2 ) z x	$&\multicolumn{1}{c}{($\bm{r} \rightarrow \bm{k}$)} \\ 
		$E_{1g}^+$
				&\{$Q_{21}^{-},Q_{21}^{+}$\},&\{$  yz , zx$\}& \\ 
				&\{$Q_{41}^{-},Q_{41}^{+}$\}&& \\ 
		$E_{2g}^+$
				&\{$Q_{22}^{-},Q_{22}^{+}$\},&\{$xy, x^2-y^2$\}& \\ 
				&\{$Q_{42}^{-},Q_{42}^{+}$\}, \{$Q_{44}^{-},Q_{44}^{+}$\}&& \\ \hline
		$A_{1u}^+$
				&$Q_{76}^{-}$&$ x y z (3 x^4 - 10 x^2 y^2+3 y^4)$&$k_z \hat{\bm{z}}, k_x\hat{\bm{x}}+k_y \hat{\bm{y}}$ \\ 
		$A_{2u}$
				&$Q_{10}^{},$&$ z 	$&$k_x \hat{\bm{y}}-k_y \hat{\bm{x}}$\\ 
				&$Q_{30}^{}$&&\\ 
		$B_{1u}^+$
				&$Q_{33}^{+}$&$ x (x^2-3y^2) 	$&$ 	k_y (3 k_x^2-k_y^2	)	 \hat{\bm{z}}, k_z(k_x^2-k_y^2)   \hat{\bm{y}}+2k_x k_y k_z  \hat{\bm{x}} 	$\\  
		$B_{2u}^+$
				&$Q_{33}^{-}$&$ y (3 x^2-y^2) 	$&$  k_x (k_x^2-3 k_y^2 )  \hat{\bm{z}}, k_z(k_x^2-k_y^2)   \hat{\bm{x}}-2k_x k_y k_z  \hat{\bm{y}} 	$\\  
		$E_{1u}^+$
				&\{$Q_{11}^{+},Q_{11}^{-}$\},&\{$x ,y $\}&\multirow{1}{*}{\{$ k_y \hat{\bm{z}} +k_z \hat{\bm{y}},k_z \hat{\bm{x}}+k_x \hat{\bm{z}}$\} }\\ 
				&\{$Q_{31}^{+},Q_{31}^{-}$\}&& \{$k_y\hat{\bm{z}}-k_z\hat{\bm{y}},  k_z\hat{\bm{x}}-k_x\hat{\bm{z}}$\} \\ 
		$E_{2u}^+$
				&\{$Q_{32}^{+},Q_{32}^{-}$\}&\{$ (x^2-y^2)z , x y z$\}&\{$ k_x \hat{\bm{y}} +k_y \hat{\bm{x}}, k_x \hat{\bm{x}}- k_y \hat{\bm{y}}$\} \\ \hline\hline
		\end{tabular}
}
		\end{table}

	\begin{table}[H]
		\caption{Time-reversal odd basis functions for irreducible representations in the $D_{6h}$ point group. Basis functions in the real space, those in the momentum space, and multipole moments are listed.}
		\label{magnetic_basis_hexagonal}
		\centering
{\renewcommand\arraystretch{1.3}
		\begin{tabular}{llll}\hline\hline
		IR &$M_{lm}$&Basis in real space&Basis in momentum space	 \\  \hline	
		$A_{1g}^-$
				&$M_{76}^- $&$ z \left(  x\hat{\bm{y}}-y \hat{\bm{x}}	\right)	$& \\ 	
		$A_{2g}^-$
				&$M_{10},  $&$\hat{\bm{z}} $&\\  
				&$M_{30}  $&& \\  
		$ B_{1g}^-$
				&$M_{33}^+  $&$ (x^2-y^2)\hat{\bm{x}} -2 x y\hat{\bm{y}}  $&\multicolumn{1}{c}{($\bm{r} \rightarrow \bm{k}$)}\\ 
		$ B_{2g}^-$
				&$M_{33}^-  $&$ (x^2-y^2)\hat{\bm{y}} +2 x y\hat{\bm{x}}  $&\\
		$ E_{1g}^-$
				&\{$M_{11}^{+},M_{11}^{-}$\},&\{$\hat{\bm{x}}, \hat{\bm{y}} $\}&\\ 
				&\{$M_{31}^{+},M_{31}^{-}$\}&&  \\  
		$E_{2g}^-	$
				&\{$M_{32}^{+},M_{32}^{-}$\}&\{$ (x^2-y^2 ) \hat{\bm{z}} ,  x y \hat{\bm{z}}  $\}, \{$ z  (x \hat{\bm{x}} -y \hat{\bm{y}}  )  ,  z ( y \hat{\bm{x}} + x\hat{\bm{y}} )	$\}& \\  \hline
		$ A_{1u}^-$
				&$M_{20}, \bm{r}\cdot{s},$&$ 2 z\hat{\bm{z}} - x\hat{\bm{x}} -y\hat{\bm{y}},x\hat{\bm{x}}+y\hat{\bm{y}}+z\hat{\bm{z}} $&\multirow{1}{*}{$k_x k_y k_z (3 k_x^4 - 10 k_x^2 k_y^2+3 k_y^4)$} \\ 
				&$M_{40} $&&\\ 
		$ A_{2u}^-$
				&$T_z, M_{66}^- $&$	 x\hat{\bm{y}}-y\hat{\bm{x}}	$&\multirow{1}{*}{$k_z $} \\ 
		$ B_{1u}^-$
				&$ M_{43}^-$&$ 	y (	3 x^2-y^2)	 \hat{\bm{z}}, z(x^2-y^2)   \hat{\bm{y}}+2x y z  \hat{\bm{x}} 	$&$k_x ( k_x^2 -3 k_y^2)    $\\ 
		$ B_{2u}^-$
				&$ M_{43}^+$&$  x (x^2-3 y^2 )  \hat{\bm{z}}, z(x^2-y^2)   \hat{\bm{x}}-2x y z  \hat{\bm{y}} 	$&$ k_y (3 k_x^2-k_y^2 )$\\
		$ E_{1u}^-$
				&\{$M_{21}^{-},M_{21}^{+}$\}, \{$T_x,T_y$\},&\{$y \hat{\bm{z}} +z \hat{\bm{y}}, z \hat{\bm{x}}+x \hat{\bm{z}}  $\}, \{$y\hat{\bm{z}}-z\hat{\bm{y}}, z\hat{\bm{x}}- x\hat{\bm{z}}$\},&\{$  k_x,   k_y $\}\\
				&\{$M_{41}^{-},M_{41}^{+}$\}&&\\ 
		$ E_{2u}^-$
				&\{$M_{22}^{-},M_{22}^{+}$\},&\{$ x \hat{\bm{y}} +y \hat{\bm{x}}, x \hat{\bm{x}}- y \hat{\bm{y}} $\}&\{$k_z (k_x^2 -k_y^2 ),   k_x k_y k_z$\}\\ 
				&\{$M_{42}^{-},M_{42}^{+}$\}, \{$M_{44}^{-},M_{44}^{+}$\}&&\\ \hline \hline
		\end{tabular}
}
		\end{table}

\twocolumngrid

		\begin{table}[htbp] 
{\renewcommand\arraystretch{1.1}
		\centering
		\caption{Basic symmetry of multipole moment and ${\bm r}$, ${\bm k}$, and ${\bm M}$. }
		\label{multipole_moment_symmetry}
		\begin{tabular}{ l *{3}{>{\centering\arraybackslash}p{5mm}}} 
		\hline \hline
		&$\mathcal{P}$ & $\mathcal{T}$ & $\mathcal{PT}$  \\ \hline
          electric multipole ($l$:even) & $+$ & $+$ & $+$ \\ 
          electric multipole ($l$:odd) & $-$ & $+$ & $-$ \\ 
          magnetic multipole ($l$:even) & $-$ & $-$ & $+$ \\ 
          magnetic multipole ($l$:odd) & $+$ & $-$ & $-$ \\ 
          ${\bm r}$ & $-$ & $+$ & $-$ \\ 
          ${\bm k}$ & $-$ & $-$ & $+$ \\ 
          ${\bm M}$ & $+$ & $-$ & $-$ \\ 
		\hline \hline
		\end{tabular}
}
		\end{table}

To understand the salient difference between even-parity and odd-parity multipole ordered states, we illustrate the basic symmetry of multipole moment. The space-inversion parity $\mathcal{P}$, time-reversal parity $\mathcal{T}$, and the combined $\mathcal{PT}$ parity are shown in Table~\ref{multipole_moment_symmetry}. When the space-inversion parity is odd, the basis function must be odd-order in terms of ${\bm r}$ and ${\bm k}$. Then, to realize the same time-reversal parity, the spin dependence must be different between the real-space and the momentum-space representations, since the time-reversal parity of ${\bm r}$ and ${\bm k}$ is even and odd, respectively. This is the origin of intriguing duality and also the reason why an unusual electronic structure appears in odd-parity multipole states. As we show in the next section (Sec.~\ref{Section3}), the seemingly unusual representations in the momentum space result in emergent electromagnetic responses in the odd-parity multipole states.

\section{Emergent responses in multipole states} \label{Section3}
\begin{figure}[htbp] 
		\centering 
		\includegraphics[width=80mm,clip]{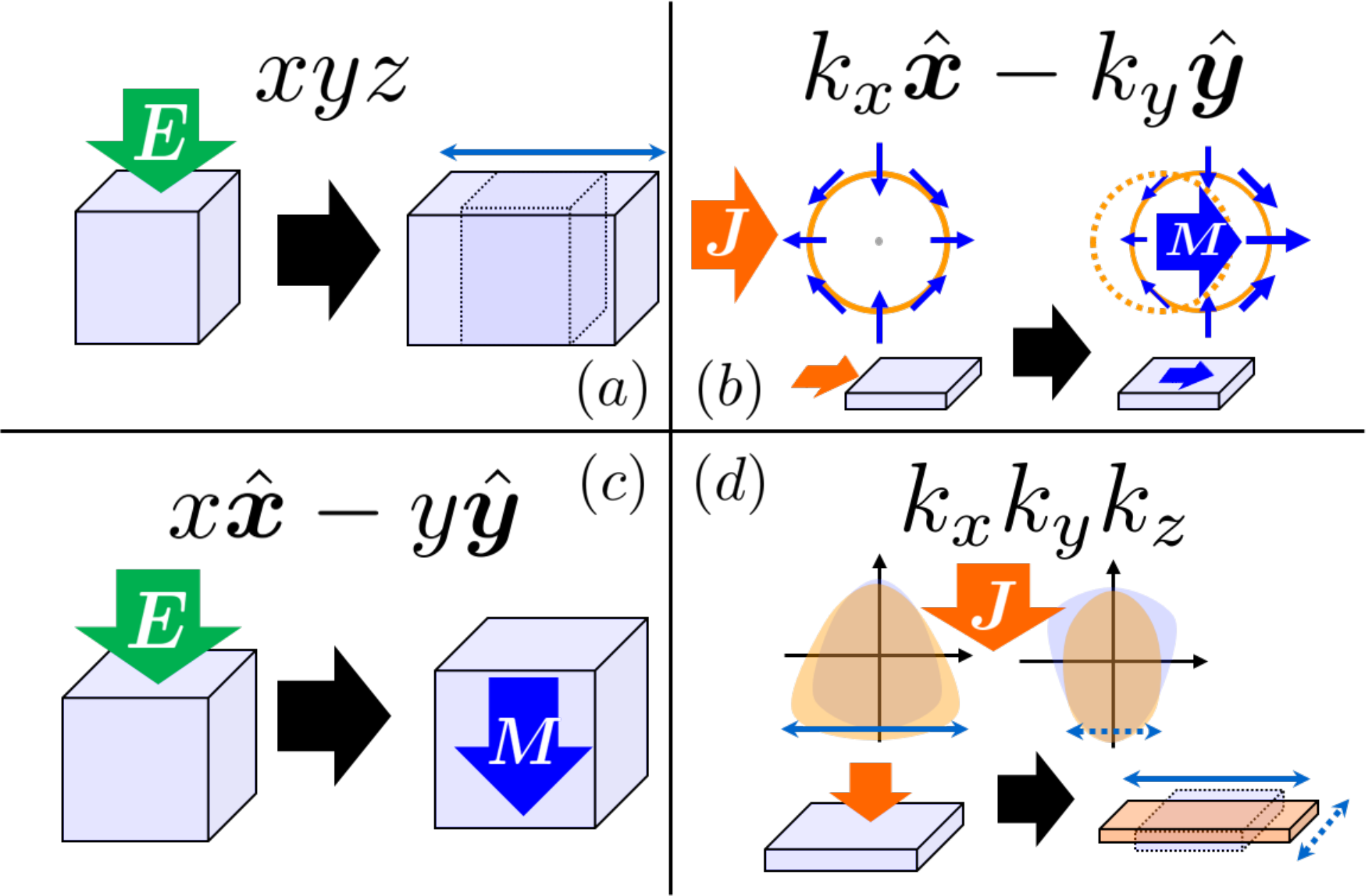} 
		\caption{Illustration of correspondence between odd-parity multipole order and cross-correlated responses. Electric multipole order gives rise to cross-correlated couplings between (a) electric fields and elasticity (piezoelectric effect), and (b) electric currents and magnetism (Edelstein effect). The magnetic counterpart induces couplings between (c) electric fields and magnetism (magnetoelectric effect), and (d) electric currents and elasticity (magnetopiezoelectric effect). Electric-field-induced phenomena [(a) and (c)] are understood by the multipole moment in the real space, while electric-current-induced phenomena [(b) and (d)] are by the corresponding representation in the momentum space.
		}
		\label{Schematic_response} 
		\end{figure}

Multipole order is accompanied by spontaneous symmetry breaking, which implies emergence of novel electromagnetic responses. For instance, cross-correlated couplings between electric, magnetic, and elastic properties have been investigated in the research field of multiferroics~\cite{Schmid2008}. Recent studies have shown large physical responses such as the anomalous Hall originating from the magnetic octupole order~\cite{Nakatsuji2015Mn3SnAnomalousHall,Ikhlas2017LargeAnomalousNernst,Higo2018largeKerr,Suzuki2017ClusterMultipole}. Similarly, an odd-parity multipole order causes a variety of emergent responses.
In this section, we demonstrate that basis functions in the classification tables~\ref{electric_basis_cubic}-\ref{magnetic_basis_hexagonal} reveal electromagnetic responses in a straightforward manner. A schematic figure illustrating the correspondence of emergent responses and multipole order is shown in Fig.~\ref{Schematic_response}, and the details are discussed below. We will obtain an intuitive and precise understanding of the emergent responses in the odd-parity multipole states. Later in Sec.~\ref{Section4_candidates}, odd-parity magnetic multipole materials are classified with use of the group theoretical analysis. Then, characteristic electromagnetic responses in candidate materials are predicted. 

\subsection{Electromagnetic cross-correlated responses} \label{Section3_magnetic_electric}

According to the classical electromagnetism in vacuum, uniform electric field and magnetic field are coupled with each other only in dynamical phenomena. The spontaneous symmetry breaking in crystals, however, allows electromagnetic coupling even in the static or stationary state. The linear responses arising from such coupling are classified into the magnetoelectric (ME) effect and the Edelstein effect. Although these two effects are seemingly similar, they are essentially different phenomena~\cite{Zelezny2017,hikaruwatanabe2017}. The following discussions clarify the nature of the ME effect and Edelstein effect from the viewpoint of multipole order. 
\\

\subsubsection{Magnetoelectric effect}

First, we discuss the ME effect, which originates from rank-2 magnetic multipole order. 
The ME effect is a static phenomenon arising from the cross-correlated coupling in the free energy
		\begin{equation}
		F_\mathrm{ME} =\sum_{\mu\nu} -\alpha_{\mu\nu} E_\mu H_{\nu}, 	
		\end{equation} 
where $\bm{E}$ and $\bm{H}$ are external electric field and magnetic field, respectively~\cite{Fiebig2005b}. Differentiating the free energy with respect to the external fields, we obtain the ME responses
		\begin{align}
		&P_{\mu} = \alpha_{\mu\nu} H_\nu, \label{ME_HtoP}\\
		&M_{\nu} = \alpha_{\mu\nu} E_\mu, \label{ME_EtoM}
		\end{align}
for the electric polarization $\bm{P}$ and magnetization $\bm{M}$. Existence of the ME effect was first pointed out by P.~Curie in 19th century~\cite{curie1894symetrie}, and later the ME effect in Cr$_2$O$_3$ was identified by theory and experiment~\cite{landau1960course,dzyaloshinskii1960magneto,astrov1960magnetoelectric,astrov1961magnetoelectric}.

The ME effect is an equilibrium phenomenon. Thus, following the Neumann's principle, the symmetry operations including the time-reversal operation restrict the form of the ME coupling tensor $\alpha_{\mu\nu}$. The ME effect at least requires inversion symmetry breaking and time-reversal symmetry breaking. The emergence of odd-parity magnetic multipole order satisfies this symmetry condition. Strictly speaking, the ME effect occurs in the odd-parity magnetic multipole state. Indeed, the ME coupling tensor $\hat{\alpha} =(\alpha_{\mu\nu})$ is decomposed as
		\begin{equation}
		\hat{\alpha} =  \frac{1}{3} \left( \rm{Tr}  \hat{\alpha} \right)  \hat{1} + \frac{1}{2}\left( \hat{\alpha} -\hat{\alpha}^T \right) 
		+ \left[ \frac{1}{2}\left( \hat{\alpha} + \hat{\alpha}^T \right) -  \frac{1}{3} \left( \rm{Tr}  \hat{\alpha} \right) \hat{1} \right],
		\end{equation}
where $\hat{1}$ and $X^T$ denote the identity matrix and the transpose of the matrix $X$, respectively. The first, second, and third terms arise from the magnetic monopole moment, toroidal moment, and quadrupole moment, respectively~\cite{Spaldin2013,Th2016}. A direct relation between the ME coupling tensor and the thermodynamically-defined magnetic multipole moment has been revealed by recent theoretical works~\cite{Gao2018spin,Shitade2018theory,gao2018theory}. These works have clarified the importance of the toroidal moments and magnetic monopole moment in crystalline materials, although both of the toroidal and monopole moments can be eliminated in Eqs.~\eqref{electric_multipole} and \eqref{magnetic_multipole}. First-principles calculations are powerful tools for a quantitative estimation of ME effects~\cite{Malashevich2012fullme,Scaramucci2012a} and magnetic multipole moments~\cite{Bultmark2009,Cricchio2010,Spaldin2013,Th2016,suzuki2018multipole,Florian2018multipole}.

Using the group-theoretical classification in Sec.~\ref{Section2}, we can identify the allowed ME effect in a straightforward manner. For example, the prototypical ME compound Cr$_2$O$_3$ crystallizes in the trigonal structure (space group No.~167, R$\bar{3}c$) and undergoes antiferromagnetic transition characterized by the $A_{1u}^-$ irreducible representation of $D_{3d}$~\cite{McGuire1956}. Following the compatibility relation, the $A_{1u}^-$ representation of $D_{3d}$ is induced to $D_{6h} \left( \supset D_{3d} \right)$ as
		\begin{equation}
		A_{1u}^- \uparrow D_{6h} =A_{1u}^- +B_{1u}^-,
		\end{equation}
where the two-fold rotation axis of $D_{3d}$ is the $x$-axis. Then, referring to Table~\ref{magnetic_basis_hexagonal}, we notice that the ME effect in Cr$_2$O$_3$ originates from the magnetic monopole moment and quadrupole moment given by
		\begin{align}
		&x\hat{\bm{x}}+y\hat{\bm{y}}+z\hat{\bm{z}}, ~~2 z\hat{\bm{z}} - x\hat{\bm{x}} -y\hat{\bm{y}}~~~\text{for $A_{1u}^-$}.
		\end{align}
The corresponding ME coupling tensor is 
			\begin{equation}
		\hat{\alpha}=
			\begin{pmatrix}
			a_1&0&0\\
			0&a_2&0\\
			0&0&a_2\\
			\end{pmatrix}
			=\begin{pmatrix}
			a&0&0\\
			0&a&0\\
			0&0&a\\
			\end{pmatrix}+
			\begin{pmatrix}
			2b&0&0\\
			0&-b&0\\
			0&0&-b\\
			\end{pmatrix},  
		\end{equation}
	where $a$ and $b$ (or $a_1$ and $a_2$) are independent of each other. The coupling constants $a$ and $b$ correspond to the magnetic monopole moment and quadrupole moment, respectively. For an intuitive understanding, an electric field $E_\mu$ induces the magnetization $M_\mu$ because the spatial coordinate takes a finite expectation value $\langle x_\mu \rangle$, and we have $x_\mu \hat{\bm{x}}_\mu \rightarrow \langle x_\mu \rangle \hat{\bm{x}}_\mu $. 

Odd-parity magnetic multipole order higher than rank-2 does not cause the ME effect. For instance, the $B_{1u}^-$ irreducible representation of $D_{6h}$ point group does not include rank-2 magnetic multipole moment in its basis. Indeed, the ME effect does not occur in the $B_{1u}^-$ magnetic hexadecapole state of hexagonal systems. On the other hand, in a tetragonal compound BaMn$_2$As$_2$, the magnetic hexadecapole order of $B_{1u}^-$ representation is admixed with the magnetic quadrupole moment~\cite{hikaruwatanabe2017}, 
		\begin{align}
		& x\hat{\bm{x}} -y\hat{\bm{y}}~~~\text{for $B_{1u}^-$ of $D_{4h}$}.
		\end{align}
Therefore, the ME effect may occur and the coupling tensor has the form,  
		\begin{equation}
		\hat{\alpha}=
		\begin{pmatrix}
			c&0&0\\
			0&-c&0\\
			0&0&0\\
		\end{pmatrix}.
		\end{equation}

Combining the classification of odd-parity magnetic multipole ordered compounds in Sec.~\ref{Section4_candidates} with the group-theoretical classification in Sec.~\ref{Section2}, 
we can identify the ME coupling tensor in the listed candidate materials. Conversely, by measuring the ME effects, we could identify the symmetry of odd-parity magnetic multipole order in the level of magnetic point group~\cite{hayami2018emergent}. Recently, such idea has been applied to identify the antiferromagnetic structure of spinel compounds~\cite{Saha2016a,Ghara2017}.

\subsubsection{Edelstein effect}

Second, we discuss the electromagnetic cross-correlated response originating from the odd-parity electric multipole. The Edelstein effect, which was theoretically clarified nearly three decades ago~\cite{Edelstein1990a}, is written as  
		\begin{equation}
			M_{\mu} = \chi_{\mu\nu} E_\nu. \label{edelstein_response}
		\end{equation}
The effect is sometimes called inverse spin-galvanic effect~\cite{Ciccarelli2015} or kinetic ME effect~\cite{levitov1985magnetoelectric}. Recently, much attention has been paid to the Edelstein effect in the research field of spintronics after the proposal of spin-orbit torque~\cite{Manchon2008,Manchon2009,Garate2009}. 
Experimental works have observed the Edelstein effect~\cite{Kato2004,Sih2005,Yang2006,Stern2006,Furukawa2017} and demonstrated switching of ferromagnetic domain by the Edelstein effect~\cite{Chernyshov2009,Miron2010,Miron2011,Ciccarelli2015}.

The response formula in Eq.~\eqref{edelstein_response} is seemingly the same as Eq.~\eqref{ME_EtoM} for the ME effect. The Edelstein effect, however, originates from the coupling between the momentum and magnetization, that is, the odd-parity magnetic multipole in the {\it momentum space} denoted by $k_\mu \hat{\bm{x}}_\nu$. This should be contrasted to the ME effect by the magnetic multipole in the real space given by $x_\mu \hat{\bm{x}}_\nu$. As we showed in Sec.~\ref{Section2_basisfunctions}, the odd-parity magnetic multipole in the momentum space corresponds to the odd-parity electric multipole in the real space. Thus, the Edelstein effect occurs in the electric multipole states with spontaneous inversion symmetry breaking.
Essential differences of the ME effect and Edelstein effect are illustrated in Table~\ref{ME_Edelstein}.

		\begin{table}[htbp] 
{\renewcommand\arraystretch{1.1}
		\centering
		\caption{Comparison of the ME effect and the Edelstein effect.}
		\label{ME_Edelstein}
		\begin{tabular}{ >{\arraybackslash}p{38mm} >{\arraybackslash}p{42mm}} 
			ME effect & Edelstein effect  \\ \hline
      Magnetic multipole state& Electric multipole state\\
      Equilibrium phenomenon & Non-equilibrium phenomenon \\
      Electric field-induced & Electric current-induced \\
      Insulators and metals & Metals \\
      Dissipationless & Dissipative \\ 
		\end{tabular}
}
		\end{table}

For instance, we here consider polar systems such as semiconductor heterostructures~\cite{Dresselhaus1992,Nitta1997}, oxide interfaces~\cite{Caviglia2010,BenShalom2010}, and bulk materials~\cite{bauer2012non}. Since the polar system is regarded as an electric dipole state, 
the Rashba coupling between the momentum $\bm{k}$ and spin $\bm{\sigma}$ appears in Hamiltonian 
		\begin{equation}
		H_{\rm Rashba}=\alpha \left(  \bm{k} \times \bm{\sigma} \right)_z =\alpha \left( k_x \sigma_y -k_y \sigma_x \right) , \label{rashba_coupling}
		\end{equation}
where we take the polar axis along the $z$-axis and $\alpha$ is the coupling constant. The non-equilibrium spin polarization along the $y$-axis ($x$-axis) is induced by the electric current along the $x$-axis ($y$-axis). We intuitively understand this effect by considering, $k_\mu \hat{\bm{x}}_\nu \rightarrow \Braket{k_\mu} \hat{\bm{x}}_\nu $, under the electric current along the $\mu$-axis. Thus, the Edelstein effect is a current-induced phenomenon while the ME effect is induced by the electric field. 
Due to the symmetry of the Rashba coupling~\eqref{rashba_coupling}, the response tensor $\hat{\chi} =(\chi_{\mu\nu})$ is given by
		\begin{equation}
         \label{Rashba-Edelstein}
		\hat{\chi}=
			\begin{pmatrix}
			0&a&0\\
			-a&0&0\\
			0&0&0\\
			\end{pmatrix}.
		\end{equation}

It is noteworthy that the time-reversal symmetry breaking is not required for the Edelstein effect while it is required for the ME effect. This is because the Edelstein effect occurs not in an equilibrium state, but in a stationary state. The essential ingredient is the broken inversion symmetry, that is, emergence of odd-parity electric multipole moment. In other words, the time-reversal symmetry breaking for the Edelstein effect arises from dissipations due to an applied electric current. To express these properties, we should rewrite the response formula as
		\begin{equation}
			M_{\mu} = \chi'_{\mu\nu} j_\nu, \label{edelstein_response_current}
		\end{equation}
instead of Eq.~\eqref{edelstein_response}, where $j_\nu$ is an electric current. As the formula implies, the Edelstein effect is a transport phenomenon in itinerant systems. This is in sharp contrast to the ME effect which is an equilibrium phenomenon in insulating systems.

	The Edelstein effect allowed in odd-parity multipole states can be identified by the classification tables in Sec.~\ref{Section2_basisfunctions}. Let us focus on cubic systems in Table~\ref{electric_multipole_cubic}. The Rashba term corresponds to the basis function of $T_{1u}^+$ representation in the momentum space. The basis in the real space, $x,~y$, and $z$, corresponds to the ferroelectric order. This correspondence reveals the Rashba-Edelstein effect [Eq.~\eqref{Rashba-Edelstein}] in polar ferroelectric-like systems, consistent with our knowledge and experimental observations~\cite{Yang2006,Miron2010,Miron2011}. 

 On the other hand, the rank-9 electric multipole order in the $A_{1u}^+$ representation breaks all the mirror symmetry (that realizes a chiral system). Then, the representation in the momentum space is magnetic monopole $\bm{k}\cdot \bm{\sigma}$, which indicates ``hedgehog'' spin-momentum locking~\cite{Fu2015}. Accordingly, the Edelstein coupling tensor is a scholar matrix, $\hat{\chi} = a \hat{1}$. Our classification of the Edelstein effect based on the representation theory is consistent with the symmetry analysis~\cite{Ciccarelli2015}.

Finally, we apply the classification of Edelstein effect to a candidate for an odd-parity electric multipole metal Cd$_2$Re$_2$O$_7$.
Cd$_2$Re$_2$O$_7$ is a superconductor unique in $\alpha$-pyrochlore oxides, crystallizing in the cubic structure (space group F$d\bar{3}m$, No.~227)~\cite{Hanawa2001,Sakai2001}. What is controversial about this compound is the structural transition at $T \sim \mr{200}{K}$ (for a brief review, see Ref.~\cite{Hiroi2018}). Several measurements including x-ray diffraction~\cite{Yamaura2002,Castellan2002,Yamaura2017a}, Raman scattering~\cite{Kendziora2005}, and non-linear optical measurement~\cite{Petersen2006b} have reported the structural phase transition characterized by the $E_u^+ \left(p \right)$ mode. The corresponding structural change is
		\begin{equation}
			\mathrm{F}d\bar{3}m~\left(\textrm{No.}~227 \right) \rightarrow \mathrm{I}\bar{4}m2~\left(\textrm{No.}~119 \right). \label{cd2re2o7_d2d_scenario}
		\end{equation}
However, a recent non-linear optical measurement by Harter \textit{et al.}~\cite{Harter2017paritybreaking} proposes that the primary order is $T_{2u}^+ \left(\gamma \right) + T_{1g}^+ \left(\hat{\bm{z}} \right)$ mode and the $E_u^+ \left(p \right)$ mode is induced as a secondary order. The corresponding structural change is written by
		\begin{equation}
			\mathrm{F}d\bar{3}m~\left(\textrm{No.}~227 \right) \rightarrow \mathrm{I}\bar{4}~\left(\textrm{No.}~82 \right). \label{cd2re2o7_s4_scenario}
		\end{equation}
Inspired by this experiment, Matteo and Norman theoretically proposed several scenarios indicating an exotic magnetic order~\cite{DiMatteo2017b}. 

Here, we compare the Edelstein effect in the two scenarios, ~\eqref{cd2re2o7_d2d_scenario} and \eqref{cd2re2o7_s4_scenario}. 
When we follow the $E_u^+ \left(p \right)$ scenario~\eqref{cd2re2o7_d2d_scenario}, the multipole order parameter is an electric dotriacontapole moment, $Q_{52}^+ = xyz(2z^2-x^2-y^2)$, which gives rise to the magnetic quadrupole moment in the momentum space, $k_x \sigma_x- k_y \sigma_y$. This spin-momentum coupling induces the longitudinal Edelstein response described by the coupling tensor
		\begin{equation}
		\hat{\chi}=
			\begin{pmatrix}
			a_{E_u}&0&0\\
			0&-a_{E_u}&0\\
			0&0&0\\
			\end{pmatrix}. \label{cd2re2o7_scenario1_Edelstein}
		\end{equation}
On the other hand, in the $T_{2u}^+ \left(\gamma \right) + T_{1g}^+ \left(\hat{\bm{z}} \right)$ scenario~\eqref{cd2re2o7_s4_scenario}, one of the primary modes $T_{2u}^+ \left(\gamma \right)$ (electric octupole order) comprises rank-2 magnetic multipole in the momentum space representation. The basis function $k_x \sigma_y+ k_y \sigma_x$ indicates the Edelstein coupling tensor
		\begin{equation}
		\hat{\chi}=
			\begin{pmatrix}
			a_{E_u}&a_{T_{2u}}&0\\
			a_{T_{2u}}&-a_{E_u}&0\\
			0&0&0\\
			\end{pmatrix}.\label{cd2re2o7_scenario2_Edelstein}
		\end{equation}
It is expected that $a_{T_{2u}} \gg a_{E_u}$ when the $T_{2u}^+ \left(\gamma \right)$ mode is a primary order. As shown by the coupling tensors~\eqref{cd2re2o7_scenario1_Edelstein} and \eqref{cd2re2o7_scenario2_Edelstein}, the two scenarios can be distinguished by the off-diagonal Edelstein effect; when the electric current is applied along the $x$-axis, the induced magnetization is  
		\begin{equation}
		\bm{M} = \left( a_{E_u} j_x,0,0  \right),
		\end{equation}
	for the $E_u^+ \left(p \right)$ scenario, whereas 
		\begin{equation}
		\bm{M} = \left( a_{E_u} j_x,a_{T_{2u}} j_x,0  \right),
		\end{equation}
	for the $T_{2u}^+ \left(\gamma \right) + T_{1g}^+ \left(\hat{\bm{z}} \right)$ scenario. Symmetry of multipole order in Cd$_2$Re$_2$O$_7$ and other candidate materials may be identified by probing magnetization under electric currents.

\subsection{Magnetopiezoelectric effect} \label{Section3_magnetopiezoelectric}

Now we clarify a non-equilibrium response of the odd-parity magnetic multipole states. 
As we show in Sec.~\ref{Section2_basisfunctions}, odd-parity magnetic multipole order is represented by the spin-independent basis functions in the momentum space, giving rise to asymmetric band distortions. Due to the peculiar symmetry of the band structures, an applied electric current may induce ``electronic strain'' $\epsilon_{\mu\nu}^{(\rm{e})}$. For instance, in the $B_{1u}^-$ state of tetragonal systems, the $k_x k_y k_z$-type asymmetry leads to the $k_x k_y$-type strain $\epsilon_{xy}^{(\rm{e})}$ under the electric current along the $z$-axis. In other words, the in-plane nematicity is induced by the out-of-plane electric current~\cite{hikaruwatanabe2017}. This response is intuitively understood by $k_x k_y k_z \rightarrow k_x k_y \langle k_z \rangle$. Since a lattice strain tensor $\epsilon_{\mu\nu}$ has the same symmetry as $\epsilon_{\mu\nu}^{(\rm{e})}$, the electric current is coupled with lattice distortions through electronic strains. 
The response is generally represented by 
		\begin{equation}
		\epsilon_{\mu\nu} = \tilde{e}_{\mu\nu \lambda} \, j_\lambda,
         \label{MPE}
		\end{equation} 
where the rank-3 tensor $\tilde{e}_{\mu\nu \lambda}$ characterizes the current-induced strain response~\cite{hikaruwatanabe2017}. 
The response resembles the well-known piezoelectricity 
		\begin{equation}
		\epsilon_{\mu\nu} = e_{\mu\nu \lambda} E_\lambda,
		\end{equation}  
which represents lattice strains induced by electric fields. For this reason, the current-induced strain described by Eq.~\eqref{MPE} is called magnetopiezoelectric (MPE) effect.

The MPE effect and piezoelectric effect are essentially different phenomena, similar to the case of the ME effect and Edelstein effect. First, the MPE effect is a non-equilibrium response, whereas the piezoelectric effect is an equilibrium response. Accordingly, the coupling tensors $\tilde{e}_{\mu\nu \lambda}$ and $e_{\mu\nu \lambda}$ have a different time-reversal parity; the time-reversal parity is even for $e_{\mu\nu \lambda}$, while it is odd for $\tilde{e}_{\mu\nu \lambda}$. 
Second, the MPE effect is caused by odd-parity magnetic multipole order, while the usual piezoelectricity occurs in odd-parity electric multipole states, as schematically illustrated in Fig.~\ref{Schematic_response}. 
For instance, the piezoelectric effect with $e_{xyz}$ occurs in the electric octupole state because $xyz \rightarrow xy \langle z \rangle$ under the electric field ${\bm E} \parallel \hat{z}$.
Third, the piezoelectricity is allowed in insulators and metals, whereas the MPE effect is realized only in metallic states and accompanied by dissipations. Thus, the MPE effect is a non-equilibrium phenomenon.

Here, we show the group-theoretical classification of the MPE effect. Table~\ref{magnetopiezo_classification} shows the allowed MPE effect determined by the crystal symmetry and the representation of magnetic multipole order. For instance, the seemingly antiferromagnetic states in some tetragonal Mn pnictides such as Ba$_{1-x}$K$_x$Mn$_2$As$_2$ are classified into the $B_{1u}^-$ representation of $D_{4h}$. Then, the MPE couplings 
		\begin{align}
		\tilde{e}_{xyz} \ne 0, \,\,\, \tilde{e}_{yzx}=\tilde{e}_{zxy} \ne 0,
		\end{align} 
indicate the current-induced nematicity~\cite{hikaruwatanabe2017}. On the other hand, GdB$_4$ undergoes magnetic transition with $A_{1u}^-$ symmetry, and then, a finite MPE coupling 
\begin{align}
\tilde{e}_{yzx}=-\tilde{e}_{zxy} \ne 0,
\end{align}
is allowed. 

Table~\ref{magnetopiezo_classification} reveals that the MPE effect is realizable in all odd-parity magnetic multipole ordered states except for the case of magnetic point group $m'\bar{3}'m'$. Therefore, most of itinerant odd-parity magnetic multipole materials potentially show the MPE effect. We can identify the symmetry of the MPE effect in candidate materials (Sec.~\ref{Section4_candidates}) by combining Tables~\ref{electric_basis_cubic}-\ref{magnetic_basis_hexagonal}, \ref{magnetopiezo_classification}, and \ref{magnetic_candidates}. 

One of the advantages is that the MPE effect is switchable by inverting the magnetic domain. In fact, the switching of domains was recently achieved in several antiferromagnetic metals~\cite{Wadley2016domain,Bodnar2018domain}. The MPE effect may be applicable in the field of the antiferromagnetic spintronics~\cite{Jungwirth2016AntiferromagneticReview,Baltz2018Antiferromagnetic,Manchon2018currentinduced}. Furthermore, the MPE materials may be useful for metal piezoelectric devices~\cite{shiomi2018AgCrSe2Piezo,shiomi2018magnetopiezo}.

\newpage

\onecolumngrid

			\begin{table}[htbp]
			\caption{List of the MPE coupling in cubic ($O_h$), tetragonal ($D_{4h}$), and hexagonal ($D_{6h}$) systems. In each odd-parity irreducible representation, the corresponding magnetic multipole moment and  symmetry-allowed MPE couplings are shown. The MPE couplings are represented by strains $\epsilon_{\mu\nu}$ and electric currents $j_\mu$.}
			\label{magnetopiezo_classification}\centering
{\renewcommand\arraystretch{1.4}
			\begin{tabular}{cccc}\hline \hline
			$\bm{G}$&IR &Multipole& Magnetopiezoelectric coupling	 \\ \hline
			$O_{h}$&$ A_{1u}^-$		&$\bm{r}\cdot \bm{s}, M_{40}, M_{44}^+, $& \\ 
			&$ A_{2u}^-$			&$M_{62}^+,M_{66}^+$&$j_x \epsilon_{yz}+j_y \epsilon_{zx}+j_z \epsilon_{xy}$ \\ 
			&$ E_{u}^-$	&$M_{20},M_{22}^+,M_{40},M_{42}^+ ,M_{44}^+ $&$\left\{ 2j_z \epsilon_{xy} -j_x \epsilon_{yz} -j_y \epsilon_{zx}, j_y \epsilon_{zx} -j_x \epsilon_{yz} \right\} $ \\ 
			&$ T_{1u}^-$	&$\bm{T}$&$\left\{j_x\epsilon_{xx},j_y\epsilon_{yy},j_z\epsilon_{zz} \right\},$ \\ 
			&&$M_{41}^\pm,M_{43}^\pm, M_{44}^-$&$\left\{j_x \left(\epsilon_{yy}+\epsilon_{zz}\right) ,j_y \left(\epsilon_{zz}+\epsilon_{xx}\right),j_z \left(\epsilon_{xx}+\epsilon_{yy}\right) \right\},$ \\ 
			&&&$\left\{j_y\epsilon_{xy}+j_z\epsilon_{zx},j_z\epsilon_{yz}+j_x\epsilon_{xy},j_x\epsilon_{zx}+j_y\epsilon_{yz} \right\}$ \\ 
			&$ T_{2u}^-$	&$ M_{21}^\pm ,M_{22}^-  $&$\left\{j_x \left( \epsilon_{yy}-\epsilon_{zz} \right),j_y \left( \epsilon_{zz}-\epsilon_{yy} \right),j_z \left( \epsilon_{xx}-\epsilon_{yy} \right) \right\},   $ \\ 
			&&$ M_{41}^\pm ,M_{42}^-, M_{43}^\pm  $&$\left\{j_y\epsilon_{xy}- j_z\epsilon_{zx},j_z\epsilon_{yz}- j_x\epsilon_{xy}, j_x\epsilon_{zx}- j_y\epsilon_{yz} \right\}   $ \\ \hline
			$D_{4h}$&$ A_{1u}^-$		&$\bm{r}\cdot \bm{s}, M_{20}, M_{40}, M_{44}^+, $&$j_x \epsilon_{yz}-j_y\epsilon_{zx} $	 \\ 
			&$ A_{2u}^-$			&$T_z,M_{44}^-$&$j_z \left(\epsilon_{xx}+\epsilon_{yy}\right), j_z \epsilon_{zz},j_y \epsilon_{yz}+j_x\epsilon_{zx}  $ \\ 
			&$ B_{1u}^-$	&$M_{22}^+,M_{42}^+ $&$j_z \epsilon_{xy},j_x \epsilon_{yz}+j_y\epsilon_{zx}   $ \\ 
			&$ B_{2u}^-$	&$M_{22}^-,M_{42}^- $&$j_z \left( \epsilon_{xx}-\epsilon_{yy}\right),j_y \epsilon_{yz}-j_x\epsilon_{zx}   $ \\ 
			&$ E_{u}^-$	&$M_{21}^\pm,T_{x},T_y $&$\left\{ j_z\epsilon_{zx},j_z\epsilon_{yz} \right\},   $ \\ 
			&			&$M_{41}^\pm,M_{43}^\pm $&$\left\{ j_x\epsilon_{zz},j_y\epsilon_{zz} \right\},\left\{ j_x \epsilon_{xx},j_y\epsilon_{xx}\right\},   $ \\ 
			&			&&$\left\{ j_y\epsilon_{xy},j_x\epsilon_{xy} \right\},\left\{ j_x \epsilon_{yy},j_y \epsilon_{yy} \right\}   $ \\ \hline 
			$D_{6h}$&$ A_{1u}^-$		&$\bm{r}\cdot \bm{s}, M_{20}, M_{40}$&$j_x \epsilon_{yz}-j_y\epsilon_{zx} $	 \\ 
			&$ A_{2u}^-$			&$T_z$&$j_z \left(\epsilon_{xx}+\epsilon_{yy}\right), j_z \epsilon_{zz},j_y \epsilon_{yz}+j_x\epsilon_{zx}  $ \\ 
			&$ B_{1u}^-$	&$M_{43}^- $&$2j_y\epsilon_{xy}- j_x \left(\epsilon_{xx}-\epsilon_{yy}\right)$ \\ 
			&$ B_{2u}^-$	&$M_{43}^+ $&$2j_x\epsilon_{xy}+ j_y \left(\epsilon_{xx}-\epsilon_{yy}\right)$ \\ 
			&$ E_{1u}^-$	&$M_{21}^\pm,T_{x},T_y $&$\left\{ j_x\epsilon_{zz},j_y\epsilon_{zz} \right\},  $ \\ 
			&			&$M_{41}^\pm $&$\left\{ 2j_y \epsilon_{xy} + j_x \left(3 \epsilon_{xx}+\epsilon_{yy}\right),2j_x \epsilon_{xy} + j_y \left( \epsilon_{xx}+3\epsilon_{yy}\right)\right\},   $ \\ 
			&			&&$\left\{ 2j_y \epsilon_{xy} - j_x \left( \epsilon_{xx}+3\epsilon_{yy}\right),2j_x \epsilon_{xy} - j_y \left(3 \epsilon_{xx}+\epsilon_{yy}\right)\right\},   $ \\ 
			&			&&$\left\{ j_z \epsilon_{zx}, j_z \epsilon_{yz} \right\}  $ \\ 
			&$ E_{2u}^-$	&$M_{22}^\pm $&$\left\{ j_z \left(\epsilon_{xx}-\epsilon_{yy}\right), j_z\epsilon_{xy}  \right\},  $ \\ 
			&			&$M_{42}^\pm, M_{44}^\pm$&$\left\{ j_x\epsilon_{zx}-j_y\epsilon_{yz}, j_y\epsilon_{zx}+j_x\epsilon_{yz} \right\}   $ \\ \hline \hline
			\end{tabular}
}
	 		\end{table}	

\twocolumngrid

\subsection{Dichromatic electron transport} \label{Section3_dichromatic}

Finally, we discuss a dichromatic transport of electrons. Within the linear response regime, the electric longitudinal conductivity is bidirectional in the sense that the conductivity is not altered by reversing either the electric fields $\bm{E}$ or magnetic fields $\bm{H}$. The bidirectional property on $\bm{H}$ is derived from the reciprocal relation~\cite{landau1980statistical}.
The bidirectional property, however, may be violated in the nonlinear regime. Here, we restrict discussions to the dichromatic longitudinal conductivity, 
		\begin{equation}
		\Delta \sigma_{\mu\mu} \equiv  \tilde{\sigma}_{\mu\mu} ( +j_\mu) - \tilde{\sigma}_{\mu\mu} ( -j_\mu), 
		\end{equation}
where $\tilde{\sigma}_{\mu\nu}(j_\mu) $ is the electric conductivity including a nonlinear component under the current $j_\mu$. Hence, $\Delta \sigma_{\mu\mu}$ depends on the electric fields. $\Delta \sigma_{\mu\mu}$ depends on $E_\mu$ or even higher-order in the electric field. 
For example, the dichromatic nonlinear electronic conductivity $\Delta \sigma_{zz}$ is allowed in polar systems with $z$ being the polar axis. 

Rikken \textit{et al.} have shown that the electronic dichroism can be tuned by magnetic fields, while the dichroism determined by the crystal polarity is not tunable.
Two setups have been studied. (i) Chiral systems such as a twisted bismuth substrate show magnetochiral anisotropy~\cite{Rikken2001}, and (ii) the cross product of external magnetic field and electric field induces magnetoelectric anisotropy~\cite{Rikken2005}. The observed dichromatic responses are 

		\begin{equation}
			\Delta \sigma_{\mu\mu} \propto a_\mathrm{\chi} j_\mu H_\mu, \label{mangetochiral_anisotropy}
		\end{equation}
for the case (i), and 
		\begin{equation}
			\Delta \sigma_{\mu\mu} \propto j_\mu \left(  \bm{E} \times \bm{H} \right)_\mu, \label{magnetoelectric_anisotropy}
		\end{equation}
for the case (ii). The coefficient $a_\mathrm{\chi}$ in Eq.~\eqref{mangetochiral_anisotropy} denotes chirality of systems.

Although the two dichromatic responses are seemingly different from each other, we may understand them in a unified way. The presence of polar axis in the \textit{momentum space} is associated with the magnetically-induced dichroism. To confirm this interpretation, we consider cubic systems as examples, and refer to the classification table of electric multipole moment (Table~\ref{electric_basis_cubic}). 

For the case (i), chiral systems are regarded as an electric multipole state in the $A_{1u}^+$ irreducible representation. Then, the hedgehog spin-momentum coupling, $\bm{k}\cdot\bm{\sigma}$, gives rise to spin-split bands. In an external magnetic field a polar axis in the momentum space emerges parallel to the field as schematically understood by 
		\begin{equation}
		\bm{k}\cdot \bm{\sigma} \xrightarrow{\bm{H}} \bm{k}\cdot \Braket{\bm{\sigma}}. 
		\end{equation}  
The anti-symmetric band dispersion gives rise to the electronic dichroism, because the nonlinear electric conductivity is determined by the anti-symmetric components of the energy spectrum in the standard Boltzmann transport theory~\cite{Ideue2017a,Yasuda2016a}. 
The direction of the momentum-space polarization is consistent with the magnetochiral anisotropy in Eq.~\eqref{mangetochiral_anisotropy}.

The case (ii) is regarded as a ferroelectric-like polar system because the electric dipole along $\bm{E}$ is induced by external electric fields. Indeed, the magnetoelectric anisotropy has recently been observed in a polar semiconductor BiTeBr~\cite{Ideue2017a} at zero electric field. These cases correspond to the $T_{1u}^+$ irreducible representation in cubic systems, and its momentum-space basis is the Rashba type spin-momentum coupling, $k_x \sigma_y- k_y \sigma_x$, for ${\bm E} \parallel \hat{z}$ [$T_{1u}^+ \left(\gamma \right)$ basis]. External magnetic fields lead to the band distortion 
		\begin{equation}
			k_x \sigma_y- k_y \sigma_x \xrightarrow{\bm{H}\parallel \hat{x}} -k_y \Braket{\sigma_x}, 
		\end{equation}
for $\bm{H} \parallel \hat{x}$, while 
		\begin{equation}
			k_x \sigma_y- k_y \sigma_x \xrightarrow{\bm{H}\parallel \hat{y}} k_x \Braket{\sigma_y}, 
		\end{equation}
for $\bm{H} \parallel \hat{y}$. Thus, the polar axis in the momentum space shows up. Then, 
the nonlinear electric current shows the magnetoelectric anisotropy in Eq.~\eqref{magnetoelectric_anisotropy}. The magnetically-induced dichroism due to the polarization in the momentum space has been microscopically clarified
by a theoretical calculation in Ref.~\onlinecite{Ideue2017a}. Furthermore, the optical dichroism has been revealed by recent theories~\cite{Norman2015,Shibata2016,Kawaguchi2016a}.

As demonstrated above, the classification tables in Sec.~\ref{Section2} provide a systematic understanding of the magnetically-induced dichroism in the longitudinal transport. Here we again discuss Cd$_2$Re$_2$O$_7$. In the $E_u^+ \left(p \right)$ scenario~\eqref{cd2re2o7_d2d_scenario}, the emergent spin-momentum coupling has been obtained as $k_x \sigma_x - k_y \sigma_y$. Then, the dichroism in the electric conductivity is obtained as
		\begin{equation}
			\left( \Delta \sigma_{xx}, \Delta \sigma_{yy} \right) = \kappa_{E_u} \left( j_x H_x, - j_y H_y \right). \label{cd2re2o7_d2d_scenario_dichroism}
		\end{equation}
Since the low-temperature structure I$\bar{4}m2$ is neither chiral nor polar, the dichromatic transport is different from magnetochiral anisotropy and magnetoelectric anisotropy. 

On the other hand, in the $T_{2u}^+ \left(\gamma \right)+T_{1g}^+ \left(\hat{\bm{z}} \right)$ scenario~\eqref{cd2re2o7_s4_scenario}, the spin-momentum coupling comprises the primarily component $k_x \sigma_y+ k_y \sigma_x$ and the secondarily component $k_x \sigma_x - k_y \sigma_y$. The primary component leads to the dichromatic response 
		\begin{equation}
			\left( \Delta \sigma_{xx}, \Delta \sigma_{yy} \right) = \kappa_{T_{2u}} \left( j_x H_y,  j_y H_x \right),
		\end{equation}
in addition to Eq.~\eqref{cd2re2o7_d2d_scenario_dichroism}. Thus, under the rotating magnetic field in the $xy$-plane, $\bm{H} =|\bm{H}| (\cos \theta, \sin \theta, 0)$, and the electric current along the $x$-axis, the electronic dichromatic response shows the field-angle dependence,  
		\begin{equation}
			\Delta \sigma_{xx} = j_x |\bm{H}| \left( \kappa_{E_u} \cos{\theta} + \kappa_{T_{2u}}  \sin{\theta} \right), \label{cd2re2o7_s4_scenario_dichroism}
		\end{equation}
from which we can evaluate $\kappa_{E_u}$ and $\kappa_{T_{2u}}$, and we may distinguish the two scenarios. When the $T_{2u}^+ \left(\gamma \right)+T_{1g}^+ \left(\hat{\bm{z}} \right)$ scenario~\eqref{cd2re2o7_s4_scenario} is correct, $\kappa_{E_u}$ and  $\kappa_{T_{2u}}$ should follow the same temperature dependence as that observed in the non-linear optical measurement by Harter \textit{et al.}~\cite{Harter2017paritybreaking} on the assumption that $\kappa_{E_u}$ and $\kappa_{T_{2u}}$ linearly depend on the corresponding order parameters.

So far we discussed magnetically-induced dichroism in electron transport. According to the above explanation of the dichroism, odd-parity magnetic multipole ordered materials may show the dichromatic transport even at zero magnetic field. The classification tables in Tables~\ref{magnetic_basis_cubic}, \ref{magnetic_basis_tetragonal}, and \ref{magnetic_basis_hexagonal} indicate polarization in the momentum space for some magnetic multipole states. In such multipole states, the dichromatic response is allowed by symmetry. This topic will be discussed elsewhere~\cite{DomainSwitch}.

In some cases, the dichromatic transport arises from more complicated asymmetric band distortions. For instance, transition metal dichalcogenide systems show the third order anti-symmetric dispersion written as $k_y \left( 3 k_x^2 -k_y^2 \right)$ under the out-of-plane magnetic field~\cite{Saito2016,Wakatsuki2017,Hoshino2018Nonreciprocal}. The normal state of transition metal dichalcogenides, however, shows a negligible dichromatic conductivity, although the effect of a superconducting fluctuation gives rise to strongly enhanced $\Delta \sigma_{\mu\mu}$~\cite{Wakatsuki2017,Hoshino2018Nonreciprocal}. Theoretical calculations showed that the contributions of spin and valley degrees of freedom are almost canceled out in the normal state~\cite{Wakatsuki2017}. It is therefore expected that such higher-order anti-symmetric distortions give a weak dichroism.

\section{Odd-parity magnetic multipole materials} \label{Section4_candidates}
 In this section, we show candidate materials of the odd-parity multipole order. For the electric multipole, parity-violating structural transitions may be regarded as odd-parity electric multipole order. Rank-1 ferroelectric order has been observed in various insulating compounds~\cite{rabe2007ferroelectric_book}. Recent studies of LiOsO$_3$~\cite{Shi2013} and carrier-doped SrTiO$_3$~\cite{Rischau2017} have explored the concept of ferroelectric-like order in metal and superconductor. 
Higher-order octupole or dotriacontapole order in Cd$_2$Re$_2$O$_7$~\cite{Hiroi2018} has been discussed in previous sections. The electric octupole order with an electronic origin has also been proposed for the hidden ordered phase of bilayer high-$T_{\rm c}$ cuprate superconductor YBa$_2$Cu$_3$O$_y$~\footnote{T. Hitomi and Y. Yanase, Submitted.}. 

The magnetic multipole order is attributed to an electronic degree of freedom. However, spin is an axial vector and space inversion parity is even. Therefore, another electronic degree of freedom has to play an essential role for the parity-violating magnetic multipole order. Recent studies pointed out the sublattice degree of freedom in locally noncentrosymmetric crystals, which allows the formation of odd-parity augmented multipole~\cite{Hitomi2014,Hitomi2016,Sumita2016,Spaldin2013,Th2016,Yanase2014,Sumita2017,Fechner2016,DiMatteo2017b,Hayami2014b,Ederer2009,Spaldin2008,Hayami2016,hikaruwatanabe2017}.  
In particular, as seen in the prototypical ME compound Cr$_2$O$_3$~\cite{dzyaloshinskii1960magneto,astrov1960magnetoelectric,astrov1961magnetoelectric}, an intra-unit-cell antiferromagnetic order in locally noncentrosymmetric crystals may be identified as an odd-parity magnetic multipole order from the viewpoint of symmetry. Accordingly, magnetic compounds realizing such spin structure may be a platform of emergent responses discussed in Sec.~\ref{Section3}. 

In this section we make a list of candidate materials for ferroic odd-parity magnetic multipole states ($\bm{Q}=\bm{0}$). Identification of multipole order is performed by two complementary methods, the magnetic representation theory and the magnetic point group analysis.

The magnetic representation theory is a standard method for magnetic structure analysis~\cite{Bertaut1968,izyumov1991neutron}. By using the method, we classify magnetic compounds into the irreducible representations of a given crystal point group. Corresponding to the odd-parity representation, the magnetic order can be identified as an odd-parity magnetic multipole order (listed in Sec.~\ref{Section2_multipole}). This method is especially useful for specifying characteristic transport properties of itinerant multipole states. Using the tables of basis functions (Sec.~\ref{Section2_basisfunctions}), we can predict some electromagnetic responses (Sec.~\ref{Section3}) in candidate materials. 

In the magnetic point group analysis, ordered phases are characterized by point group symmetry of the magnetic states. The magnetic point group gives a symmetry constraint for the equilibrium properties and transport phenomena more directly than the magnetic representation theory~\cite{birss1964symmetry,nye1985physical,Janovec2010,Litvin2014}. We can see what electromagnetic responses should occur in the magnetic states. Thus, these two approaches are complementary. 

For more information, we introduce Aizu species, which consist of a pair of symmetry of the disordered phase and ordered phase. Symmetry reduction in the magnetic phase transition is generally classified by Aizu species. The domain states are also characterized; for example, we can identify external fields suitable for switching domain states by Aizu species~\cite{Janovec2010,Litvin2014}. The concept of Aizu species is introduced in Appendix~\ref{app_aizu_species}.

Table~\ref{magnetic_candidates} shows the list of materials which are identified as ferroic odd-parity magnetic multipole states. 117 magnetic compounds are shown with their space group, Aizu species, irreducible representation, conducting properties (metal/insulator/semiconductor), N\'eel temperatures, and references. A part of the compounds has been known as ME materials~\cite{Schmid1973,Siratori1992a,Gallego2016a}, and Table~\ref{magnetic_candidates} also shows which compounds have been already identified. However, more than half of the compounds in Table~\ref{magnetic_candidates} are elucidated in this work. 
For some compounds (\textit{e.g.} CaMn$_2$Sb$_2$), experimental reports of magnetic structures or transition temperatures are contradictory, and for other compounds  (\textit{e.g.} Co$_3$O$_4$), several possibilities of magnetic structures were suggested. In such cases, we show all the proposed magnetic structures as far as possible.

We restrict our classification to the magnetic phases preserving $\mathcal{PT}$ symmetry. The $\mathcal{PT}$ symmetry is broken in the even-parity magnetic multipole phases ($\mathcal{P}$-even and $\mathcal{T}$-odd), whereas it is preserved in the odd-parity magnetic multipole phases ($\mathcal{P}$-odd and $\mathcal{T}$-odd) [see Table~\ref{multipole_moment_symmetry}]. Thus, the odd-parity (even-parity) magnetic multipole ordered states are regarded as the $\mathcal{PT}$-preserved ($\mathcal{PT}$-broken) magnetic phases. Table~\ref{magnetic_candidates} shows the \textit{pure odd-parity magnetic multipole states}.  
This constraint requires that crystal structures are centrosymmetric. Otherwise, owing to the parity violation in the paramagnetic phase, an odd-parity magnetic multipole moment should coexist with an even-parity one. In other words, an odd-parity magnetic multipole moment appears in all the magnetic phases of noncentrosymmetric crystals. An analysis of such parity-mixed magnetic states is an important future study.

\onecolumngrid

	\begin{longtable}[H]{llllclll} 
	\caption{Odd-parity magnetic multipole materials. The table lists compounds, space group, symmetry of magnetic structure denoted by Aizu species and irreducible representations ($\Gamma_\mathrm{mag}$), conducting properties (M/S/I = metal, semiconductor, and insulator), N\'eel temperatures ($T_\mathrm{N}$), and references (Ref.). In the column ``Rev.'', the symbols Sc, Si, and H specify compounds which have been clarified as parity-violating magnetic materials in the review articles by Schmid~\cite{Schmid1973}, Siratori \textit{et al.}~\cite{Siratori1992a}, and Gallego \textit{et al.}~\cite{Gallego2016a}, respectively. The numbers of the space groups and the Aizu species follow Refs.~\cite{internatinaltables} and~\cite{Litvin2008}, respectively. The blank part has not been clarified to the best of our knowledge.}
	\label{magnetic_candidates}\\
		Compounds &Space group&Aizu species&$\Gamma_\mathrm{mag}$&M/S/I&$T_\mathrm{N}$&Ref.&Rev. \\ \hline \hline
	\multicolumn{8}{l}{\textbf{Cubic systems (7)}}\\
	$\rm RbFeO_2	$&F$d\bar{3}m$ (227)&$\aizu{m\bar{3}m1'}{4'/m'm'm\Braket{xd}}$ (740)&$T_{2u}$&&$737<T<1027$&\cite{Sheptyakov2010}&\\
	$\rm CsFeO_2	$&F$d\bar{3}m$ (227)&$\aizu{m\bar{3}m1'}{4'/m'm'm\Braket{xd}}$ (740)&$T_{2u}$&&$350<T<1055$&\cite{Sheptyakov2010}&\\
	$\rm CoAl_2O_4	$&F$d\bar{3}m$ (227)&$\aizu{m\bar{3}m1'}{4'/m'm'm\Braket{xd}}$ (740)&$T_{2u}$&I&9.8&\cite{Roy2013,Ghara2017}&\\
	$\rm CoRh_2O_4	$&F$d\bar{3}m$ (227)&$\aizu{m\bar{3}m1'}{4'/m'm'm\Braket{xd}}$ (740)&$T_{2u}$&I&25&\cite{ge2017}&\\
	$\rm MnAl_2O_4	$&F$d\bar{3}m$ (227)&$\aizu{m\bar{3}m1'}{4'/m'm'm\Braket{xd}}$ (740)&$T_{2u}$&I&42&\cite{Krimmel2006}&\\
	$\rm MnGa_2O_4	$&F$d\bar{3}m$ (227)&$\aizu{m\bar{3}m1'}{\bar{3}'m'}$ (757)&$T_{2u}$&I&32&\cite{Saha2016a}&\\
	$\rm Co_3O_4	$&F$d\bar{3}m$ (227)&$\aizu{m\bar{3}m1'}{\bar{3}'m'}$ (757)&$T_{2u}$&I&30&\cite{Roth1964}&\\
	&&or~$\aizu{m\bar{3}m1'}{4'/m'm'm\Braket{xd}}$ (740)&$T_{2u}$&&&\cite{Roth1964}\vspace{5mm}&\\
	\multicolumn{8}{l}{\textbf{Tetragonal systems (53)}}\\
	$\rm BaMn_2 P_2	$&I$4/mmm$ (139)&$\aizu{4/mmm1'}{4'/m'm'm}$ (252)&$B_{1u}$&S&$>$750&\cite{Brock1994}&\\
	$\rm BaMn_2 As_2	$&I$4/mmm$ (139)&$\aizu{4/mmm1'}{4'/m'm'm}$ (252)&$B_{1u}$&S&625&\cite{Singh2009b,Singh2009c}&\\ 
	Ba$_{1-x}$K$_x$Mn$_2$As$_2$&I$4/mmm$ (139)&$\aizu{4/mmm1'}{4'/m'm'm}$ (252)&$B_{1u}$&M&&\cite{Lamsal2013}&\\ 
	$\rm BaMn_2 Sb_2	$&I$4/mmm$ (139)&$\aizu{4/mmm1'}{4'/m'm'm}$ (252)&$B_{1u}$&S&450&\cite{Sangeetha2018}&G\\
	$\rm BaMn_2 Bi_2	$&I$4/mmm$ (139)&$\aizu{4/mmm1'}{4'/m'm'm}$ (252)&$B_{1u}$&S&387&\cite{Calder2014,Saparov2013}&\\
	Ba$_{1-x}$K$_x$Mn$_2$Bi$_2$&I$4/mmm$ (139)&$\aizu{4/mmm1'}{4'/m'm'm}$ (252)&$B_{1u}$&M&&\cite{Saparov2013}&\\
	$\rm CeMn_2 Ge_2	$&I$4/mmm$ (139)&$\aizu{4/mmm1'}{mmm'\Braket{x}}$ (218)&$E_{u}$&&$318<T<417$&\cite{MdDin2015}&\\	
	$\rm EuCr_2 Si_2	$&I$4/mmm$ (139)&$\aizu{4/mmm1'}{4'/m'm'm}$ (252)&$B_{1u}$&&$2.4<T<692$&\cite{Moze2003}&\\
	$\rm HoCr_2 Si_2	$&I$4/mmm$ (139)&$\aizu{4/mmm1'}{4'/m'm'm}$ (252)&$B_{1u}$&&718&\cite{Moze2003}&\\
	$\rm TbCr_2 Si_2	$&I$4/mmm$ (139)&$\aizu{4/mmm1'}{4'/m'm'm}$ (252)&$B_{1u}$&&758&\cite{Moze2003}&\\
	$\rm EuCr_2 As_2	$&I$4/mmm$ (139)&$\aizu{4/mmm1'}{4'/m'm'm}$ (252)&$B_{1u}$&M&&\cite{Paramanik2014a}&\\
	$\rm BaCr_2 As_2	$&I$4/mmm$ (139)&$\aizu{4/mmm1'}{4'/m'm'm}$ (252)&$B_{1u}$&M&580&\cite{Singh2009d,Filsinger2017}&\\
	$\rm BaCrFe As_2	$&I$4/mmm$ (139)&$\aizu{4/mmm1'}{4'/m'm'm}$ (252)&$B_{1u}$&M&265&\cite{Filsinger2017}&\\
	$\rm SrCr_2As_2	$&I$4/mmm$ (139)&$\aizu{4/mmm1'}{4'/m'm'm}$ (252)&$B_{1u}$&M&590&\cite{Das2017a}&\\
	$\rm CaMnBi_2	$&P$4/nmm$ (123)&$\aizu{4/mmm1'}{4'/m'm'm}$ (252)&$B_{1u}$&M&300&\cite{Guo2014a,Zhang2016f}&\\
	$\rm SrMnBi_2	$&I$4/mmm$ (139)&$\aizu{4/mmm1'}{4'/m'm'm}$ (252)&$B_{1u}$&M&295&\cite{Zhang2016f}&\\
	$\rm YbMnBi_2	$&P$4/nmm$ (123)&$\aizu{4/mmm1'}{4'/m'm'm}$ (252)&$B_{1u}$&M&285&\cite{Wang2016o}&\\
	$\rm EuMnBi_2	$&I$4/mmm$ (139)&$\aizu{4/mmm1'}{4'/m'm'm}$ (252)&$B_{1u}$&M&315&\cite{Masuda2016a,Masuda2018Impact}&\\
	$\rm LaMnPO	$&P$4/nmm$ (129)&$\aizu{4/mmm1'}{4'/m'm'm}$ (252)&$B_{1u}$&I&375&\cite{Yanagi2009a,Simonson2012b}&\\
	$\rm LaMnAsO	$&P$4/nmm$ (129)&$\aizu{4/mmm1'}{4'/m'm'm}$ (252)&$B_{1u}$&S&317&\cite{Emery2010}&\\
	$\rm LaMnSbO	$&P$4/nmm$ (129)&$\aizu{4/mmm1'}{4'/m'm'm}$ (252)&$B_{1u}$&&255&\cite{Zhang2016k}&\\
	$\rm CeMnAsO	$&P$4/nmm$ (129)&$\aizu{4/mmm1'}{4'/m'm'm}$ (252)&$B_{1u}$&&$35<T<340$&\cite{Zhang2015f}&\\
	&&$\aizu{4'/m'm'm}{m'mm'\Braket{x}}$ (-)&$E_{u}$&&$7<T<35$&\cite{Zhang2015f}&\\
	&&$\aizu{m'mm'\Braket{x}}{2'/m\Braket{z}}$ (-)&$E_{u}$&&$7$&\cite{Zhang2015f}&\\
	$\rm CeMnSbO	$&P$4/nmm$ (129)&$\aizu{4/mmm1'}{4'/m'm'm}$ (252)&$B_{1u}$&&$4.5<T<240$&\cite{Zhang2016k}&\\
	&&$\aizu{4'/m'm'm}{m'mm'\Braket{x}}$ (-)&$E_{u}$&&$4.5$&\cite{Zhang2016k}&\\
	$\rm PrMnSbO	$&P$4/nmm$ (129)&$\aizu{4/mmm1'}{4'/m'm'm}$ (252)&$B_{1u}$&M&$35<T<230$&\cite{Kimber2010}&\\
					&&$\aizu{4/mmm1'}{mmm'\Braket{x}}$ (218)&$E_{u}$&M&35&\cite{Kimber2010}&\\
	$\rm NdMnAsO	$&P$4/nmm$ (129)&$\aizu{4/mmm1'}{4'/m'm'm}$ (252)&$B_{1u}$&S&$23<T<359$&\cite{Marcinkova2010,Emery2011b}&\\
	&&$\aizu{4'/m'm'm}{mmm'\Braket{x}}$ (-)&$E_{u}$&S&23&\cite{Marcinkova2010,Emery2011b}&\\
	$\rm NaMnP	$&P$4/nmm$ (129)&$\aizu{4/mmm1'}{4'/m'm'm}$ (252)&$B_{1u}$&&&\cite{Bronger1986NaMnP}&\\
	$\rm NaMnAs	$&P$4/nmm$ (129)&$\aizu{4/mmm1'}{4'/m'm'm}$ (252)&$B_{1u}$&&&\cite{Bronger1986NaMnP}&\\
	$\rm NaMnSb	$&P$4/nmm$ (129)&$\aizu{4/mmm1'}{4'/m'm'm}$ (252)&$B_{1u}$&&&\cite{Bronger1986NaMnP}&\\
	$\rm NaMnBi	$&P$4/nmm$ (129)&$\aizu{4/mmm1'}{4'/m'm'm}$ (252)&$B_{1u}$&&&\cite{Bronger1986NaMnP}&\\
	$\rm KMnSb	$&P$4/nmm$ (129)&$\aizu{4/mmm1'}{4'/m'm'm}$ (252)&$B_{1u}$&&&\cite{Schucht1999}&\\
	$\rm KMnBi	$&P$4/nmm$ (129)&$\aizu{4/mmm1'}{4'/m'm'm}$ (252)&$B_{1u}$&&&\cite{Schucht1999}&\\
	$\rm GdB_4	$&P$4/mbm$ (127)&$\aizu{4/mmm1'}{4/m'm'm'}$ (250)&$A_{1u}$&M&42&\cite{Blanco2006,Fisk1981}&\\
	$\rm DyB_4	$&P$4/mbm$ (127)&$\aizu{4/mmm1'}{mmm'\Braket{x}}$ (218)&$E_{u}$&M&$12.7<T<20.3$&\cite{Fisk1981,Will1979,Ji2007}&G\\
	$\rm ErB_4	$&P$4/mbm$ (127)&$\aizu{4/mmm1'}{mmm'\Braket{x}}$ (218)&$E_{u}$&M&13&\cite{Fisk1981,Will1979,Will1981}&\\
	$\rm TbB_4	$&P$4/mbm$ (127)&$\aizu{4/mmm1'}{4/m'm'm'}$ (250)&$A_{1u}$&M&$22<T<44$&\cite{Fisk1981,Matsumura2007}&\\
	&&$\aizu{4/m'm'm'}{m'm'm'\Braket{x}}$ (-)&$A_{1u},B_{1u}$&M&22&\cite{Fisk1981,Matsumura2007}&\\
	$\rm Ce_2 Pd Ge_3	$&P$4_2/mmc$ (131)&$\aizu{4/mmm1'}{4'/m'm'm}$ (252)&$B_{1u}$&&$2.3<T<10.7$&\cite{Bhattacharyya2016a}&\\
	$\rm EuTiO_3$&I$4_2/mcm$ (140)&$\aizu{4/mmm1'}{mmm'\Braket{d}}$ (218)&$E_u$&&5.3&\cite{Scagnoli2012}&G\\
	$\rm DyPO_4$&I$4_1/amd$ (141)&$\aizu{4/mmm1'}{4'/m'mm'}$ (252)&$B_{2u}$&&3.5&\cite{Rado1969,Thorntont1971}&Sc, Si\\
	$\rm TbPO_4$&I$4_1/amd$ (141)&$\aizu{4/mmm1'}{4'/m'mm'}$ (252)&$B_{2u}$&&2.2&\cite{Naagle1980,Rado1984}&\\
	$\rm HoPO_4$&I$4_1/amd$ (141)&$\aizu{4/mmm1'}{4'/m'mm'}$ (252)&$B_{2u}$&&1.4&\cite{Cooke1973magnetic}&Si\\
	$\rm GdVO_4$&I$4_1/amd$ (141)&$\aizu{4/mmm1'}{4'/m'mm'}$ (252)&$B_{2u}$&&2.43&\cite{Vie1974}&Sc, Si\\
	$\rm U_2 Pd_2 In$&P$4/mbm$ (127)&$\aizu{4/mmm1'}{4'/m'mm'}$ (218)&$B_{2u}$&M&36&\cite{Purwanto1994,Martin-Martin1999}&\\
	$\rm U_2 Pd_2 Sn$&P$4/mbm$ (127)&$\aizu{4/mmm1'}{4'/m'mm'}$ (218)&$B_{2u}$&M&41&\cite{Purwanto1994}&\\
	$\rm Mn_2 Au$&I$4/mmm$ (139)&$\aizu{4/mmm1'}{mmm'\Braket{d}}$ (218)&$E_{u}$&M&$>1000$&\cite{Barthem2013}&\\
	$\rm U Bi_2$&P$4/nmm$ (129)&$\aizu{4/mmm1'}{4/m'm'm'}$ (250)&$A_{1u}$&M&$183$&\cite{Barthem2013,Leciejewicz1967neutron}&\\
	$\rm U OTe$&P$4/nmm$ (129)&$\aizu{4/mmm1'}{4/m'm'm'}$ (250)&$A_{1u}$&&&\cite{murasik1969neutron}&\\
	$\rm U GeSe$&I$4/mmm$ (139)&$\aizu{4/mmm1'}{4/m'm'm'}$ (250)&$A_{1u}$&&40&\cite{Ptasiewicz1978}&\\
	$\rm Fe Sn_2$&I$4/mcm$ (140)&$\aizu{4/mmm1'}{mmm'\Braket{d}}$ (218)&$E_{u}$&M&$93<T\lesssim378$&\cite{Venturini1987,Armbruster2010}&\\
	&&$\aizu{mmm'\Braket{d}}{2'/m\Braket{z}}$ (-)&$E_{u}$&M&$93\lesssim T<378$&\cite{Venturini1987,Armbruster2010}&\\
	$\rm CuMnAs$&P$4/nmm$ (129)&$\aizu{4/mmm1'}{mmm'\Braket{x}}$ (218)&$E_{u}$&S&480&\cite{Wadley2013}&\\
	$\rm Cr_2 WO_6$&P$4_2/mnm$ (136)&$\aizu{4/mmm1'}{mmm'\Braket{x}}$ (218)&$E_{u}$&&45&\cite{Kunnmann1968magnetic,Zhu2014}&G\\
	$\rm Cr_2 TeO_6$&P$4_2/mnm$ (136)&$\aizu{4/mmm1'}{mmm'\Braket{x}}$ (218)&$E_{u}$&&93&\cite{Kunnmann1968magnetic,Zhu2014}&G\\
	$\rm Fe_2 TeO_6$&P$4_2/mnm$ (136)&$\aizu{4/mmm1'}{4/m'm'm'}$ (250)&$A_{1u}$&I&240&\cite{Kunnmann1968magnetic,Wang2014}&Sc, G\\
	Sr$_2$Ir$_{0.9}$Mn$_{0.1}$O$_4$&I$4_1/acd$ (142)&$\aizu{4/mmm1'}{4'/m'mm'}$ (252)&$B_{2u}$&I&$\sim 155$&\cite{Calder2012b}\vspace{5mm}&\\
	\multicolumn{8}{l}{\textbf{Hexagonal and trigonal systems (18)}}\\
	$\rm U_3 Ru_4 Al_{12}	$&P$6_3/mmc$ (194)&$\aizu{6/mmm1'}{mmm'\Braket{z}}$ (481)&$E_{2u},A_{2u}$&M&9.5&\cite{Pasturel2009,Troc2012}&G\\
	$\rm CaMn_2As_2	$&P$\bar{3}m1$ (164)&&&S&62&\cite{Sangeetha2016}&\\
	$\rm CaMn_2Sb_2	$&P$\bar{3}m1$ (164)&$\aizu{\bar{3}m1'}{2'/m}$ (295)&$E_{u}$&I&85&\cite{McNally2015c}&G\\
	&&$\aizu{\bar{3}m1'}{\bar{1}'}$ (286)&$A_{1u},E_{u}$&&85&\cite{Bridges2009}&\\
	$\rm CaMn_2Bi_2	$&P$\bar{3}m1$ (164)&$\aizu{\bar{3}m1'}{\bar{1}'}$ (286)&$E_{u}$&S&154&\cite{Gibson2015,Kawaguchi2018CaMn2Bi2}&\\
	$\rm SrMn_2P_2	$&P$\bar{3}m1$ (164)&&&S&53&\cite{Brock1994}&\\
	$\rm SrMn_2As_2	$&P$\bar{3}m1$ (164)&$\aizu{\bar{3}m1'}{2'/m}$ (295)&$E_{u}$&I&118&\cite{Sangeetha2016,Das2017}&\\
	$\rm SrMn_2Sb_2	$&P$\bar{3}m1$ (164)&$\aizu{\bar{3}m1'}{2'/m}$ (295)&$E_{u}$&S&110&\cite{Sangeetha2018}&\\
	$\rm EuMn_2As_2	$&P$\bar{3}m1$ (164)&&&S &142&\cite{Anand2016}&\\
	$\rm YbMn_2Sb_2	$&P$\bar{3}m1$ (164)&$\aizu{\bar{3}m1'}{\bar{1}'}$ (286)&$A_{1u},E_{u}$&&120&\cite{Morozkin2006}&\\
	$\rm U_2N_2S	$&P$\bar{3}m1$ (164)&$\aizu{\bar{3}m1'}{\bar{3}'m'}$ (313)&$A_{1u}$&&233&\cite{leciejewicz1975magnetic}&\\
	$\rm U_2N_2Se	$&P$\bar{3}m1$ (164)&$\aizu{\bar{3}m1'}{\bar{3}'m'}$ (313)&$A_{1u}$&&245&\cite{leciejewicz1975magnetic}&\\
	$\rm Cr_2O_3	$&R$\bar{3}c$ (167)&$\aizu{\bar{3}m1'}{\bar{3}'m'}$ (313)&$A_{1u}$&I&307&\cite{McGuire1956}&Sc, Si, G\\
	$\rm MnTiO_3	$&R$\bar{3}$ (148)&$\aizu{\bar{3}1'}{\bar{3}'}$ (264)&$A_{u}$&I&64&\cite{Silverstein2016,Shirane1959a}&G\\
	$\rm MnGeO_3	$&R$\bar{3}$ (148)&$\aizu{\bar{3}1'}{\bar{3}'}$ (264)&$A_{u}$&&120&\cite{Tsuzuki1974c}&G\\
	$\rm Co_4 Nb_2 O_9	$&P$\bar{3}c1$ (165)&$\aizu{\bar{3}m1'}{\bar{3}'m'}$ (313)&$A_{1u}$&I&27.4&\cite{Bertaut1961}&Sc\\
		&&$\aizu{\bar{3}m1'}{2/m'}$ (296)&$E_{u}$&I&27.2&\cite{Khanh2016a}&\\
	$\rm Mn_4 Nb_2 O_9	$&P$\bar{3}c1$ (165)&$\aizu{\bar{3}m1'}{\bar{3}'m'}$ (313)&$A_{1u}$&I&108.4&\cite{Cao2017,Bertaut1961}&Sc\\
	$\rm Co_4 Ta_2 O_9	$&P$\bar{3}c1$ (165)&$\aizu{\bar{3}m1'}{\bar{3}'m'}$ (313)&$A_{1u}$&I&21&\cite{Fischer1972,Fang2015a}&Sc\\
	$\rm Mn_4 Ta_2 O_9	$&P$\bar{3}c1$ (165)&$\aizu{\bar{3}m1'}{\bar{3}'m'}$ (313)&$A_{1u}$&I&103&\cite{Fischer1972,Liu2016a}&Sc\vspace{5mm}\\
	\multicolumn{8}{l}{\textbf{Orthorhombic systems (29)}}\\
	$\rm CaMnSb_2	$&P$nma$ (62)&&&M&302&\cite{He2017a}&\\
	$\rm NdCrTiO_5	$&Pbam (55)&$\aizu{mmm1'}{mmm'}$ (71)&$B_{1u}$&I&13&\cite{Buisson1970}&\\
	&&&&&21&\cite{Hwang2012}&G\\
	$\rm LiFePO_4	$&P$nma$ (62)&$\aizu{mmm1'}{mmm'}$ (71)&$ B_{1u} $&I&50&\cite{Santoro1967} &Sc, G\\
	&&$\aizu{mmm1'}{2/m'\Braket{z}}$ (60)&$ A_{u}, B_{1u} $&&47&\cite{Li2006,Toft-Petersen2015} &Sc\\
	$\rm LiMnPO_4	$&P$nma$ (62)&$\aizu{mmm1'}{m'm'm'}$ (73)&$ A_{u} $&I&35&\cite{PhysRev.131.38,Toft-Petersen2012} &Sc, G\\
	$\rm LiNiPO_4	$&P$nma$ (62)&$\aizu{mmm1'}{mm'm}$ (71)&$ B_{2u} $&I&20.8&\cite{Kornev2000a} &Sc\\
	$\rm LiCoPO_4	$&P$nma$ (62)&$\aizu{mmm1'}{mmm'}$ (71)&$ B_{1u} $&I&21.6&\cite{Fogh2017,santoro1966magnetic}&\\
		&&$\aizu{mmm1'}{2'\Braket{x}}$ (53)&$ B_{2g},B_{1u},B_{2u} $&&&\cite{Vaknin2002,VanAken2007observation}&\\
	$\rm Li_2Ni(SO_4)_2	$&P$bca$ (61)&$\aizu{mmm1'}{m'm'm'}$ (73)&$ A_{u} $&I&28&\cite{Reynaud2014}&\\ 
	$\rm KMn_4 (PO_4)_3	$&P$nam$ (62)&$\aizu{mmm1'}{mm'm}$ (73)&$ B_{2u} $&&$10$&\cite{Lopez2008}&G\\
	$\rm t-NaFePO_4	$&P$nma$ (62)&$\aizu{mmm1'}{mmm'}$ (71)&$ B_{1u} $&&50&\cite{Avdeev2013} &G\\
	$\rm Gd_5 Ge_4	$&P$nma$ (62)&$\aizu{mmm1'}{mmm'}$ (71)&$ B_{1u} $&M&127&\cite{Tan2005,Levin2001} &G\\
	$\rm EuZrO_3	$&P$nma$ (62)&$\aizu{mmm1'}{mm'm}$ (71)&$B_{2u}$&I&4.1&\cite{Avdeev2014}&\\
	&&$\aizu{mmm1'}{m'm'm'}$ (73)&$A_{u}$&I&4.4&\cite{Saha2016c}&G\\
	$\rm DyAlO_3	$&P$bnm$ (62)&$\aizu{mmm1'}{m'm'm'}$ (73)&$A_{u}$&I&3.48&\cite{bidaux1968etude}&Sc, Si\\
	$\rm DyCoO_3	$&P$nma$ (62)&$\aizu{mmm1'}{mm'm}$ (71)&$A_{u}$&I&3.6&\cite{Knizek2014}&\\
	$\rm TbAlO_3	$&P$bnm$ (62)&$\aizu{mmm1'}{m'm'm'}$ (73)&$A_{u}$&&3.95&\cite{Mareschal1968}&Sc, Si\\
	$\rm TbCoO_3	$&P$bnm$ (62)&$\aizu{mmm1'}{mm'm}$ (71)&$B_{2u}$&I&3.31&\cite{Knizek2014,Mareschal1968,Munoz2012}&Sc, G\\
	$\rm HoCoO_3	$&P$nma$ (62)&$\aizu{mmm1'}{m'm'm'}$ (73)&$A_{u}$&&3&\cite{Munoz2012}&\\
	$\rm GdAlO_3	$&P$bnm$ (62)&$\aizu{mmm1'}{m'm'm'}$ (73)&$A_{u}$&&3.9&\cite{Cooke1976,Quezel1982}&Sc, Si\\
	$\rm MnNb_2O_6	$&P$bcn$ (60)&$\aizu{mmm1'}{2'/m\Braket{x}}$ (59)&$B_{2u},B_{3u}$&&4.4&\cite{Jacobson1975}&\\
	$\rm CoSe_2O_5	$&P$bcn$ (60)&$\aizu{mmm1'}{m'mm}$ (71)&$B_{3u}$&&8.5&\cite{Melot2010b}&G\\
	$\rm TbGe_2	$&C$mmm$ (65)&$\aizu{mmm1'}{m'mm}$ (71)&$B_{3u}$&&41&\cite{Schobinger-Papamantellos1988}&G\\
	$\rm Ce_3Sn_7	$&C$mmm$ (65)&$\aizu{mmm1'}{m'mm}$ (71)&$B_{3u}$&M&5&\cite{Bonnet1994,Givord1989}&\\
	$\rm Sm_3 Ag_4 Sn_4	$&Immm (71)&$\aizu{mmm1'}{mmm'}$ (71)&$B_{1u}$&&8.3&\cite{Voyer2007}&\\
	&&or $\aizu{mmm1'}{mm'm}$ (71)&$B_{2u}$&&8.3&\cite{Voyer2007}&\\
	$\rm UCu_5In	$&P$nma$ (62)&$\aizu{mmm1'}{mm'm}$ (71)&$B_{2u}$&M&25&\cite{Tran2001}&\\
	$\rm KFeO_2	$&P$bca$ (61)&$\aizu{mmm1'}{m'm'm'}$ (73)&$A_{u}$&&960&\cite{Tomkowicz1977a}&\\
	&&$\aizu{mmm1'}{m'mm}$ (71)&$B_{3u}$&&$\sim 1001$&\cite{Sheptyakov2010}&\\
	$\rm RbFeO_2	$&P$bca$ (61)&$\aizu{4'/m'm'm}{m'm'm'}$ (-)&$A_{u}$&&$<$ 737&\cite{Sheptyakov2010}&\\
	$\rm CsFeO_2	$&P$bca$ (61)&$\aizu{4'/m'm'm}{m'm'm'}$ (-)&$A_{u}$&&$<$ 350&\cite{Sheptyakov2010}&\\
	$\rm CoGeO_3	$&P$bca$ (61)&$\aizu{mmm1'}{mmm'}$ (71)&$B_{1u}$&I&33.1&\cite{Redhammer2010a}&\\
	$\rm DyVO_4	$&I$mma$ (74)&$\aizu{mmm1'}{mmm'}$ (71)&$B_{1u}$&I&3.8&\cite{Search1971,Kishimoto2010}&\\
	YbAl$_{1-x}$Fe$_x$B$_{4}$&P$bam$ (55)&$\aizu{mmm1'}{m'mm}$ (71)&$B_{3u}$&M&&\cite{ybalb4}&\\
	YbAl$_{1-x}$Fe$_x$B$_{4}$&P$bam$ (55)&$\aizu{mmm1'}{m'm'm'}$ (73)&$A_{u}$&M&&\cite{ybalb4}
\vspace{5mm}&\\
	\multicolumn{8}{l}{\textbf{Monoclinic systems (10)}}\\
	$\rm Co_3 Te O_6	$&C$2/c$ (15)&$\aizu{2/m1'}{2'/m}$ (26)&$B_{u}$&&21.1&\cite{Ivanov2012}&G\\
	$\rm MnPS_3	$&C$2/m$ (12)&$\aizu{2/m1'}{2'/m}$ (26)&$B_{u}$&I&78&\cite{Kurosawa1983,Ressouche2010a}&G\\ 
	$\rm LiFeSi_2O_6	$&P$2_1/c$ (14)&$\aizu{2/m1'}{2/m'}$ (27)&$ A_{u} $& &17.8&\cite{Redhammer2009,Redhammer2001} &G\\
	&&$\aizu{2/m1'}{\bar{1}'}$ (17)&$ A_{u},B_u $& &18&\cite{Toledano2015} &\\
	$\rm LiCrSi_2O_6	$&P$2_1/c$ (14)&$\aizu{2/m1'}{2'/m}$ (26)&$ B_{u} $&&11.5&\cite{Nenert2009a,Nenert2010} &\\
	$\rm LiCrGe_2O_6	$&P$2_1/c$ (14)&$\aizu{2/m1'}{2'/m}$ (26)&$ B_{u} $&&4.8&\cite{Nenert2010} &\\
	$\rm LiVGe_2O_6	$&P$2_1/c$ (14)& &$A_u$ or $B_{u} $&&24&\cite{Lumsden2000} &\\
	$\rm NaCrSi_2O_6 	$&C$2/c$ (15)&$\aizu{2/m1'}{\bar{1}'}$ (17)&$ A_u, B_{u}$&&2.8&\cite{Nenert2010a} &\\
	$\rm CaMnGe_2O_6	$&C$2/c$ (15)&$\aizu{2/m1'}{\bar{1}'}$ (17)&$ A_{u}, B_u $&I&12&\cite{Redhammer2008} &\\
	&&$\aizu{2/m1'}{2'/m}$ (26)&$ B_u $&&15&\cite{Ding2016} &G\\
	$\rm MnGeO_3	$&C$2/c$ (15)&$\aizu{2/m1'}{2'/m}$ (26)&$ B_{u} $&&35.1&\cite{Redhammer2011} &\\
	$\rm Na_2RuO_4$&P$2_1/c$ (14)&$\aizu{2/m1'}{2/m'}$ (27)&$A_{u}$&&37.22&\cite{Mogare2006}&\\
	\end{longtable}

\twocolumngrid

	In our classification list, more than 100 magnetic materials are revealed with their symmetry and conducting property. Here we briefly discuss some intriguing examples from the viewpoint of emergent responses. 

\subsection{High-temperature ME compounds}
	Many ME compounds have already been reported~\cite{Schmid1973,Siratori1992a,Gallego2016a}, and multiferroic compounds have recently attracted much attention~\cite{Tokura2014multiferroic,Schmid2008}. The N\'eel temperature, however, is much lower than the room temperature in most cases except for a few compounds such as Cr$_2$O$_3$~\cite{astrov1961magnetoelectric}. This is unfavorable for a potential application of the ME effect to energy-saving devices. 

	On the other hand, some materials in Table~\ref{magnetic_candidates} undergo magnetic transition at high N\'eel temperatures above the room temperature. For examples, in many manganese pnictides and chromium pnictides, crystallizing in the ThCr$_2$Si$_2$ structure, odd-parity magnetic multipole order occurs at high temperatures. Furthermore, their magnetic and conducting properties are chemically controllable; \textit{e.g.} substituting pnictogen atoms with carbon group atoms. Thus, Table~\ref{magnetic_candidates} contains a new family of high-temperature ME compounds.

\subsection{Itinerant odd-parity multipole compounds}
Most of previous studies on the parity-violating magnetic order focused on insulators. It is natural to study insulating systems in the context of the multiferroics, since ferroelectricity is weakened by itinerant electrons' screening of electric polarizations.

The odd-parity magnetic multipole order in itinerant systems, however, is attracting recent attentions. The interplay between the itinerant property and parity-violating magnetic order leads to intriguing transport phenomena such as the MPE effect and dichromatic transport as we have discussed in Sec.~\ref{Section3}. Furthermore, itinerant odd-parity magnetic multipole materials may realize the Fulde-Ferrell-Larkin-Ovchinnikov state at zero external magnetic field when the superconductivity occurs in the multipole ordered state~\cite{Sumita2016,Sumita2017}. Table~\ref{magnetic_candidates} includes many metallic odd-parity magnetic multipole materials. For examples, Ba$_{1-x}$K$_x$Mn$_2$As$_2$, SrCr$_2$As$_2$, GdB$_4$, U$_2$Pd$_2$In, and others may be a platform of the exotic transport phenomena and unconventional superconductivity.

\section{Summary and Conclusions}

Previous studies clarified higher order multipole order by spontaneous ordering of atomic multipole moment, which is stemmed from the entanglement of orbital and spin angular momentum due to the strong spin-orbit coupling. Localized $d$- and $f$-electron systems have been investigated by many works~\cite{Kuramoto2009Review,Santini2009}. Then, possible symmetry is restricted to the even-parity multipole order which preserves space inversion symmetry. 

On the other hand, the augmented odd-parity multipole order has recently attracted interest. The essential ingredient is the existence of subsectors such as sublattice. The entanglement of spin, orbital, and subsector degrees of freedom may allow the formation of augmented multipole moment in a unit cell. In particular, the locally-noncentrosymmetric crystal is a platform of the odd-parity multipole order because additional negative sign due to the site permutation appears in the space inversion parity~\cite{Kotzev1982}. It has been shown that the odd-parity augmented multipole order gives rise to intriguing electromagnetic responses not only in insulators but also in metals and semiconductors~\cite{Yanase2014,Hayami2014b,Hitomi2016,hikaruwatanabe2017}. 

Inspired by the renewed interests in multipole physics, we carried out group-theoretical classification of the even-parity/odd-parity multipole order,  extending the previous result limited to tetragonal systems~\cite{hikaruwatanabe2017}. Multipole order in cubic, tetragonal, and hexagonal systems has been classified by point group symmetry in Tables~\ref{electric_basis_cubic}-\ref{magnetic_basis_hexagonal}. Classification in other crystal groups is straightforwardly obtained by compatibility relations. Even though real compounds are sometimes too complicated to see the relevant multipole moment characterizing the ordered state, we can identify the multipole order parameter by group-theoretical analysis. 

Our classification tables show the multipole moments and the basis functions in both of the real space and the momentum space. Interestingly, as for the odd-parity multipole order, the basis functions in the momentum space look quite different from the corresponding multipole moment in the real space. Indeed, they reveal unusual band structures. The momentum space representation of odd-parity magnetic multipole moment is spin-independent, implying an asymmetric band structure. Spin-dependent basis functions of odd-parity electric multipole indicate spin-momentum locking. The distinct representations in real and momentum spaces are characteristic properties of parity-violating order and result in emergent electromagnetic responses. 

We have demonstrated application of the classification theory to the emergent responses. The correspondence between the symmetry of multipole order and ME effect, Edelstein effect, piezoelectric effect, MPE effect, and dichromatic electron transport has been elucidated. The electric-field-induced responses are obtained by the basis functions in the real space, whereas the electric-current-induced responses are described by the basis functions in the momentum space. It is noteworthy that the recent studies have identified the direct relation between odd-parity magnetic multipole moments and ME effect~\cite{Gao2018spin,Shitade2018theory,gao2018theory}. As for other emergent responses, quantitative relations with odd-parity multipole moments are expected to be found in a future work.

From the classification theory, we can intuitively and precisely determine the response tensor. As an example, the parity-violating order in Cd$_2$Re$_2$O$_7$ has been discussed. The two controversial phases proposed by experiments are classified into the electric octupole and dotriacontapole states. Since the response tensors of the Edelstein effect and dichromatic electron transport are different, these effects can be used to identify the symmetry of electric multipole order in Cd$_2$Re$_2$O$_7$. On the other hand, the ME effect and MPE effect are characteristic properties of the odd-parity magnetic multipole states. The allowed MPE effect, namely, current-induced lattice distortion, has been classified on the basis of the group theory (Table~\ref{magnetopiezo_classification}). Ba$_{1-x}$K$_x$Mn$_2$As$_2$ and GdB$_4$ have been discussed as examples.

Using the group-theoretical analysis, we have identified more than 110 candidate materials for odd-parity magnetic multipole states. The list of materials contains not only well-known ME materials but also many magnetic compounds which were not clarified so far. Some of them are candidates for room-temperature ME materials, which have been searched in the research field of multiferroics. Furthermore, the list includes itinerant compounds, which may be a platform of intriguing transport properties and exotic superconductivity arising from the broken inversion and time-reversal symmetry. 

The concept of the augmented multipole with the subsector degree of freedom has extended the research field of multipole order in strongly correlated electron systems, and it has connected various research fields, such as heavy fermions, multiferroics, and spintronics. Our classification theory may provide basis for future researches. We expect that the classification tables are useful, as the classification theory of unconventional superconductivity~\cite{Sigrist1991} has been used in the research field of superconductivity. We also expect that the list of candidate materials stimulates experimental studies of odd-parity multipole order and emergent responses. We hope that our classification theory generates renewed interests in multipole physics.

\begin{acknowledgments}
We would like to thank M.~Kimata, A.~Shitade, S.~Suzuki, A.~Daido, S.~Nakatsuji, D.~Hirai, and Z.~Hiroi for fruitful discussions. This work is supported by a Grant-in-Aid for Scientific Research on Innovative Areas ``J-Physics'' (Grant No.~JP15H05884) and ``Topological Materials Science'' (Grant No.~JP16H00991,~JP18H04225) from the Japan Society for the Promotion of Science (JSPS), and by JSPS KAKENHI (Grants No.~JP15K05164, No.~JP15H05745, and No.~JP18H01178). H.W. is supported by a JSPS research fellowship and supported by JSPS KAKENHI (Grant No.~18J23115). 
\end{acknowledgments} 
\appendix

\section{Aizu species}\label{app_aizu_species}

The method of Aizu species is useful to classify possible ferroic phase transitions and resulting domain states. Aizu species was proposed by Aizu~\cite{Aizu1966,Aizu1969,Aizu1970a}, and the method has been developed in various ways~\cite{Schmid2008,Litvin2008,Hlinka2016a}. 

Aizu species is denoted as 
		\begin{equation}
		\aizu{\bm{G}}{\bm{K}},
		\end{equation} 
where $\bm{G}$ and $\bm{K}$ are (magnetic) point group of a disordered phase (high-temperature phase) and that of an ordered phase (low-temperature phase), respectively. The symbol \textit{F} represents ``Ferroic''. When the group-subgroup relation $\bm{G} \supset \bm{K}$ holds, we obtain the coset decomposition as
		\begin{equation}
	  	\bm{G} = g_1 \bm{K} +g_2 \bm{K}+\cdots g_N \bm{K}, \label{GK_decomposition}
	  \end{equation}
where $g_1 \in \bm{K}$, and $g_j \not\in \bm{K} $ for $j\neq 1$. It is indicated that the domain states $S_i~\left(i=1,2,\cdots,N \right)$ exist, and the number of domain states is $N= |\bm{G}|/|\bm{K}|$ with $|\bm{G}|$ being the order of the group $\bm{G}$.

In Aizu's classification, the domain states $S_i$ are characterized by some ferroic order parameters such as electric polarization $\bm{P}$ (ferroelectric order), magnetization $\bm{M}$ (ferromagnetic order), and strain $\hat{\epsilon}$ (ferroelastic order). For example, let us consider the pair of the cubic phase $\bm{G} = m\bar{3}m$ and the tetragonal phase $\bm{K} =4mm$. Assuming that the four-fold rotation axis is the $z$-axis, the corresponding coset decomposition is obtained as
		\begin{align}
		\begin{aligned}
	  	m\bar{3}m &= E~ 4mm + P~4mm + C_{4x}^+~4mm  \\
	  			 &+S_{4x}^+~4mm+ C_{4y}^+~4mm+ S_{4y}^+~4mm,
	  \end{aligned}
	  \end{align}
where $E,~P,~C_{4x(y)}^+,~S_{4x(y)}^+$ are identity operation, parity operation, $\pi/2$ rotation operation along the $x(y)$-axis, and $\pi/2$ rotation-inversion operation along the $x(y)$-axis, respectively.
	Accordingly, the six domain states are obtained as
		\begin{align}
		\begin{aligned}
			&S_1=E~S_1,~S_2= P~ S_1 ,~S_3= C_{4x}^+~ S_1,\\
			&S_4= S_{4x}^+~ S_1,~S_5= C_{4y}^+~ S_1,~S_6= S_{4y}^+~ S_1. 
		\end{aligned}
		\end{align}
The domain $S_1$ may have an electric polarization along the $z$-axis, $\bm{P}_1 \neq \bm{0}$. The electric polarization $\bm{P}_j$ of another domain $S_j$ is given by
			\begin{equation}
				\bm{P}_j = g_j \bm{P}_1.
			\end{equation}
The electric polarizations $\bm{P}_i$ $\left(i=1,2,\cdots,6 \right)$ are different from each other and can be distinguished completely by the conjugate field with $\bm{P}$, that is, the electric field $\bm{E}$. In this case, the Aizu species $\aizu{m\bar{3}m}{4mm}$ is ``full-electric.'' The well-known ferroelectric material BaTiO$_3$~\cite{rabe2007ferroelectric_book} belongs to this Aizu species and it is perfectly switchable by external electric fields. 

Besides, domain states given by other Aizu species are also characterized by the electric polarization. Following the notation by Aizu~\cite{Aizu1970a} and Litvin~\cite{Litvin2008}, the domain states are classified into ``full-electric'', ``partial-electric'', ``null-electric'', and ``zero-electric''. These notations are defined as follows:
		\begin{enumerate}
		\item full (F) electric: all domain states are distinguishable according to $\bm{P}$.
		\item null (N) electric: all domain states have the same $\bm{P}$, and thus, domains are not distinguishable by $\bm{P}$.
		\item zero (Z) electric: in all domain states $\bm{P}$ is zero. Therefore, domains are not distinguishable by $\bm{P}$.
		\item partial (P) electric: some domain states have the same $\bm{P}$ and are not distinguishable. However, there are several classes of domain states which are distinguishable by $\bm{P}$.
		\end{enumerate}
Accordingly, 
		\begin{enumerate}
		\item full (F) electric: all domains are switchable by $\bm{E}$.
		\item null (N) electric: domains are not switchable by $\bm{E}$.
		\item zero (Z) electric: domains are not switchable by $\bm{E}$.
		\item partial (P) electric: domains are partially switchable by $\bm{E}$. Domain states having the same $\bm{P}$ can not be controlled by $\bm{E}$.
		\end{enumerate}
Similarly, Aizu species is also characterized by magnetization $\bm{M}$, strain $\hat{\epsilon}$, and so on. 

The Aizu species is useful to classify domain states. When a pair of disordered phase and ordered phase are specified by symmetry, the corresponding Aizu species tells what ferroic order parameter characterizes domains and how domain states can be switched by external fields conjugate with the ferroic order. 
In Sec.~\ref{Section4_candidates}, we classify the odd-parity magnetic multipole order in candidate materials. The magnetic transitions in the classified list belong to Aizu species which may be called ``full odd-parity-magnetic''.

Note that Aizu species has been defined in several ways~\cite{Aizu1979}. Algebraic relation of point groups is definitely given without ambiguity. However, the coordinates can be arbitrarily taken. For example, the Aizu species $\aizu{mmm}{2mm}$ represents a nonpolar-polar phase transition in an orthorhombic system, where the polar axis of the ordered phase $\left(2mm \right)$ is not explicitly given. In the representation theory, the nonpolar-polar transition is represented by either of $B_{1u}$, $B_{2u}$ or $B_{3u}$ irreducible representation of the $D_{2h}$ point group. However, the three irreducible representations can not be distinguished by Aizu species, since they are merged into the same Aizu species $\aizu{mmm}{2mm}$. To solve this problem, we may define the Aizu species with taking into account the coordinates of the ordered phase relative to those of the disordered phase. Indeed, Aizu introduced several definitions of the Aizu species with conventions on the coordinates, which are called ``normal'', ``specific'', ``subspecific'', and ``rigorous'' definitions~\cite{Aizu1979}. In our classification theory, we adopt the subspecific definition of Aizu species so as to maintain the correspondence with the representation theory.

\newpage
\bibliography{addition,classification_library,library}

\begin{thebibliography}{275}%
\makeatletter
\providecommand \@ifxundefined [1]{%
 \@ifx{#1\undefined}
}%
\providecommand \@ifnum [1]{%
 \ifnum #1\expandafter \@firstoftwo
 \else \expandafter \@secondoftwo
 \fi
}%
\providecommand \@ifx [1]{%
 \ifx #1\expandafter \@firstoftwo
 \else \expandafter \@secondoftwo
 \fi
}%
\providecommand \natexlab [1]{#1}%
\providecommand \enquote  [1]{``#1''}%
\providecommand \bibnamefont  [1]{#1}%
\providecommand \bibfnamefont [1]{#1}%
\providecommand \citenamefont [1]{#1}%
\providecommand \href@noop [0]{\@secondoftwo}%
\providecommand \href [0]{\begingroup \@sanitize@url \@href}%
\providecommand \@href[1]{\@@startlink{#1}\@@href}%
\providecommand \@@href[1]{\endgroup#1\@@endlink}%
\providecommand \@sanitize@url [0]{\catcode `\\12\catcode `\$12\catcode
  `\&12\catcode `\#12\catcode `\^12\catcode `\_12\catcode `\%12\relax}%
\providecommand \@@startlink[1]{}%
\providecommand \@@endlink[0]{}%
\providecommand \url  [0]{\begingroup\@sanitize@url \@url }%
\providecommand \@url [1]{\endgroup\@href {#1}{\urlprefix }}%
\providecommand \urlprefix  [0]{URL }%
\providecommand \Eprint [0]{\href }%
\providecommand \doibase [0]{http://dx.doi.org/}%
\providecommand \selectlanguage [0]{\@gobble}%
\providecommand \bibinfo  [0]{\@secondoftwo}%
\providecommand \bibfield  [0]{\@secondoftwo}%
\providecommand \translation [1]{[#1]}%
\providecommand \BibitemOpen [0]{}%
\providecommand \bibitemStop [0]{}%
\providecommand \bibitemNoStop [0]{.\EOS\space}%
\providecommand \EOS [0]{\spacefactor3000\relax}%
\providecommand \BibitemShut  [1]{\csname bibitem#1\endcsname}%
\let\auto@bib@innerbib\@empty
\bibitem [{\citenamefont {Hasan}\ and\ \citenamefont {Kane}(2010)}]{Hasan2010}%
  \BibitemOpen
  \bibfield  {author} {\bibinfo {author} {\bibfnamefont {M.~Z.}\ \bibnamefont
  {Hasan}}\ and\ \bibinfo {author} {\bibfnamefont {C.~L.}\ \bibnamefont
  {Kane}},\ }\href {\doibase 10.1103/RevModPhys.82.3045} {\bibfield  {journal}
  {\bibinfo  {journal} {Rev. Mod. Phys.}\ }\textbf {\bibinfo {volume} {82}},\
  \bibinfo {pages} {3045} (\bibinfo {year} {2010})}\BibitemShut {NoStop}%
\bibitem [{\citenamefont {Qi}\ and\ \citenamefont {Zhang}(2011)}]{Qi2011}%
  \BibitemOpen
  \bibfield  {author} {\bibinfo {author} {\bibfnamefont {X.-L.}\ \bibnamefont
  {Qi}}\ and\ \bibinfo {author} {\bibfnamefont {S.-C.}\ \bibnamefont {Zhang}},\
  }\href {\doibase 10.1103/RevModPhys.83.1057} {\bibfield  {journal} {\bibinfo
  {journal} {Rev. Mod. Phys.}\ }\textbf {\bibinfo {volume} {83}},\ \bibinfo
  {pages} {1057} (\bibinfo {year} {2011})}\BibitemShut {NoStop}%
\bibitem [{\citenamefont {Witczak-Krempa}\ \emph {et~al.}(2014)\citenamefont
  {Witczak-Krempa}, \citenamefont {Chen}, \citenamefont {Kim},\ and\
  \citenamefont {Balents}}]{Witczak-Krempa2014}%
  \BibitemOpen
  \bibfield  {author} {\bibinfo {author} {\bibfnamefont {W.}~\bibnamefont
  {Witczak-Krempa}}, \bibinfo {author} {\bibfnamefont {G.}~\bibnamefont
  {Chen}}, \bibinfo {author} {\bibfnamefont {Y.~B.}\ \bibnamefont {Kim}}, \
  and\ \bibinfo {author} {\bibfnamefont {L.}~\bibnamefont {Balents}},\ }\href
  {\doibase 10.1146/annurev-conmatphys-020911-125138} {\bibfield  {journal}
  {\bibinfo  {journal} {Annu. Rev. Condens. Matter Phys.}\ }\textbf {\bibinfo
  {volume} {5}},\ \bibinfo {pages} {57} (\bibinfo {year} {2014})}\BibitemShut
  {NoStop}%
\bibitem [{\citenamefont {Bauer}\ and\ \citenamefont
  {Sigrist}(2012)}]{bauer2012non}%
  \BibitemOpen
  \bibfield  {author} {\bibinfo {author} {\bibfnamefont {E.}~\bibnamefont
  {Bauer}}\ and\ \bibinfo {author} {\bibfnamefont {M.}~\bibnamefont
  {Sigrist}},\ }\href@noop {} {\emph {\bibinfo {title} {Non-centrosymmetric
  superconductors: introduction and overview}}},\ Vol.\ \bibinfo {volume}
  {847}\ (\bibinfo  {publisher} {Springer Science \& Business Media},\ \bibinfo
  {year} {2012})\BibitemShut {NoStop}%
\bibitem [{\citenamefont {Sato}\ and\ \citenamefont
  {Ando}(2017)}]{Sato2017topological}%
  \BibitemOpen
  \bibfield  {author} {\bibinfo {author} {\bibfnamefont {M.}~\bibnamefont
  {Sato}}\ and\ \bibinfo {author} {\bibfnamefont {Y.}~\bibnamefont {Ando}},\
  }\href {http://stacks.iop.org/0034-4885/80/i=7/a=076501} {\bibfield
  {journal} {\bibinfo  {journal} {Rep. Prog. Phys.}\ }\textbf {\bibinfo
  {volume} {80}},\ \bibinfo {pages} {076501} (\bibinfo {year}
  {2017})}\BibitemShut {NoStop}%
\bibitem [{\citenamefont {Smidman}\ \emph {et~al.}(2017)\citenamefont
  {Smidman}, \citenamefont {Salamon}, \citenamefont {Yuan},\ and\ \citenamefont
  {Agterberg}}]{Smidman2016super}%
  \BibitemOpen
  \bibfield  {author} {\bibinfo {author} {\bibfnamefont {M.}~\bibnamefont
  {Smidman}}, \bibinfo {author} {\bibfnamefont {M.~B.}\ \bibnamefont
  {Salamon}}, \bibinfo {author} {\bibfnamefont {H.~Q.}\ \bibnamefont {Yuan}}, \
  and\ \bibinfo {author} {\bibfnamefont {D.~F.}\ \bibnamefont {Agterberg}},\
  }\href {http://stacks.iop.org/0034-4885/80/i=3/a=036501} {\bibfield
  {journal} {\bibinfo  {journal} {Rep. Prog. Phys.}\ }\textbf {\bibinfo
  {volume} {80}},\ \bibinfo {pages} {036501} (\bibinfo {year}
  {2017})}\BibitemShut {NoStop}%
\bibitem [{\citenamefont {Katsura}\ \emph {et~al.}(2005)\citenamefont
  {Katsura}, \citenamefont {Nagaosa},\ and\ \citenamefont
  {Balatsky}}]{Katsura2005}%
  \BibitemOpen
  \bibfield  {author} {\bibinfo {author} {\bibfnamefont {H.}~\bibnamefont
  {Katsura}}, \bibinfo {author} {\bibfnamefont {N.}~\bibnamefont {Nagaosa}}, \
  and\ \bibinfo {author} {\bibfnamefont {A.~V.}\ \bibnamefont {Balatsky}},\
  }\href {\doibase 10.1103/PhysRevLett.95.057205} {\bibfield  {journal}
  {\bibinfo  {journal} {Phys. Rev. Lett.}\ }\textbf {\bibinfo {volume} {95}},\
  \bibinfo {pages} {057205} (\bibinfo {year} {2005})}\BibitemShut {NoStop}%
\bibitem [{\citenamefont {Arima}(2007)}]{Arima2007}%
  \BibitemOpen
  \bibfield  {author} {\bibinfo {author} {\bibfnamefont {T.-h.~H.}\
  \bibnamefont {Arima}},\ }\href {\doibase 10.1143/JPSJ.76.073702} {\bibfield
  {journal} {\bibinfo  {journal} {J. Phys. Soc. Jpn.}\ }\textbf {\bibinfo
  {volume} {76}},\ \bibinfo {pages} {1} (\bibinfo {year} {2007})}\BibitemShut
  {NoStop}%
\bibitem [{\citenamefont {Kimura}\ \emph {et~al.}(2003)\citenamefont {Kimura},
  \citenamefont {Goto}, \citenamefont {Shintani}, \citenamefont {Ishizaka},
  \citenamefont {Arima},\ and\ \citenamefont {Tokura}}]{Kimura2003}%
  \BibitemOpen
  \bibfield  {author} {\bibinfo {author} {\bibfnamefont {T.}~\bibnamefont
  {Kimura}}, \bibinfo {author} {\bibfnamefont {T.}~\bibnamefont {Goto}},
  \bibinfo {author} {\bibfnamefont {H.}~\bibnamefont {Shintani}}, \bibinfo
  {author} {\bibfnamefont {K.}~\bibnamefont {Ishizaka}}, \bibinfo {author}
  {\bibfnamefont {T.}~\bibnamefont {Arima}}, \ and\ \bibinfo {author}
  {\bibfnamefont {Y.}~\bibnamefont {Tokura}},\ }\href {\doibase
  10.1038/nature02018} {\bibfield  {journal} {\bibinfo  {journal} {Nature}\
  }\textbf {\bibinfo {volume} {426}},\ \bibinfo {pages} {55} (\bibinfo {year}
  {2003})}\BibitemShut {NoStop}%
\bibitem [{\citenamefont {Fiebig}(2005)}]{Fiebig2005b}%
  \BibitemOpen
  \bibfield  {author} {\bibinfo {author} {\bibfnamefont {M.}~\bibnamefont
  {Fiebig}},\ }\href {\doibase 10.1088/0022-3727/38/8/R01} {\bibfield
  {journal} {\bibinfo  {journal} {J. Phys. D. Appl. Phys.}\ }\textbf {\bibinfo
  {volume} {38}},\ \bibinfo {pages} {R123} (\bibinfo {year}
  {2005})}\BibitemShut {NoStop}%
\bibitem [{\citenamefont {Tokura}\ \emph {et~al.}(2014)\citenamefont {Tokura},
  \citenamefont {Seki},\ and\ \citenamefont
  {Nagaosa}}]{Tokura2014multiferroic}%
  \BibitemOpen
  \bibfield  {author} {\bibinfo {author} {\bibfnamefont {Y.}~\bibnamefont
  {Tokura}}, \bibinfo {author} {\bibfnamefont {S.}~\bibnamefont {Seki}}, \ and\
  \bibinfo {author} {\bibfnamefont {N.}~\bibnamefont {Nagaosa}},\ }\href
  {http://stacks.iop.org/0034-4885/77/i=7/a=076501} {\bibfield  {journal}
  {\bibinfo  {journal} {Rep. Prog. Phys.}\ }\textbf {\bibinfo {volume} {77}},\
  \bibinfo {pages} {076501} (\bibinfo {year} {2014})}\BibitemShut {NoStop}%
\bibitem [{\citenamefont {Schmid}(2008)}]{Schmid2008}%
  \BibitemOpen
  \bibfield  {author} {\bibinfo {author} {\bibfnamefont {H.}~\bibnamefont
  {Schmid}},\ }\href {\doibase 10.1088/0953-8984/20/43/434201} {\bibfield
  {journal} {\bibinfo  {journal} {J. Phys. Condens. Matter}\ }\textbf {\bibinfo
  {volume} {20}},\ \bibinfo {pages} {434201} (\bibinfo {year}
  {2008})}\BibitemShut {NoStop}%
\bibitem [{\citenamefont {Murakami}\ \emph {et~al.}(2003)\citenamefont
  {Murakami}, \citenamefont {Nagaosa},\ and\ \citenamefont
  {Zhang}}]{Murakami2003spinhall}%
  \BibitemOpen
  \bibfield  {author} {\bibinfo {author} {\bibfnamefont {S.}~\bibnamefont
  {Murakami}}, \bibinfo {author} {\bibfnamefont {N.}~\bibnamefont {Nagaosa}}, \
  and\ \bibinfo {author} {\bibfnamefont {S.-C.}\ \bibnamefont {Zhang}},\ }\href
  {\doibase 10.1126/science.1087128} {\bibfield  {journal} {\bibinfo  {journal}
  {Science}\ }\textbf {\bibinfo {volume} {301}},\ \bibinfo {pages} {1348}
  (\bibinfo {year} {2003})}\BibitemShut {NoStop}%
\bibitem [{\citenamefont {Sinova}\ \emph {et~al.}(2004)\citenamefont {Sinova},
  \citenamefont {Culcer}, \citenamefont {Niu}, \citenamefont {Sinitsyn},
  \citenamefont {Jungwirth},\ and\ \citenamefont {MacDonald}}]{Sinova2004}%
  \BibitemOpen
  \bibfield  {author} {\bibinfo {author} {\bibfnamefont {J.}~\bibnamefont
  {Sinova}}, \bibinfo {author} {\bibfnamefont {D.}~\bibnamefont {Culcer}},
  \bibinfo {author} {\bibfnamefont {Q.}~\bibnamefont {Niu}}, \bibinfo {author}
  {\bibfnamefont {N.~A.}\ \bibnamefont {Sinitsyn}}, \bibinfo {author}
  {\bibfnamefont {T.}~\bibnamefont {Jungwirth}}, \ and\ \bibinfo {author}
  {\bibfnamefont {A.~H.}\ \bibnamefont {MacDonald}},\ }\href {\doibase
  10.1103/PhysRevLett.92.126603} {\bibfield  {journal} {\bibinfo  {journal}
  {Phys. Rev. Lett.}\ }\textbf {\bibinfo {volume} {92}},\ \bibinfo {pages}
  {126603} (\bibinfo {year} {2004})}\BibitemShut {NoStop}%
\bibitem [{\citenamefont {Manchon}\ and\ \citenamefont
  {Zhang}(2008)}]{Manchon2008}%
  \BibitemOpen
  \bibfield  {author} {\bibinfo {author} {\bibfnamefont {A.}~\bibnamefont
  {Manchon}}\ and\ \bibinfo {author} {\bibfnamefont {S.}~\bibnamefont
  {Zhang}},\ }\href {\doibase 10.1103/PhysRevB.78.212405} {\bibfield  {journal}
  {\bibinfo  {journal} {Phys. Rev. B}\ }\textbf {\bibinfo {volume} {78}},\
  \bibinfo {pages} {212405} (\bibinfo {year} {2008})}\BibitemShut {NoStop}%
\bibitem [{\citenamefont {Manchon}\ and\ \citenamefont
  {Zhang}(2009)}]{Manchon2009}%
  \BibitemOpen
  \bibfield  {author} {\bibinfo {author} {\bibfnamefont {A.}~\bibnamefont
  {Manchon}}\ and\ \bibinfo {author} {\bibfnamefont {S.}~\bibnamefont
  {Zhang}},\ }\href {\doibase 10.1103/PhysRevB.79.094422} {\bibfield  {journal}
  {\bibinfo  {journal} {Phys. Rev. B}\ }\textbf {\bibinfo {volume} {79}},\
  \bibinfo {pages} {094422} (\bibinfo {year} {2009})}\BibitemShut {NoStop}%
\bibitem [{\citenamefont {Garate}\ and\ \citenamefont
  {MacDonald}(2009)}]{Garate2009}%
  \BibitemOpen
  \bibfield  {author} {\bibinfo {author} {\bibfnamefont {I.}~\bibnamefont
  {Garate}}\ and\ \bibinfo {author} {\bibfnamefont {A.~H.}\ \bibnamefont
  {MacDonald}},\ }\href {\doibase 10.1103/PhysRevB.80.134403} {\bibfield
  {journal} {\bibinfo  {journal} {Phys. Rev. B}\ }\textbf {\bibinfo {volume}
  {80}},\ \bibinfo {pages} {134403} (\bibinfo {year} {2009})}\BibitemShut
  {NoStop}%
\bibitem [{\citenamefont {Kuramoto}\ \emph {et~al.}(2009)\citenamefont
  {Kuramoto}, \citenamefont {Kusunose},\ and\ \citenamefont
  {Kiss}}]{Kuramoto2009Review}%
  \BibitemOpen
  \bibfield  {author} {\bibinfo {author} {\bibfnamefont {Y.}~\bibnamefont
  {Kuramoto}}, \bibinfo {author} {\bibfnamefont {H.}~\bibnamefont {Kusunose}},
  \ and\ \bibinfo {author} {\bibfnamefont {A.}~\bibnamefont {Kiss}},\ }\href
  {\doibase 10.1143/JPSJ.78.072001} {\bibfield  {journal} {\bibinfo  {journal}
  {J. Phys. Soc. Jpn.}\ }\textbf {\bibinfo {volume} {78}},\ \bibinfo {pages}
  {072001} (\bibinfo {year} {2009})}\BibitemShut {NoStop}%
\bibitem [{\citenamefont {Santini}\ \emph {et~al.}(2009)\citenamefont
  {Santini}, \citenamefont {Carretta}, \citenamefont {Amoretti}, \citenamefont
  {Caciuffo}, \citenamefont {Magnani},\ and\ \citenamefont
  {Lander}}]{Santini2009}%
  \BibitemOpen
  \bibfield  {author} {\bibinfo {author} {\bibfnamefont {P.}~\bibnamefont
  {Santini}}, \bibinfo {author} {\bibfnamefont {S.}~\bibnamefont {Carretta}},
  \bibinfo {author} {\bibfnamefont {G.}~\bibnamefont {Amoretti}}, \bibinfo
  {author} {\bibfnamefont {R.}~\bibnamefont {Caciuffo}}, \bibinfo {author}
  {\bibfnamefont {N.}~\bibnamefont {Magnani}}, \ and\ \bibinfo {author}
  {\bibfnamefont {G.~H.}\ \bibnamefont {Lander}},\ }\href {\doibase
  10.1103/RevModPhys.81.807} {\bibfield  {journal} {\bibinfo  {journal} {Rev.
  Mod. Phys.}\ }\textbf {\bibinfo {volume} {81}},\ \bibinfo {pages} {807}
  (\bibinfo {year} {2009})}\BibitemShut {NoStop}%
\bibitem [{\citenamefont {Koga}\ \emph {et~al.}(2006)\citenamefont {Koga},
  \citenamefont {Matsumoto},\ and\ \citenamefont {Shiba}}]{Koga2006}%
  \BibitemOpen
  \bibfield  {author} {\bibinfo {author} {\bibfnamefont {M.}~\bibnamefont
  {Koga}}, \bibinfo {author} {\bibfnamefont {M.}~\bibnamefont {Matsumoto}}, \
  and\ \bibinfo {author} {\bibfnamefont {H.}~\bibnamefont {Shiba}},\ }\href
  {\doibase 10.1143/JPSJ.75.014709} {\bibfield  {journal} {\bibinfo  {journal}
  {J. Phys. Soc. Jpn.}\ }\textbf {\bibinfo {volume} {75}},\ \bibinfo {pages}
  {014709} (\bibinfo {year} {2006})}\BibitemShut {NoStop}%
\bibitem [{\citenamefont {Matsubayashi}\ \emph {et~al.}(2012)\citenamefont
  {Matsubayashi}, \citenamefont {Tanaka}, \citenamefont {Sakai}, \citenamefont
  {Nakatsuji}, \citenamefont {Kubo},\ and\ \citenamefont
  {Uwatoko}}]{Matsubayachi2012MultipolarSC}%
  \BibitemOpen
  \bibfield  {author} {\bibinfo {author} {\bibfnamefont {K.}~\bibnamefont
  {Matsubayashi}}, \bibinfo {author} {\bibfnamefont {T.}~\bibnamefont
  {Tanaka}}, \bibinfo {author} {\bibfnamefont {A.}~\bibnamefont {Sakai}},
  \bibinfo {author} {\bibfnamefont {S.}~\bibnamefont {Nakatsuji}}, \bibinfo
  {author} {\bibfnamefont {Y.}~\bibnamefont {Kubo}}, \ and\ \bibinfo {author}
  {\bibfnamefont {Y.}~\bibnamefont {Uwatoko}},\ }\href {\doibase
  10.1103/PhysRevLett.109.187004} {\bibfield  {journal} {\bibinfo  {journal}
  {Phys. Rev. Lett.}\ }\textbf {\bibinfo {volume} {109}},\ \bibinfo {pages}
  {187004} (\bibinfo {year} {2012})}\BibitemShut {NoStop}%
\bibitem [{\citenamefont {Cox}\ and\ \citenamefont
  {Zawadowski}(1998)}]{cox1998exotic}%
  \BibitemOpen
  \bibfield  {author} {\bibinfo {author} {\bibfnamefont {D.~L.}\ \bibnamefont
  {Cox}}\ and\ \bibinfo {author} {\bibfnamefont {A.}~\bibnamefont
  {Zawadowski}},\ }\href {\doibase 10.1080/000187398243500} {\bibfield
  {journal} {\bibinfo  {journal} {Advances in Physics}\ }\textbf {\bibinfo
  {volume} {47}},\ \bibinfo {pages} {599} (\bibinfo {year} {1998})}\BibitemShut
  {NoStop}%
\bibitem [{\citenamefont {Onimaru}\ and\ \citenamefont
  {Kusunose}(2016)}]{Onimaru2016QuadKondo}%
  \BibitemOpen
  \bibfield  {author} {\bibinfo {author} {\bibfnamefont {T.}~\bibnamefont
  {Onimaru}}\ and\ \bibinfo {author} {\bibfnamefont {H.}~\bibnamefont
  {Kusunose}},\ }\href {\doibase 10.7566/JPSJ.85.082002} {\bibfield  {journal}
  {\bibinfo  {journal} {Journal of the Physical Society of Japan}\ }\textbf
  {\bibinfo {volume} {85}},\ \bibinfo {pages} {082002} (\bibinfo {year}
  {2016})}\BibitemShut {NoStop}%
\bibitem [{\citenamefont {Yamane}\ \emph {et~al.}(2018)\citenamefont {Yamane},
  \citenamefont {Onimaru}, \citenamefont {Wakiya}, \citenamefont {Matsumoto},
  \citenamefont {Umeo},\ and\ \citenamefont
  {Takabatake}}]{Yamane2018QuadKondo}%
  \BibitemOpen
  \bibfield  {author} {\bibinfo {author} {\bibfnamefont {Y.}~\bibnamefont
  {Yamane}}, \bibinfo {author} {\bibfnamefont {T.}~\bibnamefont {Onimaru}},
  \bibinfo {author} {\bibfnamefont {K.}~\bibnamefont {Wakiya}}, \bibinfo
  {author} {\bibfnamefont {K.~T.}\ \bibnamefont {Matsumoto}}, \bibinfo {author}
  {\bibfnamefont {K.}~\bibnamefont {Umeo}}, \ and\ \bibinfo {author}
  {\bibfnamefont {T.}~\bibnamefont {Takabatake}},\ }\href {\doibase
  10.1103/PhysRevLett.121.077206} {\bibfield  {journal} {\bibinfo  {journal}
  {Phys. Rev. Lett.}\ }\textbf {\bibinfo {volume} {121}},\ \bibinfo {pages}
  {077206} (\bibinfo {year} {2018})}\BibitemShut {NoStop}%
\bibitem [{\citenamefont {Hitomi}\ and\ \citenamefont
  {Yanase}(2014)}]{Hitomi2014}%
  \BibitemOpen
  \bibfield  {author} {\bibinfo {author} {\bibfnamefont {T.}~\bibnamefont
  {Hitomi}}\ and\ \bibinfo {author} {\bibfnamefont {Y.}~\bibnamefont
  {Yanase}},\ }\href {\doibase 10.7566/JPSJ.83.114704} {\bibfield  {journal}
  {\bibinfo  {journal} {J. Phys. Soc. Jpn.}\ }\textbf {\bibinfo {volume}
  {83}},\ \bibinfo {pages} {114704} (\bibinfo {year} {2014})}\BibitemShut
  {NoStop}%
\bibitem [{\citenamefont {Hitomi}\ and\ \citenamefont
  {Yanase}(2016)}]{Hitomi2016}%
  \BibitemOpen
  \bibfield  {author} {\bibinfo {author} {\bibfnamefont {T.}~\bibnamefont
  {Hitomi}}\ and\ \bibinfo {author} {\bibfnamefont {Y.}~\bibnamefont
  {Yanase}},\ }\href {\doibase 10.7566/JPSJ.85.124702} {\bibfield  {journal}
  {\bibinfo  {journal} {J. Phys. Soc. Jpn.}\ }\textbf {\bibinfo {volume}
  {85}},\ \bibinfo {pages} {124702} (\bibinfo {year} {2016})}\BibitemShut
  {NoStop}%
\bibitem [{\citenamefont {Sumita}\ and\ \citenamefont
  {Yanase}(2016)}]{Sumita2016}%
  \BibitemOpen
  \bibfield  {author} {\bibinfo {author} {\bibfnamefont {S.}~\bibnamefont
  {Sumita}}\ and\ \bibinfo {author} {\bibfnamefont {Y.}~\bibnamefont
  {Yanase}},\ }\href {\doibase 10.1103/PhysRevB.93.224507} {\bibfield
  {journal} {\bibinfo  {journal} {Phys. Rev. B}\ }\textbf {\bibinfo {volume}
  {93}},\ \bibinfo {pages} {224507} (\bibinfo {year} {2016})}\BibitemShut
  {NoStop}%
\bibitem [{\citenamefont {Spaldin}\ \emph {et~al.}(2013)\citenamefont
  {Spaldin}, \citenamefont {Fechner}, \citenamefont {Bousquet}, \citenamefont
  {Balatsky},\ and\ \citenamefont {Nordstr{\"{o}}m}}]{Spaldin2013}%
  \BibitemOpen
  \bibfield  {author} {\bibinfo {author} {\bibfnamefont {N.~A.}\ \bibnamefont
  {Spaldin}}, \bibinfo {author} {\bibfnamefont {M.}~\bibnamefont {Fechner}},
  \bibinfo {author} {\bibfnamefont {E.}~\bibnamefont {Bousquet}}, \bibinfo
  {author} {\bibfnamefont {A.}~\bibnamefont {Balatsky}}, \ and\ \bibinfo
  {author} {\bibfnamefont {L.}~\bibnamefont {Nordstr{\"{o}}m}},\ }\href
  {\doibase 10.1103/PhysRevB.88.094429} {\bibfield  {journal} {\bibinfo
  {journal} {Phys. Rev. B}\ }\textbf {\bibinfo {volume} {88}},\ \bibinfo
  {pages} {094429} (\bibinfo {year} {2013})}\BibitemShut {NoStop}%
\bibitem [{\citenamefont {Th{\"{o}}le}\ \emph {et~al.}(2016)\citenamefont
  {Th{\"{o}}le}, \citenamefont {Fechner},\ and\ \citenamefont
  {Spaldin}}]{Th2016}%
  \BibitemOpen
  \bibfield  {author} {\bibinfo {author} {\bibfnamefont {F.}~\bibnamefont
  {Th{\"{o}}le}}, \bibinfo {author} {\bibfnamefont {M.}~\bibnamefont
  {Fechner}}, \ and\ \bibinfo {author} {\bibfnamefont {N.~A.}\ \bibnamefont
  {Spaldin}},\ }\href {\doibase 10.1103/PhysRevB.93.195167} {\bibfield
  {journal} {\bibinfo  {journal} {Phys. Rev. B}\ }\textbf {\bibinfo {volume}
  {93}},\ \bibinfo {pages} {195167} (\bibinfo {year} {2016})}\BibitemShut
  {NoStop}%
\bibitem [{\citenamefont {Yanase}(2014)}]{Yanase2014}%
  \BibitemOpen
  \bibfield  {author} {\bibinfo {author} {\bibfnamefont {Y.}~\bibnamefont
  {Yanase}},\ }\href {\doibase 10.7566/JPSJ.83.014703} {\bibfield  {journal}
  {\bibinfo  {journal} {J. Phys. Soc. Jpn.}\ }\textbf {\bibinfo {volume}
  {83}},\ \bibinfo {pages} {014703} (\bibinfo {year} {2014})}\BibitemShut
  {NoStop}%
\bibitem [{\citenamefont {Sumita}\ \emph {et~al.}(2017)\citenamefont {Sumita},
  \citenamefont {Nomoto},\ and\ \citenamefont {Yanase}}]{Sumita2017}%
  \BibitemOpen
  \bibfield  {author} {\bibinfo {author} {\bibfnamefont {S.}~\bibnamefont
  {Sumita}}, \bibinfo {author} {\bibfnamefont {T.}~\bibnamefont {Nomoto}}, \
  and\ \bibinfo {author} {\bibfnamefont {Y.}~\bibnamefont {Yanase}},\ }\href
  {\doibase 10.1103/PhysRevLett.119.027001} {\bibfield  {journal} {\bibinfo
  {journal} {Phys. Rev. Lett.}\ }\textbf {\bibinfo {volume} {119}},\ \bibinfo
  {pages} {027001} (\bibinfo {year} {2017})}\BibitemShut {NoStop}%
\bibitem [{\citenamefont {Fechner}\ \emph {et~al.}(2016)\citenamefont
  {Fechner}, \citenamefont {Fierz}, \citenamefont {Th{\"{o}}le}, \citenamefont
  {Staub},\ and\ \citenamefont {Spaldin}}]{Fechner2016}%
  \BibitemOpen
  \bibfield  {author} {\bibinfo {author} {\bibfnamefont {M.}~\bibnamefont
  {Fechner}}, \bibinfo {author} {\bibfnamefont {M.~J.~A.}\ \bibnamefont
  {Fierz}}, \bibinfo {author} {\bibfnamefont {F.}~\bibnamefont {Th{\"{o}}le}},
  \bibinfo {author} {\bibfnamefont {U.}~\bibnamefont {Staub}}, \ and\ \bibinfo
  {author} {\bibfnamefont {N.~A.}\ \bibnamefont {Spaldin}},\ }\href {\doibase
  10.1103/PhysRevB.93.174419} {\bibfield  {journal} {\bibinfo  {journal} {Phys.
  Rev. B}\ }\textbf {\bibinfo {volume} {93}},\ \bibinfo {pages} {174419}
  (\bibinfo {year} {2016})}\BibitemShut {NoStop}%
\bibitem [{\citenamefont {{Di Matteo}}\ and\ \citenamefont
  {Norman}(2017)}]{DiMatteo2017b}%
  \BibitemOpen
  \bibfield  {author} {\bibinfo {author} {\bibfnamefont {S.}~\bibnamefont {{Di
  Matteo}}}\ and\ \bibinfo {author} {\bibfnamefont {M.~R.}\ \bibnamefont
  {Norman}},\ }\href {\doibase 10.1103/PhysRevB.96.115156} {\bibfield
  {journal} {\bibinfo  {journal} {Phys. Rev. B}\ }\textbf {\bibinfo {volume}
  {96}},\ \bibinfo {pages} {115156} (\bibinfo {year} {2017})}\BibitemShut
  {NoStop}%
\bibitem [{\citenamefont {Clin}\ \emph {et~al.}(1991)\citenamefont {Clin},
  \citenamefont {Schmid}, \citenamefont {Schobinger},\ and\ \citenamefont
  {Fischer}}]{Clin1991}%
  \BibitemOpen
  \bibfield  {author} {\bibinfo {author} {\bibfnamefont {M.}~\bibnamefont
  {Clin}}, \bibinfo {author} {\bibfnamefont {H.}~\bibnamefont {Schmid}},
  \bibinfo {author} {\bibfnamefont {P.}~\bibnamefont {Schobinger}}, \ and\
  \bibinfo {author} {\bibfnamefont {P.}~\bibnamefont {Fischer}},\ }\href
  {\doibase 10.1080/01411599108207726} {\bibfield  {journal} {\bibinfo
  {journal} {Phase Transitions}\ }\textbf {\bibinfo {volume} {33}},\ \bibinfo
  {pages} {149} (\bibinfo {year} {1991})}\BibitemShut {NoStop}%
\bibitem [{\citenamefont {Popov}\ \emph {et~al.}(1998)\citenamefont {Popov},
  \citenamefont {Kadomtseva}, \citenamefont {Vorob'ev}, \citenamefont
  {Timofeeva}, \citenamefont {Ustinin}, \citenamefont {Zvezdin},\ and\
  \citenamefont {Tegeranchi}}]{Popov1998}%
  \BibitemOpen
  \bibfield  {author} {\bibinfo {author} {\bibfnamefont {Y.~F.}\ \bibnamefont
  {Popov}}, \bibinfo {author} {\bibfnamefont {A.~M.}\ \bibnamefont
  {Kadomtseva}}, \bibinfo {author} {\bibfnamefont {G.~P.}\ \bibnamefont
  {Vorob'ev}}, \bibinfo {author} {\bibfnamefont {V.~A.}\ \bibnamefont
  {Timofeeva}}, \bibinfo {author} {\bibfnamefont {D.~M.}\ \bibnamefont
  {Ustinin}}, \bibinfo {author} {\bibfnamefont {A.~K.}\ \bibnamefont
  {Zvezdin}}, \ and\ \bibinfo {author} {\bibfnamefont {M.~M.}\ \bibnamefont
  {Tegeranchi}},\ }\href {\doibase 10.1134/1.558635} {\bibfield  {journal}
  {\bibinfo  {journal} {J. Exp. Theor. Phys.}\ }\textbf {\bibinfo {volume}
  {87}},\ \bibinfo {pages} {146} (\bibinfo {year} {1998})}\BibitemShut
  {NoStop}%
\bibitem [{\citenamefont {Arima}\ \emph {et~al.}(2005)\citenamefont {Arima},
  \citenamefont {Jung}, \citenamefont {Matsubara}, \citenamefont {Kubota},
  \citenamefont {He}, \citenamefont {Kaneko},\ and\ \citenamefont
  {Tokura}}]{Arima2005}%
  \BibitemOpen
  \bibfield  {author} {\bibinfo {author} {\bibfnamefont {T.-h.}\ \bibnamefont
  {Arima}}, \bibinfo {author} {\bibfnamefont {J.-H.}\ \bibnamefont {Jung}},
  \bibinfo {author} {\bibfnamefont {M.}~\bibnamefont {Matsubara}}, \bibinfo
  {author} {\bibfnamefont {M.}~\bibnamefont {Kubota}}, \bibinfo {author}
  {\bibfnamefont {J.-P.}\ \bibnamefont {He}}, \bibinfo {author} {\bibfnamefont
  {Y.}~\bibnamefont {Kaneko}}, \ and\ \bibinfo {author} {\bibfnamefont
  {Y.}~\bibnamefont {Tokura}},\ }\href {\doibase 10.1143/JPSJ.74.1419}
  {\bibfield  {journal} {\bibinfo  {journal} {J. Phys. Soc. Jpn.}\ }\textbf
  {\bibinfo {volume} {74}},\ \bibinfo {pages} {1419} (\bibinfo {year}
  {2005})}\BibitemShut {NoStop}%
\bibitem [{\citenamefont {{Van Aken}}\ \emph {et~al.}(2007)\citenamefont {{Van
  Aken}}, \citenamefont {Rivera}, \citenamefont {Schmid},\ and\ \citenamefont
  {Fiebig}}]{VanAken2007observation}%
  \BibitemOpen
  \bibfield  {author} {\bibinfo {author} {\bibfnamefont {B.~B.}\ \bibnamefont
  {{Van Aken}}}, \bibinfo {author} {\bibfnamefont {J.-P.}\ \bibnamefont
  {Rivera}}, \bibinfo {author} {\bibfnamefont {H.}~\bibnamefont {Schmid}}, \
  and\ \bibinfo {author} {\bibfnamefont {M.}~\bibnamefont {Fiebig}},\ }\href
  {\doibase 10.1038/nature06139} {\bibfield  {journal} {\bibinfo  {journal}
  {Nature}\ }\textbf {\bibinfo {volume} {449}},\ \bibinfo {pages} {702}
  (\bibinfo {year} {2007})}\BibitemShut {NoStop}%
\bibitem [{\citenamefont {Zimmermann}\ \emph {et~al.}(2014)\citenamefont
  {Zimmermann}, \citenamefont {Meier},\ and\ \citenamefont
  {Fiebig}}]{Zimmermann2014a}%
  \BibitemOpen
  \bibfield  {author} {\bibinfo {author} {\bibfnamefont {A.~S.}\ \bibnamefont
  {Zimmermann}}, \bibinfo {author} {\bibfnamefont {D.}~\bibnamefont {Meier}}, \
  and\ \bibinfo {author} {\bibfnamefont {M.}~\bibnamefont {Fiebig}},\ }\href
  {\doibase 10.1038/ncomms5796} {\bibfield  {journal} {\bibinfo  {journal}
  {Nat. Commun.}\ }\textbf {\bibinfo {volume} {5}},\ \bibinfo {pages} {4796}
  (\bibinfo {year} {2014})}\BibitemShut {NoStop}%
\bibitem [{\citenamefont {Hayami}\ \emph {et~al.}(2014)\citenamefont {Hayami},
  \citenamefont {Kusunose},\ and\ \citenamefont {Motome}}]{Hayami2014b}%
  \BibitemOpen
  \bibfield  {author} {\bibinfo {author} {\bibfnamefont {S.}~\bibnamefont
  {Hayami}}, \bibinfo {author} {\bibfnamefont {H.}~\bibnamefont {Kusunose}}, \
  and\ \bibinfo {author} {\bibfnamefont {Y.}~\bibnamefont {Motome}},\ }\href
  {\doibase 10.1103/PhysRevB.90.024432} {\bibfield  {journal} {\bibinfo
  {journal} {Phys. Rev. B}\ }\textbf {\bibinfo {volume} {90}},\ \bibinfo
  {pages} {024432} (\bibinfo {year} {2014})}\BibitemShut {NoStop}%
\bibitem [{\citenamefont {Ederer}\ and\ \citenamefont
  {Spaldin}(2007)}]{Ederer2007}%
  \BibitemOpen
  \bibfield  {author} {\bibinfo {author} {\bibfnamefont {C.}~\bibnamefont
  {Ederer}}\ and\ \bibinfo {author} {\bibfnamefont {N.~A.}\ \bibnamefont
  {Spaldin}},\ }\href {\doibase 10.1103/PhysRevB.76.214404} {\bibfield
  {journal} {\bibinfo  {journal} {Phys. Rev. B}\ }\textbf {\bibinfo {volume}
  {76}},\ \bibinfo {pages} {214404} (\bibinfo {year} {2007})}\BibitemShut
  {NoStop}%
\bibitem [{\citenamefont {Spaldin}\ \emph {et~al.}(2008)\citenamefont
  {Spaldin}, \citenamefont {Fiebig},\ and\ \citenamefont
  {Mostovoy}}]{Spaldin2008}%
  \BibitemOpen
  \bibfield  {author} {\bibinfo {author} {\bibfnamefont {N.~A.}\ \bibnamefont
  {Spaldin}}, \bibinfo {author} {\bibfnamefont {M.}~\bibnamefont {Fiebig}}, \
  and\ \bibinfo {author} {\bibfnamefont {M.}~\bibnamefont {Mostovoy}},\ }\href
  {\doibase 10.1088/0953-8984/20/43/434203} {\bibfield  {journal} {\bibinfo
  {journal} {J. Phys. Condens. Matter}\ }\textbf {\bibinfo {volume} {20}},\
  \bibinfo {pages} {434203} (\bibinfo {year} {2008})}\BibitemShut {NoStop}%
\bibitem [{\citenamefont {Saito}\ \emph {et~al.}(2018)\citenamefont {Saito},
  \citenamefont {Uenishi}, \citenamefont {Miura}, \citenamefont {Tabata},
  \citenamefont {Hidaka}, \citenamefont {Yanagisawa},\ and\ \citenamefont
  {Amitsuka}}]{Saito2018CurrentInduced}%
  \BibitemOpen
  \bibfield  {author} {\bibinfo {author} {\bibfnamefont {H.}~\bibnamefont
  {Saito}}, \bibinfo {author} {\bibfnamefont {K.}~\bibnamefont {Uenishi}},
  \bibinfo {author} {\bibfnamefont {N.}~\bibnamefont {Miura}}, \bibinfo
  {author} {\bibfnamefont {C.}~\bibnamefont {Tabata}}, \bibinfo {author}
  {\bibfnamefont {H.}~\bibnamefont {Hidaka}}, \bibinfo {author} {\bibfnamefont
  {T.}~\bibnamefont {Yanagisawa}}, \ and\ \bibinfo {author} {\bibfnamefont
  {H.}~\bibnamefont {Amitsuka}},\ }\href {\doibase 10.7566/JPSJ.87.033702}
  {\bibfield  {journal} {\bibinfo  {journal} {J. Phys. Soc. Jpn.}\ }\textbf
  {\bibinfo {volume} {87}},\ \bibinfo {pages} {033702} (\bibinfo {year}
  {2018})}\BibitemShut {NoStop}%
\bibitem [{\citenamefont {Watanabe}\ and\ \citenamefont
  {Yanase}(2017)}]{hikaruwatanabe2017}%
  \BibitemOpen
  \bibfield  {author} {\bibinfo {author} {\bibfnamefont {H.}~\bibnamefont
  {Watanabe}}\ and\ \bibinfo {author} {\bibfnamefont {Y.}~\bibnamefont
  {Yanase}},\ }\href {\doibase 10.1103/PhysRevB.96.064432} {\bibfield
  {journal} {\bibinfo  {journal} {Phys. Rev. B}\ }\textbf {\bibinfo {volume}
  {96}},\ \bibinfo {pages} {064432} (\bibinfo {year} {2017})}\BibitemShut
  {NoStop}%
\bibitem [{\citenamefont {Fu}(2015)}]{Fu2015}%
  \BibitemOpen
  \bibfield  {author} {\bibinfo {author} {\bibfnamefont {L.}~\bibnamefont
  {Fu}},\ }\href {\doibase 10.1103/PhysRevLett.115.026401} {\bibfield
  {journal} {\bibinfo  {journal} {Phys. Rev. Lett.}\ }\textbf {\bibinfo
  {volume} {115}},\ \bibinfo {pages} {026401} (\bibinfo {year}
  {2015})}\BibitemShut {NoStop}%
\bibitem [{\citenamefont {Varshalovich}\ \emph {et~al.}(1988)\citenamefont
  {Varshalovich}, \citenamefont {Moskalev},\ and\ \citenamefont
  {Khersonskii}}]{varshalovich1988quantum}%
  \BibitemOpen
  \bibfield  {author} {\bibinfo {author} {\bibfnamefont {D.~A.}\ \bibnamefont
  {Varshalovich}}, \bibinfo {author} {\bibfnamefont {A.~N.}\ \bibnamefont
  {Moskalev}}, \ and\ \bibinfo {author} {\bibfnamefont {V.~K.}\ \bibnamefont
  {Khersonskii}},\ }\href@noop {} {\emph {\bibinfo {title} {Quantum theory of
  angular momentum}}}\ (\bibinfo  {publisher} {World Scientific},\ \bibinfo
  {year} {1988})\BibitemShut {NoStop}%
\bibitem [{\citenamefont {Dubovik}\ and\ \citenamefont
  {Tugushev}(1990)}]{Dubovik1990a}%
  \BibitemOpen
  \bibfield  {author} {\bibinfo {author} {\bibfnamefont {V.~M.}\ \bibnamefont
  {Dubovik}}\ and\ \bibinfo {author} {\bibfnamefont {V.~V.}\ \bibnamefont
  {Tugushev}},\ }\href {\doibase 10.1016/0370-1573(90)90042-Z} {\bibfield
  {journal} {\bibinfo  {journal} {Phys. Rep.}\ }\textbf {\bibinfo {volume}
  {187}},\ \bibinfo {pages} {145} (\bibinfo {year} {1990})}\BibitemShut
  {NoStop}%
\bibitem [{\citenamefont {Hayami}\ and\ \citenamefont
  {Kusunose}(2018)}]{Hayami2017MultipoleExpansion}%
  \BibitemOpen
  \bibfield  {author} {\bibinfo {author} {\bibfnamefont {S.}~\bibnamefont
  {Hayami}}\ and\ \bibinfo {author} {\bibfnamefont {H.}~\bibnamefont
  {Kusunose}},\ }\href {\doibase 10.7566/JPSJ.87.033709} {\bibfield  {journal}
  {\bibinfo  {journal} {J. Phys. Soc. Jpn.}\ }\textbf {\bibinfo {volume}
  {87}},\ \bibinfo {pages} {033709} (\bibinfo {year} {2018})}\BibitemShut
  {NoStop}%
\bibitem [{\citenamefont {Gao}\ \emph {et~al.}(2018)\citenamefont {Gao},
  \citenamefont {Vanderbilt},\ and\ \citenamefont {Xiao}}]{Gao2018spin}%
  \BibitemOpen
  \bibfield  {author} {\bibinfo {author} {\bibfnamefont {Y.}~\bibnamefont
  {Gao}}, \bibinfo {author} {\bibfnamefont {D.}~\bibnamefont {Vanderbilt}}, \
  and\ \bibinfo {author} {\bibfnamefont {D.}~\bibnamefont {Xiao}},\ }\href
  {\doibase 10.1103/PhysRevB.97.134423} {\bibfield  {journal} {\bibinfo
  {journal} {Phys. Rev. B}\ }\textbf {\bibinfo {volume} {97}},\ \bibinfo
  {pages} {134423} (\bibinfo {year} {2018})}\BibitemShut {NoStop}%
\bibitem [{\citenamefont {Shitade}\ \emph {et~al.}(2018)\citenamefont
  {Shitade}, \citenamefont {Watanabe},\ and\ \citenamefont
  {Yanase}}]{Shitade2018theory}%
  \BibitemOpen
  \bibfield  {author} {\bibinfo {author} {\bibfnamefont {A.}~\bibnamefont
  {Shitade}}, \bibinfo {author} {\bibfnamefont {H.}~\bibnamefont {Watanabe}}, \
  and\ \bibinfo {author} {\bibfnamefont {Y.}~\bibnamefont {Yanase}},\ }\href
  {\doibase 10.1103/PhysRevB.98.020407} {\bibfield  {journal} {\bibinfo
  {journal} {Phys. Rev. B}\ }\textbf {\bibinfo {volume} {98}},\ \bibinfo
  {pages} {020407} (\bibinfo {year} {2018})}\BibitemShut {NoStop}%
\bibitem [{\citenamefont {Gao}\ and\ \citenamefont
  {Xiao}(2018)}]{gao2018theory}%
  \BibitemOpen
  \bibfield  {author} {\bibinfo {author} {\bibfnamefont {Y.}~\bibnamefont
  {Gao}}\ and\ \bibinfo {author} {\bibfnamefont {D.}~\bibnamefont {Xiao}},\
  }\href {\doibase 10.1103/PhysRevB.98.060402} {\bibfield  {journal} {\bibinfo
  {journal} {Phys. Rev. B}\ }\textbf {\bibinfo {volume} {98}},\ \bibinfo
  {pages} {060402} (\bibinfo {year} {2018})}\BibitemShut {NoStop}%
\bibitem [{\citenamefont {Kusunose}(2008)}]{Kusunose2008}%
  \BibitemOpen
  \bibfield  {author} {\bibinfo {author} {\bibfnamefont {H.}~\bibnamefont
  {Kusunose}},\ }\href {\doibase 10.1143/JPSJ.77.064710} {\bibfield  {journal}
  {\bibinfo  {journal} {J. Phys. Soc. Jpn.}\ }\textbf {\bibinfo {volume}
  {77}},\ \bibinfo {pages} {064710} (\bibinfo {year} {2008})}\BibitemShut
  {NoStop}%
\bibitem [{\citenamefont {Schwartz}(1957)}]{Schwartz1957}%
  \BibitemOpen
  \bibfield  {author} {\bibinfo {author} {\bibfnamefont {C.}~\bibnamefont
  {Schwartz}},\ }\href {\doibase 10.1103/PhysRev.105.173} {\bibfield  {journal}
  {\bibinfo  {journal} {Phys. Rev.}\ }\textbf {\bibinfo {volume} {105}},\
  \bibinfo {pages} {173} (\bibinfo {year} {1957})}\BibitemShut {NoStop}%
\bibitem [{\citenamefont {Sigrist}\ and\ \citenamefont
  {Ueda}(1991)}]{Sigrist1991}%
  \BibitemOpen
  \bibfield  {author} {\bibinfo {author} {\bibfnamefont {M.}~\bibnamefont
  {Sigrist}}\ and\ \bibinfo {author} {\bibfnamefont {K.}~\bibnamefont {Ueda}},\
  }\href {\doibase 10.1103/RevModPhys.63.239} {\bibfield  {journal} {\bibinfo
  {journal} {Rev. Mod. Phys.}\ }\textbf {\bibinfo {volume} {63}},\ \bibinfo
  {pages} {239} (\bibinfo {year} {1991})}\BibitemShut {NoStop}%
\bibitem [{\citenamefont {Inui}\ \emph {et~al.}(1990)\citenamefont {Inui},
  \citenamefont {Tanabe},\ and\ \citenamefont {Onodera}}]{inui1990group}%
  \BibitemOpen
  \bibfield  {author} {\bibinfo {author} {\bibfnamefont {T.}~\bibnamefont
  {Inui}}, \bibinfo {author} {\bibfnamefont {Y.}~\bibnamefont {Tanabe}}, \ and\
  \bibinfo {author} {\bibfnamefont {Y.}~\bibnamefont {Onodera}},\ }\href@noop
  {} {\emph {\bibinfo {title} {Group theory and its applications in
  physics}}},\ Vol.~\bibinfo {volume} {78}\ (\bibinfo  {publisher} {Springer
  Science \& Business Media},\ \bibinfo {year} {1990})\BibitemShut {NoStop}%
\bibitem [{\citenamefont {Shiina}\ \emph {et~al.}(1997)\citenamefont {Shiina},
  \citenamefont {Shiba},\ and\ \citenamefont {Thalmeier}}]{Shiina1997b}%
  \BibitemOpen
  \bibfield  {author} {\bibinfo {author} {\bibfnamefont {R.}~\bibnamefont
  {Shiina}}, \bibinfo {author} {\bibfnamefont {H.}~\bibnamefont {Shiba}}, \
  and\ \bibinfo {author} {\bibfnamefont {P.}~\bibnamefont {Thalmeier}},\ }\href
  {\doibase 10.1143/JPSJ.66.1741} {\bibfield  {journal} {\bibinfo  {journal}
  {J. Phys. Soc. Jpn.}\ }\textbf {\bibinfo {volume} {66}},\ \bibinfo {pages}
  {1741} (\bibinfo {year} {1997})}\BibitemShut {NoStop}%
\bibitem [{\citenamefont {Nissinen}\ \emph {et~al.}(2016)\citenamefont
  {Nissinen}, \citenamefont {Liu}, \citenamefont {Slager}, \citenamefont {Wu},\
  and\ \citenamefont {Zaanen}}]{Liu2016GaugeTheoreticClassification}%
  \BibitemOpen
  \bibfield  {author} {\bibinfo {author} {\bibfnamefont {J.}~\bibnamefont
  {Nissinen}}, \bibinfo {author} {\bibfnamefont {K.}~\bibnamefont {Liu}},
  \bibinfo {author} {\bibfnamefont {R.-J.}\ \bibnamefont {Slager}}, \bibinfo
  {author} {\bibfnamefont {K.}~\bibnamefont {Wu}}, \ and\ \bibinfo {author}
  {\bibfnamefont {J.}~\bibnamefont {Zaanen}},\ }\href {\doibase
  10.1103/PhysRevE.94.022701} {\bibfield  {journal} {\bibinfo  {journal} {Phys.
  Rev. E}\ }\textbf {\bibinfo {volume} {94}},\ \bibinfo {pages} {022701}
  (\bibinfo {year} {2016})}\BibitemShut {NoStop}%
\bibitem [{\citenamefont {Chu}\ \emph {et~al.}(2012)\citenamefont {Chu},
  \citenamefont {Kuo}, \citenamefont {Analytis},\ and\ \citenamefont
  {Fisher}}]{Chu2012}%
  \BibitemOpen
  \bibfield  {author} {\bibinfo {author} {\bibfnamefont {J.-H.}\ \bibnamefont
  {Chu}}, \bibinfo {author} {\bibfnamefont {H.-H.}\ \bibnamefont {Kuo}},
  \bibinfo {author} {\bibfnamefont {J.~G.}\ \bibnamefont {Analytis}}, \ and\
  \bibinfo {author} {\bibfnamefont {I.~R.}\ \bibnamefont {Fisher}},\ }\href
  {\doibase 10.1126/science.1221713} {\bibfield  {journal} {\bibinfo  {journal}
  {Science}\ }\textbf {\bibinfo {volume} {337}},\ \bibinfo {pages} {710}
  (\bibinfo {year} {2012})}\BibitemShut {NoStop}%
\bibitem [{\citenamefont {Fernandes}\ \emph {et~al.}(2014)\citenamefont
  {Fernandes}, \citenamefont {Chubukov}, \citenamefont {Schmalian},\ and\
  \citenamefont {Article}}]{Fernandes2014}%
  \BibitemOpen
  \bibfield  {author} {\bibinfo {author} {\bibfnamefont {R.~M.}\ \bibnamefont
  {Fernandes}}, \bibinfo {author} {\bibfnamefont {a.~V.}\ \bibnamefont
  {Chubukov}}, \bibinfo {author} {\bibfnamefont {J.}~\bibnamefont {Schmalian}},
  \ and\ \bibinfo {author} {\bibfnamefont {R.}~\bibnamefont {Article}},\ }\href
  {\doibase 10.1038/nphys2877} {\bibfield  {journal} {\bibinfo  {journal} {Nat.
  Phys.}\ }\textbf {\bibinfo {volume} {10}},\ \bibinfo {pages} {97} (\bibinfo
  {year} {2014})}\BibitemShut {NoStop}%
\bibitem [{\citenamefont {Hayami}\ \emph {et~al.}(2016)\citenamefont {Hayami},
  \citenamefont {Kusunose},\ and\ \citenamefont {Motome}}]{Hayami2016}%
  \BibitemOpen
  \bibfield  {author} {\bibinfo {author} {\bibfnamefont {S.}~\bibnamefont
  {Hayami}}, \bibinfo {author} {\bibfnamefont {H.}~\bibnamefont {Kusunose}}, \
  and\ \bibinfo {author} {\bibfnamefont {Y.}~\bibnamefont {Motome}},\ }\href
  {\doibase 10.7566/JPSJ.85.053705} {\bibfield  {journal} {\bibinfo  {journal}
  {J. Phys. Soc. Jpn.}\ }\textbf {\bibinfo {volume} {85}},\ \bibinfo {pages}
  {053705} (\bibinfo {year} {2016})}\BibitemShut {NoStop}%
\bibitem [{\citenamefont {Nakatsuji}\ \emph {et~al.}(2015)\citenamefont
  {Nakatsuji}, \citenamefont {Kiyohara},\ and\ \citenamefont
  {Higo}}]{Nakatsuji2015Mn3SnAnomalousHall}%
  \BibitemOpen
  \bibfield  {author} {\bibinfo {author} {\bibfnamefont {S.}~\bibnamefont
  {Nakatsuji}}, \bibinfo {author} {\bibfnamefont {N.}~\bibnamefont {Kiyohara}},
  \ and\ \bibinfo {author} {\bibfnamefont {T.}~\bibnamefont {Higo}},\ }\href
  {\doibase 10.1038/nature15723} {\bibfield  {journal} {\bibinfo  {journal}
  {Nature}\ }\textbf {\bibinfo {volume} {527}},\ \bibinfo {pages} {212}
  (\bibinfo {year} {2015})}\BibitemShut {NoStop}%
\bibitem [{\citenamefont {Ikhlas}\ \emph {et~al.}(2017)\citenamefont {Ikhlas},
  \citenamefont {Tomita}, \citenamefont {Koretsune}, \citenamefont {Suzuki},
  \citenamefont {Nishio-Hamane}, \citenamefont {Arita}, \citenamefont {Otani},\
  and\ \citenamefont {Nakatsuji}}]{Ikhlas2017LargeAnomalousNernst}%
  \BibitemOpen
  \bibfield  {author} {\bibinfo {author} {\bibfnamefont {M.}~\bibnamefont
  {Ikhlas}}, \bibinfo {author} {\bibfnamefont {T.}~\bibnamefont {Tomita}},
  \bibinfo {author} {\bibfnamefont {T.}~\bibnamefont {Koretsune}}, \bibinfo
  {author} {\bibfnamefont {M.~T.}\ \bibnamefont {Suzuki}}, \bibinfo {author}
  {\bibfnamefont {D.}~\bibnamefont {Nishio-Hamane}}, \bibinfo {author}
  {\bibfnamefont {R.}~\bibnamefont {Arita}}, \bibinfo {author} {\bibfnamefont
  {Y.}~\bibnamefont {Otani}}, \ and\ \bibinfo {author} {\bibfnamefont
  {S.}~\bibnamefont {Nakatsuji}},\ }\href {\doibase 10.1038/nphys4181}
  {\bibfield  {journal} {\bibinfo  {journal} {Nat. Phys.}\ }\textbf {\bibinfo
  {volume} {13}},\ \bibinfo {pages} {1085} (\bibinfo {year}
  {2017})}\BibitemShut {NoStop}%
\bibitem [{\citenamefont {Higo}\ \emph {et~al.}(2018)\citenamefont {Higo},
  \citenamefont {Man}, \citenamefont {Gopman}, \citenamefont {Wu},
  \citenamefont {Koretsune}, \citenamefont {van't Erve}, \citenamefont
  {Kabanov}, \citenamefont {Rees}, \citenamefont {Li}, \citenamefont {Suzuki}
  \emph {et~al.}}]{Higo2018largeKerr}%
  \BibitemOpen
  \bibfield  {author} {\bibinfo {author} {\bibfnamefont {T.}~\bibnamefont
  {Higo}}, \bibinfo {author} {\bibfnamefont {H.}~\bibnamefont {Man}}, \bibinfo
  {author} {\bibfnamefont {D.~B.}\ \bibnamefont {Gopman}}, \bibinfo {author}
  {\bibfnamefont {L.}~\bibnamefont {Wu}}, \bibinfo {author} {\bibfnamefont
  {T.}~\bibnamefont {Koretsune}}, \bibinfo {author} {\bibfnamefont {O.~M.}\
  \bibnamefont {van't Erve}}, \bibinfo {author} {\bibfnamefont {Y.~P.}\
  \bibnamefont {Kabanov}}, \bibinfo {author} {\bibfnamefont {D.}~\bibnamefont
  {Rees}}, \bibinfo {author} {\bibfnamefont {Y.}~\bibnamefont {Li}}, \bibinfo
  {author} {\bibfnamefont {M.-T.}\ \bibnamefont {Suzuki}},  \emph {et~al.},\
  }\href@noop {} {\bibfield  {journal} {\bibinfo  {journal} {Nature photonics}\
  }\textbf {\bibinfo {volume} {12}},\ \bibinfo {pages} {73} (\bibinfo {year}
  {2018})}\BibitemShut {NoStop}%
\bibitem [{\citenamefont {Suzuki}\ \emph {et~al.}(2017)\citenamefont {Suzuki},
  \citenamefont {Koretsune}, \citenamefont {Ochi},\ and\ \citenamefont
  {Arita}}]{Suzuki2017ClusterMultipole}%
  \BibitemOpen
  \bibfield  {author} {\bibinfo {author} {\bibfnamefont {M.-T.}\ \bibnamefont
  {Suzuki}}, \bibinfo {author} {\bibfnamefont {T.}~\bibnamefont {Koretsune}},
  \bibinfo {author} {\bibfnamefont {M.}~\bibnamefont {Ochi}}, \ and\ \bibinfo
  {author} {\bibfnamefont {R.}~\bibnamefont {Arita}},\ }\href {\doibase
  10.1103/PhysRevB.95.094406} {\bibfield  {journal} {\bibinfo  {journal} {Phys.
  Rev. B}\ }\textbf {\bibinfo {volume} {95}},\ \bibinfo {pages} {094406}
  (\bibinfo {year} {2017})}\BibitemShut {NoStop}%
\bibitem [{\citenamefont {{\v{Z}}elezn{\'{y}}}\ \emph
  {et~al.}(2017)\citenamefont {{\v{Z}}elezn{\'{y}}}, \citenamefont {Gao},
  \citenamefont {Manchon}, \citenamefont {Freimuth}, \citenamefont {Mokrousov},
  \citenamefont {Zemen}, \citenamefont {Ma{\v{s}}ek}, \citenamefont {Sinova},\
  and\ \citenamefont {Jungwirth}}]{Zelezny2017}%
  \BibitemOpen
  \bibfield  {author} {\bibinfo {author} {\bibfnamefont {J.}~\bibnamefont
  {{\v{Z}}elezn{\'{y}}}}, \bibinfo {author} {\bibfnamefont {H.}~\bibnamefont
  {Gao}}, \bibinfo {author} {\bibfnamefont {A.}~\bibnamefont {Manchon}},
  \bibinfo {author} {\bibfnamefont {F.}~\bibnamefont {Freimuth}}, \bibinfo
  {author} {\bibfnamefont {Y.}~\bibnamefont {Mokrousov}}, \bibinfo {author}
  {\bibfnamefont {J.}~\bibnamefont {Zemen}}, \bibinfo {author} {\bibfnamefont
  {J.}~\bibnamefont {Ma{\v{s}}ek}}, \bibinfo {author} {\bibfnamefont
  {J.}~\bibnamefont {Sinova}}, \ and\ \bibinfo {author} {\bibfnamefont
  {T.}~\bibnamefont {Jungwirth}},\ }\href {\doibase 10.1103/PhysRevB.95.014403}
  {\bibfield  {journal} {\bibinfo  {journal} {Phys. Rev. B}\ }\textbf {\bibinfo
  {volume} {95}},\ \bibinfo {pages} {014403} (\bibinfo {year}
  {2017})}\BibitemShut {NoStop}%
\bibitem [{\citenamefont {Curie}(1894)}]{curie1894symetrie}%
  \BibitemOpen
  \bibfield  {author} {\bibinfo {author} {\bibfnamefont {P.}~\bibnamefont
  {Curie}},\ }\href@noop {} {\bibfield  {journal} {\bibinfo  {journal} {Journal
  de physique th{\'e}orique et appliqu{\'e}e}\ }\textbf {\bibinfo {volume}
  {3}},\ \bibinfo {pages} {393} (\bibinfo {year} {1894})}\BibitemShut {NoStop}%
\bibitem [{\citenamefont {Landau}\ and\ \citenamefont
  {Lifshitz}(1960)}]{landau1960course}%
  \BibitemOpen
  \bibfield  {author} {\bibinfo {author} {\bibfnamefont {L.~D.}\ \bibnamefont
  {Landau}}\ and\ \bibinfo {author} {\bibfnamefont {E.~M.}\ \bibnamefont
  {Lifshitz}},\ }\href@noop {} {\emph {\bibinfo {title} {Course of Theoretical
  Physics. Vol. 8: Electrodynamics of Continuous Media}}}\ (\bibinfo
  {publisher} {Oxford},\ \bibinfo {year} {1960})\BibitemShut {NoStop}%
\bibitem [{\citenamefont {Dzyaloshinskii}(1960)}]{dzyaloshinskii1960magneto}%
  \BibitemOpen
  \bibfield  {author} {\bibinfo {author} {\bibfnamefont {I.~E.}\ \bibnamefont
  {Dzyaloshinskii}},\ }\href@noop {} {\bibfield  {journal} {\bibinfo  {journal}
  {Sov. Phys. JETP}\ }\textbf {\bibinfo {volume} {10}},\ \bibinfo {pages} {628}
  (\bibinfo {year} {1960})},\ \bibinfo {note} {[I. E. Dzyaloshinskii, Zh. Exp.
  Teor. Fiz. \textbf{37}, 881 (1959)]}\BibitemShut {NoStop}%
\bibitem [{\citenamefont {Astrov}(1960)}]{astrov1960magnetoelectric}%
  \BibitemOpen
  \bibfield  {author} {\bibinfo {author} {\bibfnamefont {D.~N.}\ \bibnamefont
  {Astrov}},\ }\href@noop {} {\bibfield  {journal} {\bibinfo  {journal} {Sov.
  Phys. JETP}\ }\textbf {\bibinfo {volume} {11}},\ \bibinfo {pages} {708}
  (\bibinfo {year} {1960})},\ \bibinfo {note} {[D. N. Astrov, Zh. Exp. Teor.
  Fiz. \textbf{38}, 984 (1960)]}\BibitemShut {NoStop}%
\bibitem [{\citenamefont {Astrov}(1961)}]{astrov1961magnetoelectric}%
  \BibitemOpen
  \bibfield  {author} {\bibinfo {author} {\bibfnamefont {D.~N.}\ \bibnamefont
  {Astrov}},\ }\href@noop {} {\bibfield  {journal} {\bibinfo  {journal} {Sov.
  Phys. JETP}\ }\textbf {\bibinfo {volume} {13}},\ \bibinfo {pages} {729}
  (\bibinfo {year} {1961})},\ \bibinfo {note} {[D. N. Astrov, Zh. Exp. Teor.
  Fiz. \textbf{40}, 1035 (1961)]}\BibitemShut {NoStop}%
\bibitem [{\citenamefont {Malashevich}\ \emph {et~al.}(2012)\citenamefont
  {Malashevich}, \citenamefont {Coh}, \citenamefont {Souza},\ and\
  \citenamefont {Vanderbilt}}]{Malashevich2012fullme}%
  \BibitemOpen
  \bibfield  {author} {\bibinfo {author} {\bibfnamefont {A.}~\bibnamefont
  {Malashevich}}, \bibinfo {author} {\bibfnamefont {S.}~\bibnamefont {Coh}},
  \bibinfo {author} {\bibfnamefont {I.}~\bibnamefont {Souza}}, \ and\ \bibinfo
  {author} {\bibfnamefont {D.}~\bibnamefont {Vanderbilt}},\ }\href {\doibase
  10.1103/PhysRevB.86.094430} {\bibfield  {journal} {\bibinfo  {journal} {Phys.
  Rev. B}\ }\textbf {\bibinfo {volume} {86}},\ \bibinfo {pages} {094430}
  (\bibinfo {year} {2012})}\BibitemShut {NoStop}%
\bibitem [{\citenamefont {Scaramucci}\ \emph {et~al.}(2012)\citenamefont
  {Scaramucci}, \citenamefont {Bousquet}, \citenamefont {Fechner},
  \citenamefont {Mostovoy},\ and\ \citenamefont {Spaldin}}]{Scaramucci2012a}%
  \BibitemOpen
  \bibfield  {author} {\bibinfo {author} {\bibfnamefont {A.}~\bibnamefont
  {Scaramucci}}, \bibinfo {author} {\bibfnamefont {E.}~\bibnamefont
  {Bousquet}}, \bibinfo {author} {\bibfnamefont {M.}~\bibnamefont {Fechner}},
  \bibinfo {author} {\bibfnamefont {M.}~\bibnamefont {Mostovoy}}, \ and\
  \bibinfo {author} {\bibfnamefont {N.~A.}\ \bibnamefont {Spaldin}},\ }\href
  {\doibase 10.1103/PhysRevLett.109.197203} {\bibfield  {journal} {\bibinfo
  {journal} {Phys. Rev. Lett.}\ }\textbf {\bibinfo {volume} {109}},\ \bibinfo
  {pages} {197203} (\bibinfo {year} {2012})}\BibitemShut {NoStop}%
\bibitem [{\citenamefont {Bultmark}\ \emph {et~al.}(2009)\citenamefont
  {Bultmark}, \citenamefont {Cricchio}, \citenamefont {Gr{\aa}n{\"{a}}s},\ and\
  \citenamefont {Nordstr{\"{o}}m}}]{Bultmark2009}%
  \BibitemOpen
  \bibfield  {author} {\bibinfo {author} {\bibfnamefont {F.}~\bibnamefont
  {Bultmark}}, \bibinfo {author} {\bibfnamefont {F.}~\bibnamefont {Cricchio}},
  \bibinfo {author} {\bibfnamefont {O.}~\bibnamefont {Gr{\aa}n{\"{a}}s}}, \
  and\ \bibinfo {author} {\bibfnamefont {L.}~\bibnamefont {Nordstr{\"{o}}m}},\
  }\href {\doibase 10.1103/PhysRevB.80.035121} {\bibfield  {journal} {\bibinfo
  {journal} {Phys. Rev. B}\ }\textbf {\bibinfo {volume} {80}},\ \bibinfo
  {pages} {035121} (\bibinfo {year} {2009})}\BibitemShut {NoStop}%
\bibitem [{\citenamefont {Cricchio}\ \emph {et~al.}(2010)\citenamefont
  {Cricchio}, \citenamefont {Gr{\aa}n{\"{a}}s},\ and\ \citenamefont
  {Nordstr{\"{o}}m}}]{Cricchio2010}%
  \BibitemOpen
  \bibfield  {author} {\bibinfo {author} {\bibfnamefont {F.}~\bibnamefont
  {Cricchio}}, \bibinfo {author} {\bibfnamefont {O.}~\bibnamefont
  {Gr{\aa}n{\"{a}}s}}, \ and\ \bibinfo {author} {\bibfnamefont
  {L.}~\bibnamefont {Nordstr{\"{o}}m}},\ }\href {\doibase
  10.1103/PhysRevB.81.140403} {\bibfield  {journal} {\bibinfo  {journal} {Phys.
  Rev. B}\ }\textbf {\bibinfo {volume} {81}},\ \bibinfo {pages} {140403}
  (\bibinfo {year} {2010})}\BibitemShut {NoStop}%
\bibitem [{\citenamefont {Suzuki}\ \emph {et~al.}(2018)\citenamefont {Suzuki},
  \citenamefont {Ikeda},\ and\ \citenamefont {Oppeneer}}]{suzuki2018multipole}%
  \BibitemOpen
  \bibfield  {author} {\bibinfo {author} {\bibfnamefont {M.-T.}\ \bibnamefont
  {Suzuki}}, \bibinfo {author} {\bibfnamefont {H.}~\bibnamefont {Ikeda}}, \
  and\ \bibinfo {author} {\bibfnamefont {P.~M.}\ \bibnamefont {Oppeneer}},\
  }\href {\doibase 10.7566/JPSJ.87.041008} {\bibfield  {journal} {\bibinfo
  {journal} {J. Phys. Soc. Jpn.}\ }\textbf {\bibinfo {volume} {87}},\ \bibinfo
  {pages} {041008} (\bibinfo {year} {2018})}\BibitemShut {NoStop}%
\bibitem [{\citenamefont {Th{\"o}le}\ and\ \citenamefont
  {Spaldin}(2018)}]{Florian2018multipole}%
  \BibitemOpen
  \bibfield  {author} {\bibinfo {author} {\bibfnamefont {F.}~\bibnamefont
  {Th{\"o}le}}\ and\ \bibinfo {author} {\bibfnamefont {N.~A.}\ \bibnamefont
  {Spaldin}},\ }\href {\doibase 10.1098/rsta.2017.0450} {\bibfield  {journal}
  {\bibinfo  {journal} {Phil. Trans. R. Soc. A}\ }\textbf {\bibinfo {volume}
  {376}},\ \bibinfo {pages} {20170450} (\bibinfo {year} {2018})}\BibitemShut
  {NoStop}%
\bibitem [{\citenamefont {McGuire}\ \emph {et~al.}(1956)\citenamefont
  {McGuire}, \citenamefont {Scott},\ and\ \citenamefont
  {Grannis}}]{McGuire1956}%
  \BibitemOpen
  \bibfield  {author} {\bibinfo {author} {\bibfnamefont {T.~R.}\ \bibnamefont
  {McGuire}}, \bibinfo {author} {\bibfnamefont {E.~J.}\ \bibnamefont {Scott}},
  \ and\ \bibinfo {author} {\bibfnamefont {F.~H.}\ \bibnamefont {Grannis}},\
  }\href {\doibase 10.1103/PhysRev.102.1000} {\bibfield  {journal} {\bibinfo
  {journal} {Phys. Rev.}\ }\textbf {\bibinfo {volume} {102}},\ \bibinfo {pages}
  {1000} (\bibinfo {year} {1956})}\BibitemShut {NoStop}%
\bibitem [{\citenamefont {Hayami}\ \emph {et~al.}(2018)\citenamefont {Hayami},
  \citenamefont {Kusunose},\ and\ \citenamefont {Motome}}]{hayami2018emergent}%
  \BibitemOpen
  \bibfield  {author} {\bibinfo {author} {\bibfnamefont {S.}~\bibnamefont
  {Hayami}}, \bibinfo {author} {\bibfnamefont {H.}~\bibnamefont {Kusunose}}, \
  and\ \bibinfo {author} {\bibfnamefont {Y.}~\bibnamefont {Motome}},\ }\href
  {\doibase 10.1103/PhysRevB.97.024414} {\bibfield  {journal} {\bibinfo
  {journal} {Phys. Rev. B}\ }\textbf {\bibinfo {volume} {97}},\ \bibinfo
  {pages} {024414} (\bibinfo {year} {2018})}\BibitemShut {NoStop}%
\bibitem [{\citenamefont {Saha}\ \emph
  {et~al.}(2016{\natexlab{a}})\citenamefont {Saha}, \citenamefont {Ghara},
  \citenamefont {Suard}, \citenamefont {Jang}, \citenamefont {Kim},
  \citenamefont {Sundaresan}, \citenamefont {Ter-Oganessian},\ and\
  \citenamefont {Sundaresan}}]{Saha2016a}%
  \BibitemOpen
  \bibfield  {author} {\bibinfo {author} {\bibfnamefont {R.}~\bibnamefont
  {Saha}}, \bibinfo {author} {\bibfnamefont {S.}~\bibnamefont {Ghara}},
  \bibinfo {author} {\bibfnamefont {E.}~\bibnamefont {Suard}}, \bibinfo
  {author} {\bibfnamefont {D.~H.}\ \bibnamefont {Jang}}, \bibinfo {author}
  {\bibfnamefont {K.~H.}\ \bibnamefont {Kim}}, \bibinfo {author} {\bibfnamefont
  {A.}~\bibnamefont {Sundaresan}}, \bibinfo {author} {\bibfnamefont {N.~V.}\
  \bibnamefont {Ter-Oganessian}}, \ and\ \bibinfo {author} {\bibfnamefont
  {A.}~\bibnamefont {Sundaresan}},\ }\href {\doibase
  10.1103/PhysRevB.94.014428} {\bibfield  {journal} {\bibinfo  {journal} {Phys.
  Rev. B}\ }\textbf {\bibinfo {volume} {94}},\ \bibinfo {pages} {014428}
  (\bibinfo {year} {2016}{\natexlab{a}})}\BibitemShut {NoStop}%
\bibitem [{\citenamefont {Ghara}\ \emph {et~al.}(2017)\citenamefont {Ghara},
  \citenamefont {Ter-Oganessian},\ and\ \citenamefont
  {Sundaresan}}]{Ghara2017}%
  \BibitemOpen
  \bibfield  {author} {\bibinfo {author} {\bibfnamefont {S.}~\bibnamefont
  {Ghara}}, \bibinfo {author} {\bibfnamefont {N.~V.}\ \bibnamefont
  {Ter-Oganessian}}, \ and\ \bibinfo {author} {\bibfnamefont {A.}~\bibnamefont
  {Sundaresan}},\ }\href {\doibase 10.1103/PhysRevB.95.094404} {\bibfield
  {journal} {\bibinfo  {journal} {Phys. Rev. B}\ }\textbf {\bibinfo {volume}
  {95}},\ \bibinfo {pages} {094404} (\bibinfo {year} {2017})}\BibitemShut
  {NoStop}%
\bibitem [{\citenamefont {Edelstein}(1990)}]{Edelstein1990a}%
  \BibitemOpen
  \bibfield  {author} {\bibinfo {author} {\bibfnamefont {V.~M.}\ \bibnamefont
  {Edelstein}},\ }\href {\doibase 10.1016/0038-1098(90)90963-C} {\bibfield
  {journal} {\bibinfo  {journal} {Solid State Commun.}\ }\textbf {\bibinfo
  {volume} {73}},\ \bibinfo {pages} {233} (\bibinfo {year} {1990})}\BibitemShut
  {NoStop}%
\bibitem [{\citenamefont {Ciccarelli}\ \emph {et~al.}(2016)\citenamefont
  {Ciccarelli}, \citenamefont {Anderson}, \citenamefont {Tshitoyan},
  \citenamefont {Ferguson}, \citenamefont {Gerhard}, \citenamefont {Gould},
  \citenamefont {Molenkamp}, \citenamefont {Gayles}, \citenamefont
  {{\v{Z}}elezn{\'{y}}}, \citenamefont {{\v{S}}mejkal}, \citenamefont {Yuan},
  \citenamefont {Sinova}, \citenamefont {Freimuth},\ and\ \citenamefont
  {Jungwirth}}]{Ciccarelli2015}%
  \BibitemOpen
  \bibfield  {author} {\bibinfo {author} {\bibfnamefont {C.}~\bibnamefont
  {Ciccarelli}}, \bibinfo {author} {\bibfnamefont {L.}~\bibnamefont
  {Anderson}}, \bibinfo {author} {\bibfnamefont {V.}~\bibnamefont {Tshitoyan}},
  \bibinfo {author} {\bibfnamefont {A.~J.}\ \bibnamefont {Ferguson}}, \bibinfo
  {author} {\bibfnamefont {F.}~\bibnamefont {Gerhard}}, \bibinfo {author}
  {\bibfnamefont {C.}~\bibnamefont {Gould}}, \bibinfo {author} {\bibfnamefont
  {L.~W.}\ \bibnamefont {Molenkamp}}, \bibinfo {author} {\bibfnamefont
  {J.}~\bibnamefont {Gayles}}, \bibinfo {author} {\bibfnamefont
  {J.}~\bibnamefont {{\v{Z}}elezn{\'{y}}}}, \bibinfo {author} {\bibfnamefont
  {L.}~\bibnamefont {{\v{S}}mejkal}}, \bibinfo {author} {\bibfnamefont
  {Z.}~\bibnamefont {Yuan}}, \bibinfo {author} {\bibfnamefont {J.}~\bibnamefont
  {Sinova}}, \bibinfo {author} {\bibfnamefont {F.}~\bibnamefont {Freimuth}}, \
  and\ \bibinfo {author} {\bibfnamefont {T.}~\bibnamefont {Jungwirth}},\ }\href
  {\doibase 10.1038/nphys3772} {\bibfield  {journal} {\bibinfo  {journal} {Nat.
  Phys.}\ }\textbf {\bibinfo {volume} {12}},\ \bibinfo {pages} {855} (\bibinfo
  {year} {2016})}\BibitemShut {NoStop}%
\bibitem [{\citenamefont {Levitov}\ \emph {et~al.}(1985)\citenamefont
  {Levitov}, \citenamefont {Nazarov},\ and\ \citenamefont
  {Eliashberg}}]{levitov1985magnetoelectric}%
  \BibitemOpen
  \bibfield  {author} {\bibinfo {author} {\bibfnamefont {L.~S.}\ \bibnamefont
  {Levitov}}, \bibinfo {author} {\bibfnamefont {Y.~V.}\ \bibnamefont
  {Nazarov}}, \ and\ \bibinfo {author} {\bibfnamefont {G.~M.}\ \bibnamefont
  {Eliashberg}},\ }\href@noop {} {\bibfield  {journal} {\bibinfo  {journal}
  {Sov. Phys. JETP}\ }\textbf {\bibinfo {volume} {61}},\ \bibinfo {pages} {133}
  (\bibinfo {year} {1985})},\ \bibinfo {note} {[L. S. Levitov, Y. V. Nazarov,
  and G. M. Eliashberg, Zh. Exp. Teor. Fiz. \textbf{88}, 229
  (1985)]}\BibitemShut {NoStop}%
\bibitem [{\citenamefont {Kato}\ \emph {et~al.}(2004)\citenamefont {Kato},
  \citenamefont {Myers}, \citenamefont {Gossard},\ and\ \citenamefont
  {Awschalom}}]{Kato2004}%
  \BibitemOpen
  \bibfield  {author} {\bibinfo {author} {\bibfnamefont {Y.~K.}\ \bibnamefont
  {Kato}}, \bibinfo {author} {\bibfnamefont {R.~C.}\ \bibnamefont {Myers}},
  \bibinfo {author} {\bibfnamefont {A.~C.}\ \bibnamefont {Gossard}}, \ and\
  \bibinfo {author} {\bibfnamefont {D.~D.}\ \bibnamefont {Awschalom}},\ }\href
  {\doibase 10.1103/PhysRevLett.93.176601} {\bibfield  {journal} {\bibinfo
  {journal} {Phys. Rev. Lett.}\ }\textbf {\bibinfo {volume} {93}},\ \bibinfo
  {pages} {176601} (\bibinfo {year} {2004})}\BibitemShut {NoStop}%
\bibitem [{\citenamefont {Sih}\ \emph {et~al.}(2005)\citenamefont {Sih},
  \citenamefont {Myers}, \citenamefont {Kato}, \citenamefont {Lau},
  \citenamefont {Gossard},\ and\ \citenamefont {Awschalom}}]{Sih2005}%
  \BibitemOpen
  \bibfield  {author} {\bibinfo {author} {\bibfnamefont {V.}~\bibnamefont
  {Sih}}, \bibinfo {author} {\bibfnamefont {R.~C.}\ \bibnamefont {Myers}},
  \bibinfo {author} {\bibfnamefont {Y.~K.}\ \bibnamefont {Kato}}, \bibinfo
  {author} {\bibfnamefont {W.~H.}\ \bibnamefont {Lau}}, \bibinfo {author}
  {\bibfnamefont {A.~C.}\ \bibnamefont {Gossard}}, \ and\ \bibinfo {author}
  {\bibfnamefont {D.~D.}\ \bibnamefont {Awschalom}},\ }\href {\doibase
  10.1038/nphys009} {\bibfield  {journal} {\bibinfo  {journal} {Nat. Phys.}\
  }\textbf {\bibinfo {volume} {1}},\ \bibinfo {pages} {31} (\bibinfo {year}
  {2005})}\BibitemShut {NoStop}%
\bibitem [{\citenamefont {Yang}\ \emph {et~al.}(2006)\citenamefont {Yang},
  \citenamefont {He}, \citenamefont {Ding}, \citenamefont {Cui}, \citenamefont
  {Zeng}, \citenamefont {Wang},\ and\ \citenamefont {Ge}}]{Yang2006}%
  \BibitemOpen
  \bibfield  {author} {\bibinfo {author} {\bibfnamefont {C.~L.}\ \bibnamefont
  {Yang}}, \bibinfo {author} {\bibfnamefont {H.~T.}\ \bibnamefont {He}},
  \bibinfo {author} {\bibfnamefont {L.}~\bibnamefont {Ding}}, \bibinfo {author}
  {\bibfnamefont {L.~J.}\ \bibnamefont {Cui}}, \bibinfo {author} {\bibfnamefont
  {Y.~P.}\ \bibnamefont {Zeng}}, \bibinfo {author} {\bibfnamefont {J.~N.}\
  \bibnamefont {Wang}}, \ and\ \bibinfo {author} {\bibfnamefont {W.~K.}\
  \bibnamefont {Ge}},\ }\href {\doibase 10.1103/PhysRevLett.96.186605}
  {\bibfield  {journal} {\bibinfo  {journal} {Phys. Rev. Lett.}\ }\textbf
  {\bibinfo {volume} {96}},\ \bibinfo {pages} {186605} (\bibinfo {year}
  {2006})}\BibitemShut {NoStop}%
\bibitem [{\citenamefont {Stern}\ \emph {et~al.}(2006)\citenamefont {Stern},
  \citenamefont {Ghosh}, \citenamefont {Xiang}, \citenamefont {Zhu},
  \citenamefont {Samarth},\ and\ \citenamefont {Awschalom}}]{Stern2006}%
  \BibitemOpen
  \bibfield  {author} {\bibinfo {author} {\bibfnamefont {N.~P.}\ \bibnamefont
  {Stern}}, \bibinfo {author} {\bibfnamefont {S.}~\bibnamefont {Ghosh}},
  \bibinfo {author} {\bibfnamefont {G.}~\bibnamefont {Xiang}}, \bibinfo
  {author} {\bibfnamefont {M.}~\bibnamefont {Zhu}}, \bibinfo {author}
  {\bibfnamefont {N.}~\bibnamefont {Samarth}}, \ and\ \bibinfo {author}
  {\bibfnamefont {D.~D.}\ \bibnamefont {Awschalom}},\ }\href {\doibase
  10.1103/PhysRevLett.97.126603} {\bibfield  {journal} {\bibinfo  {journal}
  {Phys. Rev. Lett.}\ }\textbf {\bibinfo {volume} {97}},\ \bibinfo {pages}
  {126603} (\bibinfo {year} {2006})}\BibitemShut {NoStop}%
\bibitem [{\citenamefont {Furukawa}\ \emph {et~al.}(2017)\citenamefont
  {Furukawa}, \citenamefont {Shimokawa}, \citenamefont {Kobayashi},\ and\
  \citenamefont {Itou}}]{Furukawa2017}%
  \BibitemOpen
  \bibfield  {author} {\bibinfo {author} {\bibfnamefont {T.}~\bibnamefont
  {Furukawa}}, \bibinfo {author} {\bibfnamefont {Y.}~\bibnamefont {Shimokawa}},
  \bibinfo {author} {\bibfnamefont {K.}~\bibnamefont {Kobayashi}}, \ and\
  \bibinfo {author} {\bibfnamefont {T.}~\bibnamefont {Itou}},\ }\href {\doibase
  10.1038/s41467-017-01093-3} {\bibfield  {journal} {\bibinfo  {journal} {Nat.
  Commun.}\ }\textbf {\bibinfo {volume} {8}},\ \bibinfo {pages} {954} (\bibinfo
  {year} {2017})}\BibitemShut {NoStop}%
\bibitem [{\citenamefont {Chernyshov}\ \emph {et~al.}(2009)\citenamefont
  {Chernyshov}, \citenamefont {Overby}, \citenamefont {Liu}, \citenamefont
  {Furdyna}, \citenamefont {Lyanda-Geller},\ and\ \citenamefont
  {Rokhinson}}]{Chernyshov2009}%
  \BibitemOpen
  \bibfield  {author} {\bibinfo {author} {\bibfnamefont {A.}~\bibnamefont
  {Chernyshov}}, \bibinfo {author} {\bibfnamefont {M.}~\bibnamefont {Overby}},
  \bibinfo {author} {\bibfnamefont {X.}~\bibnamefont {Liu}}, \bibinfo {author}
  {\bibfnamefont {J.~K.}\ \bibnamefont {Furdyna}}, \bibinfo {author}
  {\bibfnamefont {Y.}~\bibnamefont {Lyanda-Geller}}, \ and\ \bibinfo {author}
  {\bibfnamefont {L.~P.}\ \bibnamefont {Rokhinson}},\ }\href {\doibase
  10.1038/nphys1362} {\bibfield  {journal} {\bibinfo  {journal} {Nat. Phys.}\
  }\textbf {\bibinfo {volume} {5}},\ \bibinfo {pages} {656} (\bibinfo {year}
  {2009})}\BibitemShut {NoStop}%
\bibitem [{\citenamefont {{Mihai Miron}}\ \emph {et~al.}(2010)\citenamefont
  {{Mihai Miron}}, \citenamefont {Gaudin}, \citenamefont {Auffret},
  \citenamefont {Rodmacq}, \citenamefont {Schuhl}, \citenamefont {Pizzini},
  \citenamefont {Vogel},\ and\ \citenamefont {Gambardella}}]{Miron2010}%
  \BibitemOpen
  \bibfield  {author} {\bibinfo {author} {\bibfnamefont {I.}~\bibnamefont
  {{Mihai Miron}}}, \bibinfo {author} {\bibfnamefont {G.}~\bibnamefont
  {Gaudin}}, \bibinfo {author} {\bibfnamefont {S.}~\bibnamefont {Auffret}},
  \bibinfo {author} {\bibfnamefont {B.}~\bibnamefont {Rodmacq}}, \bibinfo
  {author} {\bibfnamefont {A.}~\bibnamefont {Schuhl}}, \bibinfo {author}
  {\bibfnamefont {S.}~\bibnamefont {Pizzini}}, \bibinfo {author} {\bibfnamefont
  {J.}~\bibnamefont {Vogel}}, \ and\ \bibinfo {author} {\bibfnamefont
  {P.}~\bibnamefont {Gambardella}},\ }\href {\doibase 10.1038/nmat2613}
  {\bibfield  {journal} {\bibinfo  {journal} {Nat. Mater.}\ }\textbf {\bibinfo
  {volume} {9}},\ \bibinfo {pages} {230} (\bibinfo {year} {2010})}\BibitemShut
  {NoStop}%
\bibitem [{\citenamefont {Miron}\ \emph {et~al.}(2011)\citenamefont {Miron},
  \citenamefont {Garello}, \citenamefont {Gaudin}, \citenamefont {Zermatten},
  \citenamefont {Costache}, \citenamefont {Auffret}, \citenamefont {Bandiera},
  \citenamefont {Rodmacq}, \citenamefont {Schuhl},\ and\ \citenamefont
  {Gambardella}}]{Miron2011}%
  \BibitemOpen
  \bibfield  {author} {\bibinfo {author} {\bibfnamefont {I.~M.}\ \bibnamefont
  {Miron}}, \bibinfo {author} {\bibfnamefont {K.}~\bibnamefont {Garello}},
  \bibinfo {author} {\bibfnamefont {G.}~\bibnamefont {Gaudin}}, \bibinfo
  {author} {\bibfnamefont {P.-J.}\ \bibnamefont {Zermatten}}, \bibinfo {author}
  {\bibfnamefont {M.~V.}\ \bibnamefont {Costache}}, \bibinfo {author}
  {\bibfnamefont {S.}~\bibnamefont {Auffret}}, \bibinfo {author} {\bibfnamefont
  {S.}~\bibnamefont {Bandiera}}, \bibinfo {author} {\bibfnamefont
  {B.}~\bibnamefont {Rodmacq}}, \bibinfo {author} {\bibfnamefont
  {A.}~\bibnamefont {Schuhl}}, \ and\ \bibinfo {author} {\bibfnamefont
  {P.}~\bibnamefont {Gambardella}},\ }\href {\doibase 10.1038/nature10309}
  {\bibfield  {journal} {\bibinfo  {journal} {Nature}\ }\textbf {\bibinfo
  {volume} {476}},\ \bibinfo {pages} {189} (\bibinfo {year}
  {2011})}\BibitemShut {NoStop}%
\bibitem [{\citenamefont {Dresselhaus}\ \emph {et~al.}(1992)\citenamefont
  {Dresselhaus}, \citenamefont {Papavassiliou}, \citenamefont {Wheeler},\ and\
  \citenamefont {Sacks}}]{Dresselhaus1992}%
  \BibitemOpen
  \bibfield  {author} {\bibinfo {author} {\bibfnamefont {P.~D.}\ \bibnamefont
  {Dresselhaus}}, \bibinfo {author} {\bibfnamefont {C.~M.~A.}\ \bibnamefont
  {Papavassiliou}}, \bibinfo {author} {\bibfnamefont {R.~G.}\ \bibnamefont
  {Wheeler}}, \ and\ \bibinfo {author} {\bibfnamefont {R.~N.}\ \bibnamefont
  {Sacks}},\ }\href {\doibase 10.1103/PhysRevLett.68.106} {\bibfield  {journal}
  {\bibinfo  {journal} {Phys. Rev. Lett.}\ }\textbf {\bibinfo {volume} {68}},\
  \bibinfo {pages} {106} (\bibinfo {year} {1992})}\BibitemShut {NoStop}%
\bibitem [{\citenamefont {Nitta}\ \emph {et~al.}(1997)\citenamefont {Nitta},
  \citenamefont {Akazaki}, \citenamefont {Takayanagi},\ and\ \citenamefont
  {Enoki}}]{Nitta1997}%
  \BibitemOpen
  \bibfield  {author} {\bibinfo {author} {\bibfnamefont {J.}~\bibnamefont
  {Nitta}}, \bibinfo {author} {\bibfnamefont {T.}~\bibnamefont {Akazaki}},
  \bibinfo {author} {\bibfnamefont {H.}~\bibnamefont {Takayanagi}}, \ and\
  \bibinfo {author} {\bibfnamefont {T.}~\bibnamefont {Enoki}},\ }\href
  {\doibase 10.1103/PhysRevLett.78.1335} {\bibfield  {journal} {\bibinfo
  {journal} {Phys. Rev. Lett.}\ }\textbf {\bibinfo {volume} {78}},\ \bibinfo
  {pages} {1335} (\bibinfo {year} {1997})}\BibitemShut {NoStop}%
\bibitem [{\citenamefont {Caviglia}\ \emph {et~al.}(2010)\citenamefont
  {Caviglia}, \citenamefont {Gabay}, \citenamefont {Gariglio}, \citenamefont
  {Reyren}, \citenamefont {Cancellieri},\ and\ \citenamefont
  {Triscone}}]{Caviglia2010}%
  \BibitemOpen
  \bibfield  {author} {\bibinfo {author} {\bibfnamefont {A.~D.}\ \bibnamefont
  {Caviglia}}, \bibinfo {author} {\bibfnamefont {M.}~\bibnamefont {Gabay}},
  \bibinfo {author} {\bibfnamefont {S.}~\bibnamefont {Gariglio}}, \bibinfo
  {author} {\bibfnamefont {N.}~\bibnamefont {Reyren}}, \bibinfo {author}
  {\bibfnamefont {C.}~\bibnamefont {Cancellieri}}, \ and\ \bibinfo {author}
  {\bibfnamefont {J.-M.}\ \bibnamefont {Triscone}},\ }\href {\doibase
  10.1103/PhysRevLett.104.126803} {\bibfield  {journal} {\bibinfo  {journal}
  {Phys. Rev. Lett.}\ }\textbf {\bibinfo {volume} {104}},\ \bibinfo {pages}
  {126803} (\bibinfo {year} {2010})}\BibitemShut {NoStop}%
\bibitem [{\citenamefont {{Ben Shalom}}\ \emph {et~al.}(2010)\citenamefont
  {{Ben Shalom}}, \citenamefont {Sachs}, \citenamefont {Rakhmilevitch},
  \citenamefont {Palevski},\ and\ \citenamefont {Dagan}}]{BenShalom2010}%
  \BibitemOpen
  \bibfield  {author} {\bibinfo {author} {\bibfnamefont {M.}~\bibnamefont {{Ben
  Shalom}}}, \bibinfo {author} {\bibfnamefont {M.}~\bibnamefont {Sachs}},
  \bibinfo {author} {\bibfnamefont {D.}~\bibnamefont {Rakhmilevitch}}, \bibinfo
  {author} {\bibfnamefont {A.}~\bibnamefont {Palevski}}, \ and\ \bibinfo
  {author} {\bibfnamefont {Y.}~\bibnamefont {Dagan}},\ }\href {\doibase
  10.1103/PhysRevLett.104.126802} {\bibfield  {journal} {\bibinfo  {journal}
  {Phys. Rev. Lett.}\ }\textbf {\bibinfo {volume} {104}},\ \bibinfo {pages}
  {126802} (\bibinfo {year} {2010})}\BibitemShut {NoStop}%
\bibitem [{\citenamefont {Hanawa}\ \emph {et~al.}(2001)\citenamefont {Hanawa},
  \citenamefont {Muraoka}, \citenamefont {Tayama}, \citenamefont {Sakakibara},
  \citenamefont {Yamaura},\ and\ \citenamefont {Hiroi}}]{Hanawa2001}%
  \BibitemOpen
  \bibfield  {author} {\bibinfo {author} {\bibfnamefont {M.}~\bibnamefont
  {Hanawa}}, \bibinfo {author} {\bibfnamefont {Y.}~\bibnamefont {Muraoka}},
  \bibinfo {author} {\bibfnamefont {T.}~\bibnamefont {Tayama}}, \bibinfo
  {author} {\bibfnamefont {T.}~\bibnamefont {Sakakibara}}, \bibinfo {author}
  {\bibfnamefont {J.}~\bibnamefont {Yamaura}}, \ and\ \bibinfo {author}
  {\bibfnamefont {Z.}~\bibnamefont {Hiroi}},\ }\href {\doibase
  10.1103/PhysRevLett.87.187001} {\bibfield  {journal} {\bibinfo  {journal}
  {Phys. Rev. Lett.}\ }\textbf {\bibinfo {volume} {87}},\ \bibinfo {pages}
  {187001} (\bibinfo {year} {2001})}\BibitemShut {NoStop}%
\bibitem [{\citenamefont {Sakai}\ \emph {et~al.}(2001)\citenamefont {Sakai},
  \citenamefont {Yoshimura}, \citenamefont {Ohno}, \citenamefont {Kato},
  \citenamefont {Kambe}, \citenamefont {Walstedt}, \citenamefont {Matsuda},
  \citenamefont {Haga},\ and\ \citenamefont {Onuki}}]{Sakai2001}%
  \BibitemOpen
  \bibfield  {author} {\bibinfo {author} {\bibfnamefont {H.}~\bibnamefont
  {Sakai}}, \bibinfo {author} {\bibfnamefont {K.}~\bibnamefont {Yoshimura}},
  \bibinfo {author} {\bibfnamefont {H.}~\bibnamefont {Ohno}}, \bibinfo {author}
  {\bibfnamefont {H.}~\bibnamefont {Kato}}, \bibinfo {author} {\bibfnamefont
  {S.}~\bibnamefont {Kambe}}, \bibinfo {author} {\bibfnamefont {R.~E.}\
  \bibnamefont {Walstedt}}, \bibinfo {author} {\bibfnamefont {T.~D.}\
  \bibnamefont {Matsuda}}, \bibinfo {author} {\bibfnamefont {Y.}~\bibnamefont
  {Haga}}, \ and\ \bibinfo {author} {\bibfnamefont {Y.}~\bibnamefont {Onuki}},\
  }\href {\doibase 10.1088/0953-8984/13/33/105} {\bibfield  {journal} {\bibinfo
   {journal} {J. Phys. Condens. Matter}\ }\textbf {\bibinfo {volume} {13}},\
  \bibinfo {pages} {L785} (\bibinfo {year} {2001})}\BibitemShut {NoStop}%
\bibitem [{\citenamefont {Hiroi}\ \emph {et~al.}(2018)\citenamefont {Hiroi},
  \citenamefont {Yamaura}, \citenamefont {Kobayashi}, \citenamefont
  {Matsubayashi},\ and\ \citenamefont {Hirai}}]{Hiroi2018}%
  \BibitemOpen
  \bibfield  {author} {\bibinfo {author} {\bibfnamefont {Z.}~\bibnamefont
  {Hiroi}}, \bibinfo {author} {\bibfnamefont {J.-i.}\ \bibnamefont {Yamaura}},
  \bibinfo {author} {\bibfnamefont {T.~C.}\ \bibnamefont {Kobayashi}}, \bibinfo
  {author} {\bibfnamefont {Y.}~\bibnamefont {Matsubayashi}}, \ and\ \bibinfo
  {author} {\bibfnamefont {D.}~\bibnamefont {Hirai}},\ }\href {\doibase
  10.7566/JPSJ.87.024702} {\bibfield  {journal} {\bibinfo  {journal} {J. Phys.
  Soc. Jpn.}\ }\textbf {\bibinfo {volume} {87}},\ \bibinfo {pages} {024702}
  (\bibinfo {year} {2018})}\BibitemShut {NoStop}%
\bibitem [{\citenamefont {Yamaura}\ and\ \citenamefont
  {Hiroi}(2002)}]{Yamaura2002}%
  \BibitemOpen
  \bibfield  {author} {\bibinfo {author} {\bibfnamefont {J.-i.}\ \bibnamefont
  {Yamaura}}\ and\ \bibinfo {author} {\bibfnamefont {Z.}~\bibnamefont
  {Hiroi}},\ }\href {\doibase 10.1143/JPSJ.71.2598} {\bibfield  {journal}
  {\bibinfo  {journal} {J. Phys. Soc. Jpn.}\ }\textbf {\bibinfo {volume}
  {71}},\ \bibinfo {pages} {2598} (\bibinfo {year} {2002})}\BibitemShut
  {NoStop}%
\bibitem [{\citenamefont {Castellan}\ \emph {et~al.}(2002)\citenamefont
  {Castellan}, \citenamefont {Gaulin}, \citenamefont {van Duijn}, \citenamefont
  {Lewis}, \citenamefont {Lumsden}, \citenamefont {Jin}, \citenamefont {He},
  \citenamefont {Nagler},\ and\ \citenamefont {Mandrus}}]{Castellan2002}%
  \BibitemOpen
  \bibfield  {author} {\bibinfo {author} {\bibfnamefont {J.~P.}\ \bibnamefont
  {Castellan}}, \bibinfo {author} {\bibfnamefont {B.~D.}\ \bibnamefont
  {Gaulin}}, \bibinfo {author} {\bibfnamefont {J.}~\bibnamefont {van Duijn}},
  \bibinfo {author} {\bibfnamefont {M.~J.}\ \bibnamefont {Lewis}}, \bibinfo
  {author} {\bibfnamefont {M.~D.}\ \bibnamefont {Lumsden}}, \bibinfo {author}
  {\bibfnamefont {R.}~\bibnamefont {Jin}}, \bibinfo {author} {\bibfnamefont
  {J.}~\bibnamefont {He}}, \bibinfo {author} {\bibfnamefont {S.~E.}\
  \bibnamefont {Nagler}}, \ and\ \bibinfo {author} {\bibfnamefont
  {D.}~\bibnamefont {Mandrus}},\ }\href {\doibase 10.1103/PhysRevB.66.134528}
  {\bibfield  {journal} {\bibinfo  {journal} {Phys. Rev. B}\ }\textbf {\bibinfo
  {volume} {66}},\ \bibinfo {pages} {134528} (\bibinfo {year}
  {2002})}\BibitemShut {NoStop}%
\bibitem [{\citenamefont {Yamaura}\ \emph {et~al.}(2017)\citenamefont
  {Yamaura}, \citenamefont {Takeda}, \citenamefont {Ikeda}, \citenamefont
  {Hirao}, \citenamefont {Ohishi}, \citenamefont {Kobayashi},\ and\
  \citenamefont {Hiroi}}]{Yamaura2017a}%
  \BibitemOpen
  \bibfield  {author} {\bibinfo {author} {\bibfnamefont {J.-i.}\ \bibnamefont
  {Yamaura}}, \bibinfo {author} {\bibfnamefont {K.}~\bibnamefont {Takeda}},
  \bibinfo {author} {\bibfnamefont {Y.}~\bibnamefont {Ikeda}}, \bibinfo
  {author} {\bibfnamefont {N.}~\bibnamefont {Hirao}}, \bibinfo {author}
  {\bibfnamefont {Y.}~\bibnamefont {Ohishi}}, \bibinfo {author} {\bibfnamefont
  {T.~C.}\ \bibnamefont {Kobayashi}}, \ and\ \bibinfo {author} {\bibfnamefont
  {Z.}~\bibnamefont {Hiroi}},\ }\href {\doibase 10.1103/PhysRevB.95.020102}
  {\bibfield  {journal} {\bibinfo  {journal} {Phys. Rev. B}\ }\textbf {\bibinfo
  {volume} {95}},\ \bibinfo {pages} {020102} (\bibinfo {year}
  {2017})}\BibitemShut {NoStop}%
\bibitem [{\citenamefont {Kendziora}\ \emph {et~al.}(2005)\citenamefont
  {Kendziora}, \citenamefont {Sergienko}, \citenamefont {Jin}, \citenamefont
  {He}, \citenamefont {Keppens}, \citenamefont {Sales},\ and\ \citenamefont
  {Mandrus}}]{Kendziora2005}%
  \BibitemOpen
  \bibfield  {author} {\bibinfo {author} {\bibfnamefont {C.~A.}\ \bibnamefont
  {Kendziora}}, \bibinfo {author} {\bibfnamefont {I.~A.}\ \bibnamefont
  {Sergienko}}, \bibinfo {author} {\bibfnamefont {R.}~\bibnamefont {Jin}},
  \bibinfo {author} {\bibfnamefont {J.}~\bibnamefont {He}}, \bibinfo {author}
  {\bibfnamefont {V.}~\bibnamefont {Keppens}}, \bibinfo {author} {\bibfnamefont
  {B.~C.}\ \bibnamefont {Sales}}, \ and\ \bibinfo {author} {\bibfnamefont
  {D.}~\bibnamefont {Mandrus}},\ }\href {\doibase
  10.1103/PhysRevLett.95.125503} {\bibfield  {journal} {\bibinfo  {journal}
  {Phys. Rev. Lett.}\ }\textbf {\bibinfo {volume} {95}},\ \bibinfo {pages}
  {125503} (\bibinfo {year} {2005})}\BibitemShut {NoStop}%
\bibitem [{\citenamefont {Petersen}\ \emph {et~al.}(2006)\citenamefont
  {Petersen}, \citenamefont {Caswell}, \citenamefont {Dodge}, \citenamefont
  {Sergienko}, \citenamefont {He}, \citenamefont {Jin},\ and\ \citenamefont
  {Mandrus}}]{Petersen2006b}%
  \BibitemOpen
  \bibfield  {author} {\bibinfo {author} {\bibfnamefont {J.~C.}\ \bibnamefont
  {Petersen}}, \bibinfo {author} {\bibfnamefont {M.~D.}\ \bibnamefont
  {Caswell}}, \bibinfo {author} {\bibfnamefont {J.~S.}\ \bibnamefont {Dodge}},
  \bibinfo {author} {\bibfnamefont {I.~A.}\ \bibnamefont {Sergienko}}, \bibinfo
  {author} {\bibfnamefont {J.}~\bibnamefont {He}}, \bibinfo {author}
  {\bibfnamefont {R.}~\bibnamefont {Jin}}, \ and\ \bibinfo {author}
  {\bibfnamefont {D.}~\bibnamefont {Mandrus}},\ }\href {\doibase
  10.1038/nphys392} {\bibfield  {journal} {\bibinfo  {journal} {Nat. Phys.}\
  }\textbf {\bibinfo {volume} {2}},\ \bibinfo {pages} {605} (\bibinfo {year}
  {2006})}\BibitemShut {NoStop}%
\bibitem [{\citenamefont {Harter}\ \emph {et~al.}(2017)\citenamefont {Harter},
  \citenamefont {Zhao}, \citenamefont {Yan}, \citenamefont {Mandrus},\ and\
  \citenamefont {Hsieh}}]{Harter2017paritybreaking}%
  \BibitemOpen
  \bibfield  {author} {\bibinfo {author} {\bibfnamefont {J.~W.}\ \bibnamefont
  {Harter}}, \bibinfo {author} {\bibfnamefont {Z.~Y.}\ \bibnamefont {Zhao}},
  \bibinfo {author} {\bibfnamefont {J.-Q.}\ \bibnamefont {Yan}}, \bibinfo
  {author} {\bibfnamefont {D.~G.}\ \bibnamefont {Mandrus}}, \ and\ \bibinfo
  {author} {\bibfnamefont {D.}~\bibnamefont {Hsieh}},\ }\href {\doibase
  10.1126/science.aad1188} {\bibfield  {journal} {\bibinfo  {journal}
  {Science}\ }\textbf {\bibinfo {volume} {356}},\ \bibinfo {pages} {295}
  (\bibinfo {year} {2017})}\BibitemShut {NoStop}%
\bibitem [{\citenamefont {Wadley}\ \emph {et~al.}(2016)\citenamefont {Wadley},
  \citenamefont {Howells}, \citenamefont {Elezny}, \citenamefont {Andrews},
  \citenamefont {Hills}, \citenamefont {Campion}, \citenamefont {Novak},
  \citenamefont {Olejnik}, \citenamefont {Maccherozzi}, \citenamefont {Dhesi},
  \citenamefont {Martin}, \citenamefont {Wagner}, \citenamefont {Wunderlich},
  \citenamefont {Freimuth}, \citenamefont {Mokrousov}, \citenamefont {Kune},
  \citenamefont {Chauhan}, \citenamefont {Grzybowski}, \citenamefont
  {Rushforth}, \citenamefont {Edmonds}, \citenamefont {Gallagher},\ and\
  \citenamefont {Jungwirth}}]{Wadley2016domain}%
  \BibitemOpen
  \bibfield  {author} {\bibinfo {author} {\bibfnamefont {P.}~\bibnamefont
  {Wadley}}, \bibinfo {author} {\bibfnamefont {B.}~\bibnamefont {Howells}},
  \bibinfo {author} {\bibfnamefont {J.}~\bibnamefont {Elezny}}, \bibinfo
  {author} {\bibfnamefont {C.}~\bibnamefont {Andrews}}, \bibinfo {author}
  {\bibfnamefont {V.}~\bibnamefont {Hills}}, \bibinfo {author} {\bibfnamefont
  {R.~P.}\ \bibnamefont {Campion}}, \bibinfo {author} {\bibfnamefont
  {V.}~\bibnamefont {Novak}}, \bibinfo {author} {\bibfnamefont
  {K.}~\bibnamefont {Olejnik}}, \bibinfo {author} {\bibfnamefont
  {F.}~\bibnamefont {Maccherozzi}}, \bibinfo {author} {\bibfnamefont {S.~S.}\
  \bibnamefont {Dhesi}}, \bibinfo {author} {\bibfnamefont {S.~Y.}\ \bibnamefont
  {Martin}}, \bibinfo {author} {\bibfnamefont {T.}~\bibnamefont {Wagner}},
  \bibinfo {author} {\bibfnamefont {J.}~\bibnamefont {Wunderlich}}, \bibinfo
  {author} {\bibfnamefont {F.}~\bibnamefont {Freimuth}}, \bibinfo {author}
  {\bibfnamefont {Y.}~\bibnamefont {Mokrousov}}, \bibinfo {author}
  {\bibfnamefont {J.}~\bibnamefont {Kune}}, \bibinfo {author} {\bibfnamefont
  {J.~S.}\ \bibnamefont {Chauhan}}, \bibinfo {author} {\bibfnamefont {M.~J.}\
  \bibnamefont {Grzybowski}}, \bibinfo {author} {\bibfnamefont {A.~W.}\
  \bibnamefont {Rushforth}}, \bibinfo {author} {\bibfnamefont {K.~W.}\
  \bibnamefont {Edmonds}}, \bibinfo {author} {\bibfnamefont {B.~L.}\
  \bibnamefont {Gallagher}}, \ and\ \bibinfo {author} {\bibfnamefont
  {T.}~\bibnamefont {Jungwirth}},\ }\href {\doibase 10.1126/science.aab1031}
  {\bibfield  {journal} {\bibinfo  {journal} {Science}\ }\textbf {\bibinfo
  {volume} {351}},\ \bibinfo {pages} {587} (\bibinfo {year}
  {2016})}\BibitemShut {NoStop}%
\bibitem [{\citenamefont {Bodnar}\ \emph {et~al.}(2018)\citenamefont {Bodnar},
  \citenamefont {{\v{S}}mejkal}, \citenamefont {Turek}, \citenamefont
  {Jungwirth}, \citenamefont {Gomonay}, \citenamefont {Sinova}, \citenamefont
  {Sapozhnik}, \citenamefont {Elmers}, \citenamefont {Kl{\"{a}}ui},\ and\
  \citenamefont {Jourdan}}]{Bodnar2018domain}%
  \BibitemOpen
  \bibfield  {author} {\bibinfo {author} {\bibfnamefont {S.~Y.}\ \bibnamefont
  {Bodnar}}, \bibinfo {author} {\bibfnamefont {L.}~\bibnamefont
  {{\v{S}}mejkal}}, \bibinfo {author} {\bibfnamefont {I.}~\bibnamefont
  {Turek}}, \bibinfo {author} {\bibfnamefont {T.}~\bibnamefont {Jungwirth}},
  \bibinfo {author} {\bibfnamefont {O.}~\bibnamefont {Gomonay}}, \bibinfo
  {author} {\bibfnamefont {J.}~\bibnamefont {Sinova}}, \bibinfo {author}
  {\bibfnamefont {A.~A.}\ \bibnamefont {Sapozhnik}}, \bibinfo {author}
  {\bibfnamefont {H.-J.}\ \bibnamefont {Elmers}}, \bibinfo {author}
  {\bibfnamefont {M.}~\bibnamefont {Kl{\"{a}}ui}}, \ and\ \bibinfo {author}
  {\bibfnamefont {M.}~\bibnamefont {Jourdan}},\ }\href {\doibase
  10.1038/s41467-017-02780-x} {\bibfield  {journal} {\bibinfo  {journal} {Nat.
  Commun.}\ }\textbf {\bibinfo {volume} {9}},\ \bibinfo {pages} {348} (\bibinfo
  {year} {2018})}\BibitemShut {NoStop}%
\bibitem [{\citenamefont {Jungwirth}\ \emph {et~al.}(2016)\citenamefont
  {Jungwirth}, \citenamefont {Marti}, \citenamefont {Wadley},\ and\
  \citenamefont {Wunderlich}}]{Jungwirth2016AntiferromagneticReview}%
  \BibitemOpen
  \bibfield  {author} {\bibinfo {author} {\bibfnamefont {T.}~\bibnamefont
  {Jungwirth}}, \bibinfo {author} {\bibfnamefont {X.}~\bibnamefont {Marti}},
  \bibinfo {author} {\bibfnamefont {P.}~\bibnamefont {Wadley}}, \ and\ \bibinfo
  {author} {\bibfnamefont {J.}~\bibnamefont {Wunderlich}},\ }\href {\doibase
  10.1038/nnano.2016.18} {\bibfield  {journal} {\bibinfo  {journal} {Nat.
  Nanotechnol.}\ }\textbf {\bibinfo {volume} {11}},\ \bibinfo {pages} {231}
  (\bibinfo {year} {2016})}\BibitemShut {NoStop}%
\bibitem [{\citenamefont {Baltz}\ \emph {et~al.}(2018)\citenamefont {Baltz},
  \citenamefont {Manchon}, \citenamefont {Tsoi}, \citenamefont {Moriyama},
  \citenamefont {Ono},\ and\ \citenamefont
  {Tserkovnyak}}]{Baltz2018Antiferromagnetic}%
  \BibitemOpen
  \bibfield  {author} {\bibinfo {author} {\bibfnamefont {V.}~\bibnamefont
  {Baltz}}, \bibinfo {author} {\bibfnamefont {A.}~\bibnamefont {Manchon}},
  \bibinfo {author} {\bibfnamefont {M.}~\bibnamefont {Tsoi}}, \bibinfo {author}
  {\bibfnamefont {T.}~\bibnamefont {Moriyama}}, \bibinfo {author}
  {\bibfnamefont {T.}~\bibnamefont {Ono}}, \ and\ \bibinfo {author}
  {\bibfnamefont {Y.}~\bibnamefont {Tserkovnyak}},\ }\href {\doibase
  10.1103/RevModPhys.90.015005} {\bibfield  {journal} {\bibinfo  {journal}
  {Rev. Mod. Phys.}\ }\textbf {\bibinfo {volume} {90}},\ \bibinfo {pages}
  {015005} (\bibinfo {year} {2018})}\BibitemShut {NoStop}%
\bibitem [{\citenamefont {{Manchon}}\ \emph {et~al.}(2018)\citenamefont
  {{Manchon}}, \citenamefont {{Miron}}, \citenamefont {{Jungwirth}},
  \citenamefont {{Sinova}}, \citenamefont {{{\v{Z}}elezn{\'y}}}, \citenamefont
  {{Thiaville}}, \citenamefont {{Garello}},\ and\ \citenamefont
  {{Gambardella}}}]{Manchon2018currentinduced}%
  \BibitemOpen
  \bibfield  {author} {\bibinfo {author} {\bibfnamefont {A.}~\bibnamefont
  {{Manchon}}}, \bibinfo {author} {\bibfnamefont {I.~M.}\ \bibnamefont
  {{Miron}}}, \bibinfo {author} {\bibfnamefont {T.}~\bibnamefont
  {{Jungwirth}}}, \bibinfo {author} {\bibfnamefont {J.}~\bibnamefont
  {{Sinova}}}, \bibinfo {author} {\bibfnamefont {J.}~\bibnamefont
  {{{\v{Z}}elezn{\'y}}}}, \bibinfo {author} {\bibfnamefont {A.}~\bibnamefont
  {{Thiaville}}}, \bibinfo {author} {\bibfnamefont {K.}~\bibnamefont
  {{Garello}}}, \ and\ \bibinfo {author} {\bibfnamefont {P.}~\bibnamefont
  {{Gambardella}}},\ }\href@noop {} {\  (\bibinfo {year} {2018})},\ \Eprint
  {http://arxiv.org/abs/1801.09636} {arXiv:1801.09636} \BibitemShut {NoStop}%
\bibitem [{\citenamefont {Shiomi}\ \emph
  {et~al.}(2018{\natexlab{a}})\citenamefont {Shiomi}, \citenamefont {Akiba},
  \citenamefont {Takahashi},\ and\ \citenamefont
  {Ishiwata}}]{shiomi2018AgCrSe2Piezo}%
  \BibitemOpen
  \bibfield  {author} {\bibinfo {author} {\bibfnamefont {Y.}~\bibnamefont
  {Shiomi}}, \bibinfo {author} {\bibfnamefont {T.}~\bibnamefont {Akiba}},
  \bibinfo {author} {\bibfnamefont {H.}~\bibnamefont {Takahashi}}, \ and\
  \bibinfo {author} {\bibfnamefont {S.}~\bibnamefont {Ishiwata}},\ }\href
  {\doibase 10.1002/aelm.201800174} {\bibfield  {journal} {\bibinfo  {journal}
  {Advanced Electronic Materials}\ }\textbf {\bibinfo {volume} {0}},\ \bibinfo
  {pages} {1800174} (\bibinfo {year} {2018}{\natexlab{a}})}\BibitemShut
  {NoStop}%
\bibitem [{\citenamefont {Shiomi}\ \emph
  {et~al.}(2018{\natexlab{b}})\citenamefont {Shiomi}, \citenamefont {Watanabe},
  \citenamefont {Hidetoshi}, \citenamefont {Takahashi}, \citenamefont
  {Yanase},\ and\ \citenamefont {Ishiwata}}]{shiomi2018magnetopiezo}%
  \BibitemOpen
  \bibfield  {author} {\bibinfo {author} {\bibfnamefont {Y.}~\bibnamefont
  {Shiomi}}, \bibinfo {author} {\bibfnamefont {H.}~\bibnamefont {Watanabe}},
  \bibinfo {author} {\bibfnamefont {M.}~\bibnamefont {Hidetoshi}}, \bibinfo
  {author} {\bibfnamefont {H.}~\bibnamefont {Takahashi}}, \bibinfo {author}
  {\bibfnamefont {Y.}~\bibnamefont {Yanase}}, \ and\ \bibinfo {author}
  {\bibfnamefont {S.}~\bibnamefont {Ishiwata}},\ }\href@noop {} {\  (\bibinfo
  {year} {2018}{\natexlab{b}})},\ \Eprint {http://arxiv.org/abs/1811.08107}
  {arXiv:1811.08107} \BibitemShut {NoStop}%
\bibitem [{\citenamefont {Landau}\ \emph {et~al.}(1980)\citenamefont {Landau},
  \citenamefont {Lifshitz},\ and\ \citenamefont
  {Pitaevskii}}]{landau1980statistical}%
  \BibitemOpen
  \bibfield  {author} {\bibinfo {author} {\bibfnamefont {L.~D.}\ \bibnamefont
  {Landau}}, \bibinfo {author} {\bibfnamefont {E.}~\bibnamefont {Lifshitz}}, \
  and\ \bibinfo {author} {\bibfnamefont {L.}~\bibnamefont {Pitaevskii}},\
  }\href@noop {} {\emph {\bibinfo {title} {Statistical Physics (Course of
  Theoretical Physics, Volume 5)}}}\ (\bibinfo  {publisher} {pergamon,
  Oxford},\ \bibinfo {year} {1980})\BibitemShut {NoStop}%
\bibitem [{\citenamefont {Rikken}\ \emph {et~al.}(2001)\citenamefont {Rikken},
  \citenamefont {F{\"{o}}lling},\ and\ \citenamefont {Wyder}}]{Rikken2001}%
  \BibitemOpen
  \bibfield  {author} {\bibinfo {author} {\bibfnamefont {G.~L. J.~A.}\
  \bibnamefont {Rikken}}, \bibinfo {author} {\bibfnamefont {J.}~\bibnamefont
  {F{\"{o}}lling}}, \ and\ \bibinfo {author} {\bibfnamefont {P.}~\bibnamefont
  {Wyder}},\ }\href {\doibase 10.1103/PhysRevLett.87.236602} {\bibfield
  {journal} {\bibinfo  {journal} {Phys. Rev. Lett.}\ }\textbf {\bibinfo
  {volume} {87}},\ \bibinfo {pages} {236602} (\bibinfo {year}
  {2001})}\BibitemShut {NoStop}%
\bibitem [{\citenamefont {Rikken}\ and\ \citenamefont
  {Wyder}(2005)}]{Rikken2005}%
  \BibitemOpen
  \bibfield  {author} {\bibinfo {author} {\bibfnamefont {G.~L. J.~A.}\
  \bibnamefont {Rikken}}\ and\ \bibinfo {author} {\bibfnamefont
  {P.}~\bibnamefont {Wyder}},\ }\href {\doibase 10.1103/PhysRevLett.94.016601}
  {\bibfield  {journal} {\bibinfo  {journal} {Phys. Rev. Lett.}\ }\textbf
  {\bibinfo {volume} {94}},\ \bibinfo {pages} {016601} (\bibinfo {year}
  {2005})}\BibitemShut {NoStop}%
\bibitem [{\citenamefont {Ideue}\ \emph {et~al.}(2017)\citenamefont {Ideue},
  \citenamefont {Hamamoto}, \citenamefont {Koshikawa}, \citenamefont {Ezawa},
  \citenamefont {Shimizu}, \citenamefont {Kaneko}, \citenamefont {Tokura},
  \citenamefont {Nagaosa},\ and\ \citenamefont {Iwasa}}]{Ideue2017a}%
  \BibitemOpen
  \bibfield  {author} {\bibinfo {author} {\bibfnamefont {T.}~\bibnamefont
  {Ideue}}, \bibinfo {author} {\bibfnamefont {K.}~\bibnamefont {Hamamoto}},
  \bibinfo {author} {\bibfnamefont {S.}~\bibnamefont {Koshikawa}}, \bibinfo
  {author} {\bibfnamefont {M.}~\bibnamefont {Ezawa}}, \bibinfo {author}
  {\bibfnamefont {S.}~\bibnamefont {Shimizu}}, \bibinfo {author} {\bibfnamefont
  {Y.}~\bibnamefont {Kaneko}}, \bibinfo {author} {\bibfnamefont
  {Y.}~\bibnamefont {Tokura}}, \bibinfo {author} {\bibfnamefont
  {N.}~\bibnamefont {Nagaosa}}, \ and\ \bibinfo {author} {\bibfnamefont
  {Y.}~\bibnamefont {Iwasa}},\ }\href {\doibase 10.1038/nphys4056} {\bibfield
  {journal} {\bibinfo  {journal} {Nat. Phys.}\ }\textbf {\bibinfo {volume}
  {13}},\ \bibinfo {pages} {578} (\bibinfo {year} {2017})}\BibitemShut
  {NoStop}%
\bibitem [{\citenamefont {Yasuda}\ \emph {et~al.}(2016)\citenamefont {Yasuda},
  \citenamefont {Tsukazaki}, \citenamefont {Yoshimi}, \citenamefont
  {Takahashi}, \citenamefont {Kawasaki},\ and\ \citenamefont
  {Tokura}}]{Yasuda2016a}%
  \BibitemOpen
  \bibfield  {author} {\bibinfo {author} {\bibfnamefont {K.}~\bibnamefont
  {Yasuda}}, \bibinfo {author} {\bibfnamefont {A.}~\bibnamefont {Tsukazaki}},
  \bibinfo {author} {\bibfnamefont {R.}~\bibnamefont {Yoshimi}}, \bibinfo
  {author} {\bibfnamefont {K.~S.}\ \bibnamefont {Takahashi}}, \bibinfo {author}
  {\bibfnamefont {M.}~\bibnamefont {Kawasaki}}, \ and\ \bibinfo {author}
  {\bibfnamefont {Y.}~\bibnamefont {Tokura}},\ }\href {\doibase
  10.1103/PhysRevLett.117.127202} {\bibfield  {journal} {\bibinfo  {journal}
  {Phys. Rev. Lett.}\ }\textbf {\bibinfo {volume} {117}},\ \bibinfo {pages}
  {127202} (\bibinfo {year} {2016})}\BibitemShut {NoStop}%
\bibitem [{\citenamefont {Norman}(2015)}]{Norman2015}%
  \BibitemOpen
  \bibfield  {author} {\bibinfo {author} {\bibfnamefont {M.~R.}\ \bibnamefont
  {Norman}},\ }\href {\doibase 10.1103/PhysRevB.92.075113} {\bibfield
  {journal} {\bibinfo  {journal} {Phys. Rev. B}\ }\textbf {\bibinfo {volume}
  {92}},\ \bibinfo {pages} {075113} (\bibinfo {year} {2015})}\BibitemShut
  {NoStop}%
\bibitem [{\citenamefont {Shibata}\ \emph {et~al.}(2016)\citenamefont
  {Shibata}, \citenamefont {Takeuchi}, \citenamefont {Kohno},\ and\
  \citenamefont {Tatara}}]{Shibata2016}%
  \BibitemOpen
  \bibfield  {author} {\bibinfo {author} {\bibfnamefont {J.}~\bibnamefont
  {Shibata}}, \bibinfo {author} {\bibfnamefont {A.}~\bibnamefont {Takeuchi}},
  \bibinfo {author} {\bibfnamefont {H.}~\bibnamefont {Kohno}}, \ and\ \bibinfo
  {author} {\bibfnamefont {G.}~\bibnamefont {Tatara}},\ }\href {\doibase
  10.7566/JPSJ.85.033701} {\bibfield  {journal} {\bibinfo  {journal} {J. Phys.
  Soc. Jpn.}\ }\textbf {\bibinfo {volume} {85}},\ \bibinfo {pages} {033701}
  (\bibinfo {year} {2016})}\BibitemShut {NoStop}%
\bibitem [{\citenamefont {Kawaguchi}\ and\ \citenamefont
  {Tatara}(2016)}]{Kawaguchi2016a}%
  \BibitemOpen
  \bibfield  {author} {\bibinfo {author} {\bibfnamefont {H.}~\bibnamefont
  {Kawaguchi}}\ and\ \bibinfo {author} {\bibfnamefont {G.}~\bibnamefont
  {Tatara}},\ }\href {\doibase 10.1103/PhysRevB.94.235148} {\bibfield
  {journal} {\bibinfo  {journal} {Phys. Rev. B}\ }\textbf {\bibinfo {volume}
  {94}},\ \bibinfo {pages} {235148} (\bibinfo {year} {2016})}\BibitemShut
  {NoStop}%
\bibitem [{Dom()}]{DomainSwitch}%
  \BibitemOpen
  \href@noop {} {}\bibinfo {note} {H.~Watanabe and Y.~Yanase,
  Submitted.}\BibitemShut {Stop}%
\bibitem [{\citenamefont {Saito}\ \emph {et~al.}(2015)\citenamefont {Saito},
  \citenamefont {Nakamura}, \citenamefont {Bahramy}, \citenamefont {Kohama},
  \citenamefont {Ye}, \citenamefont {Kasahara}, \citenamefont {Nakagawa},
  \citenamefont {Onga}, \citenamefont {Tokunaga}, \citenamefont {Nojima},
  \citenamefont {Yanase},\ and\ \citenamefont {Iwasa}}]{Saito2016}%
  \BibitemOpen
  \bibfield  {author} {\bibinfo {author} {\bibfnamefont {Y.}~\bibnamefont
  {Saito}}, \bibinfo {author} {\bibfnamefont {Y.}~\bibnamefont {Nakamura}},
  \bibinfo {author} {\bibfnamefont {M.~S.}\ \bibnamefont {Bahramy}}, \bibinfo
  {author} {\bibfnamefont {Y.}~\bibnamefont {Kohama}}, \bibinfo {author}
  {\bibfnamefont {J.}~\bibnamefont {Ye}}, \bibinfo {author} {\bibfnamefont
  {Y.}~\bibnamefont {Kasahara}}, \bibinfo {author} {\bibfnamefont
  {Y.}~\bibnamefont {Nakagawa}}, \bibinfo {author} {\bibfnamefont
  {M.}~\bibnamefont {Onga}}, \bibinfo {author} {\bibfnamefont {M.}~\bibnamefont
  {Tokunaga}}, \bibinfo {author} {\bibfnamefont {T.}~\bibnamefont {Nojima}},
  \bibinfo {author} {\bibfnamefont {Y.}~\bibnamefont {Yanase}}, \ and\ \bibinfo
  {author} {\bibfnamefont {Y.}~\bibnamefont {Iwasa}},\ }\href {\doibase
  10.1038/nphys3580} {\bibfield  {journal} {\bibinfo  {journal} {Nat. Phys.}\
  }\textbf {\bibinfo {volume} {12}},\ \bibinfo {pages} {144} (\bibinfo {year}
  {2015})}\BibitemShut {NoStop}%
\bibitem [{\citenamefont {Wakatsuki}\ \emph {et~al.}(2017)\citenamefont
  {Wakatsuki}, \citenamefont {Saito}, \citenamefont {Hoshino}, \citenamefont
  {Itahashi}, \citenamefont {Ideue}, \citenamefont {Ezawa}, \citenamefont
  {Iwasa},\ and\ \citenamefont {Nagaosa}}]{Wakatsuki2017}%
  \BibitemOpen
  \bibfield  {author} {\bibinfo {author} {\bibfnamefont {R.}~\bibnamefont
  {Wakatsuki}}, \bibinfo {author} {\bibfnamefont {Y.}~\bibnamefont {Saito}},
  \bibinfo {author} {\bibfnamefont {S.}~\bibnamefont {Hoshino}}, \bibinfo
  {author} {\bibfnamefont {Y.~M.}\ \bibnamefont {Itahashi}}, \bibinfo {author}
  {\bibfnamefont {T.}~\bibnamefont {Ideue}}, \bibinfo {author} {\bibfnamefont
  {M.}~\bibnamefont {Ezawa}}, \bibinfo {author} {\bibfnamefont
  {Y.}~\bibnamefont {Iwasa}}, \ and\ \bibinfo {author} {\bibfnamefont
  {N.}~\bibnamefont {Nagaosa}},\ }\href {\doibase 10.1126/sciadv.1602390}
  {\bibfield  {journal} {\bibinfo  {journal} {Sci. Adv.}\ }\textbf {\bibinfo
  {volume} {3}},\ \bibinfo {pages} {e1602390} (\bibinfo {year}
  {2017})}\BibitemShut {NoStop}%
\bibitem [{\citenamefont {Hoshino}\ \emph {et~al.}(2018)\citenamefont
  {Hoshino}, \citenamefont {Wakatsuki}, \citenamefont {Hamamoto},\ and\
  \citenamefont {Nagaosa}}]{Hoshino2018Nonreciprocal}%
  \BibitemOpen
  \bibfield  {author} {\bibinfo {author} {\bibfnamefont {S.}~\bibnamefont
  {Hoshino}}, \bibinfo {author} {\bibfnamefont {R.}~\bibnamefont {Wakatsuki}},
  \bibinfo {author} {\bibfnamefont {K.}~\bibnamefont {Hamamoto}}, \ and\
  \bibinfo {author} {\bibfnamefont {N.}~\bibnamefont {Nagaosa}},\ }\href
  {\doibase 10.1103/PhysRevB.98.054510} {\bibfield  {journal} {\bibinfo
  {journal} {Phys. Rev. B}\ }\textbf {\bibinfo {volume} {98}},\ \bibinfo
  {pages} {054510} (\bibinfo {year} {2018})}\BibitemShut {NoStop}%
\bibitem [{\citenamefont {Rabe}\ \emph {et~al.}(2007)\citenamefont {Rabe},
  \citenamefont {Ahn},\ and\ \citenamefont
  {Triscone}}]{rabe2007ferroelectric_book}%
  \BibitemOpen
  \bibfield  {author} {\bibinfo {author} {\bibfnamefont {K.~M.}\ \bibnamefont
  {Rabe}}, \bibinfo {author} {\bibfnamefont {C.~H.}\ \bibnamefont {Ahn}}, \
  and\ \bibinfo {author} {\bibfnamefont {J.-M.}\ \bibnamefont {Triscone}},\
  }\href@noop {} {\emph {\bibinfo {title} {Physics of ferroelectrics: a modern
  perspective}}},\ Vol.\ \bibinfo {volume} {105}\ (\bibinfo  {publisher}
  {Springer Science \& Business Media},\ \bibinfo {year} {2007})\BibitemShut
  {NoStop}%
\bibitem [{\citenamefont {Shi}\ \emph {et~al.}(2013)\citenamefont {Shi},
  \citenamefont {Guo}, \citenamefont {Wang}, \citenamefont {Princep},
  \citenamefont {Khalyavin}, \citenamefont {Manuel}, \citenamefont {Michiue},
  \citenamefont {Sato}, \citenamefont {Tsuda}, \citenamefont {Yu},
  \citenamefont {Arai}, \citenamefont {Shirako}, \citenamefont {Akaogi},
  \citenamefont {Wang}, \citenamefont {Yamaura},\ and\ \citenamefont
  {Boothroyd}}]{Shi2013}%
  \BibitemOpen
  \bibfield  {author} {\bibinfo {author} {\bibfnamefont {Y.}~\bibnamefont
  {Shi}}, \bibinfo {author} {\bibfnamefont {Y.}~\bibnamefont {Guo}}, \bibinfo
  {author} {\bibfnamefont {X.}~\bibnamefont {Wang}}, \bibinfo {author}
  {\bibfnamefont {A.~J.}\ \bibnamefont {Princep}}, \bibinfo {author}
  {\bibfnamefont {D.}~\bibnamefont {Khalyavin}}, \bibinfo {author}
  {\bibfnamefont {P.}~\bibnamefont {Manuel}}, \bibinfo {author} {\bibfnamefont
  {Y.}~\bibnamefont {Michiue}}, \bibinfo {author} {\bibfnamefont
  {A.}~\bibnamefont {Sato}}, \bibinfo {author} {\bibfnamefont {K.}~\bibnamefont
  {Tsuda}}, \bibinfo {author} {\bibfnamefont {S.}~\bibnamefont {Yu}}, \bibinfo
  {author} {\bibfnamefont {M.}~\bibnamefont {Arai}}, \bibinfo {author}
  {\bibfnamefont {Y.}~\bibnamefont {Shirako}}, \bibinfo {author} {\bibfnamefont
  {M.}~\bibnamefont {Akaogi}}, \bibinfo {author} {\bibfnamefont
  {N.}~\bibnamefont {Wang}}, \bibinfo {author} {\bibfnamefont {K.}~\bibnamefont
  {Yamaura}}, \ and\ \bibinfo {author} {\bibfnamefont {A.~T.}\ \bibnamefont
  {Boothroyd}},\ }\href {\doibase 10.1038/nmat3754} {\bibfield  {journal}
  {\bibinfo  {journal} {Nat. Mater.}\ }\textbf {\bibinfo {volume} {12}},\
  \bibinfo {pages} {1024} (\bibinfo {year} {2013})}\BibitemShut {NoStop}%
\bibitem [{\citenamefont {Rischau}\ \emph {et~al.}(2017)\citenamefont
  {Rischau}, \citenamefont {Lin}, \citenamefont {Grams}, \citenamefont {Finck},
  \citenamefont {Harms}, \citenamefont {Engelmayer}, \citenamefont {Lorenz},
  \citenamefont {Gallais}, \citenamefont {Fauqu{\'{e}}}, \citenamefont
  {Hemberger},\ and\ \citenamefont {Behnia}}]{Rischau2017}%
  \BibitemOpen
  \bibfield  {author} {\bibinfo {author} {\bibfnamefont {C.~W.}\ \bibnamefont
  {Rischau}}, \bibinfo {author} {\bibfnamefont {X.}~\bibnamefont {Lin}},
  \bibinfo {author} {\bibfnamefont {C.~P.}\ \bibnamefont {Grams}}, \bibinfo
  {author} {\bibfnamefont {D.}~\bibnamefont {Finck}}, \bibinfo {author}
  {\bibfnamefont {S.}~\bibnamefont {Harms}}, \bibinfo {author} {\bibfnamefont
  {J.}~\bibnamefont {Engelmayer}}, \bibinfo {author} {\bibfnamefont
  {T.}~\bibnamefont {Lorenz}}, \bibinfo {author} {\bibfnamefont
  {Y.}~\bibnamefont {Gallais}}, \bibinfo {author} {\bibfnamefont
  {B.}~\bibnamefont {Fauqu{\'{e}}}}, \bibinfo {author} {\bibfnamefont
  {J.}~\bibnamefont {Hemberger}}, \ and\ \bibinfo {author} {\bibfnamefont
  {K.}~\bibnamefont {Behnia}},\ }\href {\doibase 10.1038/nphys4085} {\bibfield
  {journal} {\bibinfo  {journal} {Nat. Phys.}\ }\textbf {\bibinfo {volume}
  {13}},\ \bibinfo {pages} {643} (\bibinfo {year} {2017})}\BibitemShut
  {NoStop}%
\bibitem [{Note1()}]{Note1}%
  \BibitemOpen
  \bibinfo {note} {T. Hitomi and Y. Yanase, Submitted.}\BibitemShut {Stop}%
\bibitem [{\citenamefont {Ederer}(2009)}]{Ederer2009}%
  \BibitemOpen
  \bibfield  {author} {\bibinfo {author} {\bibfnamefont {C.}~\bibnamefont
  {Ederer}},\ }\href {\doibase 10.1140/epjb/e2009-00274-4} {\bibfield
  {journal} {\bibinfo  {journal} {Eur. Phys. J. B}\ }\textbf {\bibinfo {volume}
  {71}},\ \bibinfo {pages} {349} (\bibinfo {year} {2009})}\BibitemShut
  {NoStop}%
\bibitem [{\citenamefont {Bertaut}(1968)}]{Bertaut1968}%
  \BibitemOpen
  \bibfield  {author} {\bibinfo {author} {\bibfnamefont {E.~F.}\ \bibnamefont
  {Bertaut}},\ }\href {\doibase 10.1107/S0567739468000306} {\bibfield
  {journal} {\bibinfo  {journal} {Acta Crystallogr. Sect. A}\ }\textbf
  {\bibinfo {volume} {24}},\ \bibinfo {pages} {217} (\bibinfo {year}
  {1968})}\BibitemShut {NoStop}%
\bibitem [{\citenamefont {Izyumov}\ \emph {et~al.}(1991)\citenamefont
  {Izyumov}, \citenamefont {Ozerov},\ and\ \citenamefont
  {Naish}}]{izyumov1991neutron}%
  \BibitemOpen
  \bibfield  {author} {\bibinfo {author} {\bibfnamefont {Y.~A.}\ \bibnamefont
  {Izyumov}}, \bibinfo {author} {\bibfnamefont {R.}~\bibnamefont {Ozerov}}, \
  and\ \bibinfo {author} {\bibfnamefont {V.}~\bibnamefont {Naish}},\
  }\href@noop {} {\emph {\bibinfo {title} {Neutron diffraction of magnetic
  materials}}}\ (\bibinfo  {publisher} {Springer},\ \bibinfo {year}
  {1991})\BibitemShut {NoStop}%
\bibitem [{\citenamefont {Birss}\ \emph {et~al.}(1964)\citenamefont {Birss}
  \emph {et~al.}}]{birss1964symmetry}%
  \BibitemOpen
  \bibfield  {author} {\bibinfo {author} {\bibfnamefont {R.~R.}\ \bibnamefont
  {Birss}} \emph {et~al.},\ }\href@noop {} {\emph {\bibinfo {title} {Symmetry
  and magnetism}}},\ Vol.\ \bibinfo {volume} {863}\ (\bibinfo  {publisher}
  {North-Holland Amsterdam},\ \bibinfo {year} {1964})\BibitemShut {NoStop}%
\bibitem [{\citenamefont {Nye}(1985)}]{nye1985physical}%
  \BibitemOpen
  \bibfield  {author} {\bibinfo {author} {\bibfnamefont {J.~F.}\ \bibnamefont
  {Nye}},\ }\href@noop {} {\emph {\bibinfo {title} {Physical properties of
  crystals: their representation by tensors and matrices}}}\ (\bibinfo
  {publisher} {Oxford university press},\ \bibinfo {year} {1985})\BibitemShut
  {NoStop}%
\bibitem [{\citenamefont {Janovec}\ \emph {et~al.}(2010)\citenamefont
  {Janovec}, \citenamefont {{\v{C}}mel{\'{i}}k},\ and\ \citenamefont
  {Machonsk{\'{y}}}}]{Janovec2010}%
  \BibitemOpen
  \bibfield  {author} {\bibinfo {author} {\bibfnamefont {V.}~\bibnamefont
  {Janovec}}, \bibinfo {author} {\bibfnamefont {M.}~\bibnamefont
  {{\v{C}}mel{\'{i}}k}}, \ and\ \bibinfo {author} {\bibfnamefont
  {L.}~\bibnamefont {Machonsk{\'{y}}}},\ }\href {\doibase
  10.1080/01411594.2010.499494} {\bibfield  {journal} {\bibinfo  {journal}
  {Phase Transitions}\ }\textbf {\bibinfo {volume} {83}},\ \bibinfo {pages}
  {670} (\bibinfo {year} {2010})}\BibitemShut {NoStop}%
\bibitem [{\citenamefont {Litvin}\ and\ \citenamefont
  {Janovec}(2014)}]{Litvin2014}%
  \BibitemOpen
  \bibfield  {author} {\bibinfo {author} {\bibfnamefont {D.~B.}\ \bibnamefont
  {Litvin}}\ and\ \bibinfo {author} {\bibfnamefont {V.}~\bibnamefont
  {Janovec}},\ }\href {\doibase 10.1080/00150193.2014.889523} {\bibfield
  {journal} {\bibinfo  {journal} {Ferroelectrics}\ }\textbf {\bibinfo {volume}
  {461}},\ \bibinfo {pages} {10} (\bibinfo {year} {2014})}\BibitemShut
  {NoStop}%
\bibitem [{\citenamefont {Schmid}(1973)}]{Schmid1973}%
  \BibitemOpen
  \bibfield  {author} {\bibinfo {author} {\bibfnamefont {H.}~\bibnamefont
  {Schmid}},\ }\href {http://archive-ouverte.unige.ch/unige:32935} {\bibfield
  {journal} {\bibinfo  {journal} {Int. J. Magn.}\ }\textbf {\bibinfo {volume}
  {4}},\ \bibinfo {pages} {337} (\bibinfo {year} {1973})}\BibitemShut {NoStop}%
\bibitem [{\citenamefont {Siratori}\ \emph {et~al.}(1992)\citenamefont
  {Siratori}, \citenamefont {Kohn},\ and\ \citenamefont
  {Kita}}]{Siratori1992a}%
  \BibitemOpen
  \bibfield  {author} {\bibinfo {author} {\bibfnamefont {K.}~\bibnamefont
  {Siratori}}, \bibinfo {author} {\bibfnamefont {K.}~\bibnamefont {Kohn}}, \
  and\ \bibinfo {author} {\bibfnamefont {E.}~\bibnamefont {Kita}},\ }\href
  {http://psjd.icm.edu.pl/psjd/element/bwmeta1.element.bwnjournal-article-appv81z401kz}
  {\bibfield  {journal} {\bibinfo  {journal} {Acta Phys. Pol. A}\ }\textbf
  {\bibinfo {volume} {81}},\ \bibinfo {pages} {431} (\bibinfo {year}
  {1992})}\BibitemShut {NoStop}%
\bibitem [{\citenamefont {Gallego}\ \emph {et~al.}(2016)\citenamefont
  {Gallego}, \citenamefont {Perez-Mato}, \citenamefont {Elcoro}, \citenamefont
  {Tasci}, \citenamefont {Hanson}, \citenamefont {Momma}, \citenamefont
  {Aroyo},\ and\ \citenamefont {Madariaga}}]{Gallego2016a}%
  \BibitemOpen
  \bibfield  {author} {\bibinfo {author} {\bibfnamefont {S.~V.}\ \bibnamefont
  {Gallego}}, \bibinfo {author} {\bibfnamefont {J.~M.}\ \bibnamefont
  {Perez-Mato}}, \bibinfo {author} {\bibfnamefont {L.}~\bibnamefont {Elcoro}},
  \bibinfo {author} {\bibfnamefont {E.~S.}\ \bibnamefont {Tasci}}, \bibinfo
  {author} {\bibfnamefont {R.~M.}\ \bibnamefont {Hanson}}, \bibinfo {author}
  {\bibfnamefont {K.}~\bibnamefont {Momma}}, \bibinfo {author} {\bibfnamefont
  {M.~I.}\ \bibnamefont {Aroyo}}, \ and\ \bibinfo {author} {\bibfnamefont
  {G.}~\bibnamefont {Madariaga}},\ }\href {\doibase 10.1107/S1600576716012863}
  {\bibfield  {journal} {\bibinfo  {journal} {J. Appl. Crystallogr.}\ }\textbf
  {\bibinfo {volume} {49}},\ \bibinfo {pages} {1750} (\bibinfo {year}
  {2016})}\BibitemShut {NoStop}%
\bibitem [{\citenamefont {Aroyo}(2016)}]{internatinaltables}%
  \BibitemOpen
  \bibinfo {editor} {\bibfnamefont {M.~I.}\ \bibnamefont {Aroyo}},\ ed.,\ \href
  {\doibase 10.1107/97809553602060000114} {\emph {\bibinfo {title}
  {International Tables for Crystallography}}}\ (\bibinfo  {publisher}
  {International Union of Crystallography},\ \bibinfo {year}
  {2016})\BibitemShut {NoStop}%
\bibitem [{\citenamefont {Litvin}(2008)}]{Litvin2008}%
  \BibitemOpen
  \bibfield  {author} {\bibinfo {author} {\bibfnamefont {D.~B.}\ \bibnamefont
  {Litvin}},\ }\href {\doibase 10.1107/S0108767307068262} {\bibfield  {journal}
  {\bibinfo  {journal} {Acta Crystallogr. Sect. A Found. Crystallogr.}\
  }\textbf {\bibinfo {volume} {64}},\ \bibinfo {pages} {316} (\bibinfo {year}
  {2008})}\BibitemShut {NoStop}%
\bibitem [{\citenamefont {Sheptyakov}\ \emph {et~al.}(2010)\citenamefont
  {Sheptyakov}, \citenamefont {Ali},\ and\ \citenamefont
  {Jansen}}]{Sheptyakov2010}%
  \BibitemOpen
  \bibfield  {author} {\bibinfo {author} {\bibfnamefont {D.}~\bibnamefont
  {Sheptyakov}}, \bibinfo {author} {\bibfnamefont {N.~Z.}\ \bibnamefont {Ali}},
  \ and\ \bibinfo {author} {\bibfnamefont {M.}~\bibnamefont {Jansen}},\ }\href
  {\doibase 10.1088/0953-8984/22/42/426001} {\bibfield  {journal} {\bibinfo
  {journal} {J. Phys. Condens. Matter}\ }\textbf {\bibinfo {volume} {22}},\
  \bibinfo {pages} {426001} (\bibinfo {year} {2010})}\BibitemShut {NoStop}%
\bibitem [{\citenamefont {Roy}\ \emph {et~al.}(2013)\citenamefont {Roy},
  \citenamefont {Pandey}, \citenamefont {Zhang}, \citenamefont {Heitmann},
  \citenamefont {Vaknin}, \citenamefont {Johnston},\ and\ \citenamefont
  {Furukawa}}]{Roy2013}%
  \BibitemOpen
  \bibfield  {author} {\bibinfo {author} {\bibfnamefont {B.}~\bibnamefont
  {Roy}}, \bibinfo {author} {\bibfnamefont {A.}~\bibnamefont {Pandey}},
  \bibinfo {author} {\bibfnamefont {Q.}~\bibnamefont {Zhang}}, \bibinfo
  {author} {\bibfnamefont {T.~W.}\ \bibnamefont {Heitmann}}, \bibinfo {author}
  {\bibfnamefont {D.}~\bibnamefont {Vaknin}}, \bibinfo {author} {\bibfnamefont
  {D.~C.}\ \bibnamefont {Johnston}}, \ and\ \bibinfo {author} {\bibfnamefont
  {Y.}~\bibnamefont {Furukawa}},\ }\href {\doibase 10.1103/PhysRevB.88.174415}
  {\bibfield  {journal} {\bibinfo  {journal} {Phys. Rev. B}\ }\textbf {\bibinfo
  {volume} {88}},\ \bibinfo {pages} {174415} (\bibinfo {year}
  {2013})}\BibitemShut {NoStop}%
\bibitem [{\citenamefont {Ge}\ \emph {et~al.}(2017)\citenamefont {Ge},
  \citenamefont {Flynn}, \citenamefont {Paddison}, \citenamefont {Stone},
  \citenamefont {Calder}, \citenamefont {Subramanian}, \citenamefont
  {Ramirez},\ and\ \citenamefont {Mourigal}}]{ge2017}%
  \BibitemOpen
  \bibfield  {author} {\bibinfo {author} {\bibfnamefont {L.}~\bibnamefont
  {Ge}}, \bibinfo {author} {\bibfnamefont {J.}~\bibnamefont {Flynn}}, \bibinfo
  {author} {\bibfnamefont {J.~A.~M.}\ \bibnamefont {Paddison}}, \bibinfo
  {author} {\bibfnamefont {M.~B.}\ \bibnamefont {Stone}}, \bibinfo {author}
  {\bibfnamefont {S.}~\bibnamefont {Calder}}, \bibinfo {author} {\bibfnamefont
  {M.~A.}\ \bibnamefont {Subramanian}}, \bibinfo {author} {\bibfnamefont
  {A.~P.}\ \bibnamefont {Ramirez}}, \ and\ \bibinfo {author} {\bibfnamefont
  {M.}~\bibnamefont {Mourigal}},\ }\href {\doibase 10.1103/PhysRevB.96.064413}
  {\bibfield  {journal} {\bibinfo  {journal} {Phys. Rev. B}\ }\textbf {\bibinfo
  {volume} {96}},\ \bibinfo {pages} {064413} (\bibinfo {year}
  {2017})}\BibitemShut {NoStop}%
\bibitem [{\citenamefont {Krimmel}\ \emph {et~al.}(2006)\citenamefont
  {Krimmel}, \citenamefont {Tsurkan}, \citenamefont {Sheptyakov},\ and\
  \citenamefont {Loidl}}]{Krimmel2006}%
  \BibitemOpen
  \bibfield  {author} {\bibinfo {author} {\bibfnamefont {A.}~\bibnamefont
  {Krimmel}}, \bibinfo {author} {\bibfnamefont {V.}~\bibnamefont {Tsurkan}},
  \bibinfo {author} {\bibfnamefont {D.}~\bibnamefont {Sheptyakov}}, \ and\
  \bibinfo {author} {\bibfnamefont {A.}~\bibnamefont {Loidl}},\ }\href
  {\doibase 10.1016/j.physb.2006.01.413} {\bibfield  {journal} {\bibinfo
  {journal} {Phys. B Condens. Matter}\ }\textbf {\bibinfo {volume} {378-380}},\
  \bibinfo {pages} {583} (\bibinfo {year} {2006})}\BibitemShut {NoStop}%
\bibitem [{\citenamefont {Roth}(1964)}]{Roth1964}%
  \BibitemOpen
  \bibfield  {author} {\bibinfo {author} {\bibfnamefont {W.}~\bibnamefont
  {Roth}},\ }\href {\doibase 10.1016/0022-3697(64)90156-8} {\bibfield
  {journal} {\bibinfo  {journal} {J. Phys. Chem. Solids}\ }\textbf {\bibinfo
  {volume} {25}},\ \bibinfo {pages} {1} (\bibinfo {year} {1964})}\BibitemShut
  {NoStop}%
\bibitem [{\citenamefont {Brock}\ \emph {et~al.}(1994)\citenamefont {Brock},
  \citenamefont {Greedan},\ and\ \citenamefont {Kauzlarich}}]{Brock1994}%
  \BibitemOpen
  \bibfield  {author} {\bibinfo {author} {\bibfnamefont {S.~L.}\ \bibnamefont
  {Brock}}, \bibinfo {author} {\bibfnamefont {J.}~\bibnamefont {Greedan}}, \
  and\ \bibinfo {author} {\bibfnamefont {S.~M.}\ \bibnamefont {Kauzlarich}},\
  }\href {\doibase 10.1006/jssc.1994.1375} {\bibfield  {journal} {\bibinfo
  {journal} {J. Solid State Chem.}\ }\textbf {\bibinfo {volume} {113}},\
  \bibinfo {pages} {303} (\bibinfo {year} {1994})}\BibitemShut {NoStop}%
\bibitem [{\citenamefont {Singh}\ \emph
  {et~al.}(2009{\natexlab{a}})\citenamefont {Singh}, \citenamefont {Ellern},\
  and\ \citenamefont {Johnston}}]{Singh2009b}%
  \BibitemOpen
  \bibfield  {author} {\bibinfo {author} {\bibfnamefont {Y.}~\bibnamefont
  {Singh}}, \bibinfo {author} {\bibfnamefont {A.}~\bibnamefont {Ellern}}, \
  and\ \bibinfo {author} {\bibfnamefont {D.~C.}\ \bibnamefont {Johnston}},\
  }\href {\doibase 10.1103/PhysRevB.79.094519} {\bibfield  {journal} {\bibinfo
  {journal} {Phys. Rev. B}\ }\textbf {\bibinfo {volume} {79}},\ \bibinfo
  {pages} {094519} (\bibinfo {year} {2009}{\natexlab{a}})}\BibitemShut
  {NoStop}%
\bibitem [{\citenamefont {Singh}\ \emph
  {et~al.}(2009{\natexlab{b}})\citenamefont {Singh}, \citenamefont {Green},
  \citenamefont {Huang}, \citenamefont {Kreyssig}, \citenamefont {McQueeney},
  \citenamefont {Johnston},\ and\ \citenamefont {Goldman}}]{Singh2009c}%
  \BibitemOpen
  \bibfield  {author} {\bibinfo {author} {\bibfnamefont {Y.}~\bibnamefont
  {Singh}}, \bibinfo {author} {\bibfnamefont {M.~A.}\ \bibnamefont {Green}},
  \bibinfo {author} {\bibfnamefont {Q.}~\bibnamefont {Huang}}, \bibinfo
  {author} {\bibfnamefont {A.}~\bibnamefont {Kreyssig}}, \bibinfo {author}
  {\bibfnamefont {R.~J.}\ \bibnamefont {McQueeney}}, \bibinfo {author}
  {\bibfnamefont {D.~C.}\ \bibnamefont {Johnston}}, \ and\ \bibinfo {author}
  {\bibfnamefont {A.~I.}\ \bibnamefont {Goldman}},\ }\href {\doibase
  10.1103/PhysRevB.80.100403} {\bibfield  {journal} {\bibinfo  {journal} {Phys.
  Rev. B}\ }\textbf {\bibinfo {volume} {80}},\ \bibinfo {pages} {100403}
  (\bibinfo {year} {2009}{\natexlab{b}})}\BibitemShut {NoStop}%
\bibitem [{\citenamefont {Lamsal}\ \emph {et~al.}(2013)\citenamefont {Lamsal},
  \citenamefont {Tucker}, \citenamefont {Heitmann}, \citenamefont {Kreyssig},
  \citenamefont {Jesche}, \citenamefont {Pandey}, \citenamefont {Tian},
  \citenamefont {McQueeney}, \citenamefont {Johnston},\ and\ \citenamefont
  {Goldman}}]{Lamsal2013}%
  \BibitemOpen
  \bibfield  {author} {\bibinfo {author} {\bibfnamefont {J.}~\bibnamefont
  {Lamsal}}, \bibinfo {author} {\bibfnamefont {G.~S.}\ \bibnamefont {Tucker}},
  \bibinfo {author} {\bibfnamefont {T.~W.}\ \bibnamefont {Heitmann}}, \bibinfo
  {author} {\bibfnamefont {A.}~\bibnamefont {Kreyssig}}, \bibinfo {author}
  {\bibfnamefont {A.}~\bibnamefont {Jesche}}, \bibinfo {author} {\bibfnamefont
  {A.}~\bibnamefont {Pandey}}, \bibinfo {author} {\bibfnamefont
  {W.}~\bibnamefont {Tian}}, \bibinfo {author} {\bibfnamefont {R.~J.}\
  \bibnamefont {McQueeney}}, \bibinfo {author} {\bibfnamefont {D.~C.}\
  \bibnamefont {Johnston}}, \ and\ \bibinfo {author} {\bibfnamefont {A.~I.}\
  \bibnamefont {Goldman}},\ }\href {\doibase 10.1103/PhysRevB.87.144418}
  {\bibfield  {journal} {\bibinfo  {journal} {Phys. Rev. B}\ }\textbf {\bibinfo
  {volume} {87}},\ \bibinfo {pages} {144418} (\bibinfo {year}
  {2013})}\BibitemShut {NoStop}%
\bibitem [{\citenamefont {Sangeetha}\ \emph {et~al.}(2018)\citenamefont
  {Sangeetha}, \citenamefont {Smetana}, \citenamefont {Mudring},\ and\
  \citenamefont {Johnston}}]{Sangeetha2018}%
  \BibitemOpen
  \bibfield  {author} {\bibinfo {author} {\bibfnamefont {N.~S.}\ \bibnamefont
  {Sangeetha}}, \bibinfo {author} {\bibfnamefont {V.}~\bibnamefont {Smetana}},
  \bibinfo {author} {\bibfnamefont {A.-V.}\ \bibnamefont {Mudring}}, \ and\
  \bibinfo {author} {\bibfnamefont {D.~C.}\ \bibnamefont {Johnston}},\ }\href
  {\doibase 10.1103/PhysRevB.97.014402} {\bibfield  {journal} {\bibinfo
  {journal} {Phys. Rev. B}\ }\textbf {\bibinfo {volume} {97}},\ \bibinfo
  {pages} {014402} (\bibinfo {year} {2018})}\BibitemShut {NoStop}%
\bibitem [{\citenamefont {Calder}\ \emph {et~al.}(2014)\citenamefont {Calder},
  \citenamefont {Saparov}, \citenamefont {Cao}, \citenamefont {Niedziela},
  \citenamefont {Lumsden}, \citenamefont {Sefat},\ and\ \citenamefont
  {Christianson}}]{Calder2014}%
  \BibitemOpen
  \bibfield  {author} {\bibinfo {author} {\bibfnamefont {S.}~\bibnamefont
  {Calder}}, \bibinfo {author} {\bibfnamefont {B.}~\bibnamefont {Saparov}},
  \bibinfo {author} {\bibfnamefont {H.~B.}\ \bibnamefont {Cao}}, \bibinfo
  {author} {\bibfnamefont {J.~L.}\ \bibnamefont {Niedziela}}, \bibinfo {author}
  {\bibfnamefont {M.~D.}\ \bibnamefont {Lumsden}}, \bibinfo {author}
  {\bibfnamefont {A.~S.}\ \bibnamefont {Sefat}}, \ and\ \bibinfo {author}
  {\bibfnamefont {A.~D.}\ \bibnamefont {Christianson}},\ }\href {\doibase
  10.1103/PhysRevB.89.064417} {\bibfield  {journal} {\bibinfo  {journal} {Phys.
  Rev. B}\ }\textbf {\bibinfo {volume} {89}},\ \bibinfo {pages} {064417}
  (\bibinfo {year} {2014})}\BibitemShut {NoStop}%
\bibitem [{\citenamefont {Saparov}\ and\ \citenamefont
  {Sefat}(2013)}]{Saparov2013}%
  \BibitemOpen
  \bibfield  {author} {\bibinfo {author} {\bibfnamefont {B.}~\bibnamefont
  {Saparov}}\ and\ \bibinfo {author} {\bibfnamefont {A.~S.}\ \bibnamefont
  {Sefat}},\ }\href {\doibase 10.1016/j.jssc.2013.05.010} {\bibfield  {journal}
  {\bibinfo  {journal} {J. Solid State Chem.}\ }\textbf {\bibinfo {volume}
  {204}},\ \bibinfo {pages} {32} (\bibinfo {year} {2013})}\BibitemShut
  {NoStop}%
\bibitem [{\citenamefont {{Md Din}}\ \emph {et~al.}(2015)\citenamefont {{Md
  Din}}, \citenamefont {Wang}, \citenamefont {Cheng}, \citenamefont {Dou},
  \citenamefont {Kennedy}, \citenamefont {Avdeev},\ and\ \citenamefont
  {Campbell}}]{MdDin2015}%
  \BibitemOpen
  \bibfield  {author} {\bibinfo {author} {\bibfnamefont {M.~F.}\ \bibnamefont
  {{Md Din}}}, \bibinfo {author} {\bibfnamefont {J.~L.}\ \bibnamefont {Wang}},
  \bibinfo {author} {\bibfnamefont {Z.~X.}\ \bibnamefont {Cheng}}, \bibinfo
  {author} {\bibfnamefont {S.~X.}\ \bibnamefont {Dou}}, \bibinfo {author}
  {\bibfnamefont {S.~J.}\ \bibnamefont {Kennedy}}, \bibinfo {author}
  {\bibfnamefont {M.}~\bibnamefont {Avdeev}}, \ and\ \bibinfo {author}
  {\bibfnamefont {S.~J.}\ \bibnamefont {Campbell}},\ }\href {\doibase
  10.1038/srep11288} {\bibfield  {journal} {\bibinfo  {journal} {Sci. Rep.}\
  }\textbf {\bibinfo {volume} {5}},\ \bibinfo {pages} {11288} (\bibinfo {year}
  {2015})}\BibitemShut {NoStop}%
\bibitem [{\citenamefont {Moze}\ \emph {et~al.}(2003)\citenamefont {Moze},
  \citenamefont {Hofmann}, \citenamefont {Cadogan}, \citenamefont {Buschow},\
  and\ \citenamefont {Ryan}}]{Moze2003}%
  \BibitemOpen
  \bibfield  {author} {\bibinfo {author} {\bibfnamefont {O.}~\bibnamefont
  {Moze}}, \bibinfo {author} {\bibfnamefont {M.}~\bibnamefont {Hofmann}},
  \bibinfo {author} {\bibfnamefont {J.~M.}\ \bibnamefont {Cadogan}}, \bibinfo
  {author} {\bibfnamefont {K.~H.~J.}\ \bibnamefont {Buschow}}, \ and\ \bibinfo
  {author} {\bibfnamefont {D.~H.}\ \bibnamefont {Ryan}},\ }\href {\doibase
  10.1140/epjb/e2004-00006-4} {\bibfield  {journal} {\bibinfo  {journal} {Eur.
  Phys. J. B - Condens. Matter}\ }\textbf {\bibinfo {volume} {36}},\ \bibinfo
  {pages} {511} (\bibinfo {year} {2003})}\BibitemShut {NoStop}%
\bibitem [{\citenamefont {Paramanik}\ \emph {et~al.}(2014)\citenamefont
  {Paramanik}, \citenamefont {Prasad}, \citenamefont {Geibel},\ and\
  \citenamefont {Hossain}}]{Paramanik2014a}%
  \BibitemOpen
  \bibfield  {author} {\bibinfo {author} {\bibfnamefont {U.~B.}\ \bibnamefont
  {Paramanik}}, \bibinfo {author} {\bibfnamefont {R.}~\bibnamefont {Prasad}},
  \bibinfo {author} {\bibfnamefont {C.}~\bibnamefont {Geibel}}, \ and\ \bibinfo
  {author} {\bibfnamefont {Z.}~\bibnamefont {Hossain}},\ }\href {\doibase
  10.1103/PhysRevB.89.144423} {\bibfield  {journal} {\bibinfo  {journal} {Phys.
  Rev. B}\ }\textbf {\bibinfo {volume} {89}},\ \bibinfo {pages} {144423}
  (\bibinfo {year} {2014})}\BibitemShut {NoStop}%
\bibitem [{\citenamefont {Singh}\ \emph
  {et~al.}(2009{\natexlab{c}})\citenamefont {Singh}, \citenamefont {Sefat},
  \citenamefont {McGuire}, \citenamefont {Sales}, \citenamefont {Mandrus},
  \citenamefont {VanBebber},\ and\ \citenamefont {Keppens}}]{Singh2009d}%
  \BibitemOpen
  \bibfield  {author} {\bibinfo {author} {\bibfnamefont {D.~J.}\ \bibnamefont
  {Singh}}, \bibinfo {author} {\bibfnamefont {A.~S.}\ \bibnamefont {Sefat}},
  \bibinfo {author} {\bibfnamefont {M.~A.}\ \bibnamefont {McGuire}}, \bibinfo
  {author} {\bibfnamefont {B.~C.}\ \bibnamefont {Sales}}, \bibinfo {author}
  {\bibfnamefont {D.}~\bibnamefont {Mandrus}}, \bibinfo {author} {\bibfnamefont
  {L.~H.}\ \bibnamefont {VanBebber}}, \ and\ \bibinfo {author} {\bibfnamefont
  {V.}~\bibnamefont {Keppens}},\ }\href {\doibase 10.1103/PhysRevB.79.094429}
  {\bibfield  {journal} {\bibinfo  {journal} {Phys. Rev. B}\ }\textbf {\bibinfo
  {volume} {79}},\ \bibinfo {pages} {094429} (\bibinfo {year}
  {2009}{\natexlab{c}})}\BibitemShut {NoStop}%
\bibitem [{\citenamefont {Filsinger}\ \emph {et~al.}(2017)\citenamefont
  {Filsinger}, \citenamefont {Schnelle}, \citenamefont {Adler}, \citenamefont
  {Fecher}, \citenamefont {Reehuis}, \citenamefont {Hoser}, \citenamefont
  {Hoffmann}, \citenamefont {Werner}, \citenamefont {Greenblatt},\ and\
  \citenamefont {Felser}}]{Filsinger2017}%
  \BibitemOpen
  \bibfield  {author} {\bibinfo {author} {\bibfnamefont {K.~A.}\ \bibnamefont
  {Filsinger}}, \bibinfo {author} {\bibfnamefont {W.}~\bibnamefont {Schnelle}},
  \bibinfo {author} {\bibfnamefont {P.}~\bibnamefont {Adler}}, \bibinfo
  {author} {\bibfnamefont {G.~H.}\ \bibnamefont {Fecher}}, \bibinfo {author}
  {\bibfnamefont {M.}~\bibnamefont {Reehuis}}, \bibinfo {author} {\bibfnamefont
  {A.}~\bibnamefont {Hoser}}, \bibinfo {author} {\bibfnamefont {J.-U.}\
  \bibnamefont {Hoffmann}}, \bibinfo {author} {\bibfnamefont {P.}~\bibnamefont
  {Werner}}, \bibinfo {author} {\bibfnamefont {M.}~\bibnamefont {Greenblatt}},
  \ and\ \bibinfo {author} {\bibfnamefont {C.}~\bibnamefont {Felser}},\ }\href
  {\doibase 10.1103/PhysRevB.95.184414} {\bibfield  {journal} {\bibinfo
  {journal} {Phys. Rev. B}\ }\textbf {\bibinfo {volume} {95}},\ \bibinfo
  {pages} {184414} (\bibinfo {year} {2017})}\BibitemShut {NoStop}%
\bibitem [{\citenamefont {Das}\ \emph {et~al.}(2017{\natexlab{a}})\citenamefont
  {Das}, \citenamefont {Sangeetha}, \citenamefont {Lindemann}, \citenamefont
  {Heitmann}, \citenamefont {Kreyssig}, \citenamefont {Goldman}, \citenamefont
  {McQueeney}, \citenamefont {Johnston},\ and\ \citenamefont
  {Vaknin}}]{Das2017a}%
  \BibitemOpen
  \bibfield  {author} {\bibinfo {author} {\bibfnamefont {P.}~\bibnamefont
  {Das}}, \bibinfo {author} {\bibfnamefont {N.~S.}\ \bibnamefont {Sangeetha}},
  \bibinfo {author} {\bibfnamefont {G.~R.}\ \bibnamefont {Lindemann}}, \bibinfo
  {author} {\bibfnamefont {T.~W.}\ \bibnamefont {Heitmann}}, \bibinfo {author}
  {\bibfnamefont {A.}~\bibnamefont {Kreyssig}}, \bibinfo {author}
  {\bibfnamefont {A.~I.}\ \bibnamefont {Goldman}}, \bibinfo {author}
  {\bibfnamefont {R.~J.}\ \bibnamefont {McQueeney}}, \bibinfo {author}
  {\bibfnamefont {D.~C.}\ \bibnamefont {Johnston}}, \ and\ \bibinfo {author}
  {\bibfnamefont {D.}~\bibnamefont {Vaknin}},\ }\href {\doibase
  10.1103/PhysRevB.96.014411} {\bibfield  {journal} {\bibinfo  {journal} {Phys.
  Rev. B}\ }\textbf {\bibinfo {volume} {96}},\ \bibinfo {pages} {014411}
  (\bibinfo {year} {2017}{\natexlab{a}})}\BibitemShut {NoStop}%
\bibitem [{\citenamefont {Guo}\ \emph {et~al.}(2014)\citenamefont {Guo},
  \citenamefont {Princep}, \citenamefont {Zhang}, \citenamefont {Manuel},
  \citenamefont {Khalyavin}, \citenamefont {Mazin}, \citenamefont {Shi},\ and\
  \citenamefont {Boothroyd}}]{Guo2014a}%
  \BibitemOpen
  \bibfield  {author} {\bibinfo {author} {\bibfnamefont {Y.~F.}\ \bibnamefont
  {Guo}}, \bibinfo {author} {\bibfnamefont {A.~J.}\ \bibnamefont {Princep}},
  \bibinfo {author} {\bibfnamefont {X.}~\bibnamefont {Zhang}}, \bibinfo
  {author} {\bibfnamefont {P.}~\bibnamefont {Manuel}}, \bibinfo {author}
  {\bibfnamefont {D.}~\bibnamefont {Khalyavin}}, \bibinfo {author}
  {\bibfnamefont {I.~I.}\ \bibnamefont {Mazin}}, \bibinfo {author}
  {\bibfnamefont {Y.~G.}\ \bibnamefont {Shi}}, \ and\ \bibinfo {author}
  {\bibfnamefont {A.~T.}\ \bibnamefont {Boothroyd}},\ }\href {\doibase
  10.1103/PhysRevB.90.075120} {\bibfield  {journal} {\bibinfo  {journal} {Phys.
  Rev. B}\ }\textbf {\bibinfo {volume} {90}},\ \bibinfo {pages} {075120}
  (\bibinfo {year} {2014})}\BibitemShut {NoStop}%
\bibitem [{\citenamefont {Zhang}\ \emph
  {et~al.}(2016{\natexlab{a}})\citenamefont {Zhang}, \citenamefont {Liu},
  \citenamefont {Yi}, \citenamefont {Zhao}, \citenamefont {Xia}, \citenamefont
  {Ji}, \citenamefont {Shi}, \citenamefont {Yu}, \citenamefont {Wang},
  \citenamefont {Chen},\ and\ \citenamefont {Zhang}}]{Zhang2016f}%
  \BibitemOpen
  \bibfield  {author} {\bibinfo {author} {\bibfnamefont {A.}~\bibnamefont
  {Zhang}}, \bibinfo {author} {\bibfnamefont {C.}~\bibnamefont {Liu}}, \bibinfo
  {author} {\bibfnamefont {C.}~\bibnamefont {Yi}}, \bibinfo {author}
  {\bibfnamefont {G.}~\bibnamefont {Zhao}}, \bibinfo {author} {\bibfnamefont
  {T.-l.}\ \bibnamefont {Xia}}, \bibinfo {author} {\bibfnamefont
  {J.}~\bibnamefont {Ji}}, \bibinfo {author} {\bibfnamefont {Y.}~\bibnamefont
  {Shi}}, \bibinfo {author} {\bibfnamefont {R.}~\bibnamefont {Yu}}, \bibinfo
  {author} {\bibfnamefont {X.}~\bibnamefont {Wang}}, \bibinfo {author}
  {\bibfnamefont {C.}~\bibnamefont {Chen}}, \ and\ \bibinfo {author}
  {\bibfnamefont {Q.}~\bibnamefont {Zhang}},\ }\href {\doibase
  10.1038/ncomms13833} {\bibfield  {journal} {\bibinfo  {journal} {Nat.
  Commun.}\ }\textbf {\bibinfo {volume} {7}},\ \bibinfo {pages} {13833}
  (\bibinfo {year} {2016}{\natexlab{a}})}\BibitemShut {NoStop}%
\bibitem [{\citenamefont {Wang}\ \emph {et~al.}(2016)\citenamefont {Wang},
  \citenamefont {Zaliznyak}, \citenamefont {Ren}, \citenamefont {Wu},
  \citenamefont {Graf}, \citenamefont {Garlea}, \citenamefont {Warren},
  \citenamefont {Bozin}, \citenamefont {Zhu},\ and\ \citenamefont
  {Petrovic}}]{Wang2016o}%
  \BibitemOpen
  \bibfield  {author} {\bibinfo {author} {\bibfnamefont {A.}~\bibnamefont
  {Wang}}, \bibinfo {author} {\bibfnamefont {I.}~\bibnamefont {Zaliznyak}},
  \bibinfo {author} {\bibfnamefont {W.}~\bibnamefont {Ren}}, \bibinfo {author}
  {\bibfnamefont {L.}~\bibnamefont {Wu}}, \bibinfo {author} {\bibfnamefont
  {D.}~\bibnamefont {Graf}}, \bibinfo {author} {\bibfnamefont {V.~O.}\
  \bibnamefont {Garlea}}, \bibinfo {author} {\bibfnamefont {J.~B.}\
  \bibnamefont {Warren}}, \bibinfo {author} {\bibfnamefont {E.}~\bibnamefont
  {Bozin}}, \bibinfo {author} {\bibfnamefont {Y.}~\bibnamefont {Zhu}}, \ and\
  \bibinfo {author} {\bibfnamefont {C.}~\bibnamefont {Petrovic}},\ }\href
  {\doibase 10.1103/PhysRevB.94.165161} {\bibfield  {journal} {\bibinfo
  {journal} {Phys. Rev. B}\ }\textbf {\bibinfo {volume} {94}},\ \bibinfo
  {pages} {165161} (\bibinfo {year} {2016})}\BibitemShut {NoStop}%
\bibitem [{\citenamefont {Masuda}\ \emph {et~al.}(2016)\citenamefont {Masuda},
  \citenamefont {Sakai}, \citenamefont {Tokunaga}, \citenamefont {Yamasaki},
  \citenamefont {Miyake}, \citenamefont {Shiogai}, \citenamefont {Nakamura},
  \citenamefont {Awaji}, \citenamefont {Tsukazaki}, \citenamefont {Nakao},
  \citenamefont {Murakami}, \citenamefont {Arima}, \citenamefont {Tokura},\
  and\ \citenamefont {Ishiwata}}]{Masuda2016a}%
  \BibitemOpen
  \bibfield  {author} {\bibinfo {author} {\bibfnamefont {H.}~\bibnamefont
  {Masuda}}, \bibinfo {author} {\bibfnamefont {H.}~\bibnamefont {Sakai}},
  \bibinfo {author} {\bibfnamefont {M.}~\bibnamefont {Tokunaga}}, \bibinfo
  {author} {\bibfnamefont {Y.}~\bibnamefont {Yamasaki}}, \bibinfo {author}
  {\bibfnamefont {A.}~\bibnamefont {Miyake}}, \bibinfo {author} {\bibfnamefont
  {J.}~\bibnamefont {Shiogai}}, \bibinfo {author} {\bibfnamefont
  {S.}~\bibnamefont {Nakamura}}, \bibinfo {author} {\bibfnamefont
  {S.}~\bibnamefont {Awaji}}, \bibinfo {author} {\bibfnamefont
  {A.}~\bibnamefont {Tsukazaki}}, \bibinfo {author} {\bibfnamefont
  {H.}~\bibnamefont {Nakao}}, \bibinfo {author} {\bibfnamefont
  {Y.}~\bibnamefont {Murakami}}, \bibinfo {author} {\bibfnamefont {T.~H.
  T.-h.}\ \bibnamefont {Arima}}, \bibinfo {author} {\bibfnamefont
  {Y.}~\bibnamefont {Tokura}}, \ and\ \bibinfo {author} {\bibfnamefont
  {S.}~\bibnamefont {Ishiwata}},\ }\href {\doibase 10.1126/sciadv.1501117}
  {\bibfield  {journal} {\bibinfo  {journal} {Sci. Adv.}\ }\textbf {\bibinfo
  {volume} {2}},\ \bibinfo {pages} {e1501117} (\bibinfo {year}
  {2016})}\BibitemShut {NoStop}%
\bibitem [{\citenamefont {Masuda}\ \emph {et~al.}(2018)\citenamefont {Masuda},
  \citenamefont {Sakai}, \citenamefont {Tokunaga}, \citenamefont {Ochi},
  \citenamefont {Takahashi}, \citenamefont {Akiba}, \citenamefont {Miyake},
  \citenamefont {Kuroki}, \citenamefont {Tokura},\ and\ \citenamefont
  {Ishiwata}}]{Masuda2018Impact}%
  \BibitemOpen
  \bibfield  {author} {\bibinfo {author} {\bibfnamefont {H.}~\bibnamefont
  {Masuda}}, \bibinfo {author} {\bibfnamefont {H.}~\bibnamefont {Sakai}},
  \bibinfo {author} {\bibfnamefont {M.}~\bibnamefont {Tokunaga}}, \bibinfo
  {author} {\bibfnamefont {M.}~\bibnamefont {Ochi}}, \bibinfo {author}
  {\bibfnamefont {H.}~\bibnamefont {Takahashi}}, \bibinfo {author}
  {\bibfnamefont {K.}~\bibnamefont {Akiba}}, \bibinfo {author} {\bibfnamefont
  {A.}~\bibnamefont {Miyake}}, \bibinfo {author} {\bibfnamefont
  {K.}~\bibnamefont {Kuroki}}, \bibinfo {author} {\bibfnamefont
  {Y.}~\bibnamefont {Tokura}}, \ and\ \bibinfo {author} {\bibfnamefont
  {S.}~\bibnamefont {Ishiwata}},\ }\href {\doibase 10.1103/PhysRevB.98.161108}
  {\bibfield  {journal} {\bibinfo  {journal} {Phys. Rev. B}\ }\textbf {\bibinfo
  {volume} {98}},\ \bibinfo {pages} {161108} (\bibinfo {year}
  {2018})}\BibitemShut {NoStop}%
\bibitem [{\citenamefont {Yanagi}\ \emph {et~al.}(2009)\citenamefont {Yanagi},
  \citenamefont {Watanabe}, \citenamefont {Kodama}, \citenamefont {Iikubo},
  \citenamefont {Shamoto}, \citenamefont {Kamiya}, \citenamefont {Hirano},\
  and\ \citenamefont {Hosono}}]{Yanagi2009a}%
  \BibitemOpen
  \bibfield  {author} {\bibinfo {author} {\bibfnamefont {H.}~\bibnamefont
  {Yanagi}}, \bibinfo {author} {\bibfnamefont {T.}~\bibnamefont {Watanabe}},
  \bibinfo {author} {\bibfnamefont {K.}~\bibnamefont {Kodama}}, \bibinfo
  {author} {\bibfnamefont {S.}~\bibnamefont {Iikubo}}, \bibinfo {author}
  {\bibfnamefont {S.-i.~I.}\ \bibnamefont {Shamoto}}, \bibinfo {author}
  {\bibfnamefont {T.}~\bibnamefont {Kamiya}}, \bibinfo {author} {\bibfnamefont
  {M.}~\bibnamefont {Hirano}}, \ and\ \bibinfo {author} {\bibfnamefont
  {H.}~\bibnamefont {Hosono}},\ }\href {\doibase 10.1063/1.3124582} {\bibfield
  {journal} {\bibinfo  {journal} {J. Appl. Phys.}\ }\textbf {\bibinfo {volume}
  {105}},\ \bibinfo {pages} {093916} (\bibinfo {year} {2009})}\BibitemShut
  {NoStop}%
\bibitem [{\citenamefont {Simonson}\ \emph {et~al.}(2012)\citenamefont
  {Simonson}, \citenamefont {Yin}, \citenamefont {Pezzoli}, \citenamefont
  {Guo}, \citenamefont {Liu}, \citenamefont {Post}, \citenamefont {Efimenko},
  \citenamefont {Hollmann}, \citenamefont {Hu}, \citenamefont {Lin},
  \citenamefont {Chen}, \citenamefont {Marques}, \citenamefont {Leyva},
  \citenamefont {Smith}, \citenamefont {Lynn}, \citenamefont {Sun},
  \citenamefont {Kotliar}, \citenamefont {Basov}, \citenamefont {Tjeng},\ and\
  \citenamefont {Aronson}}]{Simonson2012b}%
  \BibitemOpen
  \bibfield  {author} {\bibinfo {author} {\bibfnamefont {J.~W.}\ \bibnamefont
  {Simonson}}, \bibinfo {author} {\bibfnamefont {Z.~P.}\ \bibnamefont {Yin}},
  \bibinfo {author} {\bibfnamefont {M.}~\bibnamefont {Pezzoli}}, \bibinfo
  {author} {\bibfnamefont {J.}~\bibnamefont {Guo}}, \bibinfo {author}
  {\bibfnamefont {J.}~\bibnamefont {Liu}}, \bibinfo {author} {\bibfnamefont
  {K.}~\bibnamefont {Post}}, \bibinfo {author} {\bibfnamefont {A.}~\bibnamefont
  {Efimenko}}, \bibinfo {author} {\bibfnamefont {N.}~\bibnamefont {Hollmann}},
  \bibinfo {author} {\bibfnamefont {Z.}~\bibnamefont {Hu}}, \bibinfo {author}
  {\bibfnamefont {H.-J.}\ \bibnamefont {Lin}}, \bibinfo {author} {\bibfnamefont
  {C.-T.}\ \bibnamefont {Chen}}, \bibinfo {author} {\bibfnamefont
  {C.}~\bibnamefont {Marques}}, \bibinfo {author} {\bibfnamefont
  {V.}~\bibnamefont {Leyva}}, \bibinfo {author} {\bibfnamefont
  {G.}~\bibnamefont {Smith}}, \bibinfo {author} {\bibfnamefont {J.~W.}\
  \bibnamefont {Lynn}}, \bibinfo {author} {\bibfnamefont {L.~L.}\ \bibnamefont
  {Sun}}, \bibinfo {author} {\bibfnamefont {G.}~\bibnamefont {Kotliar}},
  \bibinfo {author} {\bibfnamefont {D.~N.}\ \bibnamefont {Basov}}, \bibinfo
  {author} {\bibfnamefont {L.~H.}\ \bibnamefont {Tjeng}}, \ and\ \bibinfo
  {author} {\bibfnamefont {M.~C.}\ \bibnamefont {Aronson}},\ }\href {\doibase
  10.1073/pnas.1117366109} {\bibfield  {journal} {\bibinfo  {journal} {Proc.
  Natl. Acad. Sci.}\ }\textbf {\bibinfo {volume} {109}},\ \bibinfo {pages}
  {E1815} (\bibinfo {year} {2012})}\BibitemShut {NoStop}%
\bibitem [{\citenamefont {Emery}\ \emph {et~al.}(2010)\citenamefont {Emery},
  \citenamefont {Wildman}, \citenamefont {Skakle}, \citenamefont {Giriat},
  \citenamefont {Smith},\ and\ \citenamefont {Mclaughlin}}]{Emery2010}%
  \BibitemOpen
  \bibfield  {author} {\bibinfo {author} {\bibfnamefont {N.}~\bibnamefont
  {Emery}}, \bibinfo {author} {\bibfnamefont {E.~J.}\ \bibnamefont {Wildman}},
  \bibinfo {author} {\bibfnamefont {J.~M.~S.}\ \bibnamefont {Skakle}}, \bibinfo
  {author} {\bibfnamefont {G.}~\bibnamefont {Giriat}}, \bibinfo {author}
  {\bibfnamefont {R.~I.}\ \bibnamefont {Smith}}, \ and\ \bibinfo {author}
  {\bibfnamefont {A.~C.}\ \bibnamefont {Mclaughlin}},\ }\href {\doibase
  10.1039/c0cc01380c} {\bibfield  {journal} {\bibinfo  {journal} {Chem.
  Commun.}\ }\textbf {\bibinfo {volume} {46}},\ \bibinfo {pages} {6777}
  (\bibinfo {year} {2010})}\BibitemShut {NoStop}%
\bibitem [{\citenamefont {Zhang}\ \emph
  {et~al.}(2016{\natexlab{b}})\citenamefont {Zhang}, \citenamefont {Kumar},
  \citenamefont {Tian}, \citenamefont {Dennis}, \citenamefont {Goldman},\ and\
  \citenamefont {Vaknin}}]{Zhang2016k}%
  \BibitemOpen
  \bibfield  {author} {\bibinfo {author} {\bibfnamefont {Q.}~\bibnamefont
  {Zhang}}, \bibinfo {author} {\bibfnamefont {C.~M.~N.}\ \bibnamefont {Kumar}},
  \bibinfo {author} {\bibfnamefont {W.}~\bibnamefont {Tian}}, \bibinfo {author}
  {\bibfnamefont {K.~W.}\ \bibnamefont {Dennis}}, \bibinfo {author}
  {\bibfnamefont {A.~I.}\ \bibnamefont {Goldman}}, \ and\ \bibinfo {author}
  {\bibfnamefont {D.}~\bibnamefont {Vaknin}},\ }\href {\doibase
  10.1103/PhysRevB.93.094413} {\bibfield  {journal} {\bibinfo  {journal} {Phys.
  Rev. B}\ }\textbf {\bibinfo {volume} {93}},\ \bibinfo {pages} {094413}
  (\bibinfo {year} {2016}{\natexlab{b}})}\BibitemShut {NoStop}%
\bibitem [{\citenamefont {Zhang}\ \emph {et~al.}(2015)\citenamefont {Zhang},
  \citenamefont {Tian}, \citenamefont {Peterson}, \citenamefont {Dennis},\ and\
  \citenamefont {Vaknin}}]{Zhang2015f}%
  \BibitemOpen
  \bibfield  {author} {\bibinfo {author} {\bibfnamefont {Q.}~\bibnamefont
  {Zhang}}, \bibinfo {author} {\bibfnamefont {W.}~\bibnamefont {Tian}},
  \bibinfo {author} {\bibfnamefont {S.~G.}\ \bibnamefont {Peterson}}, \bibinfo
  {author} {\bibfnamefont {K.~W.}\ \bibnamefont {Dennis}}, \ and\ \bibinfo
  {author} {\bibfnamefont {D.}~\bibnamefont {Vaknin}},\ }\href {\doibase
  10.1103/PhysRevB.91.064418} {\bibfield  {journal} {\bibinfo  {journal} {Phys.
  Rev. B}\ }\textbf {\bibinfo {volume} {91}},\ \bibinfo {pages} {064418}
  (\bibinfo {year} {2015})}\BibitemShut {NoStop}%
\bibitem [{\citenamefont {Kimber}\ \emph {et~al.}(2010)\citenamefont {Kimber},
  \citenamefont {Hill}, \citenamefont {Zhang}, \citenamefont {Jeschke},
  \citenamefont {Valent{\'{i}}}, \citenamefont {Ritter}, \citenamefont
  {Schellenberg}, \citenamefont {Hermes}, \citenamefont {P{\"{o}}ttgen},\ and\
  \citenamefont {Argyriou}}]{Kimber2010}%
  \BibitemOpen
  \bibfield  {author} {\bibinfo {author} {\bibfnamefont {S.~A.~J.}\
  \bibnamefont {Kimber}}, \bibinfo {author} {\bibfnamefont {A.~H.}\
  \bibnamefont {Hill}}, \bibinfo {author} {\bibfnamefont {Y.-Z.~Z.}\
  \bibnamefont {Zhang}}, \bibinfo {author} {\bibfnamefont {H.~O.}\ \bibnamefont
  {Jeschke}}, \bibinfo {author} {\bibfnamefont {R.}~\bibnamefont
  {Valent{\'{i}}}}, \bibinfo {author} {\bibfnamefont {C.}~\bibnamefont
  {Ritter}}, \bibinfo {author} {\bibfnamefont {I.}~\bibnamefont
  {Schellenberg}}, \bibinfo {author} {\bibfnamefont {W.}~\bibnamefont
  {Hermes}}, \bibinfo {author} {\bibfnamefont {R.}~\bibnamefont
  {P{\"{o}}ttgen}}, \ and\ \bibinfo {author} {\bibfnamefont {D.~N.}\
  \bibnamefont {Argyriou}},\ }\href {\doibase 10.1103/PhysRevB.82.100412}
  {\bibfield  {journal} {\bibinfo  {journal} {Phys. Rev. B}\ }\textbf {\bibinfo
  {volume} {82}},\ \bibinfo {pages} {100412} (\bibinfo {year}
  {2010})}\BibitemShut {NoStop}%
\bibitem [{\citenamefont {Marcinkova}\ \emph {et~al.}(2010)\citenamefont
  {Marcinkova}, \citenamefont {Hansen}, \citenamefont {Curfs}, \citenamefont
  {Margadonna},\ and\ \citenamefont {Bos}}]{Marcinkova2010}%
  \BibitemOpen
  \bibfield  {author} {\bibinfo {author} {\bibfnamefont {A.}~\bibnamefont
  {Marcinkova}}, \bibinfo {author} {\bibfnamefont {T.~C.}\ \bibnamefont
  {Hansen}}, \bibinfo {author} {\bibfnamefont {C.}~\bibnamefont {Curfs}},
  \bibinfo {author} {\bibfnamefont {S.}~\bibnamefont {Margadonna}}, \ and\
  \bibinfo {author} {\bibfnamefont {J.~W.~G.}\ \bibnamefont {Bos}},\ }\href
  {\doibase 10.1103/PhysRevB.82.174438} {\bibfield  {journal} {\bibinfo
  {journal} {Phys. Rev. B}\ }\textbf {\bibinfo {volume} {82}},\ \bibinfo
  {pages} {174438} (\bibinfo {year} {2010})}\BibitemShut {NoStop}%
\bibitem [{\citenamefont {Emery}\ \emph {et~al.}(2011)\citenamefont {Emery},
  \citenamefont {Wildman}, \citenamefont {Skakle}, \citenamefont {Mclaughlin},
  \citenamefont {Smith},\ and\ \citenamefont {Fitch}}]{Emery2011b}%
  \BibitemOpen
  \bibfield  {author} {\bibinfo {author} {\bibfnamefont {N.}~\bibnamefont
  {Emery}}, \bibinfo {author} {\bibfnamefont {E.~J.}\ \bibnamefont {Wildman}},
  \bibinfo {author} {\bibfnamefont {J.~M.~S.}\ \bibnamefont {Skakle}}, \bibinfo
  {author} {\bibfnamefont {A.~C.}\ \bibnamefont {Mclaughlin}}, \bibinfo
  {author} {\bibfnamefont {R.~I.}\ \bibnamefont {Smith}}, \ and\ \bibinfo
  {author} {\bibfnamefont {A.~N.}\ \bibnamefont {Fitch}},\ }\href {\doibase
  10.1103/PhysRevB.83.144429} {\bibfield  {journal} {\bibinfo  {journal} {Phys.
  Rev. B}\ }\textbf {\bibinfo {volume} {83}},\ \bibinfo {pages} {144429}
  (\bibinfo {year} {2011})}\BibitemShut {NoStop}%
\bibitem [{\citenamefont {Bronger}\ \emph {et~al.}(1986)\citenamefont
  {Bronger}, \citenamefont {M{\"u}ller}, \citenamefont {H{\"o}ppner},\ and\
  \citenamefont {Schuster}}]{Bronger1986NaMnP}%
  \BibitemOpen
  \bibfield  {author} {\bibinfo {author} {\bibfnamefont {W.}~\bibnamefont
  {Bronger}}, \bibinfo {author} {\bibfnamefont {P.}~\bibnamefont {M{\"u}ller}},
  \bibinfo {author} {\bibfnamefont {R.}~\bibnamefont {H{\"o}ppner}}, \ and\
  \bibinfo {author} {\bibfnamefont {H.-U.}\ \bibnamefont {Schuster}},\ }\href
  {\doibase 10.1002/zaac.19865390816} {\bibfield  {journal} {\bibinfo
  {journal} {Zeitschrift f{\"u}r Anorg. und Allg. Chemie}\ }\textbf {\bibinfo
  {volume} {539}},\ \bibinfo {pages} {175} (\bibinfo {year}
  {1986})}\BibitemShut {NoStop}%
\bibitem [{\citenamefont {Schucht}\ \emph {et~al.}(1999)\citenamefont
  {Schucht}, \citenamefont {Dascoulidou}, \citenamefont {M{\"{u}}ller},
  \citenamefont {Jung}, \citenamefont {Schuster}, \citenamefont {Bronger},\
  and\ \citenamefont {M{\"{u}}ller}}]{Schucht1999}%
  \BibitemOpen
  \bibfield  {author} {\bibinfo {author} {\bibfnamefont {F.}~\bibnamefont
  {Schucht}}, \bibinfo {author} {\bibfnamefont {A.}~\bibnamefont
  {Dascoulidou}}, \bibinfo {author} {\bibfnamefont {R.}~\bibnamefont
  {M{\"{u}}ller}}, \bibinfo {author} {\bibfnamefont {W.}~\bibnamefont {Jung}},
  \bibinfo {author} {\bibfnamefont {H.-U.}\ \bibnamefont {Schuster}}, \bibinfo
  {author} {\bibfnamefont {W.}~\bibnamefont {Bronger}}, \ and\ \bibinfo
  {author} {\bibfnamefont {P.}~\bibnamefont {M{\"{u}}ller}},\ }\href {\doibase
  10.1002/(SICI)1521-3749(199901)625:1<31::AID-ZAAC31>3.0.CO;2-S} {\bibfield
  {journal} {\bibinfo  {journal} {Zeitschrift f{\"{u}}r Anorg. und Allg.
  Chemie}\ }\textbf {\bibinfo {volume} {625}},\ \bibinfo {pages} {31} (\bibinfo
  {year} {1999})}\BibitemShut {NoStop}%
\bibitem [{\citenamefont {Blanco}\ \emph {et~al.}(2006)\citenamefont {Blanco},
  \citenamefont {Brown}, \citenamefont {Stunault}, \citenamefont {Katsumata},
  \citenamefont {Iga},\ and\ \citenamefont {Michimura}}]{Blanco2006}%
  \BibitemOpen
  \bibfield  {author} {\bibinfo {author} {\bibfnamefont {J.~A.}\ \bibnamefont
  {Blanco}}, \bibinfo {author} {\bibfnamefont {P.~J.}\ \bibnamefont {Brown}},
  \bibinfo {author} {\bibfnamefont {A.}~\bibnamefont {Stunault}}, \bibinfo
  {author} {\bibfnamefont {K.}~\bibnamefont {Katsumata}}, \bibinfo {author}
  {\bibfnamefont {F.}~\bibnamefont {Iga}}, \ and\ \bibinfo {author}
  {\bibfnamefont {S.}~\bibnamefont {Michimura}},\ }\href {\doibase
  10.1103/PhysRevB.73.212411} {\bibfield  {journal} {\bibinfo  {journal} {Phys.
  Rev. B}\ }\textbf {\bibinfo {volume} {73}},\ \bibinfo {pages} {212411}
  (\bibinfo {year} {2006})}\BibitemShut {NoStop}%
\bibitem [{\citenamefont {Fisk}\ \emph {et~al.}(1981)\citenamefont {Fisk},
  \citenamefont {Maple}, \citenamefont {Johnston},\ and\ \citenamefont
  {Woolf}}]{Fisk1981}%
  \BibitemOpen
  \bibfield  {author} {\bibinfo {author} {\bibfnamefont {Z.}~\bibnamefont
  {Fisk}}, \bibinfo {author} {\bibfnamefont {M.~B.}\ \bibnamefont {Maple}},
  \bibinfo {author} {\bibfnamefont {D.~C.}\ \bibnamefont {Johnston}}, \ and\
  \bibinfo {author} {\bibfnamefont {L.~D.}\ \bibnamefont {Woolf}},\ }\href
  {\doibase 10.1016/0038-1098(81)91111-X} {\bibfield  {journal} {\bibinfo
  {journal} {Solid State Commun.}\ }\textbf {\bibinfo {volume} {39}},\ \bibinfo
  {pages} {1189} (\bibinfo {year} {1981})}\BibitemShut {NoStop}%
\bibitem [{\citenamefont {Will}\ and\ \citenamefont
  {Schafer}(1979)}]{Will1979}%
  \BibitemOpen
  \bibfield  {author} {\bibinfo {author} {\bibfnamefont {G.}~\bibnamefont
  {Will}}\ and\ \bibinfo {author} {\bibfnamefont {W.}~\bibnamefont {Schafer}},\
  }\href {\doibase 10.1016/0022-5088(79)90071-7} {\bibfield  {journal}
  {\bibinfo  {journal} {J. Less-Common Met.}\ }\textbf {\bibinfo {volume}
  {67}},\ \bibinfo {pages} {31} (\bibinfo {year} {1979})}\BibitemShut {NoStop}%
\bibitem [{\citenamefont {Ji}\ \emph {et~al.}(2007)\citenamefont {Ji},
  \citenamefont {Song}, \citenamefont {Koo}, \citenamefont {Park},
  \citenamefont {Park}, \citenamefont {Lee}, \citenamefont {Lee}, \citenamefont
  {Park}, \citenamefont {Kim}, \citenamefont {Cho}, \citenamefont {Hong},
  \citenamefont {Lee},\ and\ \citenamefont {Iga}}]{Ji2007}%
  \BibitemOpen
  \bibfield  {author} {\bibinfo {author} {\bibfnamefont {S.}~\bibnamefont
  {Ji}}, \bibinfo {author} {\bibfnamefont {C.}~\bibnamefont {Song}}, \bibinfo
  {author} {\bibfnamefont {J.}~\bibnamefont {Koo}}, \bibinfo {author}
  {\bibfnamefont {J.}~\bibnamefont {Park}}, \bibinfo {author} {\bibfnamefont
  {Y.~J.}\ \bibnamefont {Park}}, \bibinfo {author} {\bibfnamefont {K.-B.}\
  \bibnamefont {Lee}}, \bibinfo {author} {\bibfnamefont {S.}~\bibnamefont
  {Lee}}, \bibinfo {author} {\bibfnamefont {J.-G.}\ \bibnamefont {Park}},
  \bibinfo {author} {\bibfnamefont {J.~Y.}\ \bibnamefont {Kim}}, \bibinfo
  {author} {\bibfnamefont {B.~K.}\ \bibnamefont {Cho}}, \bibinfo {author}
  {\bibfnamefont {K.-P.}\ \bibnamefont {Hong}}, \bibinfo {author}
  {\bibfnamefont {C.-H.}\ \bibnamefont {Lee}}, \ and\ \bibinfo {author}
  {\bibfnamefont {F.}~\bibnamefont {Iga}},\ }\href {\doibase
  10.1103/PhysRevLett.99.076401} {\bibfield  {journal} {\bibinfo  {journal}
  {Phys. Rev. Lett.}\ }\textbf {\bibinfo {volume} {99}},\ \bibinfo {pages}
  {076401} (\bibinfo {year} {2007})}\BibitemShut {NoStop}%
\bibitem [{\citenamefont {Will}\ \emph {et~al.}(1981)\citenamefont {Will},
  \citenamefont {Sch{\"{a}}fer}, \citenamefont {Pfeiffer}, \citenamefont
  {Elf},\ and\ \citenamefont {Etourneau}}]{Will1981}%
  \BibitemOpen
  \bibfield  {author} {\bibinfo {author} {\bibfnamefont {G.}~\bibnamefont
  {Will}}, \bibinfo {author} {\bibfnamefont {W.}~\bibnamefont {Sch{\"{a}}fer}},
  \bibinfo {author} {\bibfnamefont {F.}~\bibnamefont {Pfeiffer}}, \bibinfo
  {author} {\bibfnamefont {F.}~\bibnamefont {Elf}}, \ and\ \bibinfo {author}
  {\bibfnamefont {J.}~\bibnamefont {Etourneau}},\ }\href {\doibase
  10.1016/0022-5088(81)90238-1} {\bibfield  {journal} {\bibinfo  {journal} {J.
  Less-Common Met.}\ }\textbf {\bibinfo {volume} {82}},\ \bibinfo {pages} {349}
  (\bibinfo {year} {1981})}\BibitemShut {NoStop}%
\bibitem [{\citenamefont {Matsumura}\ \emph {et~al.}(2007)\citenamefont
  {Matsumura}, \citenamefont {Okuyama},\ and\ \citenamefont
  {Murakami}}]{Matsumura2007}%
  \BibitemOpen
  \bibfield  {author} {\bibinfo {author} {\bibfnamefont {T.}~\bibnamefont
  {Matsumura}}, \bibinfo {author} {\bibfnamefont {D.}~\bibnamefont {Okuyama}},
  \ and\ \bibinfo {author} {\bibfnamefont {Y.}~\bibnamefont {Murakami}},\
  }\href {\doibase 10.1143/JPSJ.76.015001} {\bibfield  {journal} {\bibinfo
  {journal} {J. Phys. Soc. Jpn.}\ }\textbf {\bibinfo {volume} {76}},\ \bibinfo
  {pages} {015001} (\bibinfo {year} {2007})}\BibitemShut {NoStop}%
\bibitem [{\citenamefont {Bhattacharyya}\ \emph {et~al.}(2016)\citenamefont
  {Bhattacharyya}, \citenamefont {Ritter}, \citenamefont {Adroja},
  \citenamefont {Coomer},\ and\ \citenamefont {Strydom}}]{Bhattacharyya2016a}%
  \BibitemOpen
  \bibfield  {author} {\bibinfo {author} {\bibfnamefont {A.}~\bibnamefont
  {Bhattacharyya}}, \bibinfo {author} {\bibfnamefont {C.}~\bibnamefont
  {Ritter}}, \bibinfo {author} {\bibfnamefont {D.~T.}\ \bibnamefont {Adroja}},
  \bibinfo {author} {\bibfnamefont {F.~C.}\ \bibnamefont {Coomer}}, \ and\
  \bibinfo {author} {\bibfnamefont {A.~M.}\ \bibnamefont {Strydom}},\ }\href
  {\doibase 10.1103/PhysRevB.94.014418} {\bibfield  {journal} {\bibinfo
  {journal} {Phys. Rev. B}\ }\textbf {\bibinfo {volume} {94}},\ \bibinfo
  {pages} {014418} (\bibinfo {year} {2016})}\BibitemShut {NoStop}%
\bibitem [{\citenamefont {Scagnoli}\ \emph {et~al.}(2012)\citenamefont
  {Scagnoli}, \citenamefont {Allieta}, \citenamefont {Walker}, \citenamefont
  {Scavini}, \citenamefont {Katsufuji}, \citenamefont {Sagarna}, \citenamefont
  {Zaharko},\ and\ \citenamefont {Mazzoli}}]{Scagnoli2012}%
  \BibitemOpen
  \bibfield  {author} {\bibinfo {author} {\bibfnamefont {V.}~\bibnamefont
  {Scagnoli}}, \bibinfo {author} {\bibfnamefont {M.}~\bibnamefont {Allieta}},
  \bibinfo {author} {\bibfnamefont {H.}~\bibnamefont {Walker}}, \bibinfo
  {author} {\bibfnamefont {M.}~\bibnamefont {Scavini}}, \bibinfo {author}
  {\bibfnamefont {T.}~\bibnamefont {Katsufuji}}, \bibinfo {author}
  {\bibfnamefont {L.}~\bibnamefont {Sagarna}}, \bibinfo {author} {\bibfnamefont
  {O.}~\bibnamefont {Zaharko}}, \ and\ \bibinfo {author} {\bibfnamefont
  {C.}~\bibnamefont {Mazzoli}},\ }\href {\doibase 10.1103/PhysRevB.86.094432}
  {\bibfield  {journal} {\bibinfo  {journal} {Phys. Rev. B}\ }\textbf {\bibinfo
  {volume} {86}},\ \bibinfo {pages} {094432} (\bibinfo {year}
  {2012})}\BibitemShut {NoStop}%
\bibitem [{\citenamefont {Rado}(1969)}]{Rado1969}%
  \BibitemOpen
  \bibfield  {author} {\bibinfo {author} {\bibfnamefont {G.~T.}\ \bibnamefont
  {Rado}},\ }\href {\doibase 10.1103/PhysRevLett.23.644} {\bibfield  {journal}
  {\bibinfo  {journal} {Phys. Rev. Lett.}\ }\textbf {\bibinfo {volume} {23}},\
  \bibinfo {pages} {644} (\bibinfo {year} {1969})}\BibitemShut {NoStop}%
\bibitem [{\citenamefont {Wright}\ \emph {et~al.}(1971)\citenamefont {Wright},
  \citenamefont {Moos}, \citenamefont {Colwell}, \citenamefont {Mangum},\ and\
  \citenamefont {Thornton}}]{Thorntont1971}%
  \BibitemOpen
  \bibfield  {author} {\bibinfo {author} {\bibfnamefont {J.~C.}\ \bibnamefont
  {Wright}}, \bibinfo {author} {\bibfnamefont {H.~W.}\ \bibnamefont {Moos}},
  \bibinfo {author} {\bibfnamefont {J.~H.}\ \bibnamefont {Colwell}}, \bibinfo
  {author} {\bibfnamefont {B.~W.}\ \bibnamefont {Mangum}}, \ and\ \bibinfo
  {author} {\bibfnamefont {D.~D.}\ \bibnamefont {Thornton}},\ }\href {\doibase
  10.1103/PhysRevB.3.843} {\bibfield  {journal} {\bibinfo  {journal} {Phys.
  Rev. B}\ }\textbf {\bibinfo {volume} {3}},\ \bibinfo {pages} {843} (\bibinfo
  {year} {1971})}\BibitemShut {NoStop}%
\bibitem [{\citenamefont {N{\"a}gele}\ \emph {et~al.}(1980)\citenamefont
  {N{\"a}gele}, \citenamefont {Hohlwein},\ and\ \citenamefont
  {Domann}}]{Naagle1980}%
  \BibitemOpen
  \bibfield  {author} {\bibinfo {author} {\bibfnamefont {W.}~\bibnamefont
  {N{\"a}gele}}, \bibinfo {author} {\bibfnamefont {D.}~\bibnamefont
  {Hohlwein}}, \ and\ \bibinfo {author} {\bibfnamefont {G.}~\bibnamefont
  {Domann}},\ }\href {\doibase 10.1007/BF01305829} {\bibfield  {journal}
  {\bibinfo  {journal} {Zeitschrift f{\"u}r Physik B Condensed Matter}\
  }\textbf {\bibinfo {volume} {39}},\ \bibinfo {pages} {305} (\bibinfo {year}
  {1980})}\BibitemShut {NoStop}%
\bibitem [{\citenamefont {Rado}\ \emph {et~al.}(1984)\citenamefont {Rado},
  \citenamefont {Ferrari},\ and\ \citenamefont {Maisch}}]{Rado1984}%
  \BibitemOpen
  \bibfield  {author} {\bibinfo {author} {\bibfnamefont {G.~T.}\ \bibnamefont
  {Rado}}, \bibinfo {author} {\bibfnamefont {J.~M.}\ \bibnamefont {Ferrari}}, \
  and\ \bibinfo {author} {\bibfnamefont {W.~G.}\ \bibnamefont {Maisch}},\
  }\href {\doibase 10.1103/PhysRevB.29.4041} {\bibfield  {journal} {\bibinfo
  {journal} {Phys. Rev. B}\ }\textbf {\bibinfo {volume} {29}},\ \bibinfo
  {pages} {4041} (\bibinfo {year} {1984})}\BibitemShut {NoStop}%
\bibitem [{\citenamefont {Cooke}\ \emph {et~al.}(1973)\citenamefont {Cooke},
  \citenamefont {Swithenby},\ and\ \citenamefont {Wells}}]{Cooke1973magnetic}%
  \BibitemOpen
  \bibfield  {author} {\bibinfo {author} {\bibfnamefont {A.~H.}\ \bibnamefont
  {Cooke}}, \bibinfo {author} {\bibfnamefont {S.~J.}\ \bibnamefont
  {Swithenby}}, \ and\ \bibinfo {author} {\bibfnamefont {M.~R.}\ \bibnamefont
  {Wells}},\ }\href {http://stacks.iop.org/0022-3719/6/i=13/a=019} {\bibfield
  {journal} {\bibinfo  {journal} {J. Phys. C}\ }\textbf {\bibinfo {volume}
  {6}},\ \bibinfo {pages} {2209} (\bibinfo {year} {1973})}\BibitemShut
  {NoStop}%
\bibitem [{\citenamefont {Gorodetsky}\ \emph {et~al.}(1973)\citenamefont
  {Gorodetsky}, \citenamefont {Hornreich},\ and\ \citenamefont
  {Wanklyn}}]{Vie1974}%
  \BibitemOpen
  \bibfield  {author} {\bibinfo {author} {\bibfnamefont {G.}~\bibnamefont
  {Gorodetsky}}, \bibinfo {author} {\bibfnamefont {R.~M.}\ \bibnamefont
  {Hornreich}}, \ and\ \bibinfo {author} {\bibfnamefont {B.~M.}\ \bibnamefont
  {Wanklyn}},\ }\href {\doibase 10.1103/PhysRevB.8.2263} {\bibfield  {journal}
  {\bibinfo  {journal} {Phys. Rev. B}\ }\textbf {\bibinfo {volume} {8}},\
  \bibinfo {pages} {2263} (\bibinfo {year} {1973})}\BibitemShut {NoStop}%
\bibitem [{\citenamefont {Purwanto}\ \emph {et~al.}(1994)\citenamefont
  {Purwanto}, \citenamefont {Robinson}, \citenamefont {Havela}, \citenamefont
  {Sechovsk{\'{y}}}, \citenamefont {Svoboda}, \citenamefont {Nakotte},
  \citenamefont {Proke{\v{s}}}, \citenamefont {de~Boer}, \citenamefont {Seret},
  \citenamefont {Winand}, \citenamefont {Rebizant},\ and\ \citenamefont
  {Spirlet}}]{Purwanto1994}%
  \BibitemOpen
  \bibfield  {author} {\bibinfo {author} {\bibfnamefont {A.}~\bibnamefont
  {Purwanto}}, \bibinfo {author} {\bibfnamefont {R.~A.}\ \bibnamefont
  {Robinson}}, \bibinfo {author} {\bibfnamefont {L.}~\bibnamefont {Havela}},
  \bibinfo {author} {\bibfnamefont {V.}~\bibnamefont {Sechovsk{\'{y}}}},
  \bibinfo {author} {\bibfnamefont {P.}~\bibnamefont {Svoboda}}, \bibinfo
  {author} {\bibfnamefont {H.}~\bibnamefont {Nakotte}}, \bibinfo {author}
  {\bibfnamefont {K.}~\bibnamefont {Proke{\v{s}}}}, \bibinfo {author}
  {\bibfnamefont {F.~R.}\ \bibnamefont {de~Boer}}, \bibinfo {author}
  {\bibfnamefont {A.}~\bibnamefont {Seret}}, \bibinfo {author} {\bibfnamefont
  {J.~M.}\ \bibnamefont {Winand}}, \bibinfo {author} {\bibfnamefont
  {J.}~\bibnamefont {Rebizant}}, \ and\ \bibinfo {author} {\bibfnamefont
  {J.~C.}\ \bibnamefont {Spirlet}},\ }\href {\doibase 10.1103/PhysRevB.50.6792}
  {\bibfield  {journal} {\bibinfo  {journal} {Phys. Rev. B}\ }\textbf {\bibinfo
  {volume} {50}},\ \bibinfo {pages} {6792} (\bibinfo {year}
  {1994})}\BibitemShut {NoStop}%
\bibitem [{\citenamefont {Martin-Martin}\ \emph {et~al.}(1999)\citenamefont
  {Martin-Martin}, \citenamefont {Pereira}, \citenamefont {Lander},
  \citenamefont {Rebizant}, \citenamefont {Wastin}, \citenamefont {Spirlet},
  \citenamefont {Dervenagas},\ and\ \citenamefont {Brown}}]{Martin-Martin1999}%
  \BibitemOpen
  \bibfield  {author} {\bibinfo {author} {\bibfnamefont {A.}~\bibnamefont
  {Martin-Martin}}, \bibinfo {author} {\bibfnamefont {L.~C.~J.}\ \bibnamefont
  {Pereira}}, \bibinfo {author} {\bibfnamefont {G.~H.}\ \bibnamefont {Lander}},
  \bibinfo {author} {\bibfnamefont {J.}~\bibnamefont {Rebizant}}, \bibinfo
  {author} {\bibfnamefont {F.}~\bibnamefont {Wastin}}, \bibinfo {author}
  {\bibfnamefont {J.~C.}\ \bibnamefont {Spirlet}}, \bibinfo {author}
  {\bibfnamefont {P.}~\bibnamefont {Dervenagas}}, \ and\ \bibinfo {author}
  {\bibfnamefont {P.~J.}\ \bibnamefont {Brown}},\ }\href {\doibase
  10.1103/PhysRevB.59.11818} {\bibfield  {journal} {\bibinfo  {journal} {Phys.
  Rev. B}\ }\textbf {\bibinfo {volume} {59}},\ \bibinfo {pages} {11818}
  (\bibinfo {year} {1999})}\BibitemShut {NoStop}%
\bibitem [{\citenamefont {Barthem}\ \emph {et~al.}(2013)\citenamefont
  {Barthem}, \citenamefont {Colin}, \citenamefont {Mayaffre}, \citenamefont
  {Julien},\ and\ \citenamefont {Givord}}]{Barthem2013}%
  \BibitemOpen
  \bibfield  {author} {\bibinfo {author} {\bibfnamefont {V.~M. T.~S.}\
  \bibnamefont {Barthem}}, \bibinfo {author} {\bibfnamefont {C.~V.}\
  \bibnamefont {Colin}}, \bibinfo {author} {\bibfnamefont {H.}~\bibnamefont
  {Mayaffre}}, \bibinfo {author} {\bibfnamefont {M.-H.}\ \bibnamefont
  {Julien}}, \ and\ \bibinfo {author} {\bibfnamefont {D.}~\bibnamefont
  {Givord}},\ }\href {\doibase 10.1038/ncomms3892} {\bibfield  {journal}
  {\bibinfo  {journal} {Nat. Commun.}\ }\textbf {\bibinfo {volume} {4}},\
  \bibinfo {pages} {2892} (\bibinfo {year} {2013})}\BibitemShut {NoStop}%
\bibitem [{\citenamefont {Leciejewicz}\ \emph {et~al.}(1967)\citenamefont
  {Leciejewicz}, \citenamefont {Tro{\'{c}}}, \citenamefont {Murasik},\ and\
  \citenamefont {Zygmunt}}]{Leciejewicz1967neutron}%
  \BibitemOpen
  \bibfield  {author} {\bibinfo {author} {\bibfnamefont {J.}~\bibnamefont
  {Leciejewicz}}, \bibinfo {author} {\bibfnamefont {R.}~\bibnamefont
  {Tro{\'{c}}}}, \bibinfo {author} {\bibfnamefont {A.}~\bibnamefont {Murasik}},
  \ and\ \bibinfo {author} {\bibfnamefont {A.}~\bibnamefont {Zygmunt}},\ }\href
  {\doibase 10.1002/pssb.19670220224} {\bibfield  {journal} {\bibinfo
  {journal} {Phys. status solidi}\ }\textbf {\bibinfo {volume} {22}},\ \bibinfo
  {pages} {517} (\bibinfo {year} {1967})}\BibitemShut {NoStop}%
\bibitem [{\citenamefont {Murasik}\ \emph {et~al.}(1969)\citenamefont
  {Murasik}, \citenamefont {Leciejewicz},\ and\ \citenamefont
  {Suski}}]{murasik1969neutron}%
  \BibitemOpen
  \bibfield  {author} {\bibinfo {author} {\bibfnamefont {A.}~\bibnamefont
  {Murasik}}, \bibinfo {author} {\bibfnamefont {J.}~\bibnamefont
  {Leciejewicz}}, \ and\ \bibinfo {author} {\bibfnamefont {W.}~\bibnamefont
  {Suski}},\ }\href {\doibase 10.1002/pssb.19690340263} {\bibfield  {journal}
  {\bibinfo  {journal} {physica status solidi (b)}\ }\textbf {\bibinfo {volume}
  {34}},\ \bibinfo {pages} {K157} (\bibinfo {year} {1969})}\BibitemShut
  {NoStop}%
\bibitem [{\citenamefont {Ptasiewicz-B\c{a}k}\ \emph
  {et~al.}(1978)\citenamefont {Ptasiewicz-B\c{a}k}, \citenamefont
  {Leciejewicz},\ and\ \citenamefont {Zygmunt}}]{Ptasiewicz1978}%
  \BibitemOpen
  \bibfield  {author} {\bibinfo {author} {\bibfnamefont {H.}~\bibnamefont
  {Ptasiewicz-B\c{a}k}}, \bibinfo {author} {\bibfnamefont {J.}~\bibnamefont
  {Leciejewicz}}, \ and\ \bibinfo {author} {\bibfnamefont {A.}~\bibnamefont
  {Zygmunt}},\ }\href {\doibase 10.1002/pssa.2210470202} {\bibfield  {journal}
  {\bibinfo  {journal} {phys. status solidi}\ }\textbf {\bibinfo {volume}
  {47}},\ \bibinfo {pages} {349} (\bibinfo {year} {1978})}\BibitemShut
  {NoStop}%
\bibitem [{\citenamefont {Venturini}\ \emph {et~al.}(1987)\citenamefont
  {Venturini}, \citenamefont {Malaman}, \citenamefont {{Le Ca{\"{e}}r}},\ and\
  \citenamefont {Fruchart}}]{Venturini1987}%
  \BibitemOpen
  \bibfield  {author} {\bibinfo {author} {\bibfnamefont {G.}~\bibnamefont
  {Venturini}}, \bibinfo {author} {\bibfnamefont {B.}~\bibnamefont {Malaman}},
  \bibinfo {author} {\bibfnamefont {G.}~\bibnamefont {{Le Ca{\"{e}}r}}}, \ and\
  \bibinfo {author} {\bibfnamefont {D.}~\bibnamefont {Fruchart}},\ }\href
  {\doibase 10.1103/PhysRevB.35.7038} {\bibfield  {journal} {\bibinfo
  {journal} {Phys. Rev. B}\ }\textbf {\bibinfo {volume} {35}},\ \bibinfo
  {pages} {7038} (\bibinfo {year} {1987})}\BibitemShut {NoStop}%
\bibitem [{\citenamefont {Armbr{\"{u}}ster}\ \emph {et~al.}(2010)\citenamefont
  {Armbr{\"{u}}ster}, \citenamefont {Schnelle}, \citenamefont {Cardoso-Gil},\
  and\ \citenamefont {Grin}}]{Armbruster2010}%
  \BibitemOpen
  \bibfield  {author} {\bibinfo {author} {\bibfnamefont {M.}~\bibnamefont
  {Armbr{\"{u}}ster}}, \bibinfo {author} {\bibfnamefont {W.}~\bibnamefont
  {Schnelle}}, \bibinfo {author} {\bibfnamefont {R.}~\bibnamefont
  {Cardoso-Gil}}, \ and\ \bibinfo {author} {\bibfnamefont {Y.}~\bibnamefont
  {Grin}},\ }\href {\doibase 10.1002/chem.201001473} {\bibfield  {journal}
  {\bibinfo  {journal} {Chem. - A Eur. J.}\ }\textbf {\bibinfo {volume} {16}},\
  \bibinfo {pages} {10357} (\bibinfo {year} {2010})}\BibitemShut {NoStop}%
\bibitem [{\citenamefont {Wadley}\ \emph {et~al.}(2013)\citenamefont {Wadley},
  \citenamefont {Nov{\'{a}}k}, \citenamefont {Campion}, \citenamefont
  {Rinaldi}, \citenamefont {Mart{\'{i}}}, \citenamefont {Reichlov{\'{a}}},
  \citenamefont {{\v{Z}}elezn{\'{y}}}, \citenamefont {Gazquez}, \citenamefont
  {Roldan}, \citenamefont {Varela}, \citenamefont {Khalyavin}, \citenamefont
  {Langridge}, \citenamefont {Kriegner}, \citenamefont {M{\'{a}}ca},
  \citenamefont {Ma{\v{s}}ek}, \citenamefont {Bertacco}, \citenamefont
  {Hol{\'{y}}}, \citenamefont {Rushforth}, \citenamefont {Edmonds},
  \citenamefont {Gallagher}, \citenamefont {Foxon}, \citenamefont
  {Wunderlich},\ and\ \citenamefont {Jungwirth}}]{Wadley2013}%
  \BibitemOpen
  \bibfield  {author} {\bibinfo {author} {\bibfnamefont {P.}~\bibnamefont
  {Wadley}}, \bibinfo {author} {\bibfnamefont {V.}~\bibnamefont {Nov{\'{a}}k}},
  \bibinfo {author} {\bibfnamefont {R.}~\bibnamefont {Campion}}, \bibinfo
  {author} {\bibfnamefont {C.}~\bibnamefont {Rinaldi}}, \bibinfo {author}
  {\bibfnamefont {X.}~\bibnamefont {Mart{\'{i}}}}, \bibinfo {author}
  {\bibfnamefont {H.}~\bibnamefont {Reichlov{\'{a}}}}, \bibinfo {author}
  {\bibfnamefont {J.}~\bibnamefont {{\v{Z}}elezn{\'{y}}}}, \bibinfo {author}
  {\bibfnamefont {J.}~\bibnamefont {Gazquez}}, \bibinfo {author} {\bibfnamefont
  {M.}~\bibnamefont {Roldan}}, \bibinfo {author} {\bibfnamefont
  {M.}~\bibnamefont {Varela}}, \bibinfo {author} {\bibfnamefont
  {D.}~\bibnamefont {Khalyavin}}, \bibinfo {author} {\bibfnamefont
  {S.}~\bibnamefont {Langridge}}, \bibinfo {author} {\bibfnamefont
  {D.}~\bibnamefont {Kriegner}}, \bibinfo {author} {\bibfnamefont
  {F.}~\bibnamefont {M{\'{a}}ca}}, \bibinfo {author} {\bibfnamefont
  {J.}~\bibnamefont {Ma{\v{s}}ek}}, \bibinfo {author} {\bibfnamefont
  {R.}~\bibnamefont {Bertacco}}, \bibinfo {author} {\bibfnamefont
  {V.}~\bibnamefont {Hol{\'{y}}}}, \bibinfo {author} {\bibfnamefont
  {A.}~\bibnamefont {Rushforth}}, \bibinfo {author} {\bibfnamefont
  {K.}~\bibnamefont {Edmonds}}, \bibinfo {author} {\bibfnamefont
  {B.}~\bibnamefont {Gallagher}}, \bibinfo {author} {\bibfnamefont
  {C.}~\bibnamefont {Foxon}}, \bibinfo {author} {\bibfnamefont
  {J.}~\bibnamefont {Wunderlich}}, \ and\ \bibinfo {author} {\bibfnamefont
  {T.}~\bibnamefont {Jungwirth}},\ }\href {\doibase 10.1038/ncomms3322}
  {\bibfield  {journal} {\bibinfo  {journal} {Nat. Commun.}\ }\textbf {\bibinfo
  {volume} {4}},\ \bibinfo {pages} {2322} (\bibinfo {year} {2013})}\BibitemShut
  {NoStop}%
\bibitem [{\citenamefont {Kunnmann}\ \emph {et~al.}(1968)\citenamefont
  {Kunnmann}, \citenamefont {Placa}, \citenamefont {Corliss}, \citenamefont
  {Hastings},\ and\ \citenamefont {Banks}}]{Kunnmann1968magnetic}%
  \BibitemOpen
  \bibfield  {author} {\bibinfo {author} {\bibfnamefont {W.}~\bibnamefont
  {Kunnmann}}, \bibinfo {author} {\bibfnamefont {S.~L.}\ \bibnamefont {Placa}},
  \bibinfo {author} {\bibfnamefont {L.}~\bibnamefont {Corliss}}, \bibinfo
  {author} {\bibfnamefont {J.}~\bibnamefont {Hastings}}, \ and\ \bibinfo
  {author} {\bibfnamefont {E.}~\bibnamefont {Banks}},\ }\href
  {https://www.sciencedirect.com/science/article/pii/002236976890187X}
  {\bibfield  {journal} {\bibinfo  {journal} {J. Phys. Chem. Solids}\ }\textbf
  {\bibinfo {volume} {29}},\ \bibinfo {pages} {1359 } (\bibinfo {year}
  {1968})}\BibitemShut {NoStop}%
\bibitem [{\citenamefont {Zhu}\ \emph {et~al.}(2014)\citenamefont {Zhu},
  \citenamefont {Do}, \citenamefont {{Dela Cruz}}, \citenamefont {Dun},
  \citenamefont {Zhou}, \citenamefont {Mahanti},\ and\ \citenamefont
  {Ke}}]{Zhu2014}%
  \BibitemOpen
  \bibfield  {author} {\bibinfo {author} {\bibfnamefont {M.}~\bibnamefont
  {Zhu}}, \bibinfo {author} {\bibfnamefont {D.}~\bibnamefont {Do}}, \bibinfo
  {author} {\bibfnamefont {C.~R.}\ \bibnamefont {{Dela Cruz}}}, \bibinfo
  {author} {\bibfnamefont {Z.}~\bibnamefont {Dun}}, \bibinfo {author}
  {\bibfnamefont {H.~D.}\ \bibnamefont {Zhou}}, \bibinfo {author}
  {\bibfnamefont {S.~D.}\ \bibnamefont {Mahanti}}, \ and\ \bibinfo {author}
  {\bibfnamefont {X.}~\bibnamefont {Ke}},\ }\href {\doibase
  10.1103/PhysRevLett.113.076406} {\bibfield  {journal} {\bibinfo  {journal}
  {Phys. Rev. Lett.}\ }\textbf {\bibinfo {volume} {113}},\ \bibinfo {pages}
  {076406} (\bibinfo {year} {2014})}\BibitemShut {NoStop}%
\bibitem [{\citenamefont {Wang}\ \emph {et~al.}(2014)\citenamefont {Wang},
  \citenamefont {Santana}, \citenamefont {Wu}, \citenamefont {Karunakaran},
  \citenamefont {Wang}, \citenamefont {Dowben},\ and\ \citenamefont
  {Binek}}]{Wang2014}%
  \BibitemOpen
  \bibfield  {author} {\bibinfo {author} {\bibfnamefont {J.}~\bibnamefont
  {Wang}}, \bibinfo {author} {\bibfnamefont {J.~A.~C.}\ \bibnamefont
  {Santana}}, \bibinfo {author} {\bibfnamefont {N.}~\bibnamefont {Wu}},
  \bibinfo {author} {\bibfnamefont {C.}~\bibnamefont {Karunakaran}}, \bibinfo
  {author} {\bibfnamefont {J.}~\bibnamefont {Wang}}, \bibinfo {author}
  {\bibfnamefont {P.~A.}\ \bibnamefont {Dowben}}, \ and\ \bibinfo {author}
  {\bibfnamefont {C.}~\bibnamefont {Binek}},\ }\href {\doibase
  10.1088/0953-8984/26/5/055012} {\bibfield  {journal} {\bibinfo  {journal} {J.
  Phys. Condens. Matter}\ }\textbf {\bibinfo {volume} {26}},\ \bibinfo {pages}
  {055012} (\bibinfo {year} {2014})}\BibitemShut {NoStop}%
\bibitem [{\citenamefont {Calder}\ \emph {et~al.}(2012)\citenamefont {Calder},
  \citenamefont {Cao}, \citenamefont {Lumsden}, \citenamefont {Kim},
  \citenamefont {Gai}, \citenamefont {Sales}, \citenamefont {Mandrus},\ and\
  \citenamefont {Christianson}}]{Calder2012b}%
  \BibitemOpen
  \bibfield  {author} {\bibinfo {author} {\bibfnamefont {S.}~\bibnamefont
  {Calder}}, \bibinfo {author} {\bibfnamefont {G.-X.}\ \bibnamefont {Cao}},
  \bibinfo {author} {\bibfnamefont {M.~D.}\ \bibnamefont {Lumsden}}, \bibinfo
  {author} {\bibfnamefont {J.~W.}\ \bibnamefont {Kim}}, \bibinfo {author}
  {\bibfnamefont {Z.}~\bibnamefont {Gai}}, \bibinfo {author} {\bibfnamefont
  {B.~C.}\ \bibnamefont {Sales}}, \bibinfo {author} {\bibfnamefont
  {D.}~\bibnamefont {Mandrus}}, \ and\ \bibinfo {author} {\bibfnamefont
  {A.~D.}\ \bibnamefont {Christianson}},\ }\href {\doibase
  10.1103/PhysRevB.86.220403} {\bibfield  {journal} {\bibinfo  {journal} {Phys.
  Rev. B}\ }\textbf {\bibinfo {volume} {86}},\ \bibinfo {pages} {220403}
  (\bibinfo {year} {2012})}\BibitemShut {NoStop}%
\bibitem [{\citenamefont {Pasturel}\ \emph {et~al.}(2009)\citenamefont
  {Pasturel}, \citenamefont {Tougait}, \citenamefont {Potel}, \citenamefont
  {Roisnel}, \citenamefont {Wochowski}, \citenamefont {No{\"{e}}l},\ and\
  \citenamefont {Tro{\'{c}}}}]{Pasturel2009}%
  \BibitemOpen
  \bibfield  {author} {\bibinfo {author} {\bibfnamefont {M.}~\bibnamefont
  {Pasturel}}, \bibinfo {author} {\bibfnamefont {O.}~\bibnamefont {Tougait}},
  \bibinfo {author} {\bibfnamefont {M.}~\bibnamefont {Potel}}, \bibinfo
  {author} {\bibfnamefont {T.}~\bibnamefont {Roisnel}}, \bibinfo {author}
  {\bibfnamefont {K.}~\bibnamefont {Wochowski}}, \bibinfo {author}
  {\bibfnamefont {H.}~\bibnamefont {No{\"{e}}l}}, \ and\ \bibinfo {author}
  {\bibfnamefont {R.}~\bibnamefont {Tro{\'{c}}}},\ }\href {\doibase
  10.1088/0953-8984/21/12/125401} {\bibfield  {journal} {\bibinfo  {journal}
  {J. Phys. Condens. Matter}\ }\textbf {\bibinfo {volume} {21}},\ \bibinfo
  {pages} {125401} (\bibinfo {year} {2009})}\BibitemShut {NoStop}%
\bibitem [{\citenamefont {Tro{\'{c}}}\ \emph {et~al.}(2012)\citenamefont
  {Tro{\'{c}}}, \citenamefont {Pasturel}, \citenamefont {Tougait},
  \citenamefont {Sazonov}, \citenamefont {Gukasov}, \citenamefont
  {Su{\l}kowski},\ and\ \citenamefont {No{\"{e}}l}}]{Troc2012}%
  \BibitemOpen
  \bibfield  {author} {\bibinfo {author} {\bibfnamefont {R.}~\bibnamefont
  {Tro{\'{c}}}}, \bibinfo {author} {\bibfnamefont {M.}~\bibnamefont
  {Pasturel}}, \bibinfo {author} {\bibfnamefont {O.}~\bibnamefont {Tougait}},
  \bibinfo {author} {\bibfnamefont {A.~P.}\ \bibnamefont {Sazonov}}, \bibinfo
  {author} {\bibfnamefont {A.}~\bibnamefont {Gukasov}}, \bibinfo {author}
  {\bibfnamefont {C.}~\bibnamefont {Su{\l}kowski}}, \ and\ \bibinfo {author}
  {\bibfnamefont {H.}~\bibnamefont {No{\"{e}}l}},\ }\href {\doibase
  10.1103/PhysRevB.85.064412} {\bibfield  {journal} {\bibinfo  {journal} {Phys.
  Rev. B}\ }\textbf {\bibinfo {volume} {85}},\ \bibinfo {pages} {064412}
  (\bibinfo {year} {2012})}\BibitemShut {NoStop}%
\bibitem [{\citenamefont {Sangeetha}\ \emph {et~al.}(2016)\citenamefont
  {Sangeetha}, \citenamefont {Pandey}, \citenamefont {Benson},\ and\
  \citenamefont {Johnston}}]{Sangeetha2016}%
  \BibitemOpen
  \bibfield  {author} {\bibinfo {author} {\bibfnamefont {N.~S.}\ \bibnamefont
  {Sangeetha}}, \bibinfo {author} {\bibfnamefont {A.}~\bibnamefont {Pandey}},
  \bibinfo {author} {\bibfnamefont {Z.~A.}\ \bibnamefont {Benson}}, \ and\
  \bibinfo {author} {\bibfnamefont {D.~C.}\ \bibnamefont {Johnston}},\ }\href
  {\doibase 10.1103/PhysRevB.94.094417} {\bibfield  {journal} {\bibinfo
  {journal} {Phys. Rev. B}\ }\textbf {\bibinfo {volume} {94}},\ \bibinfo
  {pages} {094417} (\bibinfo {year} {2016})}\BibitemShut {NoStop}%
\bibitem [{\citenamefont {McNally}\ \emph {et~al.}(2015)\citenamefont
  {McNally}, \citenamefont {Simonson}, \citenamefont {Kistner-Morris},
  \citenamefont {Smith}, \citenamefont {Hassinger}, \citenamefont
  {DeBeer-Schmitt}, \citenamefont {Kolesnikov}, \citenamefont {Zaliznyak},\
  and\ \citenamefont {Aronson}}]{McNally2015c}%
  \BibitemOpen
  \bibfield  {author} {\bibinfo {author} {\bibfnamefont {D.~E.}\ \bibnamefont
  {McNally}}, \bibinfo {author} {\bibfnamefont {J.~W.}\ \bibnamefont
  {Simonson}}, \bibinfo {author} {\bibfnamefont {J.~J.}\ \bibnamefont
  {Kistner-Morris}}, \bibinfo {author} {\bibfnamefont {G.~J.}\ \bibnamefont
  {Smith}}, \bibinfo {author} {\bibfnamefont {J.~E.}\ \bibnamefont
  {Hassinger}}, \bibinfo {author} {\bibfnamefont {L.}~\bibnamefont
  {DeBeer-Schmitt}}, \bibinfo {author} {\bibfnamefont {A.~I.}\ \bibnamefont
  {Kolesnikov}}, \bibinfo {author} {\bibfnamefont {I.~A.}\ \bibnamefont
  {Zaliznyak}}, \ and\ \bibinfo {author} {\bibfnamefont {M.~C.}\ \bibnamefont
  {Aronson}},\ }\href {\doibase 10.1103/PhysRevB.91.180407} {\bibfield
  {journal} {\bibinfo  {journal} {Phys. Rev. B}\ }\textbf {\bibinfo {volume}
  {91}},\ \bibinfo {pages} {180407} (\bibinfo {year} {2015})}\BibitemShut
  {NoStop}%
\bibitem [{\citenamefont {Bridges}\ \emph {et~al.}(2009)\citenamefont
  {Bridges}, \citenamefont {Krishnamurthy}, \citenamefont {Poulton},
  \citenamefont {Paranthaman}, \citenamefont {Sales}, \citenamefont {Myers},\
  and\ \citenamefont {Bobev}}]{Bridges2009}%
  \BibitemOpen
  \bibfield  {author} {\bibinfo {author} {\bibfnamefont {C.~A.}\ \bibnamefont
  {Bridges}}, \bibinfo {author} {\bibfnamefont {V.~V.}\ \bibnamefont
  {Krishnamurthy}}, \bibinfo {author} {\bibfnamefont {S.}~\bibnamefont
  {Poulton}}, \bibinfo {author} {\bibfnamefont {M.~P.}\ \bibnamefont
  {Paranthaman}}, \bibinfo {author} {\bibfnamefont {B.~C.}\ \bibnamefont
  {Sales}}, \bibinfo {author} {\bibfnamefont {C.}~\bibnamefont {Myers}}, \ and\
  \bibinfo {author} {\bibfnamefont {S.}~\bibnamefont {Bobev}},\ }\href
  {\doibase 10.1016/j.jmmm.2009.07.015} {\bibfield  {journal} {\bibinfo
  {journal} {J. Magn. Magn. Mater.}\ }\textbf {\bibinfo {volume} {321}},\
  \bibinfo {pages} {3653} (\bibinfo {year} {2009})}\BibitemShut {NoStop}%
\bibitem [{\citenamefont {Gibson}\ \emph {et~al.}(2015)\citenamefont {Gibson},
  \citenamefont {Wu}, \citenamefont {Liang}, \citenamefont {Ali}, \citenamefont
  {Ong}, \citenamefont {Huang},\ and\ \citenamefont {Cava}}]{Gibson2015}%
  \BibitemOpen
  \bibfield  {author} {\bibinfo {author} {\bibfnamefont {Q.~D.}\ \bibnamefont
  {Gibson}}, \bibinfo {author} {\bibfnamefont {H.}~\bibnamefont {Wu}}, \bibinfo
  {author} {\bibfnamefont {T.}~\bibnamefont {Liang}}, \bibinfo {author}
  {\bibfnamefont {M.~N.}\ \bibnamefont {Ali}}, \bibinfo {author} {\bibfnamefont
  {N.~P.}\ \bibnamefont {Ong}}, \bibinfo {author} {\bibfnamefont
  {Q.}~\bibnamefont {Huang}}, \ and\ \bibinfo {author} {\bibfnamefont {R.~J.}\
  \bibnamefont {Cava}},\ }\href {\doibase 10.1103/PhysRevB.91.085128}
  {\bibfield  {journal} {\bibinfo  {journal} {Phys. Rev. B}\ }\textbf {\bibinfo
  {volume} {91}},\ \bibinfo {pages} {085128} (\bibinfo {year}
  {2015})}\BibitemShut {NoStop}%
\bibitem [{\citenamefont {Kawaguchi}\ \emph {et~al.}(2018)\citenamefont
  {Kawaguchi}, \citenamefont {Urata}, \citenamefont {Hatano}, \citenamefont
  {Iida},\ and\ \citenamefont {Ikuta}}]{Kawaguchi2018CaMn2Bi2}%
  \BibitemOpen
  \bibfield  {author} {\bibinfo {author} {\bibfnamefont {N.}~\bibnamefont
  {Kawaguchi}}, \bibinfo {author} {\bibfnamefont {T.}~\bibnamefont {Urata}},
  \bibinfo {author} {\bibfnamefont {T.}~\bibnamefont {Hatano}}, \bibinfo
  {author} {\bibfnamefont {K.}~\bibnamefont {Iida}}, \ and\ \bibinfo {author}
  {\bibfnamefont {H.}~\bibnamefont {Ikuta}},\ }\href {\doibase
  10.1103/PhysRevB.97.140403} {\bibfield  {journal} {\bibinfo  {journal} {Phys.
  Rev. B}\ }\textbf {\bibinfo {volume} {97}},\ \bibinfo {pages} {140403}
  (\bibinfo {year} {2018})}\BibitemShut {NoStop}%
\bibitem [{\citenamefont {Das}\ \emph {et~al.}(2017{\natexlab{b}})\citenamefont
  {Das}, \citenamefont {Sangeetha}, \citenamefont {Pandey}, \citenamefont
  {Benson}, \citenamefont {Heitmann}, \citenamefont {Johnston}, \citenamefont
  {Goldman},\ and\ \citenamefont {Kreyssig}}]{Das2017}%
  \BibitemOpen
  \bibfield  {author} {\bibinfo {author} {\bibfnamefont {P.}~\bibnamefont
  {Das}}, \bibinfo {author} {\bibfnamefont {N.~S.}\ \bibnamefont {Sangeetha}},
  \bibinfo {author} {\bibfnamefont {A.}~\bibnamefont {Pandey}}, \bibinfo
  {author} {\bibfnamefont {Z.~A.}\ \bibnamefont {Benson}}, \bibinfo {author}
  {\bibfnamefont {T.~W.}\ \bibnamefont {Heitmann}}, \bibinfo {author}
  {\bibfnamefont {D.~C.}\ \bibnamefont {Johnston}}, \bibinfo {author}
  {\bibfnamefont {A.~I.}\ \bibnamefont {Goldman}}, \ and\ \bibinfo {author}
  {\bibfnamefont {A.}~\bibnamefont {Kreyssig}},\ }\href {\doibase
  10.1088/0953-8984/29/3/035802} {\bibfield  {journal} {\bibinfo  {journal} {J.
  Phys. Condens. Matter}\ }\textbf {\bibinfo {volume} {29}},\ \bibinfo {pages}
  {035802} (\bibinfo {year} {2017}{\natexlab{b}})}\BibitemShut {NoStop}%
\bibitem [{\citenamefont {Anand}\ and\ \citenamefont
  {Johnston}(2016)}]{Anand2016}%
  \BibitemOpen
  \bibfield  {author} {\bibinfo {author} {\bibfnamefont {V.~K.}\ \bibnamefont
  {Anand}}\ and\ \bibinfo {author} {\bibfnamefont {D.~C.}\ \bibnamefont
  {Johnston}},\ }\href {\doibase 10.1103/PhysRevB.94.014431} {\bibfield
  {journal} {\bibinfo  {journal} {Phys. Rev. B}\ }\textbf {\bibinfo {volume}
  {94}},\ \bibinfo {pages} {014431} (\bibinfo {year} {2016})}\BibitemShut
  {NoStop}%
\bibitem [{\citenamefont {Morozkin}\ \emph {et~al.}(2006)\citenamefont
  {Morozkin}, \citenamefont {Isnard}, \citenamefont {Henry}, \citenamefont
  {Granovsky}, \citenamefont {Nirmala},\ and\ \citenamefont
  {Manfrinetti}}]{Morozkin2006}%
  \BibitemOpen
  \bibfield  {author} {\bibinfo {author} {\bibfnamefont {A.~V.}\ \bibnamefont
  {Morozkin}}, \bibinfo {author} {\bibfnamefont {O.}~\bibnamefont {Isnard}},
  \bibinfo {author} {\bibfnamefont {P.}~\bibnamefont {Henry}}, \bibinfo
  {author} {\bibfnamefont {S.}~\bibnamefont {Granovsky}}, \bibinfo {author}
  {\bibfnamefont {R.}~\bibnamefont {Nirmala}}, \ and\ \bibinfo {author}
  {\bibfnamefont {P.}~\bibnamefont {Manfrinetti}},\ }\href {\doibase
  10.1016/j.jallcom.2005.10.051} {\bibfield  {journal} {\bibinfo  {journal} {J.
  Alloys Compd.}\ }\textbf {\bibinfo {volume} {420}},\ \bibinfo {pages} {34}
  (\bibinfo {year} {2006})}\BibitemShut {NoStop}%
\bibitem [{\citenamefont {Leciejewicz}\ \emph {et~al.}(1975)\citenamefont
  {Leciejewicz}, \citenamefont {Zolnierek}, \citenamefont {Ligenza},
  \citenamefont {Troc},\ and\ \citenamefont
  {Ptasiewicz}}]{leciejewicz1975magnetic}%
  \BibitemOpen
  \bibfield  {author} {\bibinfo {author} {\bibfnamefont {J.}~\bibnamefont
  {Leciejewicz}}, \bibinfo {author} {\bibfnamefont {Z.}~\bibnamefont
  {Zolnierek}}, \bibinfo {author} {\bibfnamefont {S.}~\bibnamefont {Ligenza}},
  \bibinfo {author} {\bibfnamefont {R.}~\bibnamefont {Troc}}, \ and\ \bibinfo
  {author} {\bibfnamefont {H.}~\bibnamefont {Ptasiewicz}},\ }\href@noop {}
  {\bibfield  {journal} {\bibinfo  {journal} {J. Phys. C}\ }\textbf {\bibinfo
  {volume} {8}},\ \bibinfo {pages} {1697} (\bibinfo {year} {1975})}\BibitemShut
  {NoStop}%
\bibitem [{\citenamefont {Silverstein}\ \emph {et~al.}(2016)\citenamefont
  {Silverstein}, \citenamefont {Skoropata}, \citenamefont {Sarte},
  \citenamefont {Mauws}, \citenamefont {Aczel}, \citenamefont {Choi},
  \citenamefont {van Lierop}, \citenamefont {Wiebe},\ and\ \citenamefont
  {Zhou}}]{Silverstein2016}%
  \BibitemOpen
  \bibfield  {author} {\bibinfo {author} {\bibfnamefont {H.~J.}\ \bibnamefont
  {Silverstein}}, \bibinfo {author} {\bibfnamefont {E.}~\bibnamefont
  {Skoropata}}, \bibinfo {author} {\bibfnamefont {P.~M.}\ \bibnamefont
  {Sarte}}, \bibinfo {author} {\bibfnamefont {C.}~\bibnamefont {Mauws}},
  \bibinfo {author} {\bibfnamefont {A.~A.}\ \bibnamefont {Aczel}}, \bibinfo
  {author} {\bibfnamefont {E.~S.}\ \bibnamefont {Choi}}, \bibinfo {author}
  {\bibfnamefont {J.}~\bibnamefont {van Lierop}}, \bibinfo {author}
  {\bibfnamefont {C.~R.}\ \bibnamefont {Wiebe}}, \ and\ \bibinfo {author}
  {\bibfnamefont {H.}~\bibnamefont {Zhou}},\ }\href {\doibase
  10.1103/PhysRevB.93.054416} {\bibfield  {journal} {\bibinfo  {journal} {Phys.
  Rev. B}\ }\textbf {\bibinfo {volume} {93}},\ \bibinfo {pages} {054416}
  (\bibinfo {year} {2016})}\BibitemShut {NoStop}%
\bibitem [{\citenamefont {Shirane}\ \emph {et~al.}(1959)\citenamefont
  {Shirane}, \citenamefont {{J. Pickart}},\ and\ \citenamefont
  {Ishikawa}}]{Shirane1959a}%
  \BibitemOpen
  \bibfield  {author} {\bibinfo {author} {\bibfnamefont {G.}~\bibnamefont
  {Shirane}}, \bibinfo {author} {\bibfnamefont {S.}~\bibnamefont {{J.
  Pickart}}}, \ and\ \bibinfo {author} {\bibfnamefont {Y.}~\bibnamefont
  {Ishikawa}},\ }\href {\doibase 10.1143/JPSJ.14.1352} {\bibfield  {journal}
  {\bibinfo  {journal} {J. Phys. Soc. Jpn.}\ }\textbf {\bibinfo {volume}
  {14}},\ \bibinfo {pages} {1352} (\bibinfo {year} {1959})}\BibitemShut
  {NoStop}%
\bibitem [{\citenamefont {Tsuzuki}\ \emph {et~al.}(1974)\citenamefont
  {Tsuzuki}, \citenamefont {Ishikawa}, \citenamefont {Watanabe},\ and\
  \citenamefont {Akimoto}}]{Tsuzuki1974c}%
  \BibitemOpen
  \bibfield  {author} {\bibinfo {author} {\bibfnamefont {K.}~\bibnamefont
  {Tsuzuki}}, \bibinfo {author} {\bibfnamefont {Y.}~\bibnamefont {Ishikawa}},
  \bibinfo {author} {\bibfnamefont {N.}~\bibnamefont {Watanabe}}, \ and\
  \bibinfo {author} {\bibfnamefont {S.}~\bibnamefont {Akimoto}},\ }\href
  {\doibase 10.1143/JPSJ.37.1242} {\bibfield  {journal} {\bibinfo  {journal}
  {J. Phys. Soc. Jpn.}\ }\textbf {\bibinfo {volume} {37}},\ \bibinfo {pages}
  {1242} (\bibinfo {year} {1974})}\BibitemShut {NoStop}%
\bibitem [{\citenamefont {Bertaut}\ \emph {et~al.}(1961)\citenamefont
  {Bertaut}, \citenamefont {Corliss}, \citenamefont {Forrat}, \citenamefont
  {Aleonard},\ and\ \citenamefont {Pauthenet}}]{Bertaut1961}%
  \BibitemOpen
  \bibfield  {author} {\bibinfo {author} {\bibfnamefont {E.}~\bibnamefont
  {Bertaut}}, \bibinfo {author} {\bibfnamefont {L.}~\bibnamefont {Corliss}},
  \bibinfo {author} {\bibfnamefont {F.}~\bibnamefont {Forrat}}, \bibinfo
  {author} {\bibfnamefont {R.}~\bibnamefont {Aleonard}}, \ and\ \bibinfo
  {author} {\bibfnamefont {R.}~\bibnamefont {Pauthenet}},\ }\href {\doibase
  10.1016/0022-3697(61)90103-2} {\bibfield  {journal} {\bibinfo  {journal} {J.
  Phys. Chem. Solids}\ }\textbf {\bibinfo {volume} {21}},\ \bibinfo {pages}
  {234} (\bibinfo {year} {1961})}\BibitemShut {NoStop}%
\bibitem [{\citenamefont {Khanh}\ \emph {et~al.}(2016)\citenamefont {Khanh},
  \citenamefont {Abe}, \citenamefont {Sagayama}, \citenamefont {Nakao},
  \citenamefont {Hanashima}, \citenamefont {Kiyanagi}, \citenamefont
  {Tokunaga},\ and\ \citenamefont {Arima}}]{Khanh2016a}%
  \BibitemOpen
  \bibfield  {author} {\bibinfo {author} {\bibfnamefont {N.~D.}\ \bibnamefont
  {Khanh}}, \bibinfo {author} {\bibfnamefont {N.}~\bibnamefont {Abe}}, \bibinfo
  {author} {\bibfnamefont {H.}~\bibnamefont {Sagayama}}, \bibinfo {author}
  {\bibfnamefont {A.}~\bibnamefont {Nakao}}, \bibinfo {author} {\bibfnamefont
  {T.}~\bibnamefont {Hanashima}}, \bibinfo {author} {\bibfnamefont
  {R.}~\bibnamefont {Kiyanagi}}, \bibinfo {author} {\bibfnamefont
  {Y.}~\bibnamefont {Tokunaga}}, \ and\ \bibinfo {author} {\bibfnamefont
  {T.}~\bibnamefont {Arima}},\ }\href {\doibase 10.1103/PhysRevB.93.075117}
  {\bibfield  {journal} {\bibinfo  {journal} {Phys. Rev. B}\ }\textbf {\bibinfo
  {volume} {93}},\ \bibinfo {pages} {075117} (\bibinfo {year}
  {2016})}\BibitemShut {NoStop}%
\bibitem [{\citenamefont {Cao}\ \emph {et~al.}(2017)\citenamefont {Cao},
  \citenamefont {Xiang}, \citenamefont {Feng}, \citenamefont {Kang},
  \citenamefont {Zhang}, \citenamefont {Guiblin}, \citenamefont {Ren},
  \citenamefont {Dkhil},\ and\ \citenamefont {Cao}}]{Cao2017}%
  \BibitemOpen
  \bibfield  {author} {\bibinfo {author} {\bibfnamefont {Y.}~\bibnamefont
  {Cao}}, \bibinfo {author} {\bibfnamefont {M.}~\bibnamefont {Xiang}}, \bibinfo
  {author} {\bibfnamefont {Z.}~\bibnamefont {Feng}}, \bibinfo {author}
  {\bibfnamefont {B.}~\bibnamefont {Kang}}, \bibinfo {author} {\bibfnamefont
  {J.}~\bibnamefont {Zhang}}, \bibinfo {author} {\bibfnamefont
  {N.}~\bibnamefont {Guiblin}}, \bibinfo {author} {\bibfnamefont
  {W.}~\bibnamefont {Ren}}, \bibinfo {author} {\bibfnamefont {B.}~\bibnamefont
  {Dkhil}}, \ and\ \bibinfo {author} {\bibfnamefont {S.}~\bibnamefont {Cao}},\
  }\href {\doibase 10.1039/C6RA26231G} {\bibfield  {journal} {\bibinfo
  {journal} {RSC Adv.}\ }\textbf {\bibinfo {volume} {7}},\ \bibinfo {pages}
  {13846} (\bibinfo {year} {2017})}\BibitemShut {NoStop}%
\bibitem [{\citenamefont {Fischer}\ \emph {et~al.}(1972)\citenamefont
  {Fischer}, \citenamefont {Gorodetsky},\ and\ \citenamefont
  {Hornreich}}]{Fischer1972}%
  \BibitemOpen
  \bibfield  {author} {\bibinfo {author} {\bibfnamefont {E.}~\bibnamefont
  {Fischer}}, \bibinfo {author} {\bibfnamefont {G.}~\bibnamefont {Gorodetsky}},
  \ and\ \bibinfo {author} {\bibfnamefont {R.~M.}\ \bibnamefont {Hornreich}},\
  }\href {\doibase 10.1016/0038-1098(72)90927-1} {\bibfield  {journal}
  {\bibinfo  {journal} {Solid State Commun.}\ }\textbf {\bibinfo {volume}
  {10}},\ \bibinfo {pages} {1127} (\bibinfo {year} {1972})}\BibitemShut
  {NoStop}%
\bibitem [{\citenamefont {Fang}\ \emph {et~al.}(2015)\citenamefont {Fang},
  \citenamefont {Yan}, \citenamefont {Zhang}, \citenamefont {Han},
  \citenamefont {Qian}, \citenamefont {Wang},\ and\ \citenamefont
  {Du}}]{Fang2015a}%
  \BibitemOpen
  \bibfield  {author} {\bibinfo {author} {\bibfnamefont {Y.}~\bibnamefont
  {Fang}}, \bibinfo {author} {\bibfnamefont {S.}~\bibnamefont {Yan}}, \bibinfo
  {author} {\bibfnamefont {L.}~\bibnamefont {Zhang}}, \bibinfo {author}
  {\bibfnamefont {Z.}~\bibnamefont {Han}}, \bibinfo {author} {\bibfnamefont
  {B.}~\bibnamefont {Qian}}, \bibinfo {author} {\bibfnamefont {D.}~\bibnamefont
  {Wang}}, \ and\ \bibinfo {author} {\bibfnamefont {Y.}~\bibnamefont {Du}},\
  }\href {\doibase 10.1111/jace.13651} {\bibfield  {journal} {\bibinfo
  {journal} {J. Am. Ceram. Soc.}\ }\textbf {\bibinfo {volume} {98}},\ \bibinfo
  {pages} {2005} (\bibinfo {year} {2015})}\BibitemShut {NoStop}%
\bibitem [{\citenamefont {Liu}\ \emph {et~al.}(2016)\citenamefont {Liu},
  \citenamefont {Fang}, \citenamefont {Han}, \citenamefont {Yan}, \citenamefont
  {Zhou}, \citenamefont {Qian}, \citenamefont {Wang},\ and\ \citenamefont
  {Du}}]{Liu2016a}%
  \BibitemOpen
  \bibfield  {author} {\bibinfo {author} {\bibfnamefont {B.~B.}\ \bibnamefont
  {Liu}}, \bibinfo {author} {\bibfnamefont {Y.}~\bibnamefont {Fang}}, \bibinfo
  {author} {\bibfnamefont {Z.~D.}\ \bibnamefont {Han}}, \bibinfo {author}
  {\bibfnamefont {S.~M.}\ \bibnamefont {Yan}}, \bibinfo {author} {\bibfnamefont
  {W.~P.}\ \bibnamefont {Zhou}}, \bibinfo {author} {\bibfnamefont
  {B.}~\bibnamefont {Qian}}, \bibinfo {author} {\bibfnamefont {D.~H.}\
  \bibnamefont {Wang}}, \ and\ \bibinfo {author} {\bibfnamefont {Y.~W.}\
  \bibnamefont {Du}},\ }\href {\doibase 10.1016/j.matlet.2015.11.025}
  {\bibfield  {journal} {\bibinfo  {journal} {Mater. Lett.}\ }\textbf {\bibinfo
  {volume} {164}},\ \bibinfo {pages} {425} (\bibinfo {year}
  {2016})}\BibitemShut {NoStop}%
\bibitem [{\citenamefont {He}\ \emph {et~al.}(2017)\citenamefont {He},
  \citenamefont {Fu}, \citenamefont {Zhao}, \citenamefont {Liang},
  \citenamefont {Chen}, \citenamefont {Leng}, \citenamefont {Wang},
  \citenamefont {Li}, \citenamefont {Zhang}, \citenamefont {Xue}, \citenamefont
  {Li}, \citenamefont {Zhang}, \citenamefont {Ren},\ and\ \citenamefont
  {Chen}}]{He2017a}%
  \BibitemOpen
  \bibfield  {author} {\bibinfo {author} {\bibfnamefont {J.~B.}\ \bibnamefont
  {He}}, \bibinfo {author} {\bibfnamefont {Y.}~\bibnamefont {Fu}}, \bibinfo
  {author} {\bibfnamefont {L.~X.}\ \bibnamefont {Zhao}}, \bibinfo {author}
  {\bibfnamefont {H.}~\bibnamefont {Liang}}, \bibinfo {author} {\bibfnamefont
  {D.}~\bibnamefont {Chen}}, \bibinfo {author} {\bibfnamefont {Y.~M.}\
  \bibnamefont {Leng}}, \bibinfo {author} {\bibfnamefont {X.~M.}\ \bibnamefont
  {Wang}}, \bibinfo {author} {\bibfnamefont {J.}~\bibnamefont {Li}}, \bibinfo
  {author} {\bibfnamefont {S.}~\bibnamefont {Zhang}}, \bibinfo {author}
  {\bibfnamefont {M.~Q.}\ \bibnamefont {Xue}}, \bibinfo {author} {\bibfnamefont
  {C.~H.}\ \bibnamefont {Li}}, \bibinfo {author} {\bibfnamefont
  {P.}~\bibnamefont {Zhang}}, \bibinfo {author} {\bibfnamefont {Z.~A.}\
  \bibnamefont {Ren}}, \ and\ \bibinfo {author} {\bibfnamefont {G.~F.}\
  \bibnamefont {Chen}},\ }\href {\doibase 10.1103/PhysRevB.95.045128}
  {\bibfield  {journal} {\bibinfo  {journal} {Phys. Rev. B}\ }\textbf {\bibinfo
  {volume} {95}},\ \bibinfo {pages} {045128} (\bibinfo {year}
  {2017})}\BibitemShut {NoStop}%
\bibitem [{\citenamefont {Buisson}(1970)}]{Buisson1970}%
  \BibitemOpen
  \bibfield  {author} {\bibinfo {author} {\bibfnamefont {G.}~\bibnamefont
  {Buisson}},\ }\href {\doibase 10.1016/0022-3697(70)90326-4} {\bibfield
  {journal} {\bibinfo  {journal} {J. Phys. Chem. Solids}\ }\textbf {\bibinfo
  {volume} {31}},\ \bibinfo {pages} {1171} (\bibinfo {year}
  {1970})}\BibitemShut {NoStop}%
\bibitem [{\citenamefont {Hwang}\ \emph {et~al.}(2012)\citenamefont {Hwang},
  \citenamefont {Choi}, \citenamefont {Zhou}, \citenamefont {Lu},\ and\
  \citenamefont {Schlottmann}}]{Hwang2012}%
  \BibitemOpen
  \bibfield  {author} {\bibinfo {author} {\bibfnamefont {J.}~\bibnamefont
  {Hwang}}, \bibinfo {author} {\bibfnamefont {E.~S.}\ \bibnamefont {Choi}},
  \bibinfo {author} {\bibfnamefont {H.~D.}\ \bibnamefont {Zhou}}, \bibinfo
  {author} {\bibfnamefont {J.}~\bibnamefont {Lu}}, \ and\ \bibinfo {author}
  {\bibfnamefont {P.}~\bibnamefont {Schlottmann}},\ }\href {\doibase
  10.1103/PhysRevB.85.024415} {\bibfield  {journal} {\bibinfo  {journal} {Phys.
  Rev. B}\ }\textbf {\bibinfo {volume} {85}},\ \bibinfo {pages} {024415}
  (\bibinfo {year} {2012})}\BibitemShut {NoStop}%
\bibitem [{\citenamefont {Santoro}\ and\ \citenamefont
  {Newnham}(1967)}]{Santoro1967}%
  \BibitemOpen
  \bibfield  {author} {\bibinfo {author} {\bibfnamefont {R.~P.}\ \bibnamefont
  {Santoro}}\ and\ \bibinfo {author} {\bibfnamefont {R.~E.}\ \bibnamefont
  {Newnham}},\ }\href {\doibase 10.1107/S0365110X67000672} {\bibfield
  {journal} {\bibinfo  {journal} {Acta Crystallogr.}\ }\textbf {\bibinfo
  {volume} {22}},\ \bibinfo {pages} {344} (\bibinfo {year} {1967})}\BibitemShut
  {NoStop}%
\bibitem [{\citenamefont {Li}\ \emph {et~al.}(2006)\citenamefont {Li},
  \citenamefont {Garlea}, \citenamefont {Zarestky},\ and\ \citenamefont
  {Vaknin}}]{Li2006}%
  \BibitemOpen
  \bibfield  {author} {\bibinfo {author} {\bibfnamefont {J.}~\bibnamefont
  {Li}}, \bibinfo {author} {\bibfnamefont {V.~O.}\ \bibnamefont {Garlea}},
  \bibinfo {author} {\bibfnamefont {J.~L.}\ \bibnamefont {Zarestky}}, \ and\
  \bibinfo {author} {\bibfnamefont {D.}~\bibnamefont {Vaknin}},\ }\href
  {\doibase 10.1103/PhysRevB.73.024410} {\bibfield  {journal} {\bibinfo
  {journal} {Phys. Rev. B}\ }\textbf {\bibinfo {volume} {73}},\ \bibinfo
  {pages} {024410} (\bibinfo {year} {2006})}\BibitemShut {NoStop}%
\bibitem [{\citenamefont {Toft-Petersen}\ \emph {et~al.}(2015)\citenamefont
  {Toft-Petersen}, \citenamefont {Reehuis}, \citenamefont {Jensen},
  \citenamefont {Andersen}, \citenamefont {Li}, \citenamefont {Le},
  \citenamefont {Laver}, \citenamefont {Niedermayer}, \citenamefont {Klemke},
  \citenamefont {Lefmann},\ and\ \citenamefont {Vaknin}}]{Toft-Petersen2015}%
  \BibitemOpen
  \bibfield  {author} {\bibinfo {author} {\bibfnamefont {R.}~\bibnamefont
  {Toft-Petersen}}, \bibinfo {author} {\bibfnamefont {M.}~\bibnamefont
  {Reehuis}}, \bibinfo {author} {\bibfnamefont {T.~B.~S.}\ \bibnamefont
  {Jensen}}, \bibinfo {author} {\bibfnamefont {N.~H.}\ \bibnamefont
  {Andersen}}, \bibinfo {author} {\bibfnamefont {J.}~\bibnamefont {Li}},
  \bibinfo {author} {\bibfnamefont {M.~D.}\ \bibnamefont {Le}}, \bibinfo
  {author} {\bibfnamefont {M.}~\bibnamefont {Laver}}, \bibinfo {author}
  {\bibfnamefont {C.}~\bibnamefont {Niedermayer}}, \bibinfo {author}
  {\bibfnamefont {B.}~\bibnamefont {Klemke}}, \bibinfo {author} {\bibfnamefont
  {K.}~\bibnamefont {Lefmann}}, \ and\ \bibinfo {author} {\bibfnamefont
  {D.}~\bibnamefont {Vaknin}},\ }\href {\doibase 10.1103/PhysRevB.92.024404}
  {\bibfield  {journal} {\bibinfo  {journal} {Phys. Rev. B}\ }\textbf {\bibinfo
  {volume} {92}},\ \bibinfo {pages} {024404} (\bibinfo {year}
  {2015})}\BibitemShut {NoStop}%
\bibitem [{\citenamefont {Mays}(1963)}]{PhysRev.131.38}%
  \BibitemOpen
  \bibfield  {author} {\bibinfo {author} {\bibfnamefont {J.~M.}\ \bibnamefont
  {Mays}},\ }\href {\doibase 10.1103/PhysRev.131.38} {\bibfield  {journal}
  {\bibinfo  {journal} {Phys. Rev.}\ }\textbf {\bibinfo {volume} {131}},\
  \bibinfo {pages} {38} (\bibinfo {year} {1963})}\BibitemShut {NoStop}%
\bibitem [{\citenamefont {Toft-Petersen}\ \emph {et~al.}(2012)\citenamefont
  {Toft-Petersen}, \citenamefont {Andersen}, \citenamefont {Li}, \citenamefont
  {Li}, \citenamefont {Tian}, \citenamefont {Bud'ko}, \citenamefont {Jensen},
  \citenamefont {Niedermayer}, \citenamefont {Laver}, \citenamefont {Zaharko},
  \citenamefont {Lynn},\ and\ \citenamefont {Vaknin}}]{Toft-Petersen2012}%
  \BibitemOpen
  \bibfield  {author} {\bibinfo {author} {\bibfnamefont {R.}~\bibnamefont
  {Toft-Petersen}}, \bibinfo {author} {\bibfnamefont {N.~H.}\ \bibnamefont
  {Andersen}}, \bibinfo {author} {\bibfnamefont {H.}~\bibnamefont {Li}},
  \bibinfo {author} {\bibfnamefont {J.}~\bibnamefont {Li}}, \bibinfo {author}
  {\bibfnamefont {W.}~\bibnamefont {Tian}}, \bibinfo {author} {\bibfnamefont
  {S.~L.}\ \bibnamefont {Bud'ko}}, \bibinfo {author} {\bibfnamefont {T.~B.~S.}\
  \bibnamefont {Jensen}}, \bibinfo {author} {\bibfnamefont {C.}~\bibnamefont
  {Niedermayer}}, \bibinfo {author} {\bibfnamefont {M.}~\bibnamefont {Laver}},
  \bibinfo {author} {\bibfnamefont {O.}~\bibnamefont {Zaharko}}, \bibinfo
  {author} {\bibfnamefont {J.~W.}\ \bibnamefont {Lynn}}, \ and\ \bibinfo
  {author} {\bibfnamefont {D.}~\bibnamefont {Vaknin}},\ }\href {\doibase
  10.1103/PhysRevB.85.224415} {\bibfield  {journal} {\bibinfo  {journal} {Phys.
  Rev. B}\ }\textbf {\bibinfo {volume} {85}},\ \bibinfo {pages} {224415}
  (\bibinfo {year} {2012})}\BibitemShut {NoStop}%
\bibitem [{\citenamefont {Kornev}\ \emph {et~al.}(2000)\citenamefont {Kornev},
  \citenamefont {Bichurin}, \citenamefont {Rivera}, \citenamefont {Gentil},
  \citenamefont {Schmid}, \citenamefont {Jansen},\ and\ \citenamefont
  {Wyder}}]{Kornev2000a}%
  \BibitemOpen
  \bibfield  {author} {\bibinfo {author} {\bibfnamefont {I.}~\bibnamefont
  {Kornev}}, \bibinfo {author} {\bibfnamefont {M.}~\bibnamefont {Bichurin}},
  \bibinfo {author} {\bibfnamefont {J.-P.}\ \bibnamefont {Rivera}}, \bibinfo
  {author} {\bibfnamefont {S.}~\bibnamefont {Gentil}}, \bibinfo {author}
  {\bibfnamefont {H.}~\bibnamefont {Schmid}}, \bibinfo {author} {\bibfnamefont
  {A.~G.~M.}\ \bibnamefont {Jansen}}, \ and\ \bibinfo {author} {\bibfnamefont
  {P.}~\bibnamefont {Wyder}},\ }\href {\doibase 10.1103/PhysRevB.62.12247}
  {\bibfield  {journal} {\bibinfo  {journal} {Phys. Rev. B}\ }\textbf {\bibinfo
  {volume} {62}},\ \bibinfo {pages} {12247} (\bibinfo {year}
  {2000})}\BibitemShut {NoStop}%
\bibitem [{\citenamefont {Fogh}\ \emph {et~al.}(2017)\citenamefont {Fogh},
  \citenamefont {Toft-Petersen}, \citenamefont {Ressouche}, \citenamefont
  {Niedermayer}, \citenamefont {Holm}, \citenamefont {Bartkowiak},
  \citenamefont {Prokhnenko}, \citenamefont {Sloth}, \citenamefont {Isaksen},
  \citenamefont {Vaknin},\ and\ \citenamefont {Christensen}}]{Fogh2017}%
  \BibitemOpen
  \bibfield  {author} {\bibinfo {author} {\bibfnamefont {E.}~\bibnamefont
  {Fogh}}, \bibinfo {author} {\bibfnamefont {R.}~\bibnamefont {Toft-Petersen}},
  \bibinfo {author} {\bibfnamefont {E.}~\bibnamefont {Ressouche}}, \bibinfo
  {author} {\bibfnamefont {C.}~\bibnamefont {Niedermayer}}, \bibinfo {author}
  {\bibfnamefont {S.~L.}\ \bibnamefont {Holm}}, \bibinfo {author}
  {\bibfnamefont {M.}~\bibnamefont {Bartkowiak}}, \bibinfo {author}
  {\bibfnamefont {O.}~\bibnamefont {Prokhnenko}}, \bibinfo {author}
  {\bibfnamefont {S.}~\bibnamefont {Sloth}}, \bibinfo {author} {\bibfnamefont
  {F.~W.}\ \bibnamefont {Isaksen}}, \bibinfo {author} {\bibfnamefont
  {D.}~\bibnamefont {Vaknin}}, \ and\ \bibinfo {author} {\bibfnamefont {N.~B.}\
  \bibnamefont {Christensen}},\ }\href {\doibase 10.1103/PhysRevB.96.104420}
  {\bibfield  {journal} {\bibinfo  {journal} {Phys. Rev. B}\ }\textbf {\bibinfo
  {volume} {96}},\ \bibinfo {pages} {104420} (\bibinfo {year}
  {2017})}\BibitemShut {NoStop}%
\bibitem [{\citenamefont {Santoro}\ \emph {et~al.}(1966)\citenamefont
  {Santoro}, \citenamefont {Segal},\ and\ \citenamefont
  {Newnham}}]{santoro1966magnetic}%
  \BibitemOpen
  \bibfield  {author} {\bibinfo {author} {\bibfnamefont {R.}~\bibnamefont
  {Santoro}}, \bibinfo {author} {\bibfnamefont {D.}~\bibnamefont {Segal}}, \
  and\ \bibinfo {author} {\bibfnamefont {R.}~\bibnamefont {Newnham}},\ }\href
  {http://www.sciencedirect.com/science/article/pii/0022369766900977}
  {\bibfield  {journal} {\bibinfo  {journal} {J. Phys. Chem. Solids}\ }\textbf
  {\bibinfo {volume} {27}},\ \bibinfo {pages} {1192 } (\bibinfo {year}
  {1966})}\BibitemShut {NoStop}%
\bibitem [{\citenamefont {Vaknin}\ \emph {et~al.}(2002)\citenamefont {Vaknin},
  \citenamefont {Zarestky}, \citenamefont {Miller}, \citenamefont {Rivera},\
  and\ \citenamefont {Schmid}}]{Vaknin2002}%
  \BibitemOpen
  \bibfield  {author} {\bibinfo {author} {\bibfnamefont {D.}~\bibnamefont
  {Vaknin}}, \bibinfo {author} {\bibfnamefont {J.~L.}\ \bibnamefont
  {Zarestky}}, \bibinfo {author} {\bibfnamefont {L.~L.}\ \bibnamefont
  {Miller}}, \bibinfo {author} {\bibfnamefont {J.-P.}\ \bibnamefont {Rivera}},
  \ and\ \bibinfo {author} {\bibfnamefont {H.}~\bibnamefont {Schmid}},\ }\href
  {\doibase 10.1103/PhysRevB.65.224414} {\bibfield  {journal} {\bibinfo
  {journal} {Phys. Rev. B}\ }\textbf {\bibinfo {volume} {65}},\ \bibinfo
  {pages} {224414} (\bibinfo {year} {2002})}\BibitemShut {NoStop}%
\bibitem [{\citenamefont {Reynaud}\ \emph {et~al.}(2014)\citenamefont
  {Reynaud}, \citenamefont {Rodr{\'{i}}guez-Carvajal}, \citenamefont {Chotard},
  \citenamefont {Tarascon},\ and\ \citenamefont {Rousse}}]{Reynaud2014}%
  \BibitemOpen
  \bibfield  {author} {\bibinfo {author} {\bibfnamefont {M.}~\bibnamefont
  {Reynaud}}, \bibinfo {author} {\bibfnamefont {J.}~\bibnamefont
  {Rodr{\'{i}}guez-Carvajal}}, \bibinfo {author} {\bibfnamefont {J.-N.}\
  \bibnamefont {Chotard}}, \bibinfo {author} {\bibfnamefont {J.-M.}\
  \bibnamefont {Tarascon}}, \ and\ \bibinfo {author} {\bibfnamefont
  {G.}~\bibnamefont {Rousse}},\ }\href {\doibase 10.1103/PhysRevB.89.104419}
  {\bibfield  {journal} {\bibinfo  {journal} {Phys. Rev. B}\ }\textbf {\bibinfo
  {volume} {89}},\ \bibinfo {pages} {104419} (\bibinfo {year}
  {2014})}\BibitemShut {NoStop}%
\bibitem [{\citenamefont {L{\'{o}}pez}\ \emph {et~al.}(2008)\citenamefont
  {L{\'{o}}pez}, \citenamefont {Daidouh}, \citenamefont {Pico}, \citenamefont
  {Rodr{\'{i}}guez-Carvajal},\ and\ \citenamefont {Veiga}}]{Lopez2008}%
  \BibitemOpen
  \bibfield  {author} {\bibinfo {author} {\bibfnamefont {M.~L.}\ \bibnamefont
  {L{\'{o}}pez}}, \bibinfo {author} {\bibfnamefont {A.}~\bibnamefont
  {Daidouh}}, \bibinfo {author} {\bibfnamefont {C.}~\bibnamefont {Pico}},
  \bibinfo {author} {\bibfnamefont {J.}~\bibnamefont
  {Rodr{\'{i}}guez-Carvajal}}, \ and\ \bibinfo {author} {\bibfnamefont {M.~L.}\
  \bibnamefont {Veiga}},\ }\href {\doibase 10.1002/chem.200800763} {\bibfield
  {journal} {\bibinfo  {journal} {Chem. - A Eur. J.}\ }\textbf {\bibinfo
  {volume} {14}},\ \bibinfo {pages} {10829} (\bibinfo {year}
  {2008})}\BibitemShut {NoStop}%
\bibitem [{\citenamefont {Avdeev}\ \emph {et~al.}(2013)\citenamefont {Avdeev},
  \citenamefont {Mohamed}, \citenamefont {Ling}, \citenamefont {Lu},
  \citenamefont {Tamaru}, \citenamefont {Yamada},\ and\ \citenamefont
  {Barpanda}}]{Avdeev2013}%
  \BibitemOpen
  \bibfield  {author} {\bibinfo {author} {\bibfnamefont {M.}~\bibnamefont
  {Avdeev}}, \bibinfo {author} {\bibfnamefont {Z.}~\bibnamefont {Mohamed}},
  \bibinfo {author} {\bibfnamefont {C.~D.}\ \bibnamefont {Ling}}, \bibinfo
  {author} {\bibfnamefont {J.}~\bibnamefont {Lu}}, \bibinfo {author}
  {\bibfnamefont {M.}~\bibnamefont {Tamaru}}, \bibinfo {author} {\bibfnamefont
  {A.}~\bibnamefont {Yamada}}, \ and\ \bibinfo {author} {\bibfnamefont
  {P.}~\bibnamefont {Barpanda}},\ }\href {\doibase 10.1021/ic400870x}
  {\bibfield  {journal} {\bibinfo  {journal} {Inorg. Chem.}\ }\textbf {\bibinfo
  {volume} {52}},\ \bibinfo {pages} {8685} (\bibinfo {year}
  {2013})}\BibitemShut {NoStop}%
\bibitem [{\citenamefont {Tan}\ \emph {et~al.}(2005)\citenamefont {Tan},
  \citenamefont {Kreyssig}, \citenamefont {Kim}, \citenamefont {Goldman},
  \citenamefont {McQueeney}, \citenamefont {Wermeille}, \citenamefont {Sieve},
  \citenamefont {Lograsso}, \citenamefont {Schlagel}, \citenamefont {Budko},
  \citenamefont {Pecharsky},\ and\ \citenamefont {Gschneidner}}]{Tan2005}%
  \BibitemOpen
  \bibfield  {author} {\bibinfo {author} {\bibfnamefont {L.}~\bibnamefont
  {Tan}}, \bibinfo {author} {\bibfnamefont {A.}~\bibnamefont {Kreyssig}},
  \bibinfo {author} {\bibfnamefont {J.~W.}\ \bibnamefont {Kim}}, \bibinfo
  {author} {\bibfnamefont {A.~I.}\ \bibnamefont {Goldman}}, \bibinfo {author}
  {\bibfnamefont {R.~J.}\ \bibnamefont {McQueeney}}, \bibinfo {author}
  {\bibfnamefont {D.}~\bibnamefont {Wermeille}}, \bibinfo {author}
  {\bibfnamefont {B.}~\bibnamefont {Sieve}}, \bibinfo {author} {\bibfnamefont
  {T.~A.}\ \bibnamefont {Lograsso}}, \bibinfo {author} {\bibfnamefont {D.~L.}\
  \bibnamefont {Schlagel}}, \bibinfo {author} {\bibfnamefont {S.~L.}\
  \bibnamefont {Budko}}, \bibinfo {author} {\bibfnamefont {V.~K.}\ \bibnamefont
  {Pecharsky}}, \ and\ \bibinfo {author} {\bibfnamefont {K.~A.}\ \bibnamefont
  {Gschneidner}},\ }\href {\doibase 10.1103/PhysRevB.71.214408} {\bibfield
  {journal} {\bibinfo  {journal} {Phys. Rev. B}\ }\textbf {\bibinfo {volume}
  {71}},\ \bibinfo {pages} {214408} (\bibinfo {year} {2005})}\BibitemShut
  {NoStop}%
\bibitem [{\citenamefont {Levin}\ \emph {et~al.}(2001)\citenamefont {Levin},
  \citenamefont {Pecharsky}, \citenamefont {Gschneidner},\ and\ \citenamefont
  {Miller}}]{Levin2001}%
  \BibitemOpen
  \bibfield  {author} {\bibinfo {author} {\bibfnamefont {E.~M.}\ \bibnamefont
  {Levin}}, \bibinfo {author} {\bibfnamefont {V.~K.}\ \bibnamefont
  {Pecharsky}}, \bibinfo {author} {\bibfnamefont {K.~A.}\ \bibnamefont
  {Gschneidner}}, \ and\ \bibinfo {author} {\bibfnamefont {G.~J.}\ \bibnamefont
  {Miller}},\ }\href {\doibase 10.1103/PhysRevB.64.235103} {\bibfield
  {journal} {\bibinfo  {journal} {Phys. Rev. B}\ }\textbf {\bibinfo {volume}
  {64}},\ \bibinfo {pages} {235103} (\bibinfo {year} {2001})}\BibitemShut
  {NoStop}%
\bibitem [{\citenamefont {Avdeev}\ \emph {et~al.}(2014)\citenamefont {Avdeev},
  \citenamefont {Kennedy},\ and\ \citenamefont {Kolodiazhnyi}}]{Avdeev2014}%
  \BibitemOpen
  \bibfield  {author} {\bibinfo {author} {\bibfnamefont {M.}~\bibnamefont
  {Avdeev}}, \bibinfo {author} {\bibfnamefont {B.~J.}\ \bibnamefont {Kennedy}},
  \ and\ \bibinfo {author} {\bibfnamefont {T.}~\bibnamefont {Kolodiazhnyi}},\
  }\href {\doibase 10.1088/0953-8984/26/9/095401} {\bibfield  {journal}
  {\bibinfo  {journal} {J. Phys. Condens. Matter}\ }\textbf {\bibinfo {volume}
  {26}},\ \bibinfo {pages} {095401} (\bibinfo {year} {2014})}\BibitemShut
  {NoStop}%
\bibitem [{\citenamefont {Saha}\ \emph
  {et~al.}(2016{\natexlab{b}})\citenamefont {Saha}, \citenamefont {Sundaresan},
  \citenamefont {Sanyal}, \citenamefont {Rao}, \citenamefont {Orlandi},
  \citenamefont {Manuel},\ and\ \citenamefont {Langridge}}]{Saha2016c}%
  \BibitemOpen
  \bibfield  {author} {\bibinfo {author} {\bibfnamefont {R.}~\bibnamefont
  {Saha}}, \bibinfo {author} {\bibfnamefont {A.}~\bibnamefont {Sundaresan}},
  \bibinfo {author} {\bibfnamefont {M.~K.}\ \bibnamefont {Sanyal}}, \bibinfo
  {author} {\bibfnamefont {C.~N.~R.}\ \bibnamefont {Rao}}, \bibinfo {author}
  {\bibfnamefont {F.}~\bibnamefont {Orlandi}}, \bibinfo {author} {\bibfnamefont
  {P.}~\bibnamefont {Manuel}}, \ and\ \bibinfo {author} {\bibfnamefont
  {S.}~\bibnamefont {Langridge}},\ }\href {\doibase 10.1103/PhysRevB.93.014409}
  {\bibfield  {journal} {\bibinfo  {journal} {Phys. Rev. B}\ }\textbf {\bibinfo
  {volume} {93}},\ \bibinfo {pages} {014409} (\bibinfo {year}
  {2016}{\natexlab{b}})}\BibitemShut {NoStop}%
\bibitem [{\citenamefont {Bidaux}\ and\ \citenamefont
  {M{\'e}riel}(1968)}]{bidaux1968etude}%
  \BibitemOpen
  \bibfield  {author} {\bibinfo {author} {\bibfnamefont {R.}~\bibnamefont
  {Bidaux}}\ and\ \bibinfo {author} {\bibfnamefont {P.}~\bibnamefont
  {M{\'e}riel}},\ }\href@noop {} {\bibfield  {journal} {\bibinfo  {journal}
  {Journal de Physique}\ }\textbf {\bibinfo {volume} {29}},\ \bibinfo {pages}
  {220} (\bibinfo {year} {1968})}\BibitemShut {NoStop}%
\bibitem [{\citenamefont {Kn{\'{i}}{\v{z}}ek}\ \emph
  {et~al.}(2014)\citenamefont {Kn{\'{i}}{\v{z}}ek}, \citenamefont
  {Jir{\'{a}}k}, \citenamefont {Nov{\'{a}}k},\ and\ \citenamefont {de~la
  Cruz}}]{Knizek2014}%
  \BibitemOpen
  \bibfield  {author} {\bibinfo {author} {\bibfnamefont {K.}~\bibnamefont
  {Kn{\'{i}}{\v{z}}ek}}, \bibinfo {author} {\bibfnamefont {Z.}~\bibnamefont
  {Jir{\'{a}}k}}, \bibinfo {author} {\bibfnamefont {P.}~\bibnamefont
  {Nov{\'{a}}k}}, \ and\ \bibinfo {author} {\bibfnamefont {C.}~\bibnamefont
  {de~la Cruz}},\ }\href {\doibase 10.1016/j.solidstatesciences.2013.12.001}
  {\bibfield  {journal} {\bibinfo  {journal} {Solid State Sci.}\ }\textbf
  {\bibinfo {volume} {28}},\ \bibinfo {pages} {26} (\bibinfo {year}
  {2014})}\BibitemShut {NoStop}%
\bibitem [{\citenamefont {Mareschal}\ \emph {et~al.}(1968)\citenamefont
  {Mareschal}, \citenamefont {Sivardi{\`{e}}re}, \citenamefont {{De Vries}},\
  and\ \citenamefont {Bertaut}}]{Mareschal1968}%
  \BibitemOpen
  \bibfield  {author} {\bibinfo {author} {\bibfnamefont {J.}~\bibnamefont
  {Mareschal}}, \bibinfo {author} {\bibfnamefont {J.}~\bibnamefont
  {Sivardi{\`{e}}re}}, \bibinfo {author} {\bibfnamefont {G.~F.}\ \bibnamefont
  {{De Vries}}}, \ and\ \bibinfo {author} {\bibfnamefont {E.~F.}\ \bibnamefont
  {Bertaut}},\ }\href {\doibase 10.1063/1.1656306} {\bibfield  {journal}
  {\bibinfo  {journal} {J. Appl. Phys.}\ }\textbf {\bibinfo {volume} {39}},\
  \bibinfo {pages} {1364} (\bibinfo {year} {1968})}\BibitemShut {NoStop}%
\bibitem [{\citenamefont {Mu{\~{n}}oz}\ \emph {et~al.}(2012)\citenamefont
  {Mu{\~{n}}oz}, \citenamefont {Mart{\'{i}}nez-Lope}, \citenamefont {Alonso},\
  and\ \citenamefont {Fern{\'{a}}ndez-D{\'{i}}az}}]{Munoz2012}%
  \BibitemOpen
  \bibfield  {author} {\bibinfo {author} {\bibfnamefont {A.}~\bibnamefont
  {Mu{\~{n}}oz}}, \bibinfo {author} {\bibfnamefont {M.~J.}\ \bibnamefont
  {Mart{\'{i}}nez-Lope}}, \bibinfo {author} {\bibfnamefont {J.~A.}\
  \bibnamefont {Alonso}}, \ and\ \bibinfo {author} {\bibfnamefont {M.~T.}\
  \bibnamefont {Fern{\'{a}}ndez-D{\'{i}}az}},\ }\href {\doibase
  10.1002/ejic.201200811} {\bibfield  {journal} {\bibinfo  {journal} {Eur. J.
  Inorg. Chem.}\ }\textbf {\bibinfo {volume} {2012}},\ \bibinfo {pages} {5825}
  (\bibinfo {year} {2012})}\BibitemShut {NoStop}%
\bibitem [{\citenamefont {Cooke}\ \emph {et~al.}(1976)\citenamefont {Cooke},
  \citenamefont {England}, \citenamefont {Preston}, \citenamefont {Swithenby},\
  and\ \citenamefont {Wells}}]{Cooke1976}%
  \BibitemOpen
  \bibfield  {author} {\bibinfo {author} {\bibfnamefont {A.~H.}\ \bibnamefont
  {Cooke}}, \bibinfo {author} {\bibfnamefont {N.~J.}\ \bibnamefont {England}},
  \bibinfo {author} {\bibfnamefont {N.~F.}\ \bibnamefont {Preston}}, \bibinfo
  {author} {\bibfnamefont {S.~J.}\ \bibnamefont {Swithenby}}, \ and\ \bibinfo
  {author} {\bibfnamefont {M.~R.}\ \bibnamefont {Wells}},\ }\href {\doibase
  10.1016/0038-1098(76)90336-7} {\bibfield  {journal} {\bibinfo  {journal}
  {Solid State Commun.}\ }\textbf {\bibinfo {volume} {18}},\ \bibinfo {pages}
  {545} (\bibinfo {year} {1976})}\BibitemShut {NoStop}%
\bibitem [{\citenamefont {Quezel}\ \emph {et~al.}(1982)\citenamefont {Quezel},
  \citenamefont {Rossat-Mignod},\ and\ \citenamefont {Tcheou}}]{Quezel1982}%
  \BibitemOpen
  \bibfield  {author} {\bibinfo {author} {\bibfnamefont {S.}~\bibnamefont
  {Quezel}}, \bibinfo {author} {\bibfnamefont {J.}~\bibnamefont
  {Rossat-Mignod}}, \ and\ \bibinfo {author} {\bibfnamefont {F.}~\bibnamefont
  {Tcheou}},\ }\href@noop {} {\bibfield  {journal} {\bibinfo  {journal} {Solid
  State Commun.}\ }\textbf {\bibinfo {volume} {42}},\ \bibinfo {pages} {103}
  (\bibinfo {year} {1982})}\BibitemShut {NoStop}%
\bibitem [{\citenamefont {Nielsen}\ \emph {et~al.}(1976)\citenamefont
  {Nielsen}, \citenamefont {Lebech}, \citenamefont {Larsen}, \citenamefont
  {Holmes},\ and\ \citenamefont {Ballman}}]{Jacobson1975}%
  \BibitemOpen
  \bibfield  {author} {\bibinfo {author} {\bibfnamefont {O.~V.}\ \bibnamefont
  {Nielsen}}, \bibinfo {author} {\bibfnamefont {B.}~\bibnamefont {Lebech}},
  \bibinfo {author} {\bibfnamefont {F.~K.}\ \bibnamefont {Larsen}}, \bibinfo
  {author} {\bibfnamefont {L.~M.}\ \bibnamefont {Holmes}}, \ and\ \bibinfo
  {author} {\bibfnamefont {A.~A.}\ \bibnamefont {Ballman}},\ }\href {\doibase
  10.1088/0022-3719/9/12/023} {\bibfield  {journal} {\bibinfo  {journal} {J.
  Phys. C Solid State Phys.}\ }\textbf {\bibinfo {volume} {9}},\ \bibinfo
  {pages} {2401} (\bibinfo {year} {1976})}\BibitemShut {NoStop}%
\bibitem [{\citenamefont {Melot}\ \emph {et~al.}(2010)\citenamefont {Melot},
  \citenamefont {Paden}, \citenamefont {Seshadri}, \citenamefont {Suard},
  \citenamefont {N{\'{e}}nert}, \citenamefont {Dixit},\ and\ \citenamefont
  {Lawes}}]{Melot2010b}%
  \BibitemOpen
  \bibfield  {author} {\bibinfo {author} {\bibfnamefont {B.~C.}\ \bibnamefont
  {Melot}}, \bibinfo {author} {\bibfnamefont {B.}~\bibnamefont {Paden}},
  \bibinfo {author} {\bibfnamefont {R.}~\bibnamefont {Seshadri}}, \bibinfo
  {author} {\bibfnamefont {E.}~\bibnamefont {Suard}}, \bibinfo {author}
  {\bibfnamefont {G.}~\bibnamefont {N{\'{e}}nert}}, \bibinfo {author}
  {\bibfnamefont {A.}~\bibnamefont {Dixit}}, \ and\ \bibinfo {author}
  {\bibfnamefont {G.}~\bibnamefont {Lawes}},\ }\href {\doibase
  10.1103/PhysRevB.82.014411} {\bibfield  {journal} {\bibinfo  {journal} {Phys.
  Rev. B}\ }\textbf {\bibinfo {volume} {82}},\ \bibinfo {pages} {014411}
  (\bibinfo {year} {2010})}\BibitemShut {NoStop}%
\bibitem [{\citenamefont {Schobinger-Papamantellos}\ \emph
  {et~al.}(1988)\citenamefont {Schobinger-Papamantellos}, \citenamefont {{De
  Mooij}},\ and\ \citenamefont {Buschow}}]{Schobinger-Papamantellos1988}%
  \BibitemOpen
  \bibfield  {author} {\bibinfo {author} {\bibfnamefont {P.}~\bibnamefont
  {Schobinger-Papamantellos}}, \bibinfo {author} {\bibfnamefont
  {D.}~\bibnamefont {{De Mooij}}}, \ and\ \bibinfo {author} {\bibfnamefont
  {K.}~\bibnamefont {Buschow}},\ }\href {\doibase 10.1016/0022-5088(88)90140-3}
  {\bibfield  {journal} {\bibinfo  {journal} {J. Less Common Met.}\ }\textbf
  {\bibinfo {volume} {144}},\ \bibinfo {pages} {265} (\bibinfo {year}
  {1988})}\BibitemShut {NoStop}%
\bibitem [{\citenamefont {Bonnet}\ \emph {et~al.}(1994)\citenamefont {Bonnet},
  \citenamefont {Boucherle}, \citenamefont {Givord}, \citenamefont {Lapierre},
  \citenamefont {Lejay}, \citenamefont {Odin}, \citenamefont {Murani},
  \citenamefont {Schweizer},\ and\ \citenamefont {Stunault}}]{Bonnet1994}%
  \BibitemOpen
  \bibfield  {author} {\bibinfo {author} {\bibfnamefont {M.}~\bibnamefont
  {Bonnet}}, \bibinfo {author} {\bibfnamefont {J.~X.}\ \bibnamefont
  {Boucherle}}, \bibinfo {author} {\bibfnamefont {F.}~\bibnamefont {Givord}},
  \bibinfo {author} {\bibfnamefont {F.}~\bibnamefont {Lapierre}}, \bibinfo
  {author} {\bibfnamefont {P.}~\bibnamefont {Lejay}}, \bibinfo {author}
  {\bibfnamefont {J.}~\bibnamefont {Odin}}, \bibinfo {author} {\bibfnamefont
  {A.~P.}\ \bibnamefont {Murani}}, \bibinfo {author} {\bibfnamefont
  {J.}~\bibnamefont {Schweizer}}, \ and\ \bibinfo {author} {\bibfnamefont
  {A.}~\bibnamefont {Stunault}},\ }\href {\doibase
  http://dx.doi.org/10.1016/0304-8853(94)90324-7} {\bibfield  {journal}
  {\bibinfo  {journal} {J. Magn. Magn. Mater.}\ }\textbf {\bibinfo {volume}
  {132}},\ \bibinfo {pages} {289} (\bibinfo {year} {1994})}\BibitemShut
  {NoStop}%
\bibitem [{\citenamefont {Givord}\ \emph {et~al.}(1989)\citenamefont {Givord},
  \citenamefont {Lejay}, \citenamefont {Ressouche}, \citenamefont {Schweizer},\
  and\ \citenamefont {Stunault}}]{Givord1989}%
  \BibitemOpen
  \bibfield  {author} {\bibinfo {author} {\bibfnamefont {F.}~\bibnamefont
  {Givord}}, \bibinfo {author} {\bibfnamefont {P.}~\bibnamefont {Lejay}},
  \bibinfo {author} {\bibfnamefont {E.}~\bibnamefont {Ressouche}}, \bibinfo
  {author} {\bibfnamefont {J.}~\bibnamefont {Schweizer}}, \ and\ \bibinfo
  {author} {\bibfnamefont {A.}~\bibnamefont {Stunault}},\ }\href {\doibase
  10.1016/0921-4526(89)90799-0} {\bibfield  {journal} {\bibinfo  {journal}
  {Phys. B Condens. Matter}\ }\textbf {\bibinfo {volume} {156-157}},\ \bibinfo
  {pages} {805} (\bibinfo {year} {1989})}\BibitemShut {NoStop}%
\bibitem [{\citenamefont {Voyer}\ \emph {et~al.}(2007)\citenamefont {Voyer},
  \citenamefont {Ryan}, \citenamefont {Cadogan}, \citenamefont {Cranswick},
  \citenamefont {Napoletano}, \citenamefont {Riani},\ and\ \citenamefont
  {Canepa}}]{Voyer2007}%
  \BibitemOpen
  \bibfield  {author} {\bibinfo {author} {\bibfnamefont {C.~J.}\ \bibnamefont
  {Voyer}}, \bibinfo {author} {\bibfnamefont {D.~H.}\ \bibnamefont {Ryan}},
  \bibinfo {author} {\bibfnamefont {J.~M.}\ \bibnamefont {Cadogan}}, \bibinfo
  {author} {\bibfnamefont {L.~M.~D.}\ \bibnamefont {Cranswick}}, \bibinfo
  {author} {\bibfnamefont {M.}~\bibnamefont {Napoletano}}, \bibinfo {author}
  {\bibfnamefont {P.}~\bibnamefont {Riani}}, \ and\ \bibinfo {author}
  {\bibfnamefont {F.}~\bibnamefont {Canepa}},\ }\href {\doibase
  10.1088/0953-8984/19/43/436205} {\bibfield  {journal} {\bibinfo  {journal}
  {J. Phys. Condens. Matter}\ }\textbf {\bibinfo {volume} {19}},\ \bibinfo
  {pages} {436205} (\bibinfo {year} {2007})}\BibitemShut {NoStop}%
\bibitem [{\citenamefont {Tran}\ \emph {et~al.}(2001)\citenamefont {Tran},
  \citenamefont {Kaczorowski}, \citenamefont {Tro{\'{c}}}, \citenamefont
  {Andr{\'{e}}}, \citenamefont {Bour{\'{e}}e},\ and\ \citenamefont
  {Zaremba}}]{Tran2001}%
  \BibitemOpen
  \bibfield  {author} {\bibinfo {author} {\bibfnamefont {V.~H.}\ \bibnamefont
  {Tran}}, \bibinfo {author} {\bibfnamefont {D.}~\bibnamefont {Kaczorowski}},
  \bibinfo {author} {\bibfnamefont {R.}~\bibnamefont {Tro{\'{c}}}}, \bibinfo
  {author} {\bibfnamefont {G.}~\bibnamefont {Andr{\'{e}}}}, \bibinfo {author}
  {\bibfnamefont {F.}~\bibnamefont {Bour{\'{e}}e}}, \ and\ \bibinfo {author}
  {\bibfnamefont {V.}~\bibnamefont {Zaremba}},\ }\href {\doibase
  10.1016/S0038-1098(00)00520-2} {\bibfield  {journal} {\bibinfo  {journal}
  {Solid State Commun.}\ }\textbf {\bibinfo {volume} {117}},\ \bibinfo {pages}
  {527} (\bibinfo {year} {2001})}\BibitemShut {NoStop}%
\bibitem [{\citenamefont {Tomkowicz}\ and\ \citenamefont
  {Szytuea}(1977)}]{Tomkowicz1977a}%
  \BibitemOpen
  \bibfield  {author} {\bibinfo {author} {\bibfnamefont {Z.}~\bibnamefont
  {Tomkowicz}}\ and\ \bibinfo {author} {\bibfnamefont {a.}~\bibnamefont
  {Szytuea}},\ }\href {\doibase 10.1016/0022-3697(77)90037-3} {\bibfield
  {journal} {\bibinfo  {journal} {J. Phys. Chem. Solids}\ }\textbf {\bibinfo
  {volume} {38}},\ \bibinfo {pages} {1117} (\bibinfo {year}
  {1977})}\BibitemShut {NoStop}%
\bibitem [{\citenamefont {Redhammer}\ \emph {et~al.}(2010)\citenamefont
  {Redhammer}, \citenamefont {Senyshyn}, \citenamefont {Tippelt}, \citenamefont
  {Pietzonka}, \citenamefont {Roth},\ and\ \citenamefont
  {Amthauer}}]{Redhammer2010a}%
  \BibitemOpen
  \bibfield  {author} {\bibinfo {author} {\bibfnamefont {G.~J.}\ \bibnamefont
  {Redhammer}}, \bibinfo {author} {\bibfnamefont {A.}~\bibnamefont {Senyshyn}},
  \bibinfo {author} {\bibfnamefont {G.}~\bibnamefont {Tippelt}}, \bibinfo
  {author} {\bibfnamefont {C.}~\bibnamefont {Pietzonka}}, \bibinfo {author}
  {\bibfnamefont {G.}~\bibnamefont {Roth}}, \ and\ \bibinfo {author}
  {\bibfnamefont {G.}~\bibnamefont {Amthauer}},\ }\href {\doibase
  10.1007/s00269-009-0335-x} {\bibfield  {journal} {\bibinfo  {journal} {Phys.
  Chem. Miner.}\ }\textbf {\bibinfo {volume} {37}},\ \bibinfo {pages} {311}
  (\bibinfo {year} {2010})}\BibitemShut {NoStop}%
\bibitem [{\citenamefont {Will}\ and\ \citenamefont
  {Schafer}(1971)}]{Search1971}%
  \BibitemOpen
  \bibfield  {author} {\bibinfo {author} {\bibfnamefont {G.}~\bibnamefont
  {Will}}\ and\ \bibinfo {author} {\bibfnamefont {W.}~\bibnamefont {Schafer}},\
  }\href {\doibase 10.1088/0022-3719/4/7/005} {\bibfield  {journal} {\bibinfo
  {journal} {J. Phys. C Solid State Phys.}\ }\textbf {\bibinfo {volume} {4}},\
  \bibinfo {pages} {811} (\bibinfo {year} {1971})}\BibitemShut {NoStop}%
\bibitem [{\citenamefont {Kishimoto}\ \emph {et~al.}(2010)\citenamefont
  {Kishimoto}, \citenamefont {Ishikura}, \citenamefont {Nakamura},
  \citenamefont {Wakabayashi},\ and\ \citenamefont {Kimura}}]{Kishimoto2010}%
  \BibitemOpen
  \bibfield  {author} {\bibinfo {author} {\bibfnamefont {K.}~\bibnamefont
  {Kishimoto}}, \bibinfo {author} {\bibfnamefont {T.}~\bibnamefont {Ishikura}},
  \bibinfo {author} {\bibfnamefont {H.}~\bibnamefont {Nakamura}}, \bibinfo
  {author} {\bibfnamefont {Y.}~\bibnamefont {Wakabayashi}}, \ and\ \bibinfo
  {author} {\bibfnamefont {T.}~\bibnamefont {Kimura}},\ }\href {\doibase
  10.1103/PhysRevB.82.012103} {\bibfield  {journal} {\bibinfo  {journal} {Phys.
  Rev. B}\ }\textbf {\bibinfo {volume} {82}},\ \bibinfo {pages} {012103}
  (\bibinfo {year} {2010})}\BibitemShut {NoStop}%
\bibitem [{yba()}]{ybalb4}%
  \BibitemOpen
  \href@noop {} {}\bibinfo {note} {S.~Suzuki and S.~Nakatsuji, private
  communication.}\BibitemShut {Stop}%
\bibitem [{\citenamefont {Ivanov}\ \emph {et~al.}(2012)\citenamefont {Ivanov},
  \citenamefont {Tellgren}, \citenamefont {Ritter}, \citenamefont {Nordblad},
  \citenamefont {Mathieu}, \citenamefont {Andr{\'{e}}}, \citenamefont
  {Golubko}, \citenamefont {Politova},\ and\ \citenamefont
  {Weil}}]{Ivanov2012}%
  \BibitemOpen
  \bibfield  {author} {\bibinfo {author} {\bibfnamefont {S.}~\bibnamefont
  {Ivanov}}, \bibinfo {author} {\bibfnamefont {R.}~\bibnamefont {Tellgren}},
  \bibinfo {author} {\bibfnamefont {C.}~\bibnamefont {Ritter}}, \bibinfo
  {author} {\bibfnamefont {P.}~\bibnamefont {Nordblad}}, \bibinfo {author}
  {\bibfnamefont {R.}~\bibnamefont {Mathieu}}, \bibinfo {author} {\bibfnamefont
  {G.}~\bibnamefont {Andr{\'{e}}}}, \bibinfo {author} {\bibfnamefont
  {N.}~\bibnamefont {Golubko}}, \bibinfo {author} {\bibfnamefont
  {E.}~\bibnamefont {Politova}}, \ and\ \bibinfo {author} {\bibfnamefont
  {M.}~\bibnamefont {Weil}},\ }\href {\doibase
  10.1016/j.materresbull.2011.10.003} {\bibfield  {journal} {\bibinfo
  {journal} {Mater. Res. Bull.}\ }\textbf {\bibinfo {volume} {47}},\ \bibinfo
  {pages} {63} (\bibinfo {year} {2012})}\BibitemShut {NoStop}%
\bibitem [{\citenamefont {Kurosawa}\ \emph {et~al.}(1983)\citenamefont
  {Kurosawa}, \citenamefont {Saito},\ and\ \citenamefont
  {Yamaguchi}}]{Kurosawa1983}%
  \BibitemOpen
  \bibfield  {author} {\bibinfo {author} {\bibfnamefont {K.}~\bibnamefont
  {Kurosawa}}, \bibinfo {author} {\bibfnamefont {S.}~\bibnamefont {Saito}}, \
  and\ \bibinfo {author} {\bibfnamefont {Y.}~\bibnamefont {Yamaguchi}},\ }\href
  {\doibase 10.1143/JPSJ.52.3919} {\bibfield  {journal} {\bibinfo  {journal}
  {J. Phys. Soc. Jpn.}\ }\textbf {\bibinfo {volume} {52}},\ \bibinfo {pages}
  {3919} (\bibinfo {year} {1983})}\BibitemShut {NoStop}%
\bibitem [{\citenamefont {Ressouche}\ \emph {et~al.}(2010)\citenamefont
  {Ressouche}, \citenamefont {Loire}, \citenamefont {Simonet}, \citenamefont
  {Ballou}, \citenamefont {Stunault},\ and\ \citenamefont
  {Wildes}}]{Ressouche2010a}%
  \BibitemOpen
  \bibfield  {author} {\bibinfo {author} {\bibfnamefont {E.}~\bibnamefont
  {Ressouche}}, \bibinfo {author} {\bibfnamefont {M.}~\bibnamefont {Loire}},
  \bibinfo {author} {\bibfnamefont {V.}~\bibnamefont {Simonet}}, \bibinfo
  {author} {\bibfnamefont {R.}~\bibnamefont {Ballou}}, \bibinfo {author}
  {\bibfnamefont {A.}~\bibnamefont {Stunault}}, \ and\ \bibinfo {author}
  {\bibfnamefont {A.}~\bibnamefont {Wildes}},\ }\href {\doibase
  10.1103/PhysRevB.82.100408} {\bibfield  {journal} {\bibinfo  {journal} {Phys.
  Rev. B}\ }\textbf {\bibinfo {volume} {82}},\ \bibinfo {pages} {100408}
  (\bibinfo {year} {2010})}\BibitemShut {NoStop}%
\bibitem [{\citenamefont {Redhammer}\ \emph {et~al.}(2009)\citenamefont
  {Redhammer}, \citenamefont {Roth}, \citenamefont {Treutmann}, \citenamefont
  {Hoelzel}, \citenamefont {Paulus}, \citenamefont {Andr{\'{e}}}, \citenamefont
  {Pietzonka},\ and\ \citenamefont {Amthauer}}]{Redhammer2009}%
  \BibitemOpen
  \bibfield  {author} {\bibinfo {author} {\bibfnamefont {G.~J.}\ \bibnamefont
  {Redhammer}}, \bibinfo {author} {\bibfnamefont {G.}~\bibnamefont {Roth}},
  \bibinfo {author} {\bibfnamefont {W.}~\bibnamefont {Treutmann}}, \bibinfo
  {author} {\bibfnamefont {M.}~\bibnamefont {Hoelzel}}, \bibinfo {author}
  {\bibfnamefont {W.}~\bibnamefont {Paulus}}, \bibinfo {author} {\bibfnamefont
  {G.}~\bibnamefont {Andr{\'{e}}}}, \bibinfo {author} {\bibfnamefont
  {C.}~\bibnamefont {Pietzonka}}, \ and\ \bibinfo {author} {\bibfnamefont
  {G.}~\bibnamefont {Amthauer}},\ }\href {\doibase 10.1016/j.jssc.2009.06.013}
  {\bibfield  {journal} {\bibinfo  {journal} {J. Solid State Chem.}\ }\textbf
  {\bibinfo {volume} {182}},\ \bibinfo {pages} {2374} (\bibinfo {year}
  {2009})}\BibitemShut {NoStop}%
\bibitem [{\citenamefont {Redhammer}\ \emph {et~al.}(2001)\citenamefont
  {Redhammer}, \citenamefont {Roth}, \citenamefont {Paulus}, \citenamefont
  {Andr{\'{e}}}, \citenamefont {Lottermoser}, \citenamefont {Amthauer},
  \citenamefont {Treutmann},\ and\ \citenamefont
  {Koppelhuber-Bitschnau}}]{Redhammer2001}%
  \BibitemOpen
  \bibfield  {author} {\bibinfo {author} {\bibfnamefont {G.~J.}\ \bibnamefont
  {Redhammer}}, \bibinfo {author} {\bibfnamefont {G.}~\bibnamefont {Roth}},
  \bibinfo {author} {\bibfnamefont {W.}~\bibnamefont {Paulus}}, \bibinfo
  {author} {\bibfnamefont {G.}~\bibnamefont {Andr{\'{e}}}}, \bibinfo {author}
  {\bibfnamefont {W.}~\bibnamefont {Lottermoser}}, \bibinfo {author}
  {\bibfnamefont {G.}~\bibnamefont {Amthauer}}, \bibinfo {author}
  {\bibfnamefont {W.}~\bibnamefont {Treutmann}}, \ and\ \bibinfo {author}
  {\bibfnamefont {B.}~\bibnamefont {Koppelhuber-Bitschnau}},\ }\href {\doibase
  10.1007/s002690100159} {\bibfield  {journal} {\bibinfo  {journal} {Phys.
  Chem. Miner.}\ }\textbf {\bibinfo {volume} {28}},\ \bibinfo {pages} {337}
  (\bibinfo {year} {2001})}\BibitemShut {NoStop}%
\bibitem [{\citenamefont {Tol{\'{e}}dano}\ \emph {et~al.}(2015)\citenamefont
  {Tol{\'{e}}dano}, \citenamefont {Ackermann}, \citenamefont {Bohat{\'{y}}},
  \citenamefont {Becker}, \citenamefont {Lorenz}, \citenamefont {Leo},\ and\
  \citenamefont {Fiebig}}]{Toledano2015}%
  \BibitemOpen
  \bibfield  {author} {\bibinfo {author} {\bibfnamefont {P.}~\bibnamefont
  {Tol{\'{e}}dano}}, \bibinfo {author} {\bibfnamefont {M.}~\bibnamefont
  {Ackermann}}, \bibinfo {author} {\bibfnamefont {L.}~\bibnamefont
  {Bohat{\'{y}}}}, \bibinfo {author} {\bibfnamefont {P.}~\bibnamefont
  {Becker}}, \bibinfo {author} {\bibfnamefont {T.}~\bibnamefont {Lorenz}},
  \bibinfo {author} {\bibfnamefont {N.}~\bibnamefont {Leo}}, \ and\ \bibinfo
  {author} {\bibfnamefont {M.}~\bibnamefont {Fiebig}},\ }\href {\doibase
  10.1103/PhysRevB.92.094431} {\bibfield  {journal} {\bibinfo  {journal} {Phys.
  Rev. B}\ }\textbf {\bibinfo {volume} {92}},\ \bibinfo {pages} {094431}
  (\bibinfo {year} {2015})}\BibitemShut {NoStop}%
\bibitem [{\citenamefont {N{\'{e}}nert}\ \emph {et~al.}(2009)\citenamefont
  {N{\'{e}}nert}, \citenamefont {Isobe}, \citenamefont {Ritter}, \citenamefont
  {Isnard}, \citenamefont {Vasiliev},\ and\ \citenamefont
  {Ueda}}]{Nenert2009a}%
  \BibitemOpen
  \bibfield  {author} {\bibinfo {author} {\bibfnamefont {G.}~\bibnamefont
  {N{\'{e}}nert}}, \bibinfo {author} {\bibfnamefont {M.}~\bibnamefont {Isobe}},
  \bibinfo {author} {\bibfnamefont {C.}~\bibnamefont {Ritter}}, \bibinfo
  {author} {\bibfnamefont {O.}~\bibnamefont {Isnard}}, \bibinfo {author}
  {\bibfnamefont {A.~N.}\ \bibnamefont {Vasiliev}}, \ and\ \bibinfo {author}
  {\bibfnamefont {Y.}~\bibnamefont {Ueda}},\ }\href {\doibase
  10.1103/PhysRevB.79.064416} {\bibfield  {journal} {\bibinfo  {journal} {Phys.
  Rev. B}\ }\textbf {\bibinfo {volume} {79}},\ \bibinfo {pages} {064416}
  (\bibinfo {year} {2009})}\BibitemShut {NoStop}%
\bibitem [{\citenamefont {N{\'{e}}nert}\ \emph
  {et~al.}(2010{\natexlab{a}})\citenamefont {N{\'{e}}nert}, \citenamefont
  {Isobe}, \citenamefont {Kim}, \citenamefont {Ritter}, \citenamefont {Colin},
  \citenamefont {Vasiliev}, \citenamefont {Kim},\ and\ \citenamefont
  {Ueda}}]{Nenert2010}%
  \BibitemOpen
  \bibfield  {author} {\bibinfo {author} {\bibfnamefont {G.}~\bibnamefont
  {N{\'{e}}nert}}, \bibinfo {author} {\bibfnamefont {M.}~\bibnamefont {Isobe}},
  \bibinfo {author} {\bibfnamefont {I.}~\bibnamefont {Kim}}, \bibinfo {author}
  {\bibfnamefont {C.}~\bibnamefont {Ritter}}, \bibinfo {author} {\bibfnamefont
  {C.~V.}\ \bibnamefont {Colin}}, \bibinfo {author} {\bibfnamefont {A.~N.}\
  \bibnamefont {Vasiliev}}, \bibinfo {author} {\bibfnamefont {K.~H.}\
  \bibnamefont {Kim}}, \ and\ \bibinfo {author} {\bibfnamefont
  {Y.}~\bibnamefont {Ueda}},\ }\href {\doibase 10.1103/PhysRevB.82.024429}
  {\bibfield  {journal} {\bibinfo  {journal} {Phys. Rev. B}\ }\textbf {\bibinfo
  {volume} {82}},\ \bibinfo {pages} {024429} (\bibinfo {year}
  {2010}{\natexlab{a}})}\BibitemShut {NoStop}%
\bibitem [{\citenamefont {Lumsden}\ \emph {et~al.}(2000)\citenamefont
  {Lumsden}, \citenamefont {Granroth}, \citenamefont {Mandrus}, \citenamefont
  {Nagler}, \citenamefont {Thompson}, \citenamefont {Castellan},\ and\
  \citenamefont {Gaulin}}]{Lumsden2000}%
  \BibitemOpen
  \bibfield  {author} {\bibinfo {author} {\bibfnamefont {M.~D.}\ \bibnamefont
  {Lumsden}}, \bibinfo {author} {\bibfnamefont {G.~E.}\ \bibnamefont
  {Granroth}}, \bibinfo {author} {\bibfnamefont {D.}~\bibnamefont {Mandrus}},
  \bibinfo {author} {\bibfnamefont {S.~E.}\ \bibnamefont {Nagler}}, \bibinfo
  {author} {\bibfnamefont {J.~R.}\ \bibnamefont {Thompson}}, \bibinfo {author}
  {\bibfnamefont {J.~P.}\ \bibnamefont {Castellan}}, \ and\ \bibinfo {author}
  {\bibfnamefont {B.~D.}\ \bibnamefont {Gaulin}},\ }\href {\doibase
  10.1103/PhysRevB.62.R9244} {\bibfield  {journal} {\bibinfo  {journal} {Phys.
  Rev. B}\ }\textbf {\bibinfo {volume} {62}},\ \bibinfo {pages} {R9244}
  (\bibinfo {year} {2000})}\BibitemShut {NoStop}%
\bibitem [{\citenamefont {N{\'{e}}nert}\ \emph
  {et~al.}(2010{\natexlab{b}})\citenamefont {N{\'{e}}nert}, \citenamefont
  {Kim}, \citenamefont {Isobe}, \citenamefont {Ritter}, \citenamefont
  {Vasiliev}, \citenamefont {Kim},\ and\ \citenamefont {Ueda}}]{Nenert2010a}%
  \BibitemOpen
  \bibfield  {author} {\bibinfo {author} {\bibfnamefont {G.}~\bibnamefont
  {N{\'{e}}nert}}, \bibinfo {author} {\bibfnamefont {I.}~\bibnamefont {Kim}},
  \bibinfo {author} {\bibfnamefont {M.}~\bibnamefont {Isobe}}, \bibinfo
  {author} {\bibfnamefont {C.}~\bibnamefont {Ritter}}, \bibinfo {author}
  {\bibfnamefont {A.~N.}\ \bibnamefont {Vasiliev}}, \bibinfo {author}
  {\bibfnamefont {K.~H.}\ \bibnamefont {Kim}}, \ and\ \bibinfo {author}
  {\bibfnamefont {Y.}~\bibnamefont {Ueda}},\ }\href {\doibase
  10.1103/PhysRevB.81.184408} {\bibfield  {journal} {\bibinfo  {journal} {Phys.
  Rev. B}\ }\textbf {\bibinfo {volume} {81}},\ \bibinfo {pages} {184408}
  (\bibinfo {year} {2010}{\natexlab{b}})}\BibitemShut {NoStop}%
\bibitem [{\citenamefont {Redhammer}\ \emph {et~al.}(2008)\citenamefont
  {Redhammer}, \citenamefont {Roth}, \citenamefont {Treutmann}, \citenamefont
  {Paulus}, \citenamefont {Andr{\'{e}}}, \citenamefont {Pietzonka},\ and\
  \citenamefont {Amthauer}}]{Redhammer2008}%
  \BibitemOpen
  \bibfield  {author} {\bibinfo {author} {\bibfnamefont {G.~J.}\ \bibnamefont
  {Redhammer}}, \bibinfo {author} {\bibfnamefont {G.}~\bibnamefont {Roth}},
  \bibinfo {author} {\bibfnamefont {W.}~\bibnamefont {Treutmann}}, \bibinfo
  {author} {\bibfnamefont {W.}~\bibnamefont {Paulus}}, \bibinfo {author}
  {\bibfnamefont {G.}~\bibnamefont {Andr{\'{e}}}}, \bibinfo {author}
  {\bibfnamefont {C.}~\bibnamefont {Pietzonka}}, \ and\ \bibinfo {author}
  {\bibfnamefont {G.}~\bibnamefont {Amthauer}},\ }\href {\doibase
  10.1016/j.jssc.2008.08.014} {\bibfield  {journal} {\bibinfo  {journal} {J.
  Solid State Chem.}\ }\textbf {\bibinfo {volume} {181}},\ \bibinfo {pages}
  {3163} (\bibinfo {year} {2008})}\BibitemShut {NoStop}%
\bibitem [{\citenamefont {Ding}\ \emph {et~al.}(2016)\citenamefont {Ding},
  \citenamefont {Colin}, \citenamefont {Darie}, \citenamefont {Robert},
  \citenamefont {Gay},\ and\ \citenamefont {Bordet}}]{Ding2016}%
  \BibitemOpen
  \bibfield  {author} {\bibinfo {author} {\bibfnamefont {L.}~\bibnamefont
  {Ding}}, \bibinfo {author} {\bibfnamefont {C.~V.}\ \bibnamefont {Colin}},
  \bibinfo {author} {\bibfnamefont {C.}~\bibnamefont {Darie}}, \bibinfo
  {author} {\bibfnamefont {J.}~\bibnamefont {Robert}}, \bibinfo {author}
  {\bibfnamefont {F.}~\bibnamefont {Gay}}, \ and\ \bibinfo {author}
  {\bibfnamefont {P.}~\bibnamefont {Bordet}},\ }\href {\doibase
  10.1103/PhysRevB.93.064423} {\bibfield  {journal} {\bibinfo  {journal} {Phys.
  Rev. B}\ }\textbf {\bibinfo {volume} {93}},\ \bibinfo {pages} {064423}
  (\bibinfo {year} {2016})}\BibitemShut {NoStop}%
\bibitem [{\citenamefont {Redhammer}\ \emph {et~al.}(2011)\citenamefont
  {Redhammer}, \citenamefont {Senyshyn}, \citenamefont {Tippelt},\ and\
  \citenamefont {Roth}}]{Redhammer2011}%
  \BibitemOpen
  \bibfield  {author} {\bibinfo {author} {\bibfnamefont {G.~J.}\ \bibnamefont
  {Redhammer}}, \bibinfo {author} {\bibfnamefont {A.}~\bibnamefont {Senyshyn}},
  \bibinfo {author} {\bibfnamefont {G.}~\bibnamefont {Tippelt}}, \ and\
  \bibinfo {author} {\bibfnamefont {G.}~\bibnamefont {Roth}},\ }\href {\doibase
  10.1088/0953-8984/23/25/254202} {\bibfield  {journal} {\bibinfo  {journal}
  {J. Phys. Condens. Matter}\ }\textbf {\bibinfo {volume} {23}},\ \bibinfo
  {pages} {254202} (\bibinfo {year} {2011})}\BibitemShut {NoStop}%
\bibitem [{\citenamefont {Mogare}\ \emph {et~al.}(2006)\citenamefont {Mogare},
  \citenamefont {Sheptyakov}, \citenamefont {Bircher}, \citenamefont
  {G{\"{u}}del},\ and\ \citenamefont {Jansen}}]{Mogare2006}%
  \BibitemOpen
  \bibfield  {author} {\bibinfo {author} {\bibfnamefont {K.~M.}\ \bibnamefont
  {Mogare}}, \bibinfo {author} {\bibfnamefont {D.~V.}\ \bibnamefont
  {Sheptyakov}}, \bibinfo {author} {\bibfnamefont {R.}~\bibnamefont {Bircher}},
  \bibinfo {author} {\bibfnamefont {H.-U.}\ \bibnamefont {G{\"{u}}del}}, \ and\
  \bibinfo {author} {\bibfnamefont {M.}~\bibnamefont {Jansen}},\ }\href
  {\doibase 10.1140/epjb/e2006-00316-5} {\bibfield  {journal} {\bibinfo
  {journal} {Eur. Phys. J. B}\ }\textbf {\bibinfo {volume} {52}},\ \bibinfo
  {pages} {371} (\bibinfo {year} {2006})}\BibitemShut {NoStop}%
\bibitem [{\citenamefont {Litvin}\ \emph {et~al.}(1982)\citenamefont {Litvin},
  \citenamefont {Kotzev},\ and\ \citenamefont {Birman}}]{Kotzev1982}%
  \BibitemOpen
  \bibfield  {author} {\bibinfo {author} {\bibfnamefont {D.~B.}\ \bibnamefont
  {Litvin}}, \bibinfo {author} {\bibfnamefont {J.~N.}\ \bibnamefont {Kotzev}},
  \ and\ \bibinfo {author} {\bibfnamefont {J.~L.}\ \bibnamefont {Birman}},\
  }\href {\doibase 10.1103/PhysRevB.26.6947} {\bibfield  {journal} {\bibinfo
  {journal} {Phys. Rev. B}\ }\textbf {\bibinfo {volume} {26}},\ \bibinfo
  {pages} {6947} (\bibinfo {year} {1982})}\BibitemShut {NoStop}%
\bibitem [{\citenamefont {Aizu}(1966)}]{Aizu1966}%
  \BibitemOpen
  \bibfield  {author} {\bibinfo {author} {\bibfnamefont {K.}~\bibnamefont
  {Aizu}},\ }\href {\doibase 10.1103/PhysRev.146.423} {\bibfield  {journal}
  {\bibinfo  {journal} {Phys. Rev.}\ }\textbf {\bibinfo {volume} {146}},\
  \bibinfo {pages} {423} (\bibinfo {year} {1966})}\BibitemShut {NoStop}%
\bibitem [{\citenamefont {Aizu}(1969)}]{Aizu1969}%
  \BibitemOpen
  \bibfield  {author} {\bibinfo {author} {\bibfnamefont {K.}~\bibnamefont
  {Aizu}},\ }\href {\doibase 10.1143/JPSJ.27.387} {\bibfield  {journal}
  {\bibinfo  {journal} {J. Phys. Soc. Jpn.}\ }\textbf {\bibinfo {volume}
  {27}},\ \bibinfo {pages} {387} (\bibinfo {year} {1969})}\BibitemShut
  {NoStop}%
\bibitem [{\citenamefont {Aizu}(1970)}]{Aizu1970a}%
  \BibitemOpen
  \bibfield  {author} {\bibinfo {author} {\bibfnamefont {K.}~\bibnamefont
  {Aizu}},\ }\href {\doibase 10.1103/PhysRevB.2.754} {\bibfield  {journal}
  {\bibinfo  {journal} {Phys. Rev. B}\ }\textbf {\bibinfo {volume} {2}},\
  \bibinfo {pages} {754} (\bibinfo {year} {1970})}\BibitemShut {NoStop}%
\bibitem [{\citenamefont {Hlinka}\ \emph {et~al.}(2016)\citenamefont {Hlinka},
  \citenamefont {Privratska}, \citenamefont {Ondrejkovic},\ and\ \citenamefont
  {Janovec}}]{Hlinka2016a}%
  \BibitemOpen
  \bibfield  {author} {\bibinfo {author} {\bibfnamefont {J.}~\bibnamefont
  {Hlinka}}, \bibinfo {author} {\bibfnamefont {J.}~\bibnamefont {Privratska}},
  \bibinfo {author} {\bibfnamefont {P.}~\bibnamefont {Ondrejkovic}}, \ and\
  \bibinfo {author} {\bibfnamefont {V.}~\bibnamefont {Janovec}},\ }\href
  {\doibase 10.1103/PhysRevLett.116.177602} {\bibfield  {journal} {\bibinfo
  {journal} {Phys. Rev. Lett.}\ }\textbf {\bibinfo {volume} {116}},\ \bibinfo
  {pages} {177602} (\bibinfo {year} {2016})}\BibitemShut {NoStop}%
\bibitem [{\citenamefont {Aizu}(1979)}]{Aizu1979}%
  \BibitemOpen
  \bibfield  {author} {\bibinfo {author} {\bibfnamefont {K.}~\bibnamefont
  {Aizu}},\ }\href {\doibase 10.1143/JPSJ.46.1716} {\bibfield  {journal}
  {\bibinfo  {journal} {J. Phys. Soc. Jpn.}\ }\textbf {\bibinfo {volume}
  {46}},\ \bibinfo {pages} {1716} (\bibinfo {year} {1979})}\BibitemShut
  {NoStop}%
\end{thebibliography}%

\end{document}